\documentclass[12pt,a4paper]{article}
\pdfoutput=1
\usepackage{jheppub}
\usepackage{hyperref}
\usepackage[utf8]{inputenc}
\usepackage{amsmath}
\usepackage{amssymb}
\usepackage{graphicx}
\usepackage{booktabs}
\usepackage{array}
\usepackage{paralist}
\usepackage{verbatim}
\usepackage{xspace}
\usepackage{lscape}
\usepackage{adjustbox}
\usepackage{slashed}
\usepackage{pslatex}
\usepackage{ulem}

\graphicspath{{./}{./plots/ADJP/}}

\def\Ref#1{Ref.~\cite{#1}}
\def\Refs#1{Refs.~\cite{#1}}
\def\Eq#1{Eq.~(\ref{#1})}
\def\Eqs#1{Eqs.~(\ref{#1})}
\def\Fig#1{Fig.~\ref{#1}}
\def\Figs#1{Figs.~\ref{#1}}
\def\Tab#1{Tab.~\ref{#1}}
\providecommand{\main}{.}

\extrafloats{20}
\title{Phenomenology of $t\bar{t}j + X$ production at the LHC}

\author[a]{Simone Alioli,}
\author[b]{Juan~Fuster,}
\author[c]{Maria~Vittoria~Garzelli,}
\author[a]{Alessandro Gavardi,}
\author[b]{Adrian~Irles,}
\author[d]{Davide~Melini,}
\author[c]{Sven-Olaf~Moch,}
\author[e]{Peter~Uwer,}
\author[f,c]{Katharina~Vo\ss}

\affiliation[a]{Dipartimento di Fisica “G. Occhialini”, Universit\`a degli Studi di Milano-Bicocca and INFN,  Sezione di Milano Bicocca, \\ Piazza della Scienza 3, I~--~20126 Milano, Italy}
\affiliation[b]{IFIC, Universitat de Val\`encia and CSIC, \\
  Catedr\'atico Jose Beltr\'an 2, E~--~46980 Paterna, Spain}
\affiliation[c]{II. Institut f\"ur Theoretische Physik, Universit\"at
  Hamburg, \\ Luruper Chaussee 149, D~--~22761 Hamburg, Germany}
 \affiliation[d]{Physics Department, Technion–Institute of Technology, \\ 
 Haifa 3200003, Israel}
\affiliation[e]{Institut f\"ur Physik, Humboldt-Universit\"at zu
  Berlin, \\ Newtonstra{\ss}e 15, D~--~12489 Berlin, Germany}
  \affiliation[f]{Center For Particle Physics Siegen, Department Physik,  Universit\"at Siegen, \\
  Walter Flex Str. 3, D~--~57068 Siegen, Germany}

\emailAdd{simone.alioli@unimib.it, fuster@ific.uv.es,
  maria.vittoria.garzelli@desy.de, 
  a.gavardi@campus.unimib.it, adrian.irles@ific.uv.es, davide.melini@cern.ch, 
  sven-olaf.moch@desy.de, peter.uwer@physik.hu-berlin.de, katharina.voss@desy.de}

\abstract{ We present phenomenological results for
  $t\bar{t}j + X$ production at the Large Hadron Collider, of interest
  for designing forthcoming experimental analyses of this process. We
  focus on those cases where the $t\bar{t}j + X$ process is considered
  as a signal. We discuss present theoretical uncertainties and 
  the dependence on relevant input parameters entering the computation.
  For the ${\cal R}$
  distribution, which depends on the invariant mass of the 
  $t\bar{t}j$-system, we present reference
  predictions in the on-shell, $\overline{\mbox{MS}}$
  and MSR top-quark mass
  renormalization schemes, applying the latter scheme to this process
  for the first time.  Our conclusions are particularly
  interesting for those analyses aiming at extracting the top-quark
  mass from cross-section measurements.}


\preprint{DESY-22-030, HU-EP-22/05, IFIC/22-05}

\begin{document}
\maketitle

\newpage

\section{Introduction and motivations for this work}
\label{sec:intro}

The $t\bar{t}j + X$ production process at the Large Hadron Collider (LHC) is very
interesting both as a signal and as a background in numerous experimental
analyses. When con\-si\-de\-red as a signal, it can be used to extract precise
values of the top-quark mass from measurements of differential cross sections
particularly sensitive to the value of this Standard Model (SM)
parameter.
In particular, in \Ref{Alioli:2013mxa} it has been argued that the additional
jet activity in the ${\ensuremath{t\bar t j + X}\xspace}$ process due to gluon radiation can lead
to an enhanced mass sensitivity of the respective inclusive and differential
cross sections, compared to the corresponding inclusive and differential cross sections for the $t\bar{t} +X$ process.
This may allow for a more precise measurement, while keeping the advantage 
of uniquely fixing the renormalization scheme of the extracted mass value
through higher-order theoretical predictions.

More precisely, \Ref{Alioli:2013mxa} introduced the following observable
\begin{equation}
  {\cal R}(m_t^R,\rho_s)= 
  \frac{1}{{\ensuremath{\sigma_{t\bar t + \textnormal{\scriptsize 1-jet}}}\xspace}} 
  \frac{d{\ensuremath{\sigma_{t\bar t + \textnormal{\scriptsize 1-jet}}}\xspace}}{d\rho_s}(m_t^R, \rho_s),
\end{equation}
with $\rho_s$ given by
\begin{equation}
  \rho_s \,=\, \frac{2 m_0} {\sqrt{{\ensuremath{s_{t\bar t j}}\xspace}}} \, .
\end{equation}
In the above definition, $\sqrt{s_{t\bar{t}j}}$
is the invariant mass of the $t\bar{t}j$ system, built from the top-quark pair
and the hardest jet satisfying typical transverse momentum and pseudorapidity
cuts depending on the experimental analysis, and $m_t^R$
is used to denote the top-quark mass in the renormalization scheme $R$. 
We have not explicitly specified the renormalization scheme for the top-quark mass, as
different schemes can be and have been employed. The `mass' $m_0$
occurring in the definition of $\rho_s$ is an arbitrary scale to make
$\rho_s$ dimensionless. In practical applications it is fixed to a value
of the order of the top-quark mass, e.g. $m_0=170$ GeV that we will
use in the following. As a proof of concept, the measurement of
${\cal R}$ was used by the ATLAS collaboration to
determine the top-quark mass in the on-shell scheme using LHC data collected
at 7~TeV \cite{ATLAS:2015pfy}. 
As a follow-up, the aforementioned published experimental results were used in
\Ref{Fuster:2017rev} to determine the top-quark mass in the
${\overline{\mbox{MS}}}$ 
renormalization scheme. Later, also LHC data collected at 8~TeV were
used by both the ATLAS and CMS collaborations~\cite{CMS:2016khu,ATLAS:2019guf} 
to extract the top-quark mass, whereas new analyses
exploiting the large statistics accumulated at $\sqrt{S}$ = 13~TeV are in
preparation. 

Motivated by the experimental needs for the ongoing and forthcoming analyses,
we present in this work theoretical predictions for the $t\bar{t}j + X$ process, 
discuss their main theoretical uncertainties and study their dependence on the
choice of the input parameters and on some of the experimental cuts 
for the jet reconstruction procedure. 
We also investigate the role of dif\-fe\-rent top-quark mass renormalization schemes.
Besides the on-shell and the ${\overline{\mbox{MS}}}$ schemes, already used in previous papers,
in this work we apply, for the first time, the MSR top-quark mass
renormalization scheme~\cite{Hoang:2008yj,Hoang:2017suc} 
in deriving predictions for this process.  

We limit our discussion to fixed-order predictions in the stable top-quark case,
consi\-dering the sophisticated techniques that have been developed by the
experimental collaborations to reconstruct top quarks from their decay
products and to unfold their results from the particle to the parton level (see for example \Refs{ATLAS:2019guf,CMS:2019esx}).

We provide reference predictions at next-to-leading order (NLO) accuracy, for dif\-fe\-rent top-quark mass
renormalization schemes and state-of-the-art choices for other input
quantities, with the intent of providing useful information and concrete
numerical results to the experimental collaborations in view of their ongoing
$t\bar{t}j +X$ studies and top-quark mass extractions. In fact, to determine the top-quark mass in the experimental analyses in preparation, as well as in those already published so far,
the unfolded measurements are compared with theoretical predicitions with this accuracy in QCD.

The manuscript is organized as follows. 
In Section~\ref{sec:compu} we  briefly describe the theoretical framework. 
Theoretical predictions for fiducial inclusive and differential cross sections
and their uncertainties are discussed in Section~\ref{sec:pheno},  
where we show their dependence on va\-rious inputs, including renormalization and
factorization scale choices, parton distribution functions (PDFs) and the jet algorithm $R$ parameter.
In Section~\ref{sec:mass} we focus on the ${\cal R}$ distribution, providing detailed predictions
useful to the experimental collaborations to extract the top-quark mass value 
in different mass renormalization schemes. Furthermore we investigate the impact  of
off-shell effects and non-resonant/non-factorizable contributions.
We conclude in Section~\ref{sec:conclu}. 
Further material and Tables with our reference numerical cross sections are collected in Appendix~\ref{sec:appA}.

\section{Computational framework}
\label{sec:compu}

NLO QCD corrections for $t\bar{t}j + X$ hadroproduction at the LHC have been
computed and discussed in a number of papers, using multiple methods. In
particular, \Refs{Dittmaier:2007wz,Dittmaier:2008uj} consider the
top quark as stable, \Refs{Melnikov:2010iu} and ~\cite{Melnikov:2011qx}
complement NLO QCD production with top-quark decays at LO and NLO,
respectively, by working in the narrow width approximation (NWA) retaining
spin correlations~\footnote{In Ref.~\cite{Melnikov:2011qx} jet radiation from
  the top-quark decay products is also included.}, 
whereas \Refs{Bevilacqua:2015qha,Bevilacqua:2016jfk} include full
off-shellness and resonance effects for the cases where the top and
antitop quarks decay leptonically.  
One-loop amplitudes for the parton-parton $\rightarrow t\bar{t}j$ process were first obtained analytically in
\Refs{Dittmaier:2007wz,Dittmaier:2008uj}. \Refs{Melnikov:2010iu,Melnikov:2011qx}
used instead generalized $D$-dimensional unitarity~\cite{Ellis:2011cr} in a
numerical implementation, whereas \Refs{Bevilacqua:2015qha,Bevilacqua:2016jfk} are based on numerical results from
{\textsc{Helac-nlo}}~\cite{Bevilacqua:2011xh}, which provides one-loop
amplitudes through a numerical implementation of the OPP
method~\cite{Ossola:2006us} 
complemented by effective Feynman rules for the calculation of the
contribution due to the $R_2$ rational terms~\cite{Draggiotis:2009yb}. 
Nowadays, the same amplitudes can also be
ea\-si\-ly obtained, still numerically, by making use of various other automated
tools (the so-called one-loop providers), such as
\textsc{MadLoop}~\cite{Hirschi:2011pa},
\textsc{GoSam}~\cite{Cullen:2011ac,Cullen:2014yla},
\textsc{OpenLoops}~\cite{Cascioli:2011va,Buccioni:2019sur} and
\textsc{Recola}~\cite{Actis:2016mpe,Denner:2017wsf}. Real emission
amplitudes can also be ea\-si\-ly obtained numerically, and multiple automatic
frameworks for the cancellation of the infrared singularities when combining
real and virtual contributions at NLO have also been developed, working
according to either the Catani-Seymour (CS) dipole subtraction method for
calculations with massive partons~\cite{Catani:2002hc}, or the
Frixione-Kunst-Signer (FKS) method~\cite{Frixione:1995ms}, or the Nagy-Soper
subtraction formalism~\cite{Nagy:2008eq,Bevilacqua:2013iha}. These
developments have allowed for the automatic computation of fixed-order NLO cross sections with a variety of different 
tools, leading to predictions fully consistent among each
other~\footnote{
  The fully local schemes for the subtraction of infrared
  divergences employed at NLO are a necessary prerequisite 
  for the consistency of the predictions.
  This is not guaranteed in different regularization frameworks, which use
  non-local methods for the subtraction, as pointed out recently 
  in case of the Drell-Yan process at NNLO in \Refs{Ebert:2019zkb,Alekhin:2021xcu}. 
  For the $t\bar{t}j$ process NNLO predictions are not yet available.}. 
Additio\-nally, one-loop
providers have been interfaced to computational frameworks providing NLO QCD
matching to parton shower (PS) approaches, considering both the
POWHEG~\cite{Nason:2004rx,Frixione:2007vw}  and the
MC@NLO~\cite{Frixione:2002ik} matching methods. 
In particular, first predictions with NLO QCD~+~PS accuracy for $t\bar{t}j$ production were obtained in the
\textsc{PowHel} framework~\cite{Kardos:2011qa}, using numerical matrix elements from
\textsc{Helac-nlo} as input to the \textsc{Powheg-Box}
implementation~\cite{Alioli:2010xd}, interfaced to Shower Monte Carlo (SMC)
codes. Top-quark decays as well as PS emissions besides the first one and
hadronization effects were taken care of by the interface of the produced events
at the first radiation emission level to the
\textsc{Pythia}~\cite{Sjostrand:2006za} and
\textsc{Herwig}~\cite{Corcella:2002jc} SMC. Subsequently, the analytical
amplitudes of \Refs{Dittmaier:2007wz,Dittmaier:2008uj} were also used as
input to the \textsc{Powheg-Box}, producing, in case of stable top quarks,
predictions~\cite{Alioli:2011as} consistent with the previous ones. 
As an alternative to the top-quark decay by the SMC code, an in-house implementation
of top-quark decays including spin correlations in the NWA was considered in
Ref.~\cite{Alioli:2011as}. 
On the other hand, one-loop amplitudes as automatically obtained by
\textsc{MadLoop} are at the core of the $t\bar{t}j$ NLO QCD +  PS computation
in the \textsc{Madgraph5\_aMC@NLO} framework~\cite{Alwall:2014hca}, which
provides PS-dependent matching terms for various versions of the
\textsc{Pythia} and \textsc{Herwig} SMC
approaches~\cite{Sjostrand:2019zhc,Bellm:2017bvx,Bellm:2015jjp,Sjostrand:2006za,Corcella:2002jc}. 
The MC@NLO matching method has also been implemented in the $t\bar{t}j$ NLO QCD + PS implementation
of Ref.~\cite{Czakon:2015cla}, using amplitudes from \textsc{Helac-nlo}, in
association with the \textsc{Deductor} PS approach~\cite{Nagy:2014mqa,Nagy:2015hwa}. 
Furthermore, NLO electroweak (EW) corrections have been recently computed in
Ref.~\cite{Gutschow:2018tuk}, and embedded in the multi-jet merging technique 
MEPS@NLO~\cite{Hoeche:2012yf}, giving rise to merged predictions for
$t\bar{t}$ and $t\bar{t}j$ including NLO QCD + EW radiative
corrections~\footnote{
  In this implementation the dominant virtual electroweak
  corrections are incorporated exactly, whereas the NLO QED bremsstrahlung is
  accounted for in an approximate way.} 
together with PS and hadronization effects as available in the \textsc{Sherpa} SMC~\cite{Sherpa:2019gpd}. 

\begin{figure}
\includegraphics[width=0.49\textwidth]{\main/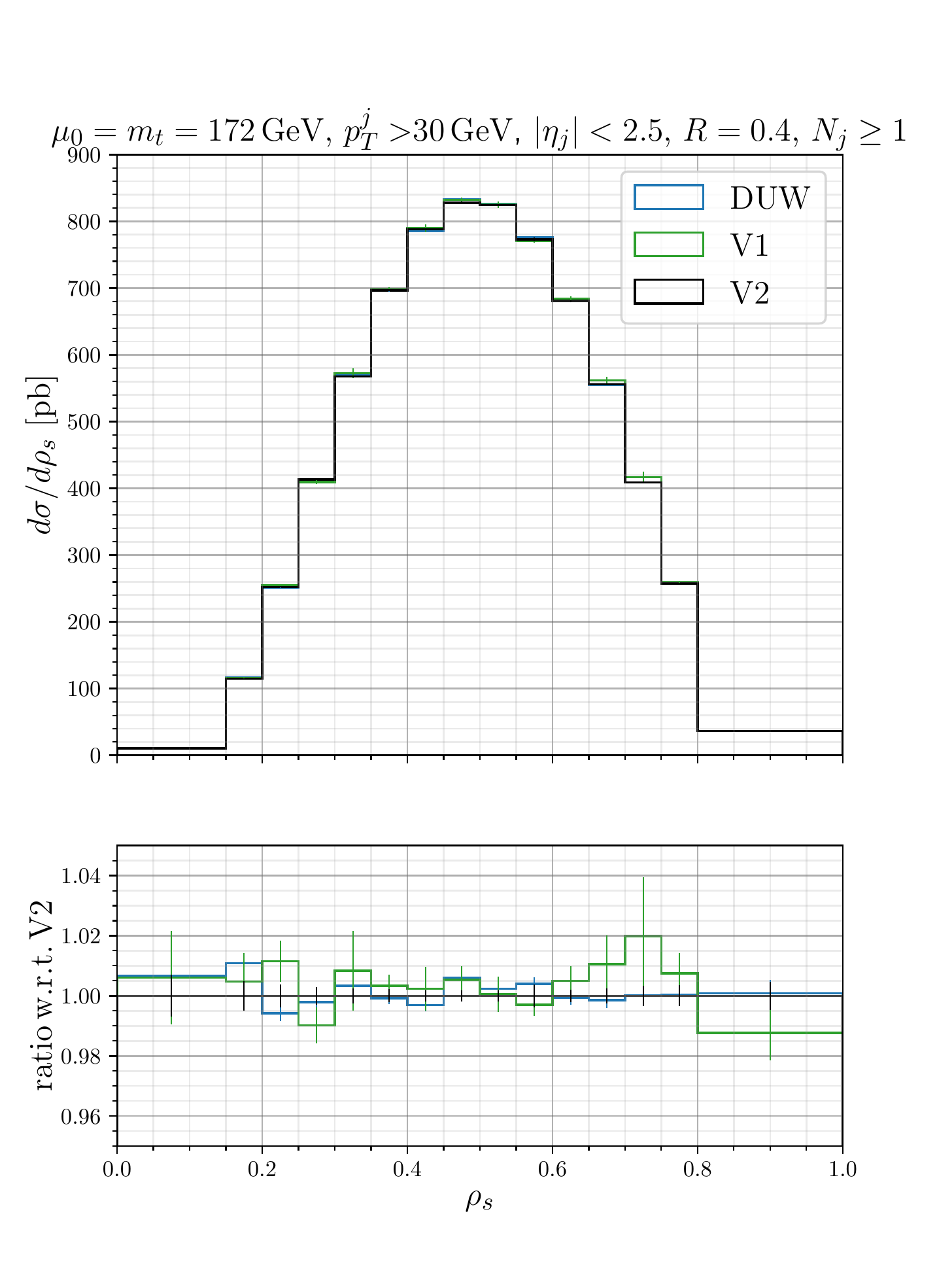}
\includegraphics[width=0.49\textwidth]{\main/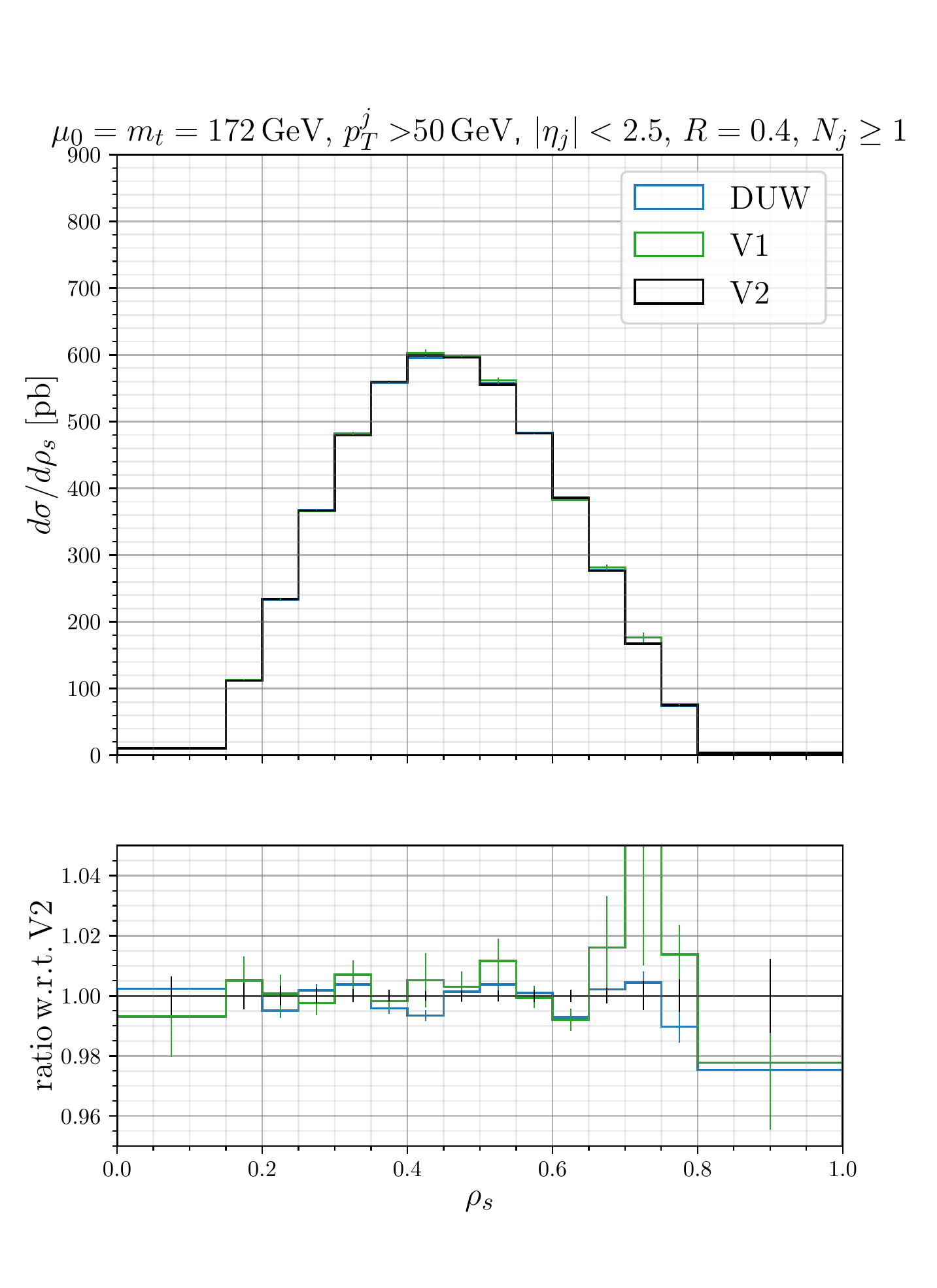}
\caption{\label{fig:cross1}
Comparison of predictions of the $\rho_s$ distribution for $t\bar{t}j + X$ production in $pp$ collisions at
$\sqrt{S} = 13\,$TeV using different frameworks described in the text:  
\textsc{Powheg-Box-v2} (black) vs. \textsc{Powheg-Box-v1} (green) vs. the in-house NLO
implementation of Dittmaier-Uwer-Weinzierl (blue) in an analysis setup where
at least one light jet reconstructed with the \mbox{anti-$k_T$} algorithm with $R =
0.4$ is required, with $|\eta_j| < 2.5$ and a $p_T^j$-cut of at least 30~GeV
(left panel) and 50~GeV (right panel). The renormalization and factorization scales are fixed to the value of the top-quark mass $m_t = 172$ GeV. 
}  
\end{figure}

In this work we have used various frameworks to compute predictions for the
$t\bar{t}j$ process at NLO QCD accuracy: i) the implementation as presented in \Refs{Dittmaier:2007wz,Dittmaier:2008uj}  
and ii) the \textsc{Powheg-Box},
which makes use of the FKS method for the infrared subtraction as numerically
implemented inside the code, together with either the analytical amplitudes of
\Refs{Dittmaier:2007wz,Dittmaier:2008uj}, implemented in a C++ library,
or the numerical amplitudes from \textsc{OpenLoops$\,\,$2}. In particular, the
first ones have been interfaced in the \textsc{Powheg-Box-v1} framework, whereas the 
second ones have been interfaced to the \textsc{Powheg-Box-v2},
respectively.  
We have extensively cross-checked these frameworks 
by comparing predictions for many differential distributions among each other, 
using also different systems of cuts. In all cases we have obtained perfect agreement. 
An illustration of these cross-checks can be found in \Fig{fig:cross1},
using as an example the $\rho_s$ distribution introduced in Section~\ref{sec:intro}.

The computations of $t\bar{t}j + X$ hadroproduction have all been performed
considering five active flavors at the scales relevant for this
process, an assumption commonly adopted for most of the top quark related
production processes~\footnote{Processes involving bottom-quarks as final states,
  like e.g. single-top and $t\bar{t}b\bar{b}$ production, besides in schemes
  with five active flavors, have also been computed in the decoupling
  factorization and renormalization scheme with four active flavors,
  i.e. considering massive $b$-quarks (see e.g. \Refs{Frederix:2012dh,Bevilacqua:2017cru}).
  For $t\bar{t}b\bar{b}$ the two descriptions have been shown to be compatible
  within uncertainties~\cite{Bevilacqua:2017cru}, 
  at least for the experimental set of analysis cuts considered in that work.  
  In $t\bar{t}j$ production at hadron colliders, the final-state jet $j$ can
  also be a $b$-jet. This case is however suppressed with respect to the cases
  where $j$ is a light jet, also due to the smallness of the $b$-quark PDF 
  with respect to the ones for other partons. 
  Therefore, given that $b$-initiated contributions  
  (absent when $b$ is considered as a massive quark)
  are indeed small when $b$ is taken to be massless, and 
  that $u$, $d$, $c$, $s$ quarks are massless in both cases,
  we do not expect a large difference in the description of the 
  $t\bar{t}j$ production using either four or five active flavors. 
  On the other hand, when considering scales well above $m_t$, as appropriate e.g. for the
  high-$p_T^t$ and high-$m_{t\bar{t}j}$ tails, it might be worth to
  investigate the $t\bar{t}j$ process even in scheme with six active flavors. 
  This may be done by using appropriate PDFs with six active flavors above the top-quark threshold. 
  Most of the PDF fits available nowadays are however limited to a maximum of five
  active flavors.}. 
Various modern PDF sets have been taken as input. In particular, in the present work we make use of the
ABMP16~\cite{Alekhin:2018pai},
CT18~\cite{Hou:2019efy},
MMHT2014~\cite{Harland-Lang:2014zoa},
MSHT20~\cite{Bailey:2020ooq} and
NNPDF3.1~\cite{NNPDF:2017mvq} NLO PDF sets,
together with their associated $\alpha_s (M_Z)$ value, and
$\alpha_s$ evolution at two-loops according to the \textsc{Lhapdf}
interface~\cite{Buckley:2014ana} for the phenomenological predictions presented
in Section~\ref{sec:pheno}.

PDF uncertainties are computed according to the specific prescriptions associated to each set. 
As we will show in the following, for the distributions and the kinematical regions in which we are
most interested, it turns out that PDF uncertainties obtained in
NLO hadronic computations involving NLO PDFs (+ $\alpha_s$) and NLO partonic
cross~sections have approximately the same relative size as PDF
uncertainties obtained in LO hadronic computations involving NLO PDFs (+ $\alpha_s$)
and LO partonic cross~sections. Therefore, considering that the second
approach is computationally much faster than the first one, we make use of it
in Section~\ref{sec:pheno} in order to compute the NLO PDF uncertainty bands for a
number of different PDF fits. We still make use of the first approach only
when computing PDF uncertainties in association with our nominal PDF set (CT18
NLO). 
Additionally, we present some considerations concerning the effect of the 
simultaneous variation of $\alpha_s(M_Z)$ and PDFs, using as a basis a set of
ABMP16 NLO PDF~+~$\alpha_S(M_Z)$ fits, which fully preserve the correlations
between these quantities. 

The predictions presented in Section~\ref{sec:pheno} refer to the case when the top-quark mass 
is renormalized in the on-shell scheme, i.e. $m_t = m_t^{\rm{pole}}$. On the other hand,
in Section~\ref{sec:mass} we discuss how to obtain predictions with the top-quark mass renormalized in the MSR and $\overline{\rm MS}$ schemes,
while we always used the latter scheme for $\alpha_s$ renormalization,
and present the corresponding results for selected distributions.

As further input, in Section~\ref{sec:pheno} we consider various choices for the renormalization and
facto\-ri\-za\-tion scales $\mu_R$ and $\mu_F$. 
To that end, we define the quantities $H_T^B$ and $m_{t\bar{t}j}^B$ as
\begin{eqnarray}
  \label{eq:HT-scale-def}
  H_T^B &=& \left( \sqrt{{p^{t,\, B}_{T}}^2 + m_t^2} + \sqrt{{p^{\bar{t},\, B}_{T}}^{2} + m_t^2} + p^{j,\, B}_{T} \right)
  \, ,
  \\
  \label{eq:mttj-scale-def}
  m_{t\bar{t}j}^B &=& \sqrt{(p^B_t + p^B_{\bar{t}} + p^B_j)^2}
  \, ,
\end{eqnarray}
which, as emphasized by the ``$B$'' superscript, are computed using the four-momenta of the outgoing particles 
of the underlying Born phase-space configuration in the \textsc{Powheg-Box} framework. 
The quantities $H_T$ and $m_{t\bar{t}j}$ are defined in analogy to \Eqs{eq:HT-scale-def} and (\ref{eq:mttj-scale-def}) 
using instead the real emission kinematics at NLO for the four-momenta of the
outgoing particles, i.e. $p_t$, $p_{\bar{t}}$ and $p_j$. 
Central predictions are obtained by fixing $\mu_R = \mu_F = \mu_0$, with $\mu_0$ being one of the
following options: 
\begin{eqnarray}
  \label{eq:scale-choices}
1)\quad \mu_0 = m_t\, ,\qquad
2)\quad \mu_0 = H_T^B/2\, ,\qquad 
3)\quad \mu_0 = H_T^B/4\, ,\qquad 
4)\quad \mu_0 = m_{t\bar{t}j}^B/2\, .\qquad 
\end{eqnarray}
The scale uncertainties are obtained by the usual seven-point prescription, i.e. by
varying independently $\mu_R$ and $\mu_F$ by factors of [1/2, 2] around their
central value, excluding the extreme configurations ($\mu_R$, $\mu_F$)~=~(2, 0.5)$\,\mu_0$ and (0.5, 2)$\,\mu_0$.

We reconstruct jets following the $E$-recombination scheme, according to which the energy and the momentum
of a jet are defined as the sums of the energies and the momenta of its constituents,
by using the \mbox{anti-$k_T$} jet clustering algorithm~\cite{Cacciari:2008gp},
with two different values for the jet radius parameter, $R = 0.4$ and $R = 0.8$. The
first value is the present default value adopted by both the ATLAS and CMS
collaborations in their analyses. 
We consider the second value as a possible alternative, in order to study the sensitivity of the predictions and their
uncertainties in dependence on $R$. 

A general discussion of the effects of these inputs on predictions for total
and dif\-fe\-ren\-tial cross sections is included in Section~\ref{sec:pheno}.  

\section{Phenomenology of $t\bar{t}j$ production at the LHC}
\label{sec:pheno}

If not specified otherwise, the predictions presented in the following refer to
our default configuration. This corresponds to a center-of-mass energy of
$\sqrt{S}=13$~TeV. The top-quark mass renormalized in the on-shell scheme is set to $m_t^{\rm{pole}} = 172$~GeV. 
In our fixed-order computation the top quarks are con\-si\-dered as 
stable and the CT18 NLO PDF set is used as default. At the phenomenological analysis level, at
least one jet is required with a transverse momentum $p_T^j > 30\,$GeV and
an absolute pseudorapidity $|\eta_j| < 2.4$. 
Jets are reconstructed using the \mbox{anti-$k_T$} jet clustering algorithm from
\textsc{FastJet}~\cite{Cacciari:2011ma} (version 3.3.4) with $R = 0.4$ and the
$E$-recombination scheme. This system of analysis cuts closely resembles
typical systems of cuts used by the ATLAS and CMS 
experimental collaborations.  

\subsection{Options for scale choice and scale uncertainty evaluation}
\label{subsec:scale}

Before studying uncertainties related to PDF and $R$ variation, as explained in
Subsection~\ref{subsec:pdf} and~\ref{subsec:r}, respectively,
we investigated the influence of different central scale de\-fi\-ni\-tions on
differential cross sections at NLO, considering many different observables. 
As mentioned in \Eq{eq:scale-choices}, 
we have considered four central scale choices, a static scale and three dynamical scales. 
The investigation of different $\mu_0$ choices was motivated by observations
presented in Ref.~\cite{Bevilacqua:2016jfk}, where a more stable behavior of the scale
variation uncertainty was found when using a dynamical scale with respect to
the static scale case. However, while in the present manuscript the top quarks
are considered as stable and the effects of relatively loose systems of cuts
are investigated, in line with the procedures applied in ongoing experimental
$t\bar{t}j + X $ analyses for the determination of the top-quark mass, in Ref.~\cite{Bevilacqua:2016jfk}
the $t\bar{t}j + X$ process was studied in presence of fully leptonic top-quark
decays and the analyzed events fulfilled more specific analysis cuts, which
differ substantially from the more inclusive cuts used here. 
This difference motivates the work presented in this Subsection. 

\begin{figure}
  \begin{center}
    \includegraphics[width=0.9\textwidth]{\main/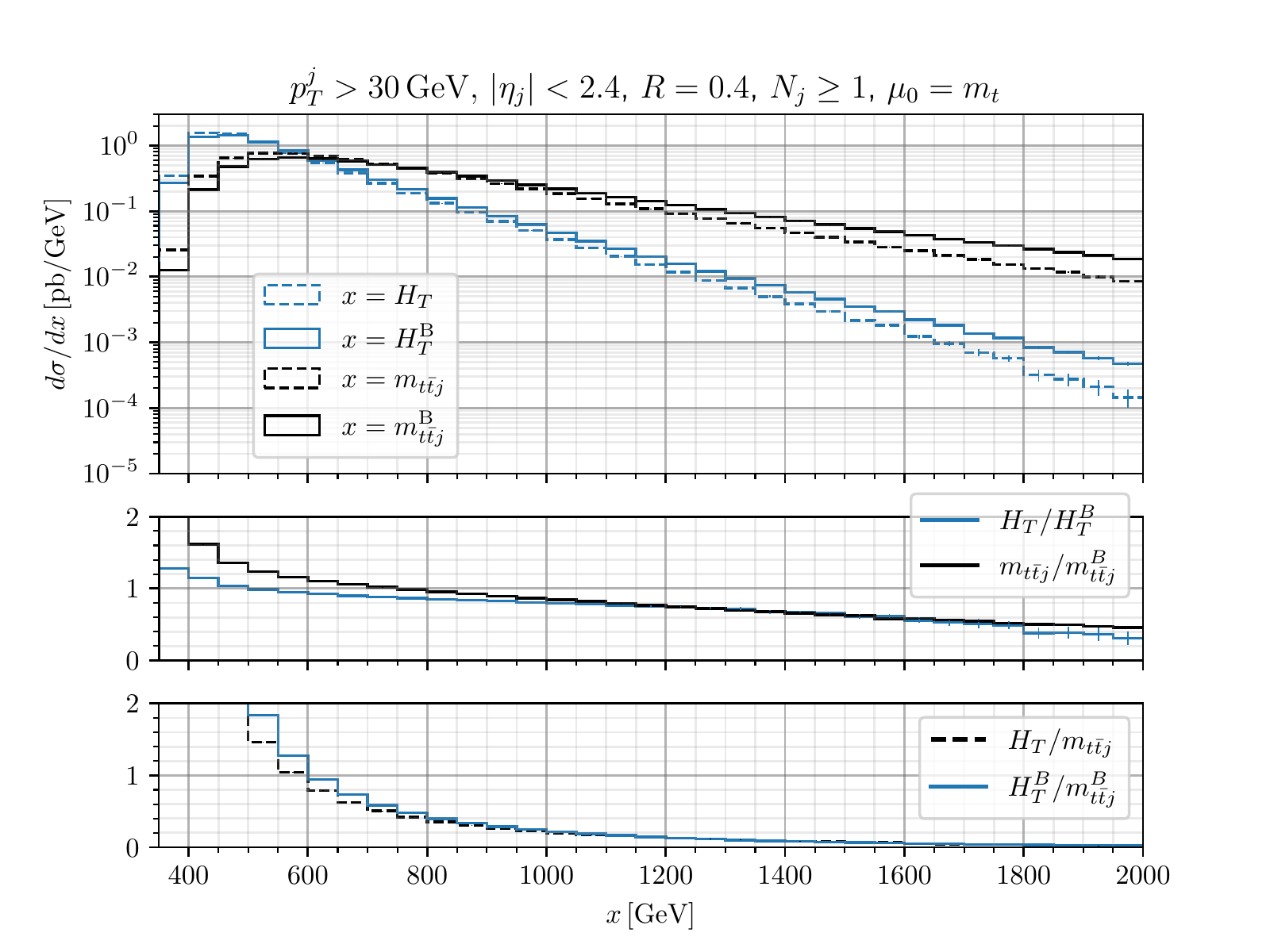}
  \end{center}
  \caption{
  \label{fig:HTB_MTTJB_DISTR}
Differential cross section as a function of $H_T^B$ (blue solid), $H_T$ (blue
dashed), $m_{t\bar{t}j}^B$ (black solid) and $m_{t\bar{t}j}$ (black dashed),
calculated for the $pp \rightarrow t\bar{t}j+X$ process at a center of mass
energy of $\sqrt{S}=13$~TeV, using as input a central scale $\mu_0$ set to
$\mu_0=m_t$.
}
\end{figure}

In \Fig{fig:HTB_MTTJB_DISTR} the NLO distributions in the variables  
$H_T^B$ (blue solid), $m_{t\bar{t}j}^B$ (black solid), 
$H_T$ (blue dashed) and $m_{t\bar{t}j}$ (black dashed) 
as defined in \Eqs{eq:HT-scale-def} and (\ref{eq:mttj-scale-def}) are shown for comparison.
All these differential cross sections were obtained by setting 
the central scale to the fixed value $\mu_0 = m_t$. 
In the lower panels ratios between various pairs of these distributions are shown. 
It is evident that the $m_{t\bar{t}j}^B$ and $m_{t\bar{t}j}$ spectra are harder than the $H_T^B$ and $H_T$ ones. 
This leads to reduced integrated and differential cross sections when using
the scales $m_{t\bar{t}j}^B$- or $m_{t\bar{t}j}$-based on the invariant-mass of the $t\bar{t}j$-system 
compared to the $H_T^B$- or $H_T$-based scales 
as the coupling $\alpha_s (\mu_R)$ decreases for increasing scales $\mu_R$.
The same considerations also apply when comparing predictions obtained with
$\mu_0=H_T^B/2$ and $\mu_0=H_T^B/4$, as we will discuss next. 
The dynamical central scale choices $\mu_0 = H_T^B/2$ and $\mu_0 = m_{t\bar{t}j}^B/2$ 
lead in general to larger numerical values 
for the renormalization and factorization scales, than the static scale choice $\mu_0 = m_t$. 
They are exactly equal to the latter only at threshold. 
Comparing the dashed and solid histograms in \Fig{fig:HTB_MTTJB_DISTR}, 
we also expect that the slightly different options $H_T$ and $m_{t\bar{t}j}$ for the dynamical scales, 
i.e. building them from the real emission kinematics instead of from the underlying Born configuration, 
will lead to slightly larger cross-section values. 

\begin{table}[h]
  \begin{center}\renewcommand{\arraystretch}{1.6}
    \begin{tabular}{c|c|c|c|c|c}
      \hline
      $\mu_0$  & $\sigma^{\text{LO}}\,$[pb] & $\delta_{\text{scale}}\,$[pb] & $\sigma^{\text{NLO}}\,$[pb] & $\delta_{\text{scale}}\,$[pb] & $\mathcal{K}=\sigma^{\text{NLO}}/\sigma^{\text{LO}}$ \\
      \hline
      $m_t$ & 294.3 & $^{+138.0 \, (+47\%)}_{-87.5\,(-30\%)}$ & 359.4 & $^{+15.3\, (+4\%)}_{- 41.5\, (-12\%)}$ & 1.22\\
      $H_T^B/2$ & 231.08 & $^{+100.30\,(+43\%)}_{-65.42\,(-28\%)}$  & 331.9 & $^{+34.9\, (+11\%)}_{-47.2\,(-14\%)}$ & 1.44\\
      $H_T^B/4$ & 331.4 & $^{+159.5 (+48\%)} _{-100.3 (-30\%)}$  &366.8 & $^{+3.3(+1\%)}_{-34.9 (-10\%)}$ & 1.11 \\      
      $m_{t\bar{t}j}^B/2$ & 202.17 & $^{+83.91\,(+42\%)}_{-55.61\,(-28\%)}$ & 302.5 & $^{+35\, (+12\%)}_{-43.2\, (-14\%)}$ & 1.50
      \\
      \hline
    \end{tabular}
  \end{center}
  \caption{
  \label{tab:RHORES_TotalXsection_ptj}
    Integrated cross section at LO and NLO of the process $pp\rightarrow
    t\bar{t}j + X$ with the analysis cuts $N_j\geq 1$, $p_T^j>30\,$GeV and
    $|\eta_j|<2.4$, where the \mbox{anti-$k_T$} jet clustering algorithm with $R=0.4$
    and the CT18 NLO PDF set were used. 
    The scale variation uncertainty $\delta_{\text{scale}}$ is
    obtained according to the seven-point scale variation procedure described
    in Section~\ref{sec:compu}.
    In each $\delta_{\text{scale}}$ column, the first number refers to the absolute uncertainty due to scale variation, whereas the second number, indicated in parentheses, gives the same uncertainty in a percentage
  format.
}
\end{table}

The values of the integrated cross sections at LO and NLO under our default
system of cuts described at the beginning of this Section are presented in
Table~\ref{tab:RHORES_TotalXsection_ptj} for the four central scale choices in \Eq{eq:scale-choices}. 
Thereby the seven-point scale variation was used to estimate the scale
uncertainty $\delta_{\text{scale}}$. The resulting $\mathcal{K}$-factors,
given by the ratio of the NLO and LO integrated cross sections evaluated with
each central scale, are presented in the last column. Both for the NLO and LO computations the CT18 NLO PDF set was used, accompanied by the associated default
$\alpha_s^{\text{NLO}}(M_Z)$ value, equal to $0.118$. Typical values of the strong
coupling accompanying LO PDF sets are in general larger, amounting to
$\alpha_s^{\text{LO}}(M_Z) \simeq 0.13$~\footnote{On the basis of this
  consideration, the LO cross sections, obtained using
  as input NLO (PDFs + $\alpha_s(M_Z)$) values and two-loop $\alpha_s$ evolution,
  have generally smaller values than those obtained
  with a LO (PDF + $\alpha_s(M_Z)$) set.}. 

Using the static scale choice leads to larger values of the integrated cross
section than the corresponding calculation with the dynamical scales
$\mu_0=H_T^B/2$ and $\mu_0=m_{t\bar{t}j}^B/2$. On the other hand, the cross
section obtained with the static scale is smaller than that obtained with the
dynamical scale $\mu_0=H_T^B/4$. In the threshold region, $\mu_0=H_T^B/4$ is
close to $m_t/2$, which leads to a larger $\alpha_s$ value in the bulk of the
phase-space at the LHC, as compared to the static scale choice $\mu_0 = m_t$, 
and, with this, to a larger integrated cross section.
The generally smaller cross sections obtained with
$\mu_0=m_{t\bar{t}j}^B/2$ compared to $\mu_0=H_T^B/2$ can be explained by the
harder spectrum of the $m^B_{t\bar{t}j}$ distribution (see
\Fig{fig:HTB_MTTJB_DISTR}), implying that events with large
$m^B_{t\bar{t}j}$, leading to small values of $\alpha_S(\mu_R = m_{t\bar{t}j}^B/2)$, 
are more frequent than events with large $H_T^B$. 

As expected the LO predictions suffer from larger scale uncertainties compared
to the NLO results. A slightly asymmetrical scale variation uncertainty is
observed at LO using all considered scale definitions, the size of the upper
uncertainty band being larger than that of the lower one.
The symmetrized scale uncertainties result in $\pm (38-39)\%$ when using the
static scale $\mu_0=m_t$ and the dynamical scale $\mu_0=H_T^B/4$, while they amount to $\pm (35-36)\%$ for the dynamical scales $\mu_0=H_T^B/2$ and $\mu_0=m_{t\bar{t}j}^B/2$.

For NLO predictions with the static scale, the downward scale variation is 
dominant, leading to a clearly asymmetrical scale uncertainty of 
$^{+4\,\%}_{-12\,\%}$ around the central value. This is also observed using 
the dynamical scale $\mu_0=H_T^B/4$ with a scale uncertainty of the NLO integrated 
cross section under the analysis cuts amounting to $^{+1\,\%}_{-10\,\%}$. Using other 
dynamical scale definitions, this asymmetry at NLO is strongly reduced, with 
scale uncertainties of $^{+11\,\%}_{-14\,\%}$  in case of 
$\mu_0=H_T^B/2$ and $^{+12\,\%}_{-14\,\%}$ in case of $\mu_0=m_{t\bar{t}j}^B/2$, 
around the respective central values.

The symmetrized scale uncertainty at NLO is larger when using the dynamical 
scales $\mu_0=H_T^B/2$ and $\mu_0=m_{t\bar{t}j}^B/2$, amounting to $\sim \pm \, 12 - 13\, \%$ 
than for the static scale case, where it amounts to $\sim \pm \, 8\,\%$. 
This can be explained as an artefact of the crossing of the scale variation 
graphs corresponding to different multiples ($K_R$, $K_F$) of the central 
renormalization and factorization scale, occurring only when using the static 
scale, as discussed in the following, which leads to an artificial reduction 
of the scale uncertainty in some kinematical regions. 
The smallest symmetrized scale uncertainty at NLO, of $\pm 5.5\%$, is found 
using the dynamical scale $\mu_0=H_T^B/4$. 
As far as the inclusive cross section is concerned the static scale as well as 
the dynamical scale $\mu_0=H_T^B/4$ show a moderate $\cal K$-factor suggesting 
a well behaved perturbative expansion. Furthermore, both scale settings give, 
within the uncertainty, consistent results. The two other dynamical scales lead 
to a larger $\cal K$-factor suggesting that higher-order corrections are 
sizable. This is consistent with the observation that for these scales the 
predictions are 10\% or even 20\% smaller than using the static scale or $H_T^B/4$. 

\begin{figure}
  \begin{center}
      \includegraphics[width=0.74\textwidth]{\main/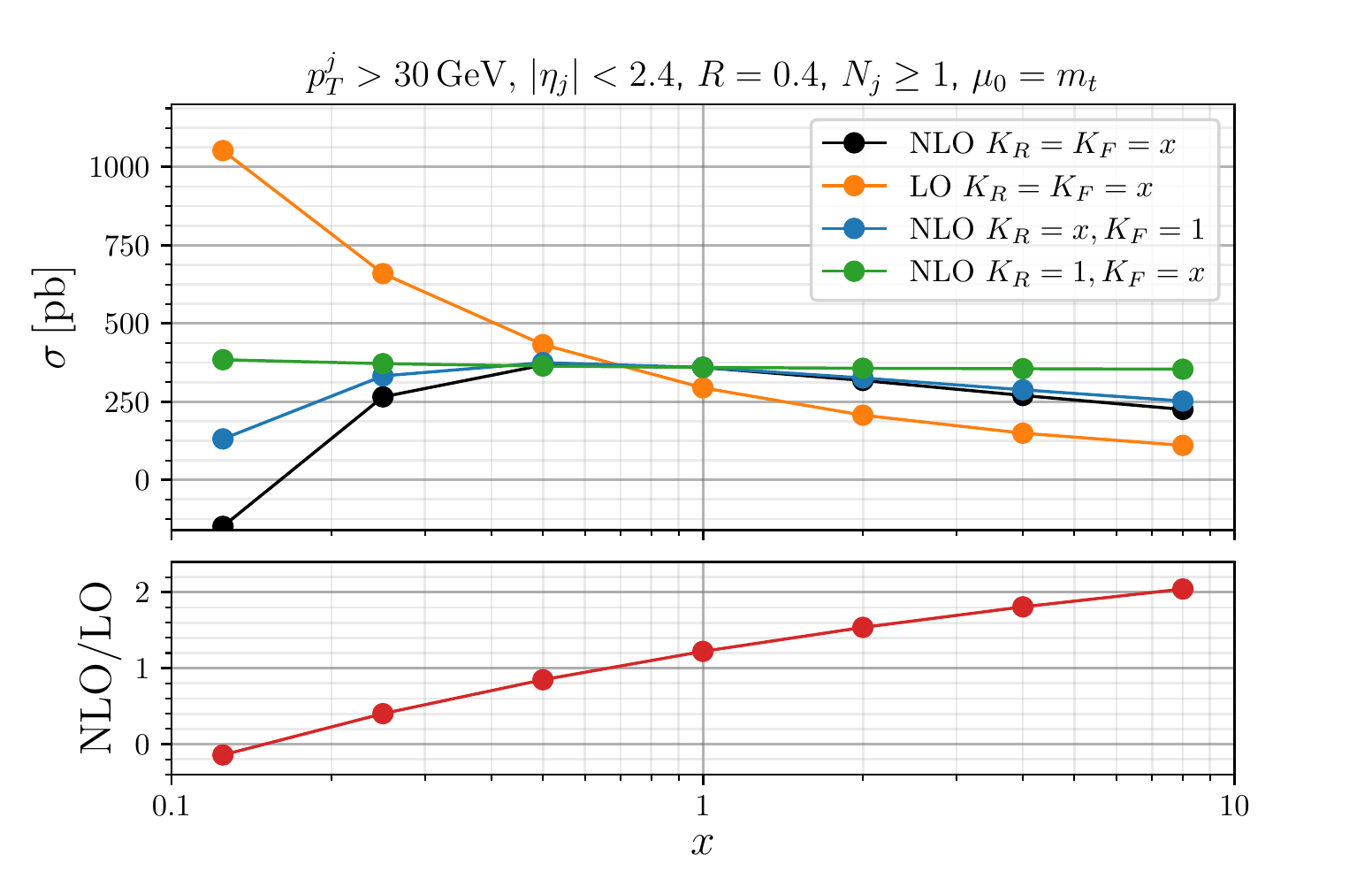}
      \label{FIG_xsec_mt}
      \\
      \includegraphics[width=0.74\textwidth]{\main/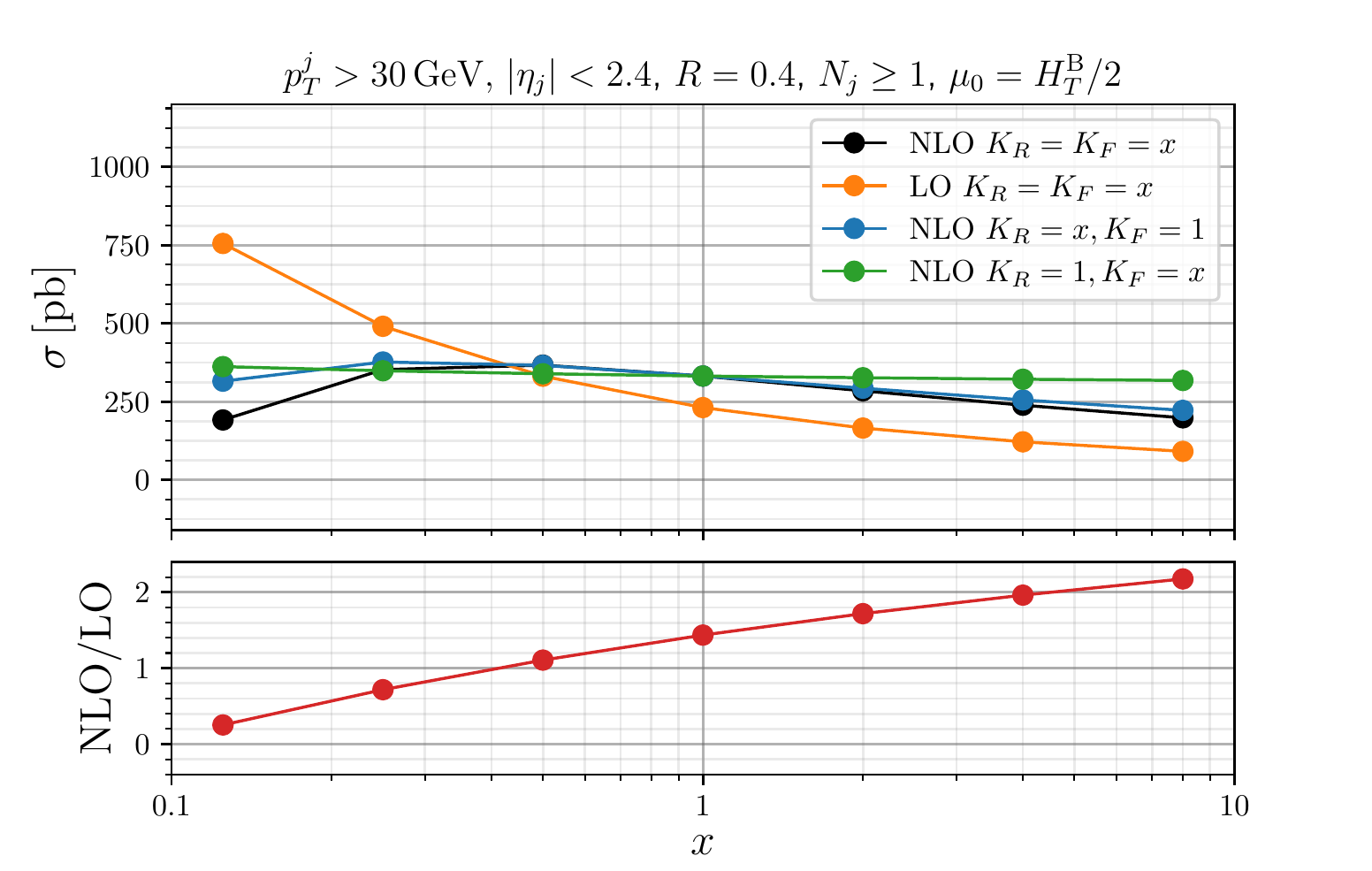}
      \label{FIG_xsec_ht}
      \\
      \includegraphics[width=0.74\textwidth]{\main/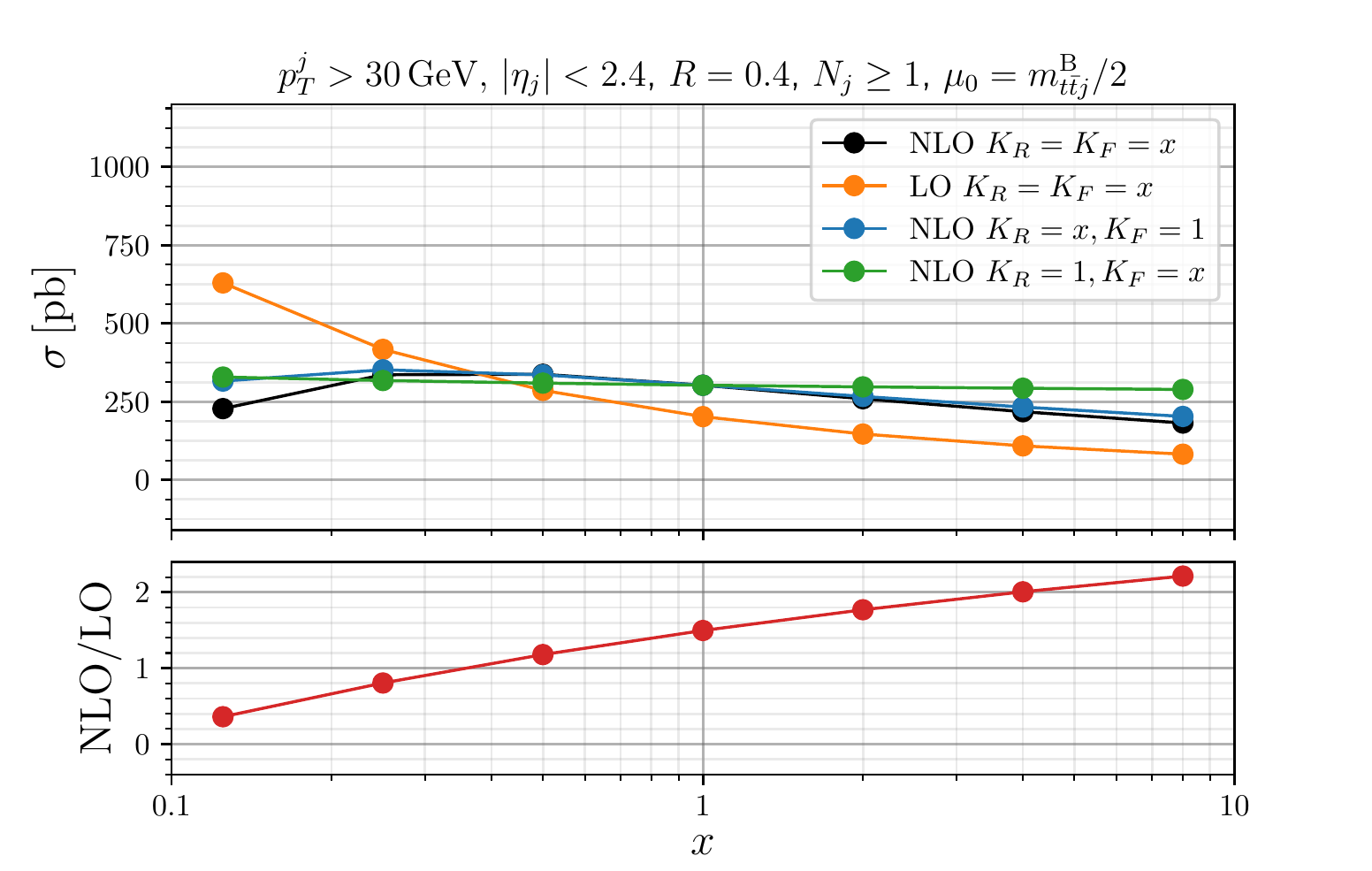}
      \label{FIG_xsec_mttj}
  \end{center}
  \caption{Integrated cross section of the $pp\rightarrow t\bar{t}j+X$ process after our default system of ana\-ly\-sis cuts,  using multiple ($K_R$, $K_F$) of the static scale definition $\mu_0 = m_t$ (upper panel) and of the dynamical scale definitions $\mu_0 = H_T^B/2$ (central panel)  and $\mu_0 = m^B_{t\bar{t}j}/2$ (lower panel).}
  \label{fig:xsec}
\end{figure}

In \Fig{fig:xsec} the NLO QCD integrated fiducial cross sections are plotted, which were 
obtained by varying the
factorization and renormalization scales simultaneously by the factor 
$x \in \{ 0.125, 0.25, 0.5, 1, 2, 4, 8\}$ (black) 
or by only varying $\mu_F$ (green) or $\mu_R$ (blue)
while keeping the other scale ($\mu_R$ or $\mu_F$, respectively) fixed.
Following Ref.~\cite{Bevilacqua:2016jfk}, 
a broad range of $x$ values is considered, in order to check the stability of
predictions with respect to scale variation. The standard seven-point scale variation procedure gives rise to a subset of these predictions,  corresponding to the points $x \in \{ 0.5, 1, 2\}$. In   \Fig{fig:xsec} it can be seen that, as also noticed in~\cite{Bevilacqua:2016jfk}, 
the ($\mu_R$, $\mu_F$) scale uncertainty band
strongly depends 
on
the variation of $\mu_R$ for all
 central
scale
choices considered in this paper, since the integrated cross section varies
only marginally when changing the value of the facto\-rization scale, if the
renormalization scale is kept fixed (green). 
Thereby the influence of either
renormalization or factorization scale variation can vary, when considering
different central scale choices. 
For example, when using $\mu_0=H_T^B/4$ the
$\mu_R$ dependence is less pronounced compared to using $\mu_0=H_T^B/2$ in the
seven-point scale variation, whereas the $\mu_F$ scale dependence is extremely
mild in both cases. This can be seen by comparing the integrated cross
sections in the intermediate panel in \Fig{fig:xsec} for values of $x \in [0.5,1,2]$
and $x \in [0.25,0.5,1]$ corresponding to the seven-point scale variation
using $\mu_0=H_T^B/2$ and $\mu_0=H_T^B/4$, respectively. The strong influence
of the renormalization scale on the 
scale uncertainty is also evident from the fact that the $K_F = K_R = x$
results (black) follow closely the $K_F=1, K_R=x$ ones (blue), with the
exception of the very extreme case of $x = 0.125$. For extreme choices as 
$K_F = K_R = 0.125$ the cross section turns out to be negative, i.e. unphysical, for the static scale choice, but not for the dynamical scale choices. This was also noticed in Ref.~\cite{Bevilacqua:2016jfk}.

The much larger width of the LO scale uncertainty band with respect to the NLO one
can also be inferred from \Fig{fig:xsec}, being already evident when the factorization
and renormalization scales are varied simultaneously by the factor $K_R=K_F=x$ (orange). 
In the lower inset of each panel of \Fig{fig:xsec} the ratio between the
NLO and LO predictions is shown. This ratio is computed from integrated
cross sections obtained by varying $\mu_R$ and $\mu_F$ simultaneously 
by $x$ in both the LO and NLO case. Thereby it
can be seen that the smallest $\mathcal{K}$-factors using the dynamical scales
are found for $\mu_0=H_T^B/4$ and $\mu_0=m_{t\bar{t}j}^B/4$. 
This motivates further investigation of these central scales.
In the sequel only the central scale $\mu_0=H_T^B/4$ is additionally studied,
since the differential cross sections obtained using $\mu_0=m_{t\bar{t}j}^B/4$
showed larger scale uncertainties, as better clarified in the following (see discussion of Fig.~\ref{fig:RHOmodified}).

\begin{figure}
  \begin{center}
    \includegraphics[width=1.\textwidth]{\main/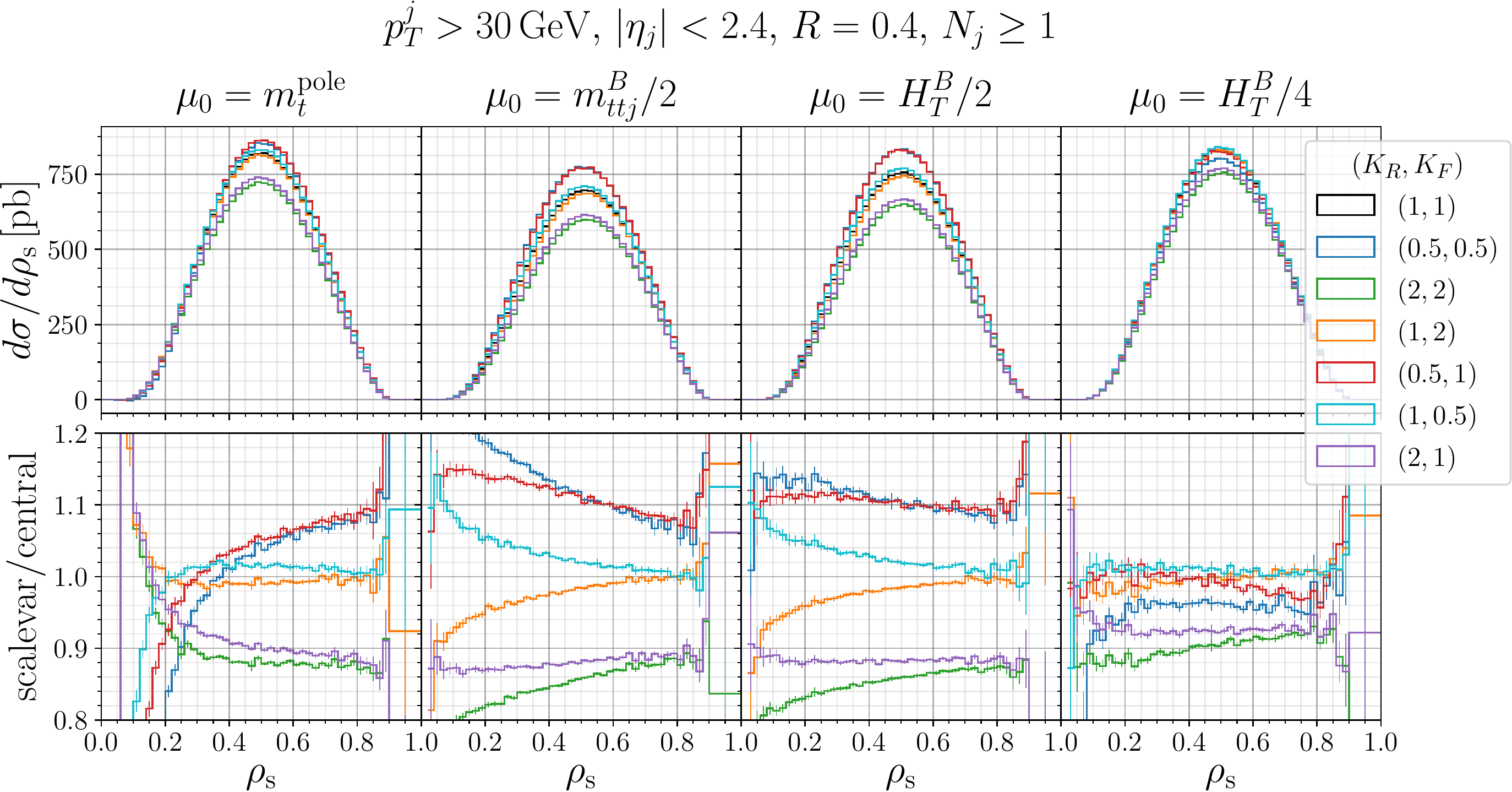}
  \end{center}
  \caption{
NLO differential cross section as a function of $\rho_s$ for the $pp \rightarrow
t\bar{t}j+X$ process at $\sqrt{S}=13\,$TeV using as central scales the static
scale $\mu_0=m_t$ (left panel) and the dynamical scales $\mu_0\in \{m_{t\bar{t}j}^B/2, H_T^B/2,
H_T^B/4\}$. Thereby the seven-point scale variation graphs
are explicitly drawn and the ratio between these and the central scale predictions
are shown in the lower ratio plot. Due to low statistics for large values of
$\rho_s$ the distribution was rebinned, such that the region $\rho_s \in
[0.9,1]$ corresponds to one bin. 
}
  \label{fig:RHO}  
\end{figure}

\begin{figure}
  \begin{center}
    \includegraphics[width=1.\textwidth]{\main/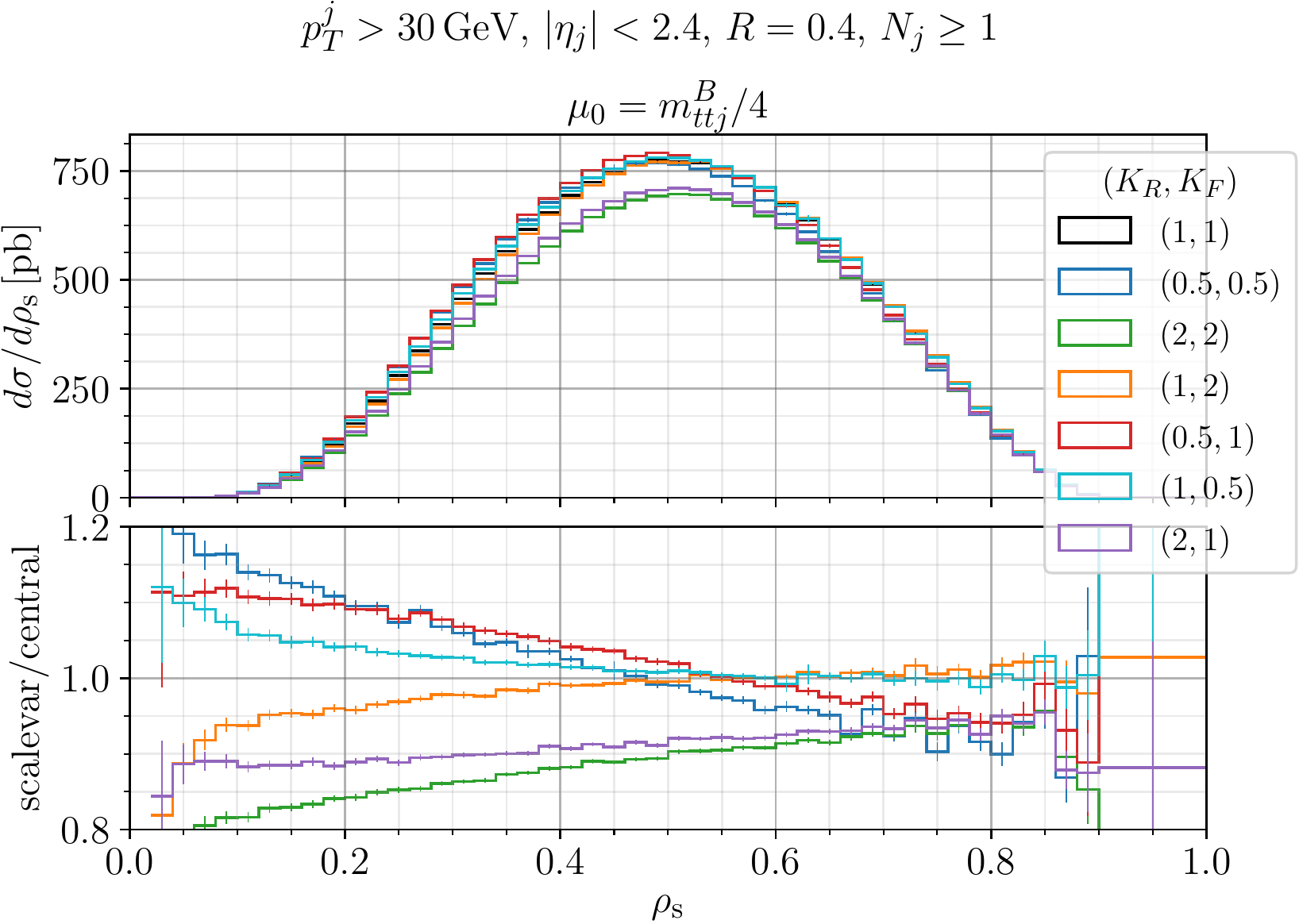}
  \end{center}
 \caption{Same as the rightmost panel of \Fig{fig:RHO}, but using the scale $\mu_0= m_{t\bar{t}j}^B/4$, instead of $H_T^B/4$.}
  \label{fig:RHOmodified}  
\end{figure}


As a first differential cross section, the $\rho_s$ distribution is
investigated.
In the most left panel of \Fig{fig:RHO}  the central scale prediction is
shown obtained by setting $\mu_R=\mu_F=m_t$, while the graphs resulting from
the various ($K_R$, $K_F$) combinations corresponding to
seven-point scale variation, performed as discussed in
Section~\ref{sec:compu}, are also drawn individually.
The scale variation
uncertainty is rapidly increasing in the high energy tails of the
distribution signaling that this region is not well described by the static
scale. The high energy tail corresponds thereby to the region of low $\rho_s$
va\-lues, since this implies large values of $m_{t\bar{t}j}$. Furthermore, a
crossing of the scale variation graphs is observed for values of 
$0.1 \lesssim \rho_s \lesssim 0.3$. 
Also a clearly asymmetrical scale variation uncertainty band is
found, in which the downwards variation is much larger than the upwards
one for values of $0.4 \lesssim \rho_s \lesssim 0.7$.
As already observed in case of integrated fiducial
cross sections, also for the $\rho_s$ differential distribution the renormalization scale dependence is stronger than the facto\-ri\-zation scale one. 
This can be directly observed in the left panel, as graphs obtained with each fixed value of $K_R$ and different values of $K_F$ are grouped together. 
This behavior is especially clear when looking e.g. at the graphs with $K_R = 2$, noticing that for most of the $\rho_s$ values 
they represent the minimum of the ($\mu_R$, $\mu_F$) uncertainty band. 
This behavior is also noticeable in the scale variation graphs
obtained with the dynamical scales $\mu_0=m_{t\bar{t}j}^B/2$ and $\mu_0=H_T^B/2$ (second and third panels of \Fig{fig:RHO}).
Using the central scale $\mu_0=H_T^B/4$ (fourth panel) the stronger
$\mu_R$ dependence with respect to  the $\mu_F$ dependence is reduced, as was already noticed when comparing the integrated cross sections in \Fig{fig:xsec}.

In contrast to the static scale prediction in \Fig{fig:RHO} (left panel), the predictions with dynamical scales (right panels) show a more uniform width of the scale variation
uncertainty band in the high energy tails and in the bulk of the $\rho_s$ distribution,
especially when using the scales $\mu_0=H_T^B/2$ and $\mu_0=H_T^B/4$. 
In addition, only a reduced crossing of the scale variation graphs among each
other is observed when using the dynamical scale. This is especially
true for the scale choice $\mu_0 = H_T^B/4$ (last panel),
for which the scale variation induces a nearly uniform shift of the central
scale result, with the exception of the large $\rho_s$ tail, where statistics
is however very limited and experimental measurements are thus not possible,
at least at present. When considering the normalized $\rho_s$ distribution,
this leads to strongly reduced scale variation uncertainties, if each scale
variation graph is normalized by its respective integrated cross section. But
even considering the differential cross section shown in \Fig{fig:RHO}, it is
clearly visible that using the dynamical scale choice $\mu_0=H_T^B/4$ the
smallest scale variation uncertainties are obtained in the experimentally
explored region. 
On the other hand, the differential cross sections obtained using
$\mu_0=m_{t\bar{t}j}^B/4$ show larger scale variation uncertainties and the
differential cross sections evaluated with dif\-fe\-rent $K_R$ and $K_F$ factors were found to cross each other, as displayed in \Fig{fig:RHOmodified}. Therefore, in the following of this paper, we will not consider anymore the latter scale choice.  

\begin{figure}
  \begin{center}
    \includegraphics[width=0.49\textwidth]{\main/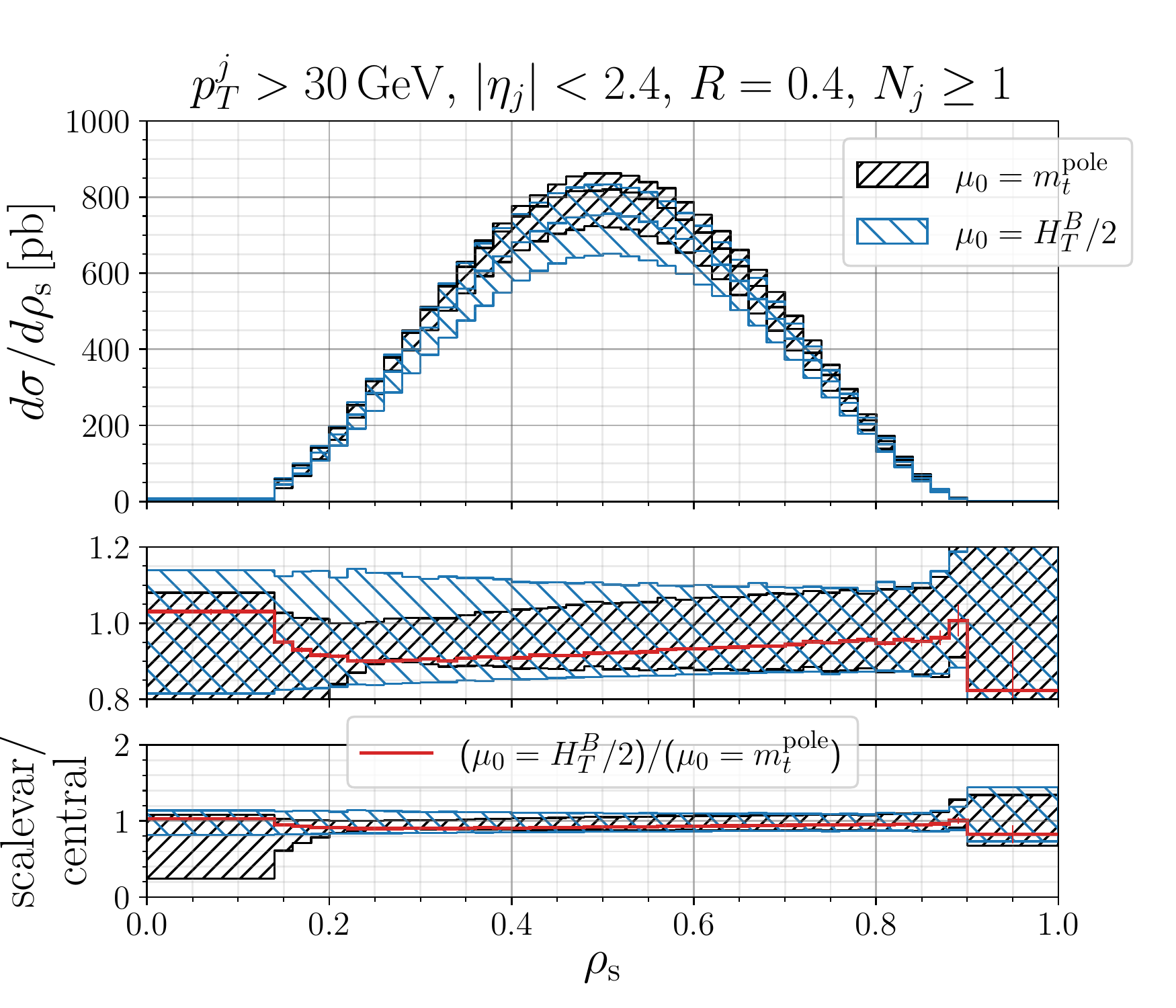}
    \includegraphics[width=0.49\textwidth]{\main/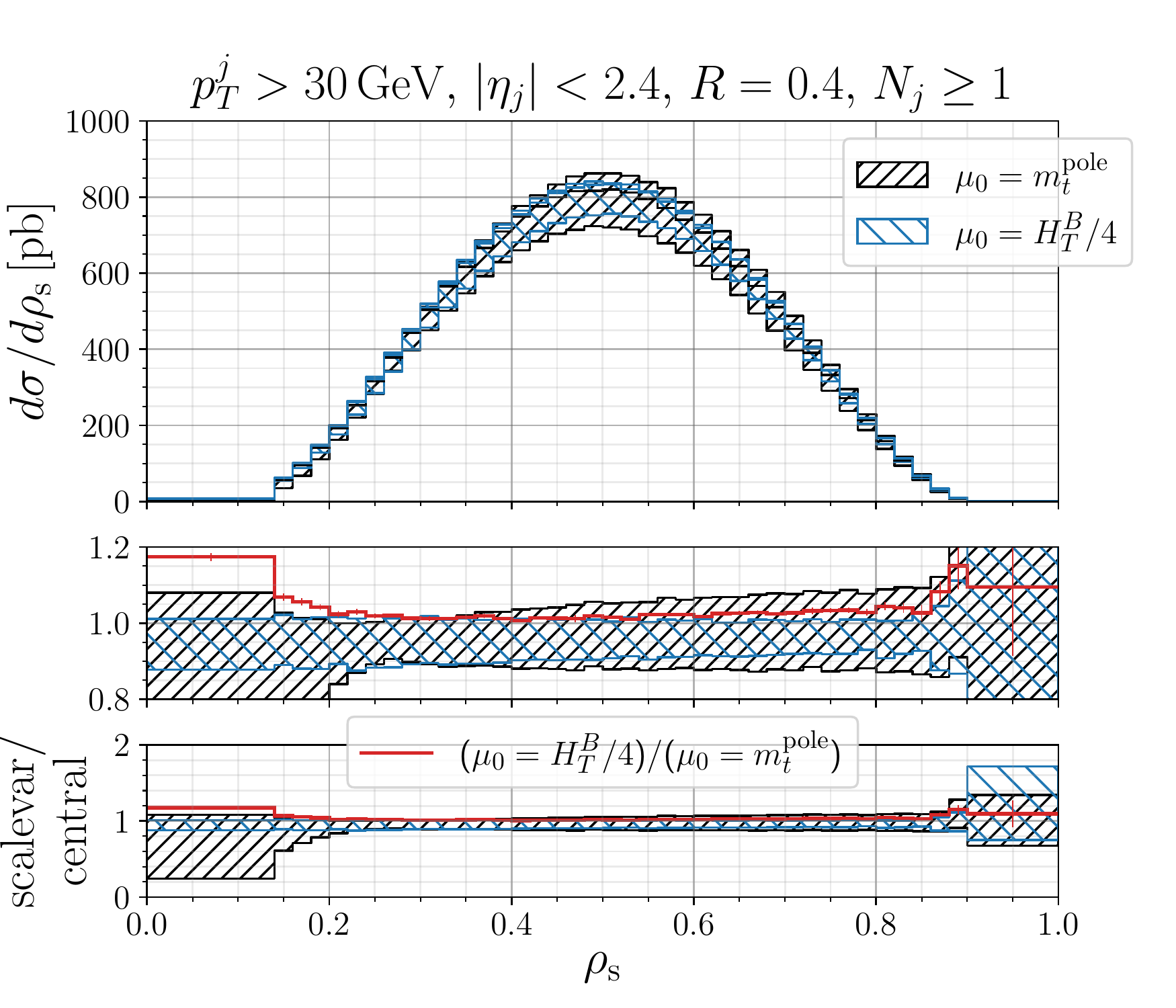}
  \end{center}
  \caption{NLO differential cross section as a function of $\rho_s$ 
    for the $pp \rightarrow t\bar{t}j+X$ process at $\sqrt{S}=13\,$TeV
    calculated
    using
    the static central scale $\mu_0=m_t$ (black) and the dynamical central scale
    $\mu_0=H_T^B/2$ (blue) (left panel) or the dynamical central scale
    $\mu_0=H_T^B/4$ (blue) (right panel). Thereby the envelopes of the seven-point
    scale variation and the central scale predictions are shown. In the two
    lower insets the ratios of the scale variation bands with the corresponding
    central scale predictions are shown with two different ranges on the
    $y$-axis, respectively. Additionally, the ratio of the central prediction
    obtained with the dynamical scale
    with the one obtained with $\mu_0 = m_t$ is shown with the red line. 
}
  \label{fig:RHO_MT_HT_OVERLAID}
\end{figure}
To better compare the distributions computed with different central scale
choices, the envelope of
the seven-point scale variation and the central scale
prediction
obtained by using
either the static $\mu_0=m_t$ (black) or the dynamical scale $\mu_0=H_T^B/2$
(blue) are overlaid in the left panel of \Fig{fig:RHO_MT_HT_OVERLAID}.  
Thereby in the lower two ratio plots the scale variation bands are rescaled by
the corresponding central scale result. The relative distortion between the two central
scale results is depicted with the red line, which shows the ratio between the
distribution obtained with $\mu_R=\mu_F=H_T^B/2$ and the one obtained with
$\mu_R=\mu_F=m_t$. From this ratio it can be seen that the differential cross
section obtained with the dynamical scale evaluates in general to lower values
compared to the one obtained with the static scale, excluding thereby the
high-energy tail, in which an unstable behavior of the $\rho_s$~distribution
using the static scale was found. This can be explained due to the larger
values of the scale (and, consequently, smaller values of $\alpha_s$), which
are obtained by applying the dynamical scale definition instead of       the
static one. 
The effect of the variation of $\alpha_s$ with the renormalization scale seems
to be dominant compared to the effect of the explicit scale logarithm in the
NLO corrections.  

For large values of $\rho_s$ the
scale variation uncertainties
for the distributions computed with $\mu_0=H_T^B/2$ and $\mu_0=m_t$
become very similar to each other. 
This is expected, since in this kinematic region of the $\rho_s$ distribution 
close to threshold this dynamical and the static scale are becoming equal. As shown in the right panel of figure 6, with the choice $\mu_0=H_T^B/4$ larger differential cross sections are found in the threshold region compared to using the scale $\mu_0=m_t$, since the dynamical scale evaluates to lower values close to the threshold region.  

Further it is observed that the scale variation uncertainty using the static
scale is in fact reduced in the region of $0.2 \lesssim \rho_s \lesssim 0.4$
compared to the uncertainty obtained with the dynamical scale definition
$\mu_0=H_T^B/2$. This can be explained with the crossing of the scale
variation graphs among each other in the static scale case in this region of
the \mbox{$\rho_s$~distribution}, as also seen in \Fig{fig:RHO}, leading to an
artificial reduction of the scale variation uncertainty. For
the lowest values of $\rho_s$ the scale variation uncertainty is strongly
reduced using the dynamical scale $\mu_0=H_T^B/2$, as is evident from the lowest ratio
in the left panel of \Fig{fig:RHO}. In contrast, using the
static scale, a steep increase in the scale variation uncertainty was found in this
region.

Looking at the right panel of \Fig{fig:RHO_MT_HT_OVERLAID}, 
it is evident that this dynamical scale choice leads to central predictions 
more similar to those obtained with the static scale than the previously 
discussed $H_T^B/2$ choice (left panel). 
Additionally, they lead to a smaller uncertainty band size, as already
discussed when commenting \Fig{fig:RHO}.  

\subsection{Comparison of NLO and LO differential cross sections}

One of the indications 
of how well a scale definition is suited to describe a
specific ob\-ser\-va\-ble is the apparent convergence of the predictions, when
including higher orders in the perturbative expansion. This applies for those cases where the same initial-state partonic channels are present at both
orders~\footnote{Differently from the case of $t\bar{t}$ production, for
  $t\bar{t}j$ production the $qg$, $gq$, $\bar{q}g$ and $g\bar{q}$ channels contribute already 
  at LO, together with the $gg$ and $q\bar{q}$ channels.}.

\begin{figure}[h!]
  \begin{center}
    \includegraphics[width=\textwidth]{\main/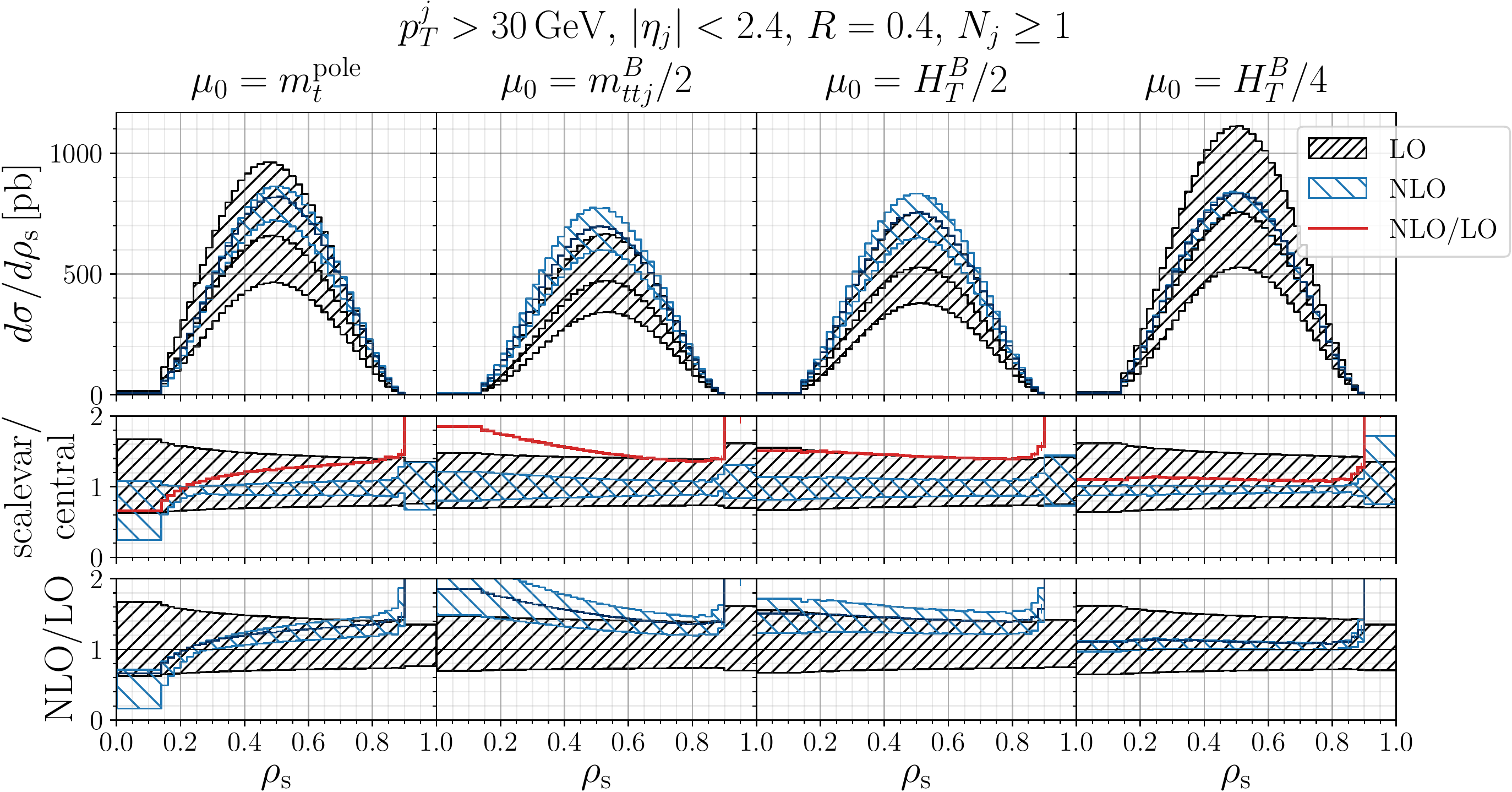}
  \end{center}
\caption{Central predictions for the $\rho_s$ distribution at NLO (blue) and
  LO (black), together the corresponding NLO and LO seven-point scale
  variation uncertainty bands, using as a central scale $\mu_0=m_t$,
  $m_{t\bar{t}j}^B/2$, $H_T^B/2$ and $H_T^B/4$ (from left to right). In the
  middle ratio plot the scale variation uncertainty at either NLO or LO is
  rescaled by the central scale prediction at NLO or LO respectively. To
  visualize the difference between the NLO and LO differential cross section
  the ratio of the NLO and LO central scale predictions is shown in red. In
  the bottom ratio insets both the NLO and LO scale variation uncertainty
  bands are rescaled by the LO central scale prediction.} 
\label{fig:scale_NLO_LO_rho}
\end{figure}

To this end we compare the LO and NLO predictions obtained with the static and
the dynamical scale choices~\footnote{Both NLO and LO predictions are obtained
  by using the CT18 NLO PDF set with its own $\alpha_S(M_Z)~=~0.118 $ value
  and two-loop $\alpha_s$ evolution, see also Sections~\ref{subsec:scale}
  and~\ref{subsec:pdf}.}.
In \Fig{fig:scale_NLO_LO_rho} the LO and NLO differential cross sections
in $\rho_s$ are compared among each other, using the scale choices in \Eq{eq:scale-choices}.
In the middle ratio plot, below the $\rho_s$ distribution, the scale variation bands, normalized
to the result obtained with $K_R=K_F=1$ at either LO (black) or NLO (blue), are shown. 
The red line in the ratio plot indicates the differential $\mathcal{K}$-factor
$(d\sigma^{\text{NLO}}/d\rho_s) \bigg / (d\sigma^{\text{LO}}/d\rho_s)$. In contrast, 
in the lowest ratio plot the scale variation bands are both rescaled by
the LO result. This way it can be easily checked if the NLO and LO scale
variation bands overlap. 

As expected, the NLO scale variation band is generally smaller compared to the
LO one. It turns out that, for $\rho_s < 0.9$, the NLO predictions obtained
with the dynamical scales $\mu_0=H^B_T/2$ and $H_T^B/4$ are shifted to higher
values with respect to  the LO predictions, by an almost uniform differential
$\mathcal{K}$-factor of $\sim 1.5$ and $1.1$, respectively, as visible in
\Fig{fig:scale_NLO_LO_rho}. The scale uncertainty band is thereby reduced
when going from LO to NLO and shows a nearly uniform width over the whole
range of the $\rho_s$~distribution, amounting to about ${\pm 10\,\%}$  for
$\mu_0=H_T^B/2$ and to about
( $+1\,\%$,  $-10\,\%$) for $\mu_0=H_T^B/4$. 
On the other hand, the same distribution evaluated with the scale
$\mu_0=m_{t\bar{t}j}^B/2$, as shown in the second panel of
\Fig{fig:scale_NLO_LO_rho}, is characterized by a larger differential
$\mathcal{K}$-factor for low values of $\rho_s$, where
$(d\sigma^{\text{NLO}}/d\rho_s)/(d\sigma^{\text{LO}}/d\rho_s) \sim 1.9$, than
for high values of $\rho_s$, where the differential $\mathcal{K}$-factor is
found to be $\sim 1.4$ (excluding the very last bin near threshold). 
Applying the scale definition $\mu_0=m_{t\bar{t}j}^B/2$, the LO and NLO scale
variation bands depart from each other in the low $\rho_s$ region, signaling
that this region is not well described. The aforementioned behaviour can be understood from the fact that the process is a multi-scale problem. While for inclusive quantities, the invariant mass of the final states at the underlying Born level $m_{t\bar{t}j}^B/2$ gives an appropriate scale setting, the scale relevant for the emission of the additional jet can be much lower. As a consequence, one expects that observables relying significantly on the additional emission are not well described using the $m_{t\bar{t}j}^B/2$ scale. This implies a large $\mathcal{K}$-factor using this setting as at NLO accuracy the scale dependence should be reduced with respect to LO and differences in the predictions on the basis of different choices should be attenuated as well.

\begin{figure}[h!]
  \begin{center}
    \includegraphics[width=\textwidth]{\main/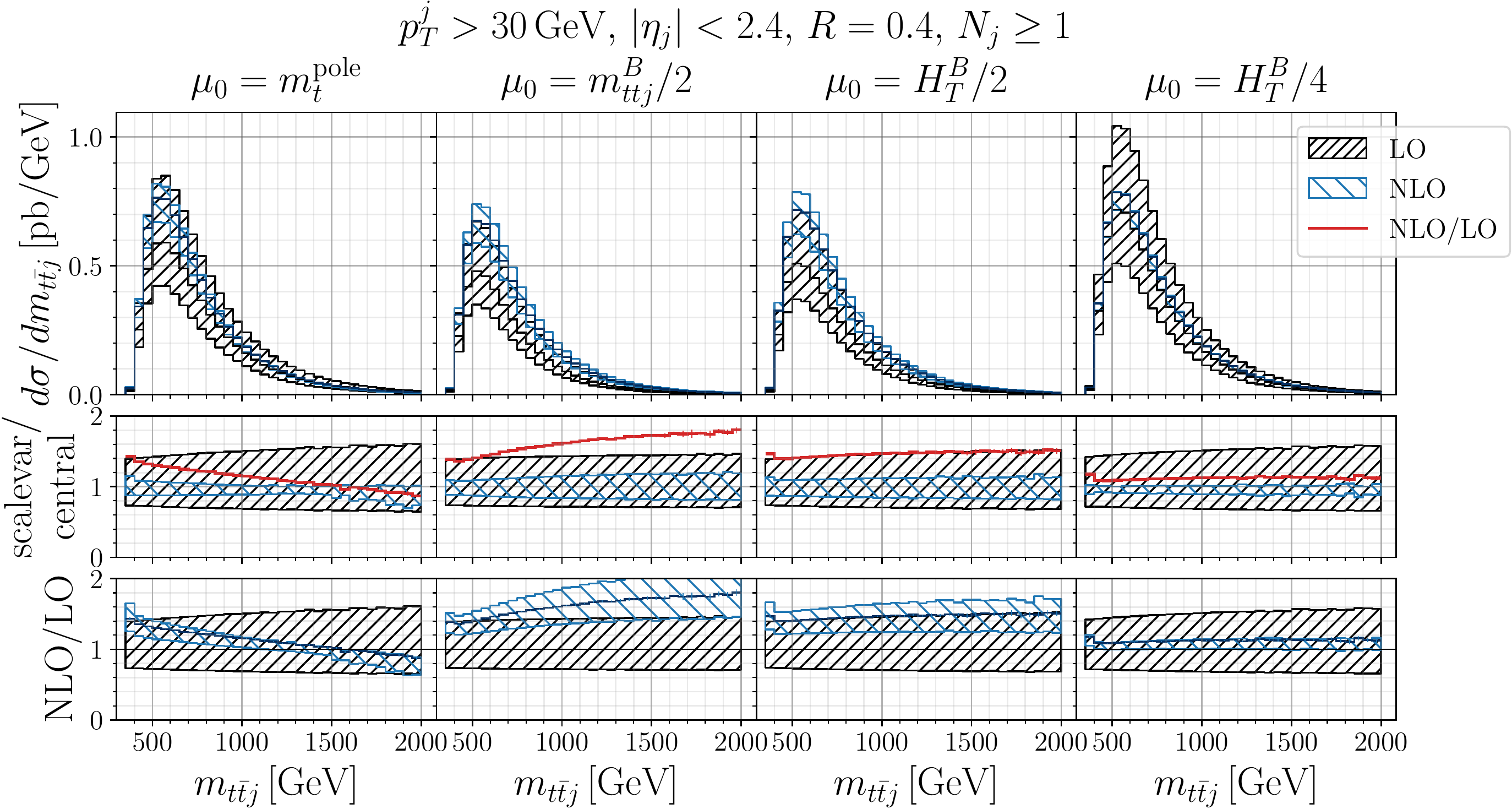}
  \end{center}
  \caption{
    \label{fig:scale_NLO_LO_mttj}
    Same as in \Fig{fig:scale_NLO_LO_rho}, but for the $m_{t\bar{t}j}$ distribution.
  }  
\end{figure}

In case of the static scale $\mu_0=m_t$, the central values of the NLO and LO
distributions are strongly distorted one with respect to  each other, as shown in the
first panel of \Fig{fig:scale_NLO_LO_rho}. 
A crossing of the NLO and LO distributions evaluated with $K_R=K_F=1$ occurs in
the same region of the $\rho_s$ distribution, namely at $\rho_s \sim 0.2$, 
where the NLO predictions, evaluated at the scales that allow to build the seven-point scale variation band, cross among each others (see the left panel of \Fig{fig:RHO}). 
Additionally, with the static scale definition,
the NLO and LO scale variation bands are only marginally overlapping in the
high $\rho_s$ region. As follows from the relation between $\rho_s$ and
$m_{t\bar{t}j}$ in the definition of $\rho_s$, the features of the $\rho_s$
distribution at low $\rho_s$ calculated with the different scales are of
course reflected in the $m_{t\bar{t}j}$ distributions for large values of
$m_{t\bar{t}j}$, shown in \Fig{fig:scale_NLO_LO_mttj}. 

\begin{figure}[h!]
  \begin{center}
    \includegraphics[width=\textwidth]{\main/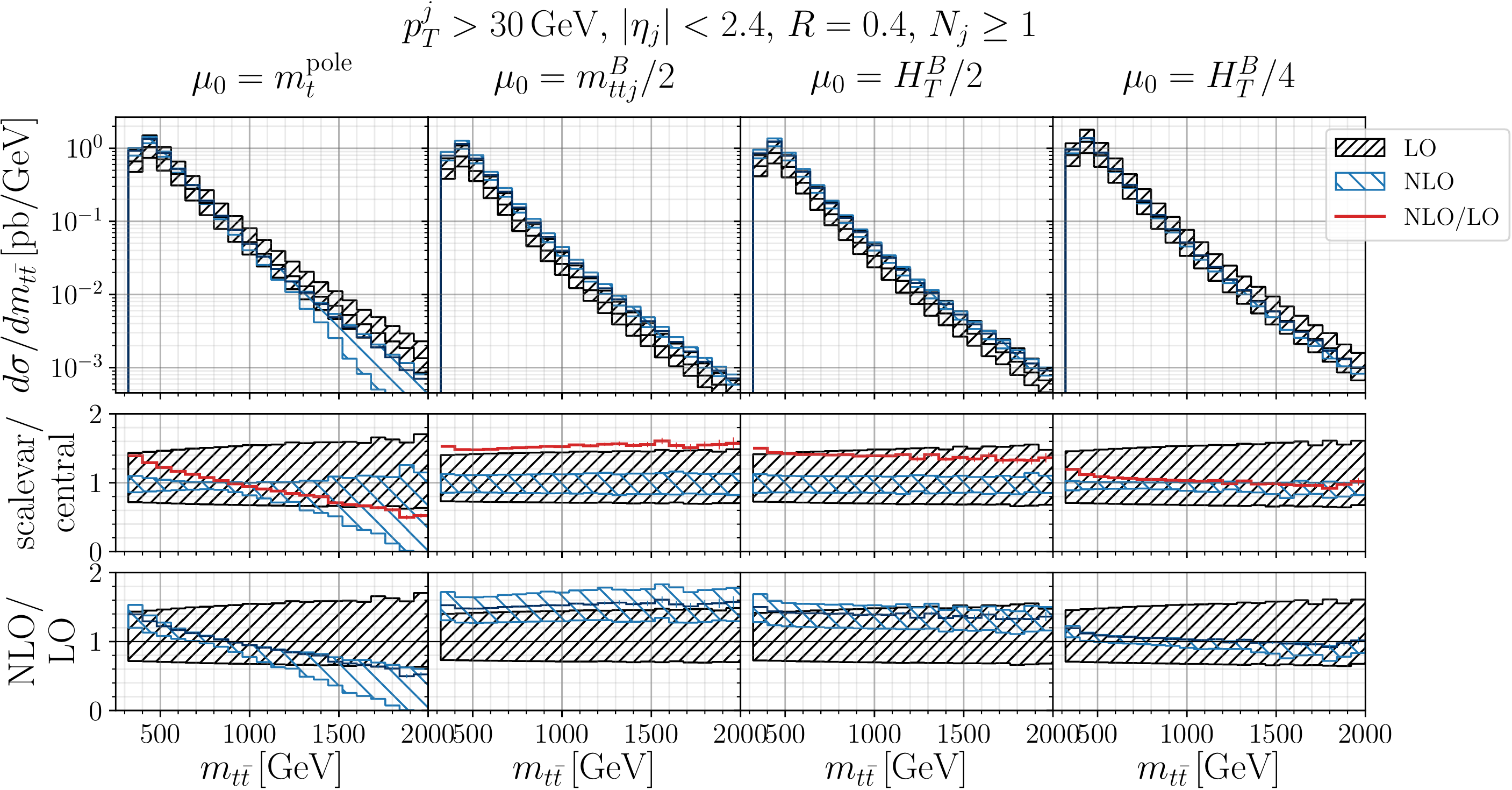}
  \end{center}
  \caption{Same as in \Fig{fig:scale_NLO_LO_rho}, but for the $m_{t\bar{t}}$ distribution.   }
\label{fig:scale_NLO_LO_mttbar}
\end{figure}

\begin{figure}[h!]
  \begin{center}
    \includegraphics[width=\textwidth]{\main/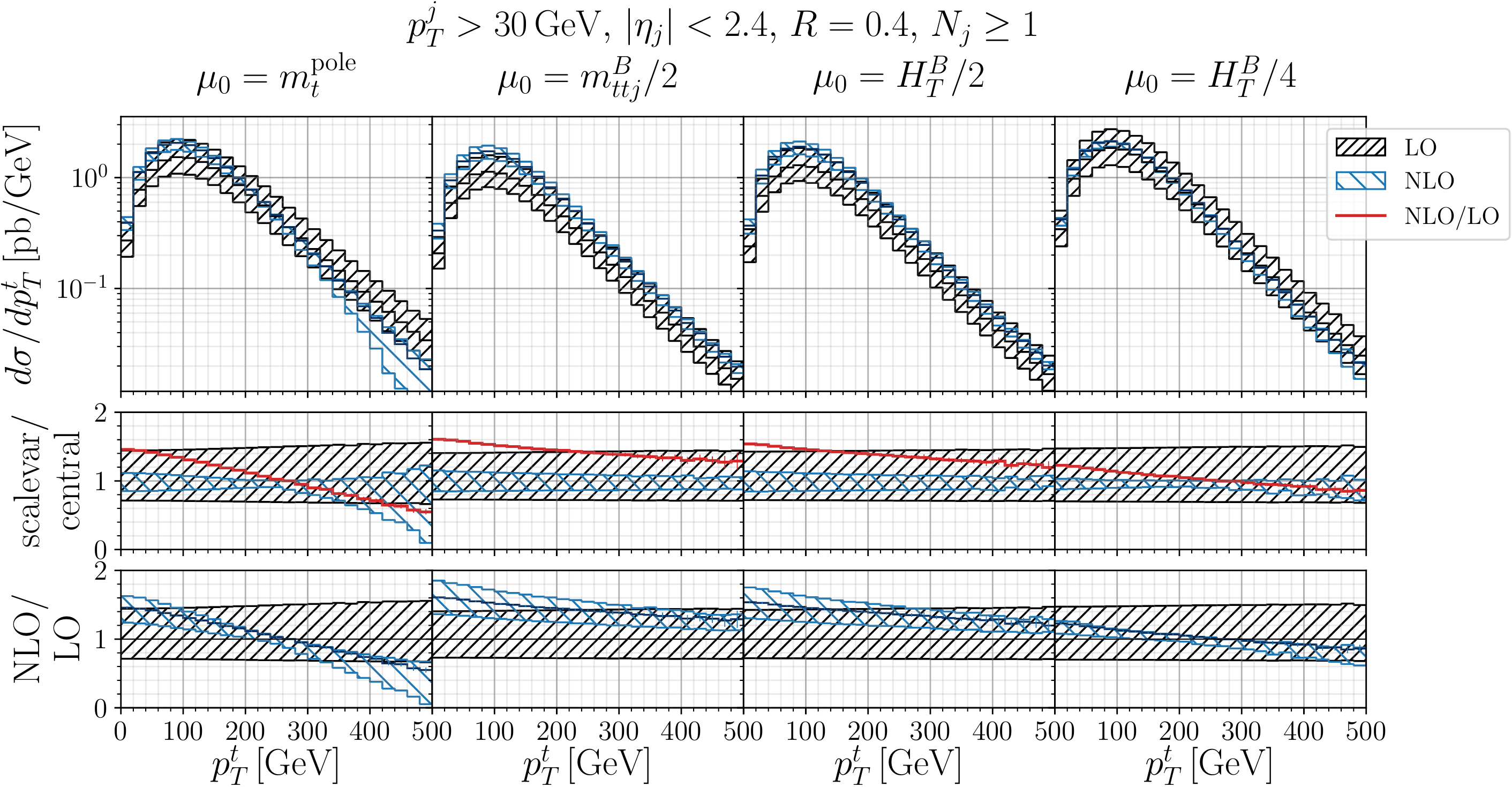}
  \end{center}
\caption{Same as in \Fig{fig:scale_NLO_LO_rho}, but for the $p_T^t$ distribution.}
\label{fig:scale_NLO_LO_pt_t}
\end{figure}

\begin{figure}[h!]
  \begin{center}
    \includegraphics[width=\textwidth]{\main/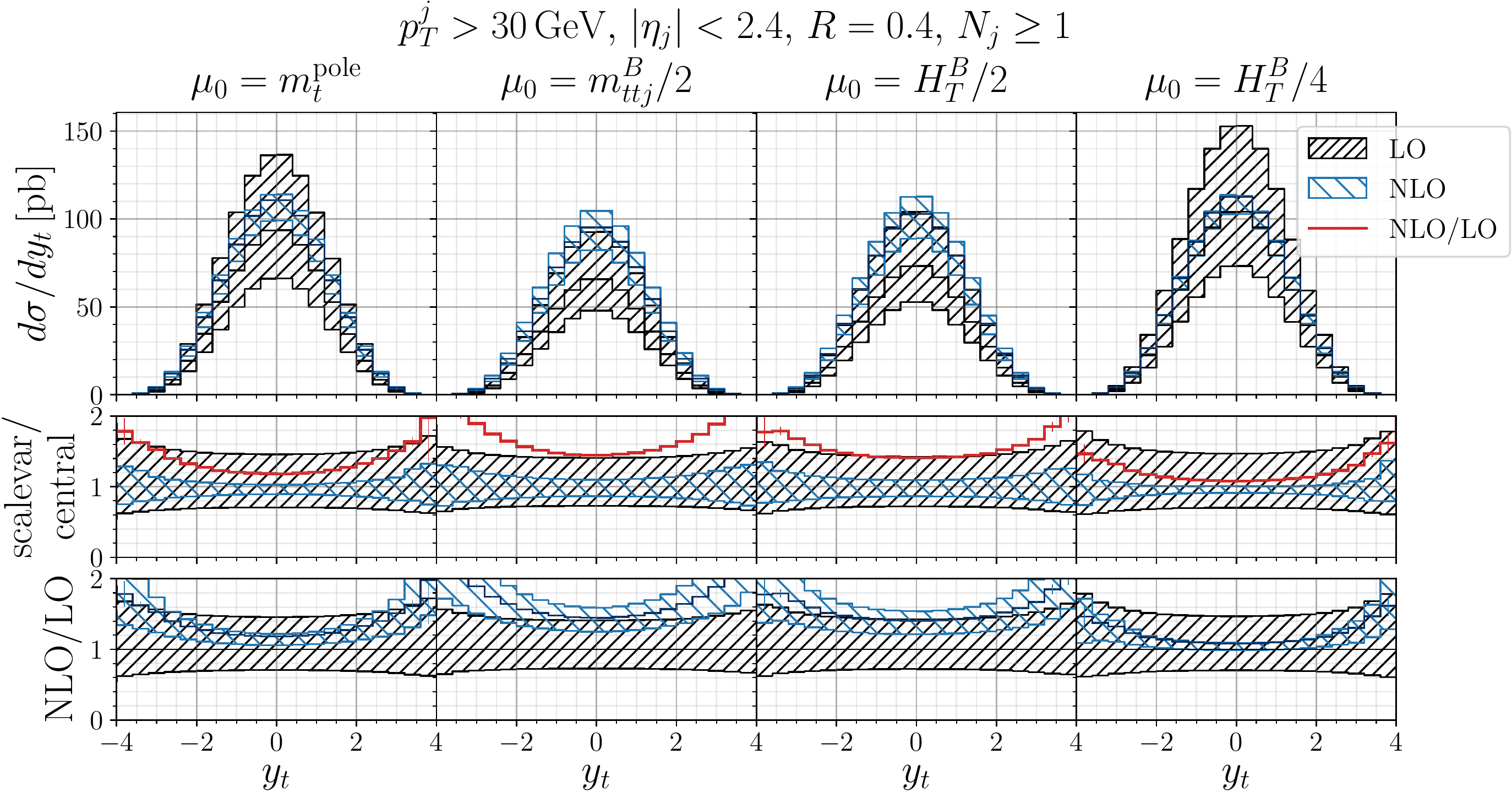}
  \end{center}
\caption{Same as in \Fig{fig:scale_NLO_LO_rho}, but for the $y_t$ distribution.}
\label{fig:scale_NLO_LO_yt}
\end{figure}

\begin{figure}[h!]
  \begin{center}
    \includegraphics[width=\textwidth]{\main/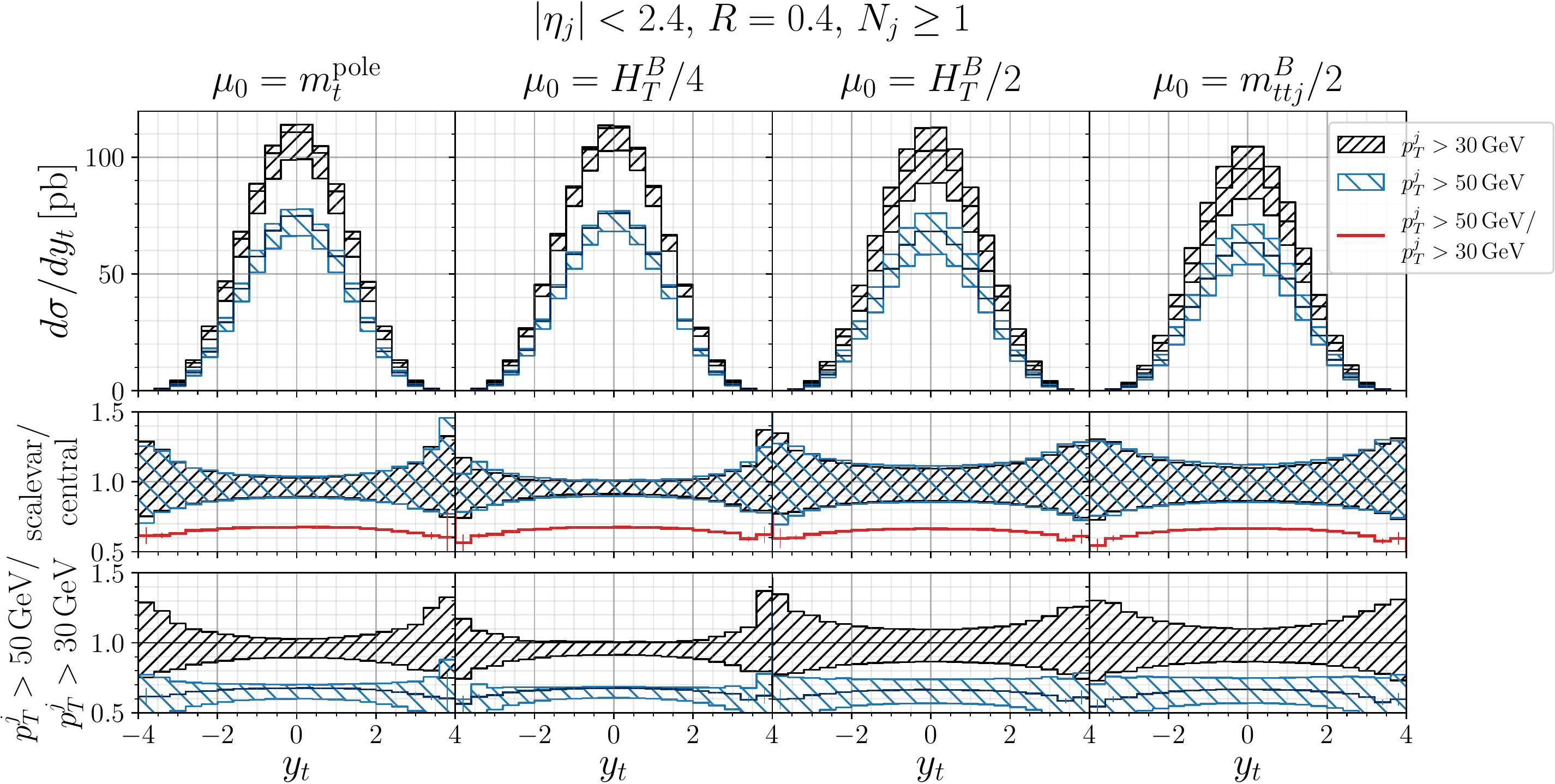}
  \end{center}
\caption{NLO central predictions for the $y_t$ distribution with a
  $p_T^j$-analysis cut of 30~GeV (black) and 50~GeV (blue), accompanied
  by the corresponding seven-point scale variation uncertainty bands, obtained
  using as a central scale $\mu_0=m_t$, $H_T^B/4$, $H_T^B/2$ and $m_{t\bar{t}j}^B/2$
   (from left to right). In the middle ratio plot the scale variation
  uncertainty band with either $p_T^j$-cut is rescaled by the central scale
  prediction with the same analysis cut. To visualize the difference between
  the dif\-fe\-ren\-tial cross section obtained with the different $p_T^j$-cuts the
  ratio of the central scale predictions is shown in red. In the bottom ratio
  plot the scale variation uncertainty bands using either analysis cut are
  rescaled by the central scale result using the requirement $p_T^j>30\,$GeV.} 
\label{fig:scale_ptj_yt}
\end{figure}

\begin{figure}[h!]
  \begin{center}
    \includegraphics[width=\textwidth]{\main/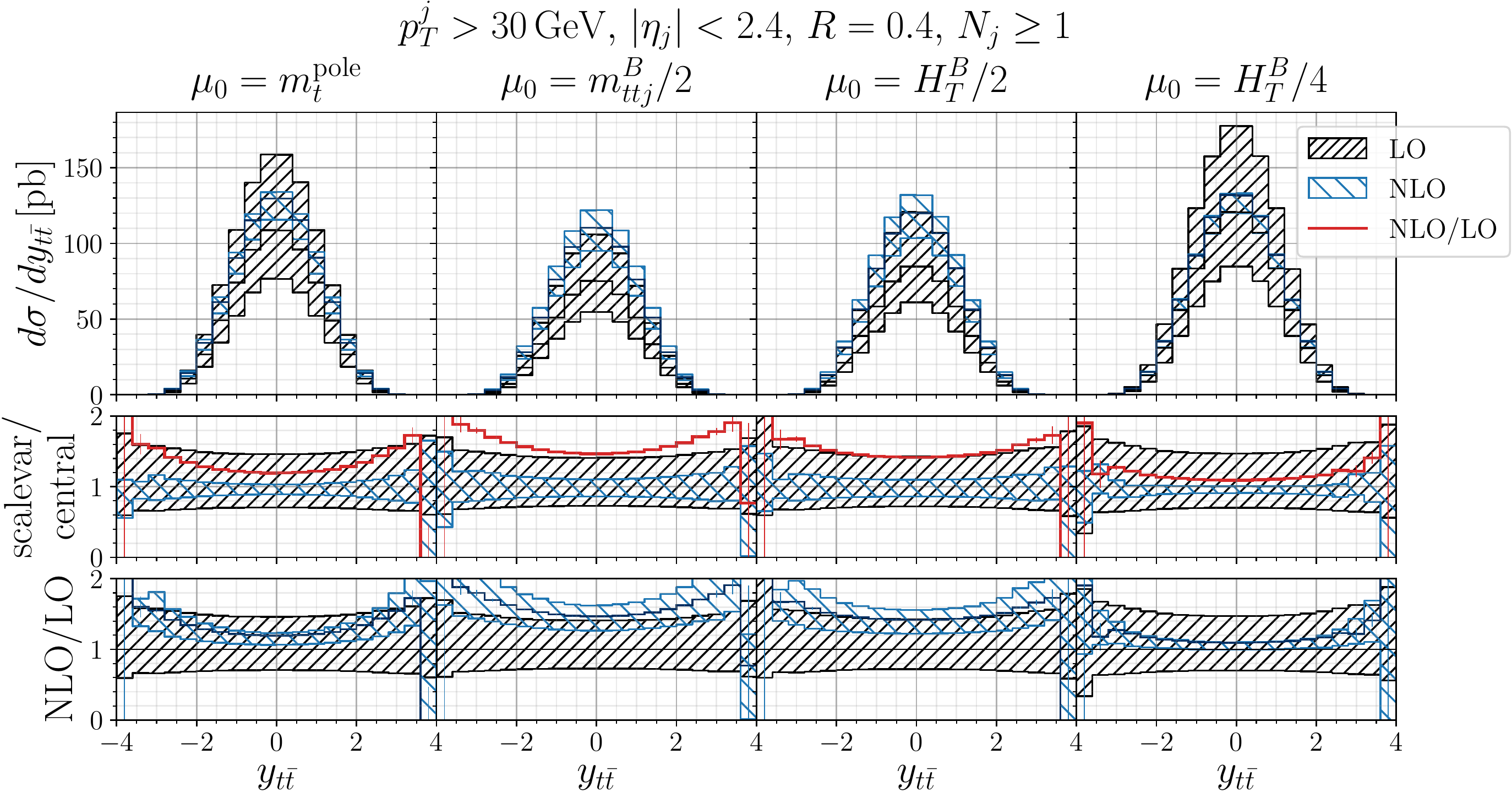}
  \end{center}
  \caption{
\label{fig:scale_NLO_LO_ytt}
    Same as in \Fig{fig:scale_NLO_LO_rho}, but for the $y_{t\bar{t}}$ distribution.}
\end{figure}

\begin{figure}[h!]
  \begin{center}
    \includegraphics[width=\textwidth]{\main/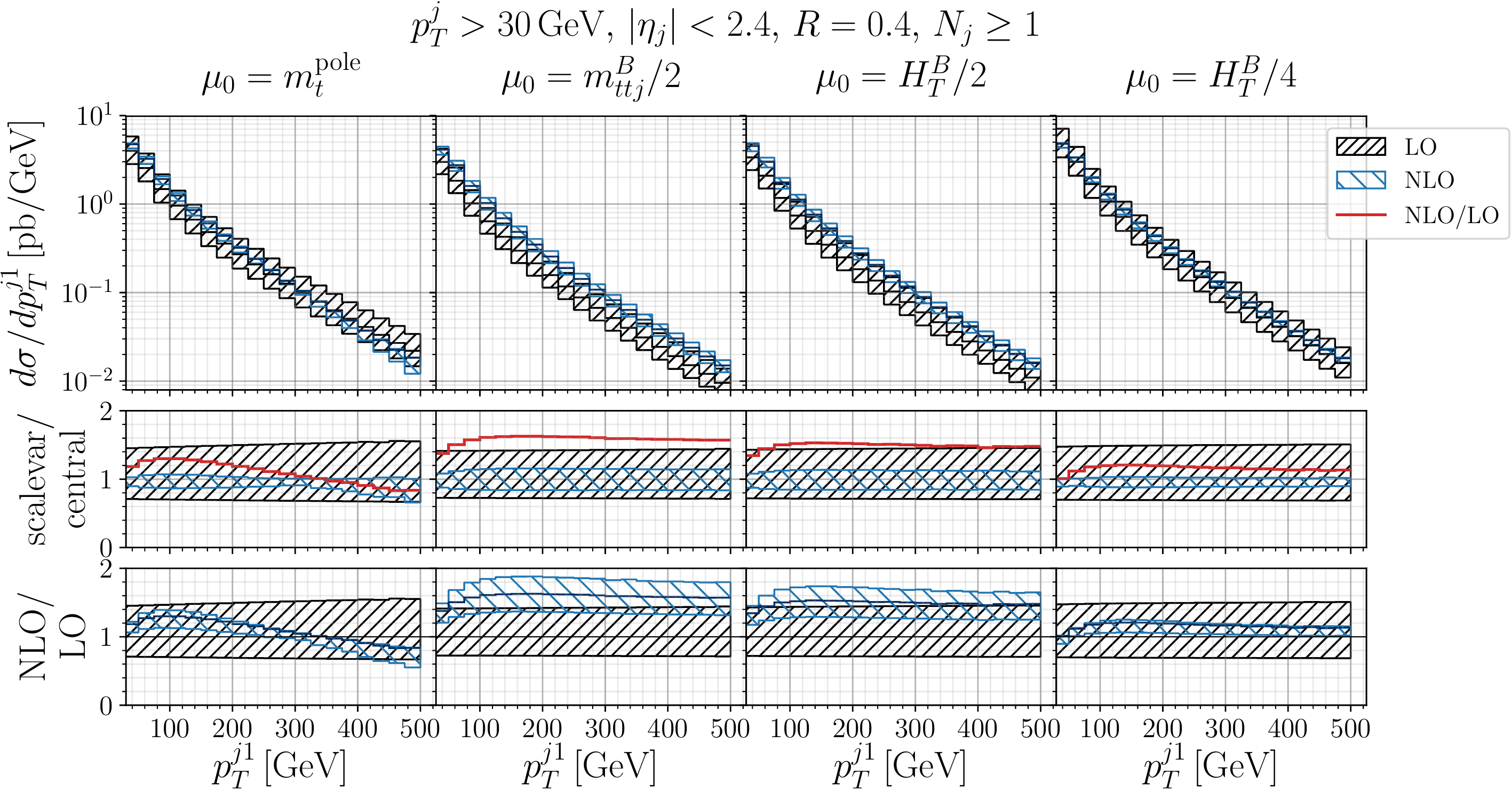}
  \end{center}
  \caption{
\label{fig:scale_NLO_LO_ptj1}
Same as in \Fig{fig:scale_NLO_LO_rho}, but for the $p_T^{j_1}$ distribution.
  }
\end{figure}

\begin{figure}[h!]
  \begin{center}
    \includegraphics[width=\textwidth]{\main/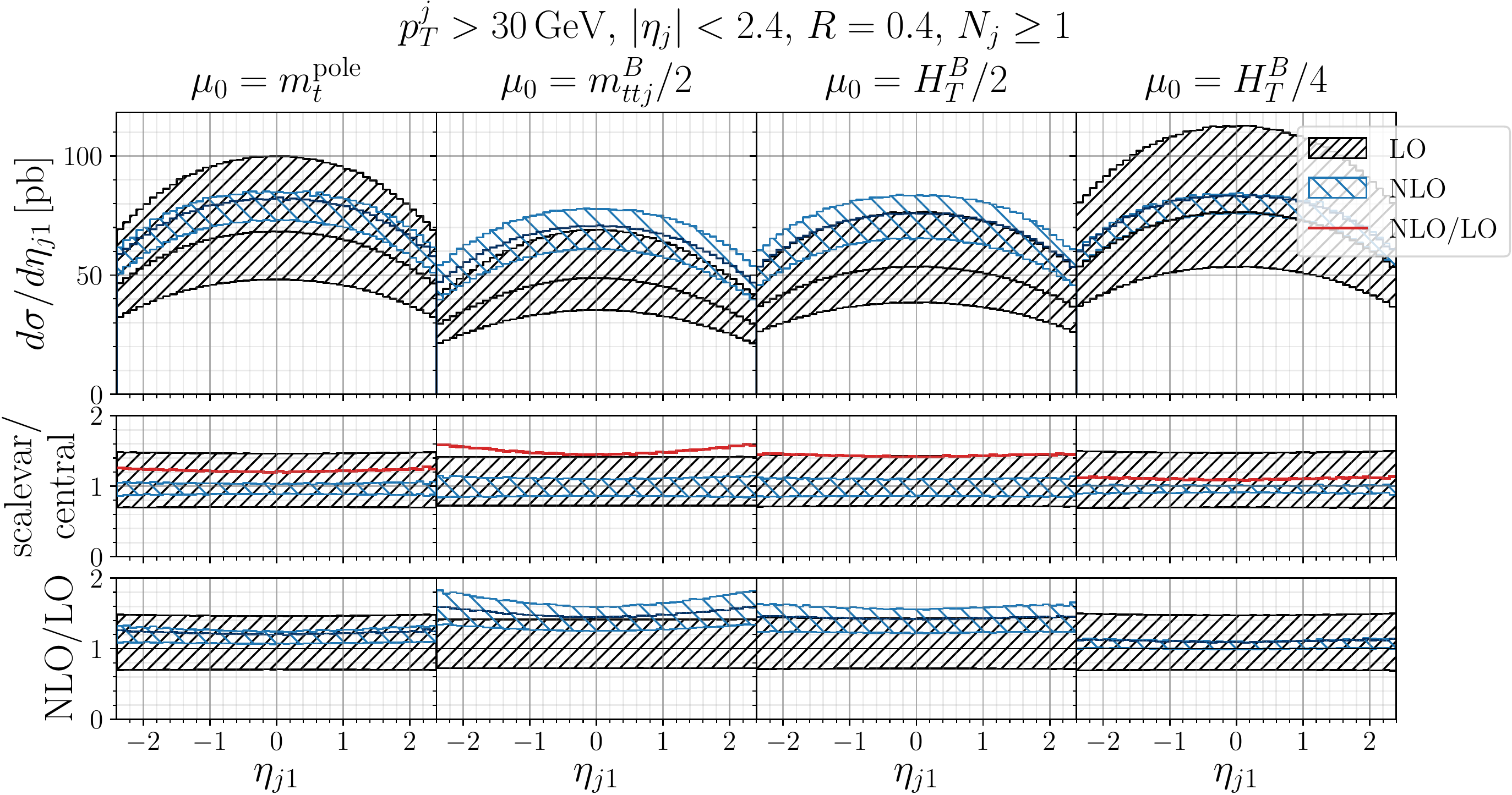}
  \end{center}
  \caption{
\label{fig:scale_NLO_LO_etaj1}
Same as in \Fig{fig:scale_NLO_LO_rho}, but for the $\eta_{j_1}$ distribution.
  }
\end{figure}

\begin{figure}[h!]
  \begin{center}
    \includegraphics[width=\textwidth]{\main/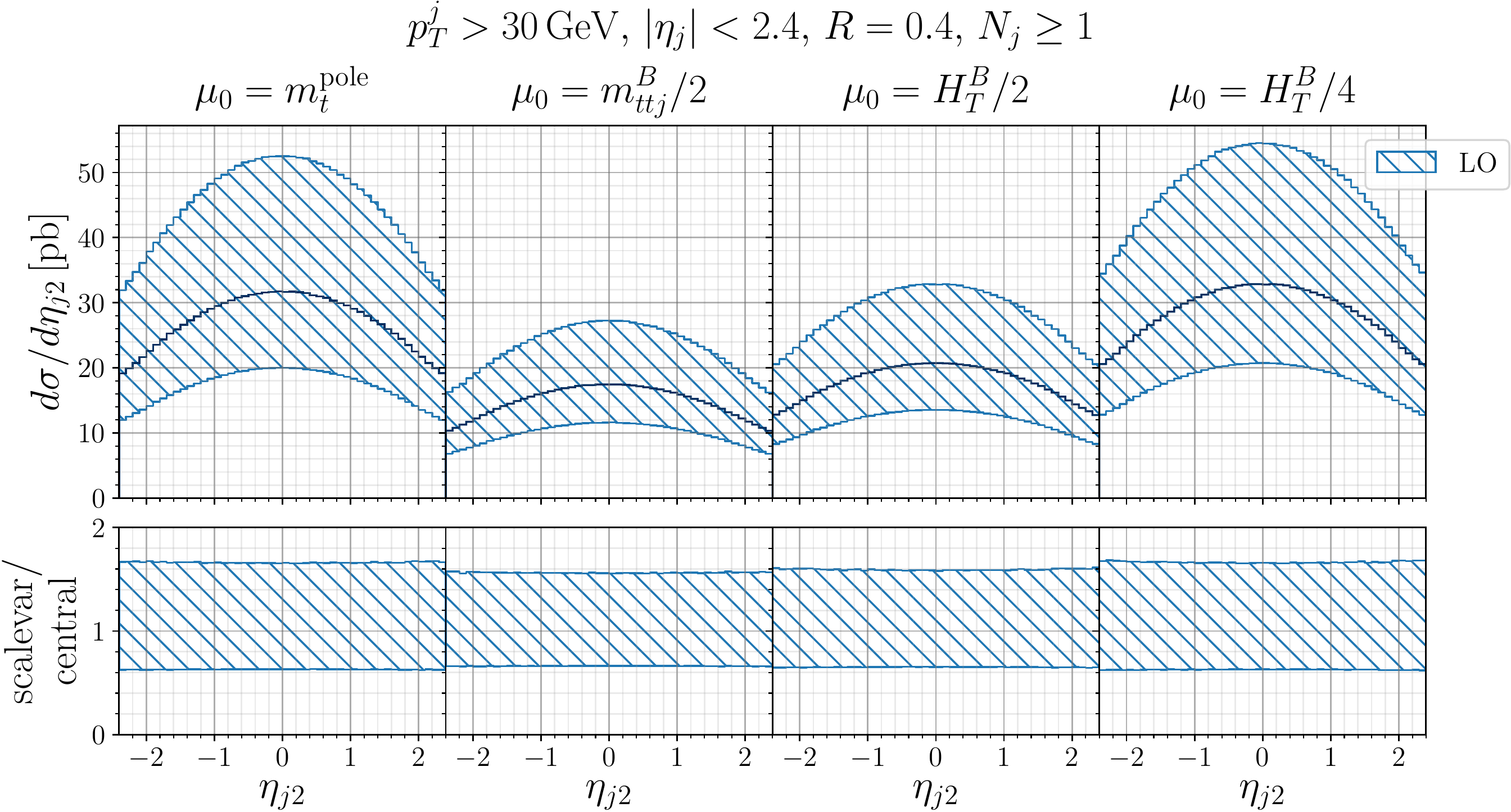}
  \end{center}
  \caption{
    \label{fig:scale_NLO_LO_etaj2}
    Same as in \Fig{fig:scale_NLO_LO_rho}, but for the $\eta_{j_2}$ distribution. The content of the panels
    reflect the fact that the second jet appears for the first time in NLO calculations and, therefore, the accuracy of the
    predictions for it is limited to LO. 
  }
\end{figure}

Another distribution, in which the more stable behavior   in the high-energy
tail with the dynamical scale choices 
is visible, is the invariant mass of the
top-quark pair $m_{t\bar{t}}=\sqrt{(p_t + p_{\bar{t}})^2}$, shown in
\Fig{fig:scale_NLO_LO_mttbar}. In case of the static scale assignment
(left panel), for large values of the invariant mass of the top-quark pair,
the scale variation band at NLO departs from the LO one. This signals again
an instability in the high energy tail of the distribution. 
The width of the NLO scale uncertainty band is smaller with respect to  the LO one
at small values of $m_{t\bar{t}}$, whereas in the high energy tail the NLO
scale variation band shows a strong increase, being of similar size as the LO
band for $m_{t\bar{t}}$ = 1.5 TeV. Also in this distribution a distortion of
the central NLO prediction with respect to the corresponding LO one is visible, the NLO
values going from $+40\%$ of the LO values in the region of small $m_{t\bar{t}}$ (large values
of $\rho_s$) to $-50\%$ in the region of high $m_{t\bar{t}}$. 
The predictions obtained with the dynamical scale show a much smaller
distortion of the central values when going from LO to NLO, while the width of
the scale variation bands are quite stable over the whole range of explored
$m_{t\bar{t}}$ values.

Similar features as for the behavior of the NLO uncertainty bands can also be
observed in e.g. the transverse momentum distribution of the top-quark
$p_T^t$, shown in \Fig{fig:scale_NLO_LO_pt_t}, although in that case a
distortion of NLO predictions with respect to LO ones is present even when
using the dynamical scales. 

In \Fig{fig:scale_NLO_LO_yt} the differential cross section is shown as a
function of the rapidity of the top-quark $y_t$. The scale variation bands in
the static and dynamical scale cases have a more similar width. While using
$\mu_0=H_T^B/4$ leads to the smallest size of the scale uncertainty band.
For the $y_t$ distribution the dynamical scale choice does not
lead to an improvement in the stability of the prediction. In fact, as also
observed in Ref.~\cite{Bevilacqua:2016jfk}, the observable $y_t$ receives
contributions from all phase-space regions, with most of them
from the bulk, where the differences between dynamical and static scales are
reduced. The shape of the $y_t$ distribution is almost independent on the
$p_T^j$ cut applied in the reconstruction of the hardest jet. 
This is shown in \Fig{fig:scale_ptj_yt}, where this observable is plotted
for two different $p_T^j$ cuts, namely $p_T^j>30$~GeV and
$p_T^j>50$~GeV. The integrated cross section becomes lower when applying a
stronger $p_T^j$-cut, but the ratio of the distributions
$\left(d\sigma/dy_t(p_T^j>50\,\text{GeV})\right) / \left(d\sigma/dy_t(p_T^j>30\,\text{GeV})\right)$
is almost uniform, showing that
no region in the $y_t$~distribution has a much stronger dependence on this
analysis cut compared to the rest of the distribution.

As far as the differential contributions are concerned, the dynamical scale
$\mu_0=H_T^B/4$ gives in most cases a flat $\cal K$-factor close to 1,
suggesting that potentially large logarithms are
effectively 
absorbed when adopting
this particular
choice. 

While the two other dynamical scales also show a flat $\cal K$-factor for most distributions, the $\cal K$-factor value itself is much larger.
Uncalculated higher
orders could thus be important. This interpretation is supported by the larger
scale uncertainty observed for these two dynamical scales. For most distributions using the static
scale leads to a $\cal K$-factor which depends
significantly on the phase space. This is in particular true for the
distributions where the individual bins set different energy scales. This is
not surprising as a static treatment cannot absorb corresponding
logarithms. However, as long as the tails of the distributions are avoided, the
static scale choice leads to a smaller $\cal K$-factor then
the two dynamical
scales $m_{t\bar{t}j}^B/2$ and $H_T^B/2$. We can 
thus conclude that the static
scale is  still a reasonable choice for many distributions, at least
for the time being. 
An improved understanding of the perturbative series for the 
$t\bar{t}j$ process will be possible  when higher-order
calculations beyond NLO will start to appear.

Since the integrated cross section is dominated by the threshold region, in
which the static and dynamical scales $\mu_0=H_T^B/2$ and
$\mu_0=m_{t\bar{t}j}^B/2$ give similar predictions, no large diffe\-ren\-ces
in the size of the scale variation bands of the $y_t$ predictions when using
either one or the other scale are expected, as observed in
\Fig{fig:scale_NLO_LO_yt}. It can also be seen in
\Fig{fig:scale_NLO_LO_yt} that the LO result varies much stronger going
from the static to the dynamical scales, than the NLO result. 
In the region of large absolute rapidity of the top quark the NLO scale
variation band starts to depart from the LO one. This is observed using all
four scale choices. The $t$-channel diagrams contributing in this region lead
to final state particles closely following the direction of the initial state
particles, such that these final state particles have a large rapidity.  
Removing the $\eta_j$-cut
would result in a $t$-channel singularity explaining
the aforementioned behavior observed at large $|y_t|$. In case of $t\bar{t}$
production, and more generally heavy-quark production, 
this was already inve\-sti\-ga\-ted
in Ref.~\cite{Catani:1990eg} and~\cite{Ball:2001pq}.
We also observe that similar trends as for the $y_t$
distribution occur even for the $y_{t\bar{t}}$ distribution, plotted for
various scale choices in \Fig{fig:scale_NLO_LO_ytt}. 

Additionally, we looked at the differential distributions of the hardest and
second hardest jet. The transverse momentum of the hardest jet is shown in
\Fig{fig:scale_NLO_LO_ptj1}, where the high-energy tail seems again to be
better described using a dynamical scale definition, than in case of a static
scale. 
Looking at the pseudorapidity of the hardest jet $\eta_{j_1}$ and second hardest
jet $\eta_{j_2}$, a relatively flat differential $\mathcal{K}$-factor is found
using either central scale, comparing \Figs{fig:scale_NLO_LO_etaj1} 
and~\ref{fig:scale_NLO_LO_etaj2}, respectively. Predictions for the second
hardest jet are only available in the NLO $t\bar{t}j$ calculation, since at LO
only one light parton is present in the final state. As expected from accuracy
considerations, these predictions
are characterized by larger scale uncertainty bands than
those           for the first hardest jet. 
The size of the scale uncertainty bands for
the $\eta_{j_2}$ distribution in
\Fig{fig:scale_NLO_LO_etaj2} is even larger than the size of
the LO scale uncertainty band for $\eta_{j1}$
in \Fig{fig:scale_NLO_LO_etaj1}. As visible in each panel of both
figures, for all central scale choices, the size of scale uncertainties
accompanying both the $\eta_{j_1}$ and $\eta_{j_2}$ distributions does not
depend on the pseudorapidity value, i.e. it is uniform, differently from the
size of scale uncertainties accompanying the top-quark $y_t$ distribution,
which increase at large absolute values of rapidity, as discussed above.

 Some of the described differential cross sections were
 also discussed in Ref.~\cite{Bevilacqua:2016jfk},
 where similar trends were observed, 
 although using more exclusive
cuts (and different scales)  as mentioned at the
beginning of this Section. 


\subsection{Effects of PDF + $\alpha_s(M_z)$ variation}
\label{subsec:pdf}


\begin{figure}
  \begin{center}
    \includegraphics[width=0.49\textwidth]{\main/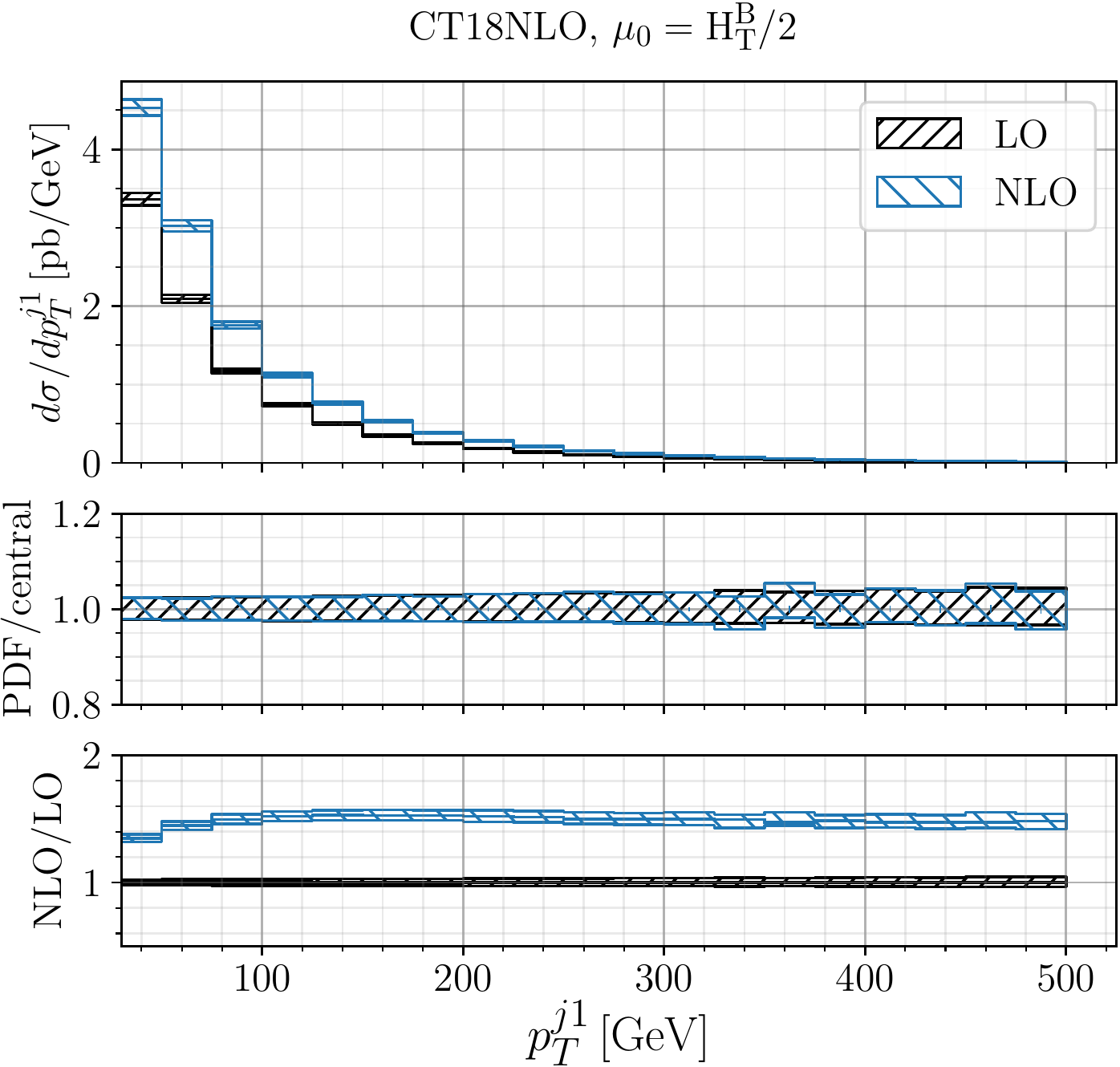}
    \includegraphics[width=0.49\textwidth]{\main/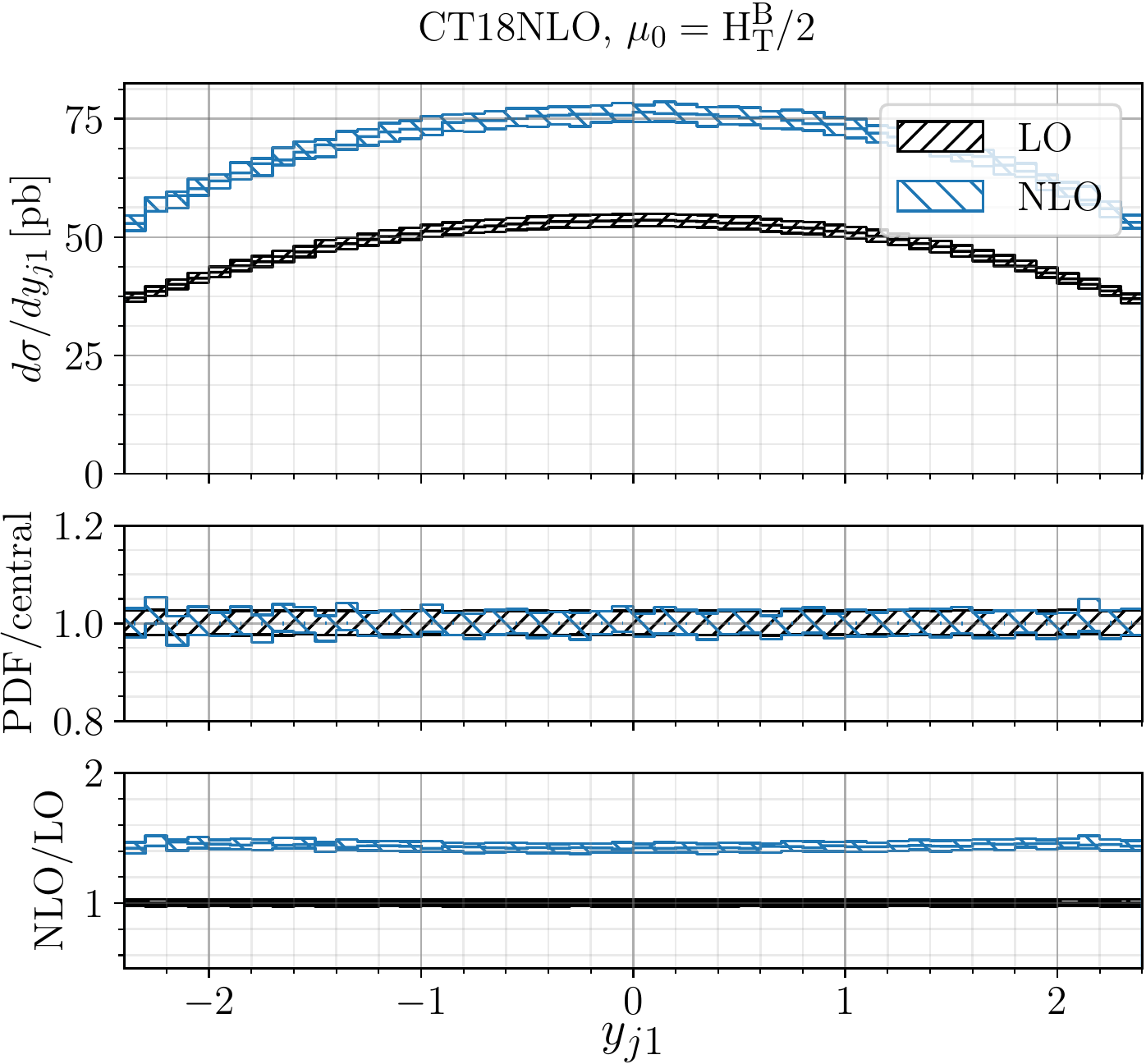}
  \end{center}
\caption{Central predictions for the $p_T^{j_1}$ (left) and $y_{j_1}$ (right)
  distributions in a full NLO computation (blue) and in a LO computation
  (black), using as input in both cases the CT18 NLO PDFs + $\alpha_s(M_Z)$
  value, two-loop evolution of $\alpha_s$ and the scale $\mu_0 = H_T^B/2$. 
  PDF uncertainty bands computed according to the CT18 prescription  
  and rescaled to the 68\% C.L. are also shown.
\label{fig:PDF_NLOLO_j1}
}
\end{figure}

\begin{figure}
  \begin{center}
    \includegraphics[width=0.49\textwidth]{\main/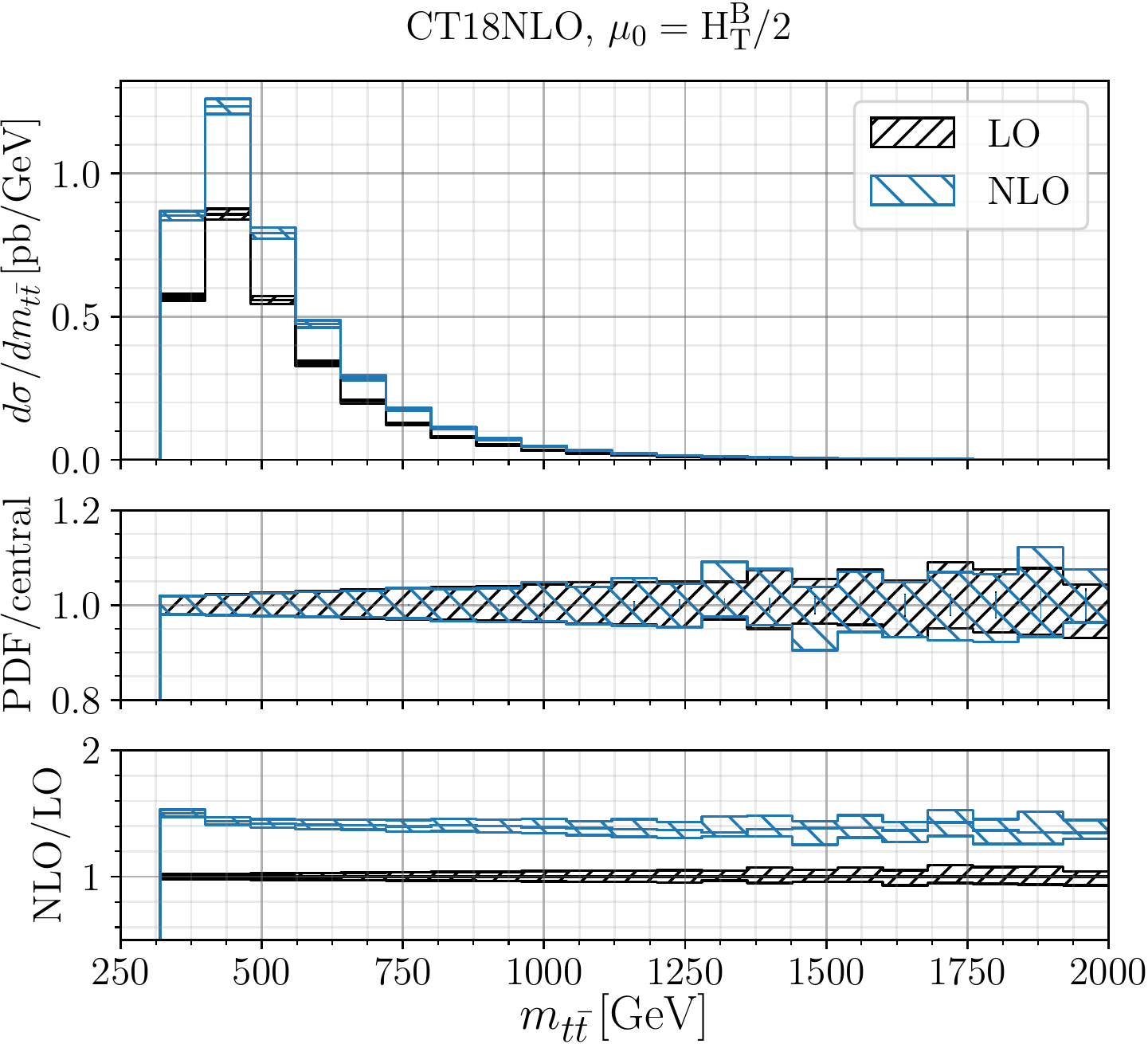}
    \includegraphics[width=0.49\textwidth]{\main/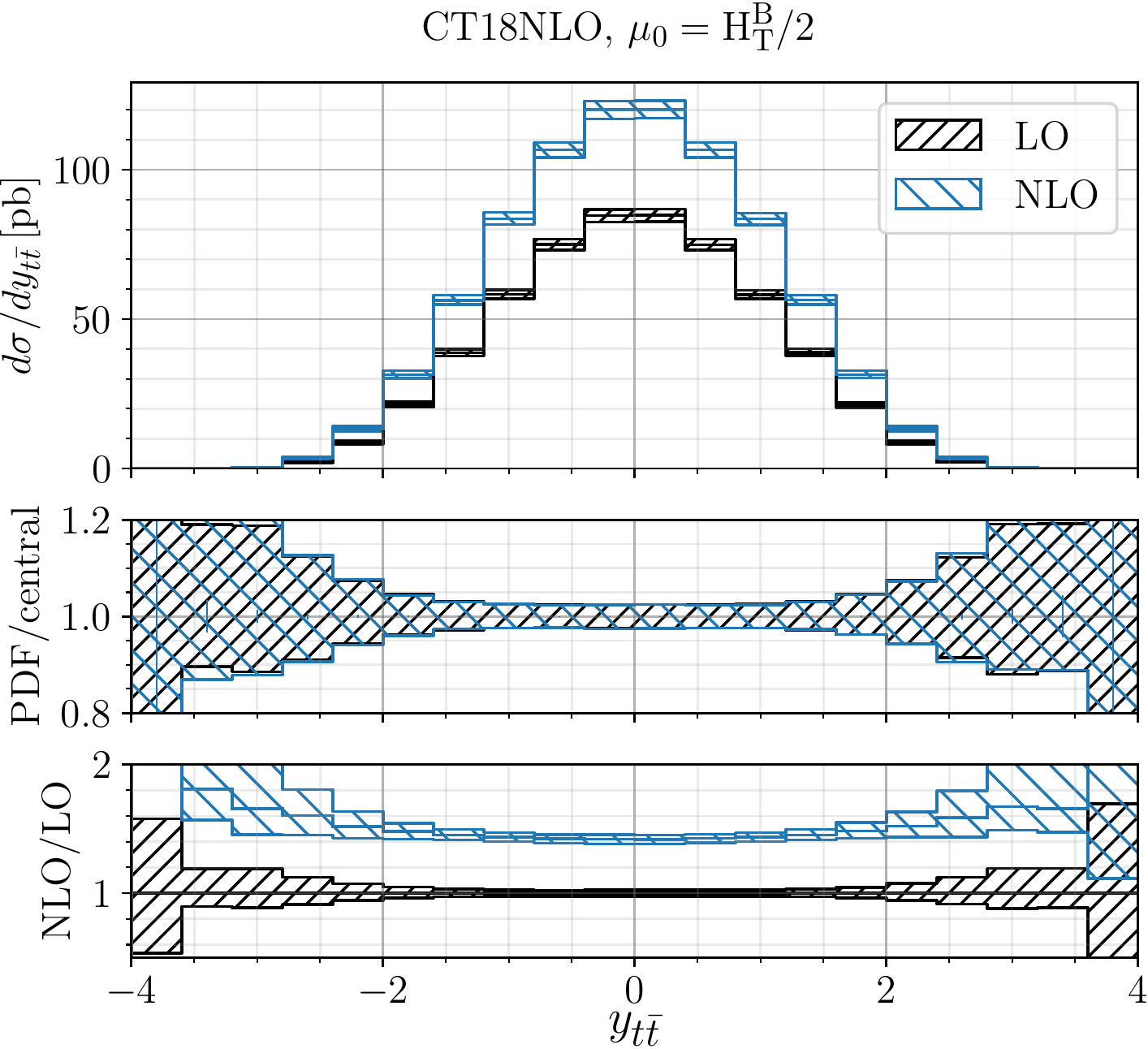}
  \end{center}
\caption{Same as \Fig{fig:PDF_NLOLO_j1}, but for the $m_{t\bar{t}}$ (left) and the $y_{t\bar{t}}$ (right) distributions.
\label{fig:PDF_NLOLO_ttbar}
}
\end{figure}

\begin{figure}
  \begin{center}
    \includegraphics[width=0.49\textwidth]{\main/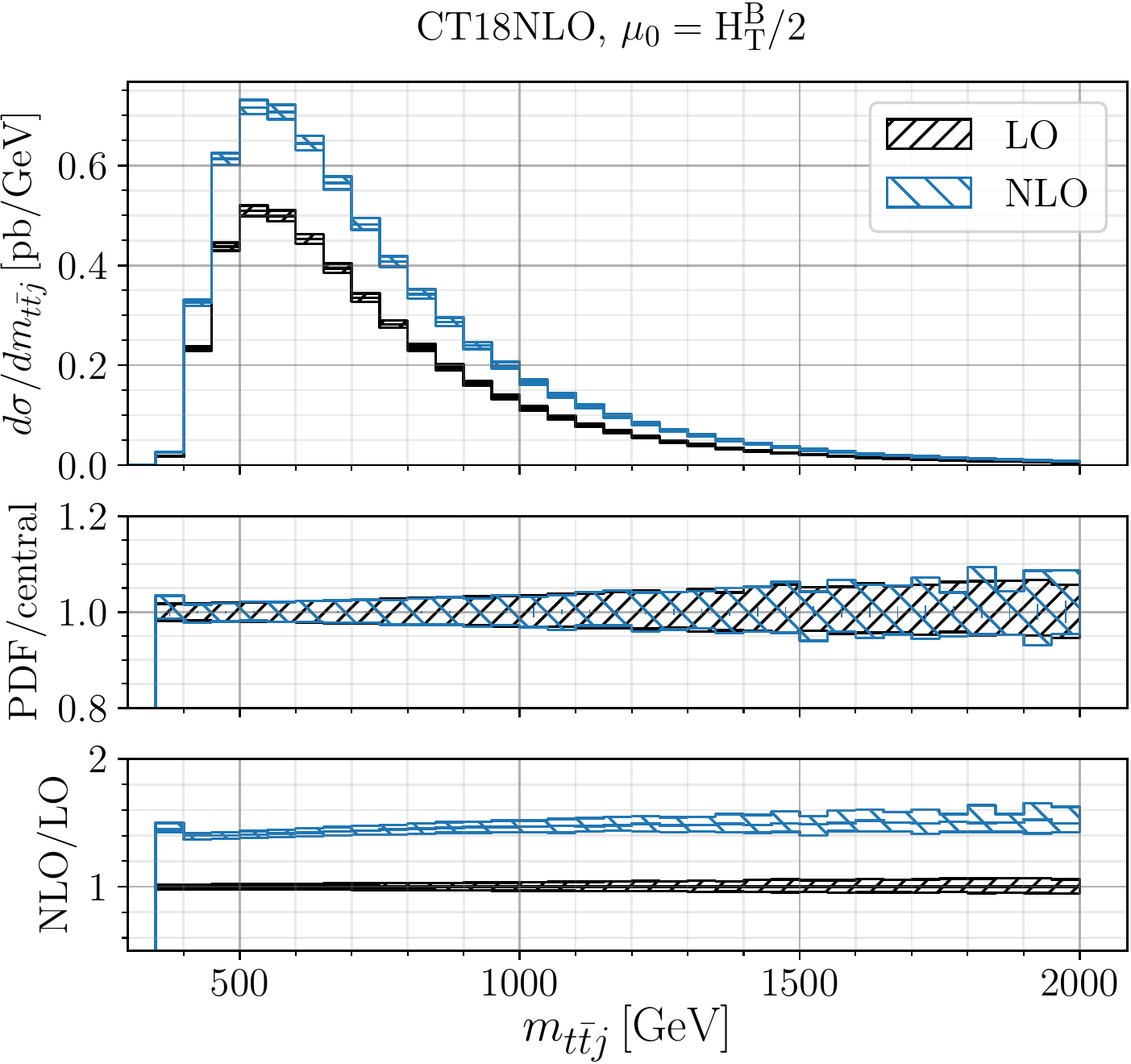}
    \includegraphics[width=0.49\textwidth]{\main/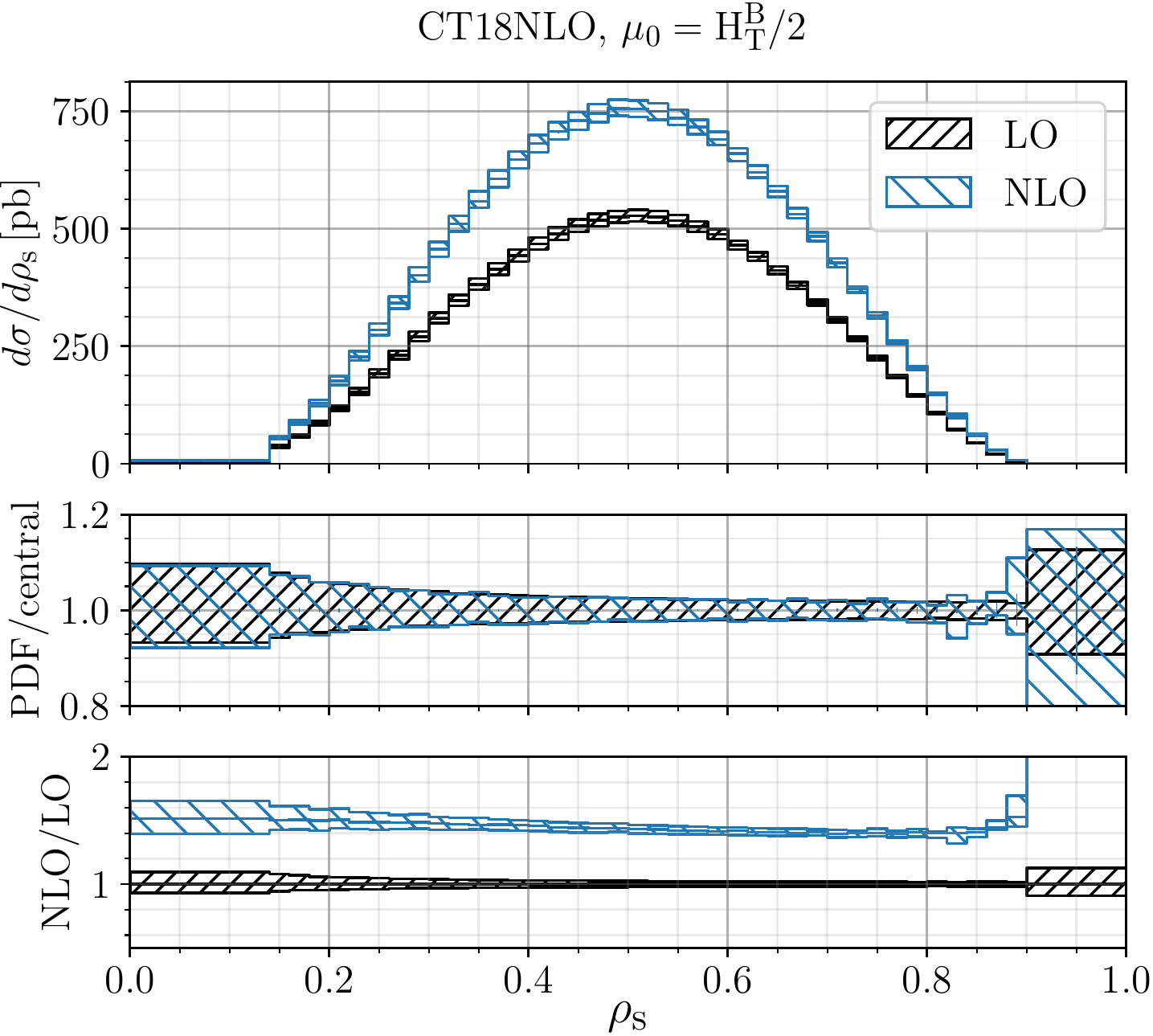}
  \end{center}
\caption{Same as \Fig{fig:PDF_NLOLO_j1}, but for the $m_{t\bar{t}j}$ (left) and the $\rho_s$ (right) distributions.
\label{fig:PDF_NLOLO_rho}
}
\end{figure}

\begin{figure}
  \begin{center}
    \includegraphics[width=0.49\textwidth]{\main/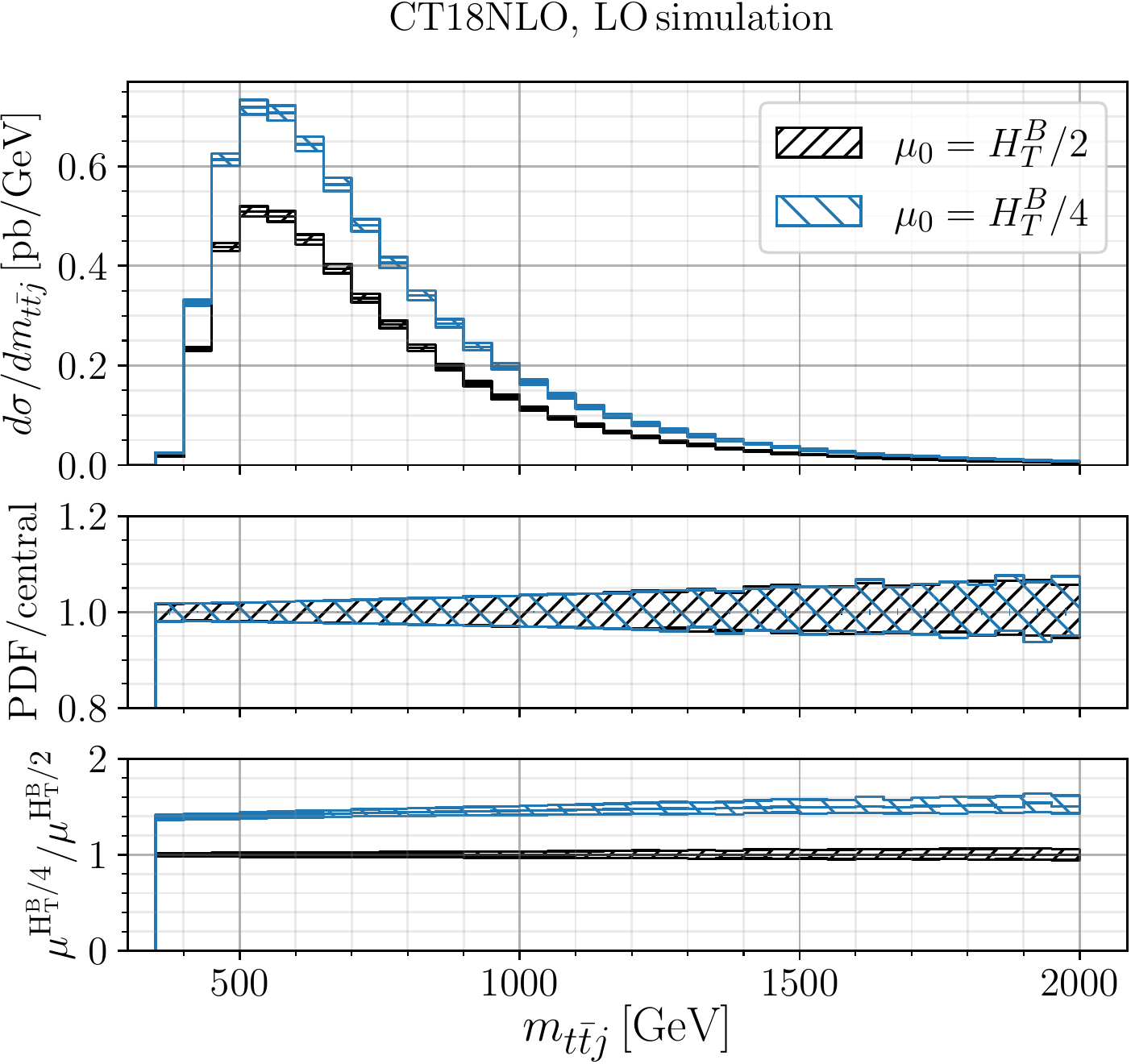}
    \includegraphics[width=0.49\textwidth]{\main/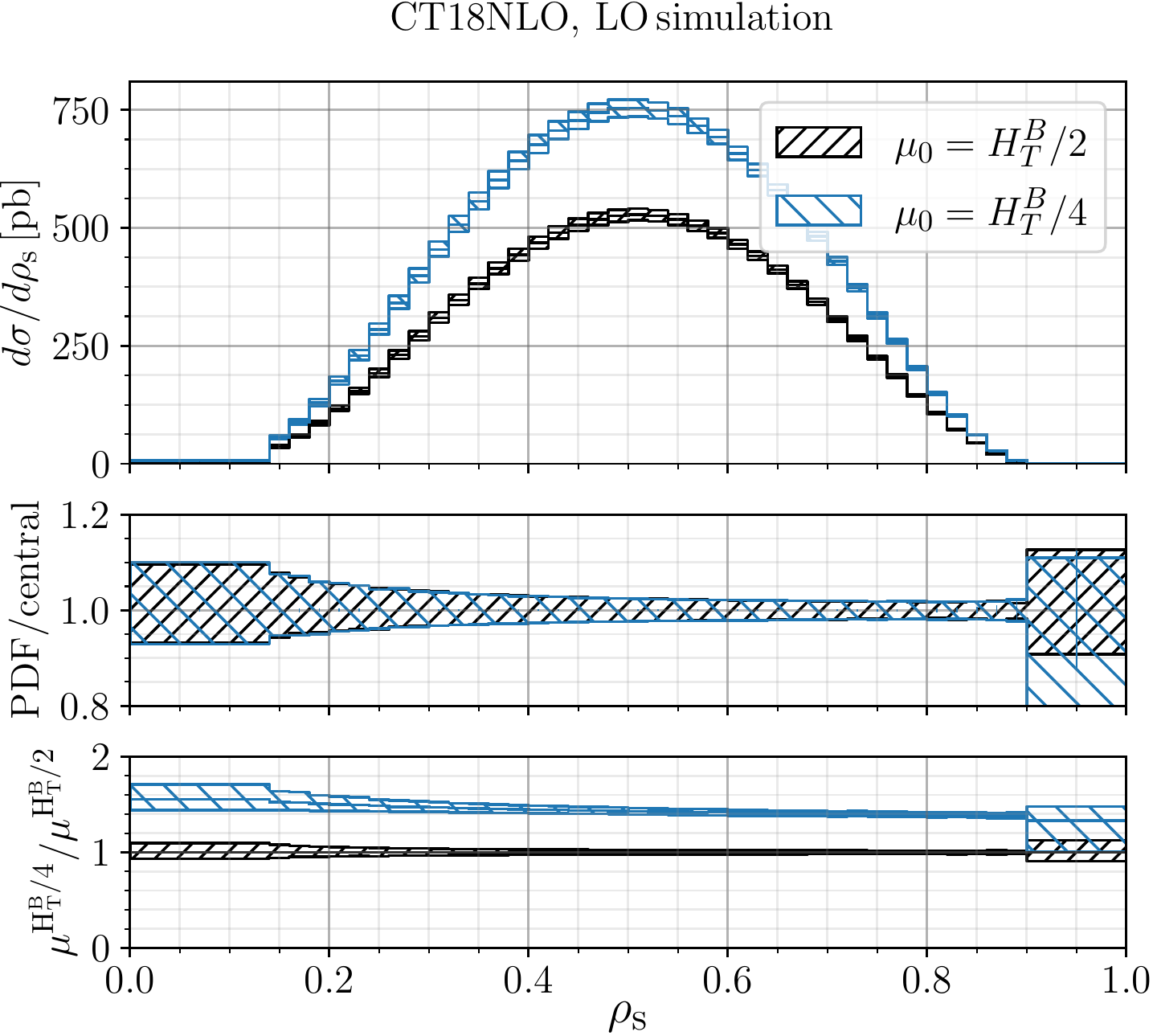}
  \end{center}
\caption{Central predictions for the $m_{t\bar{t}j}$ (left) and $\rho_s$ (right)
  distributions in a LO computation with the scales $\mu_0 = H_T^B/2$ (black)
  and $\mu_0 = H_T^B/4$ (blue), using as input in both cases the CT18 NLO PDFs + $\alpha_s(M_Z)$ 
  value and two-loop evolution of $\alpha_s$.
  PDF uncertainty bands computed according to the CT18 prescription and rescaled to 68\%
  C.L. are also shown.  
\label{fig:PDF_TWOSCALES_rho}
} 
\end{figure}

\begin{figure}
\includegraphics[width=0.49\textwidth]{\main/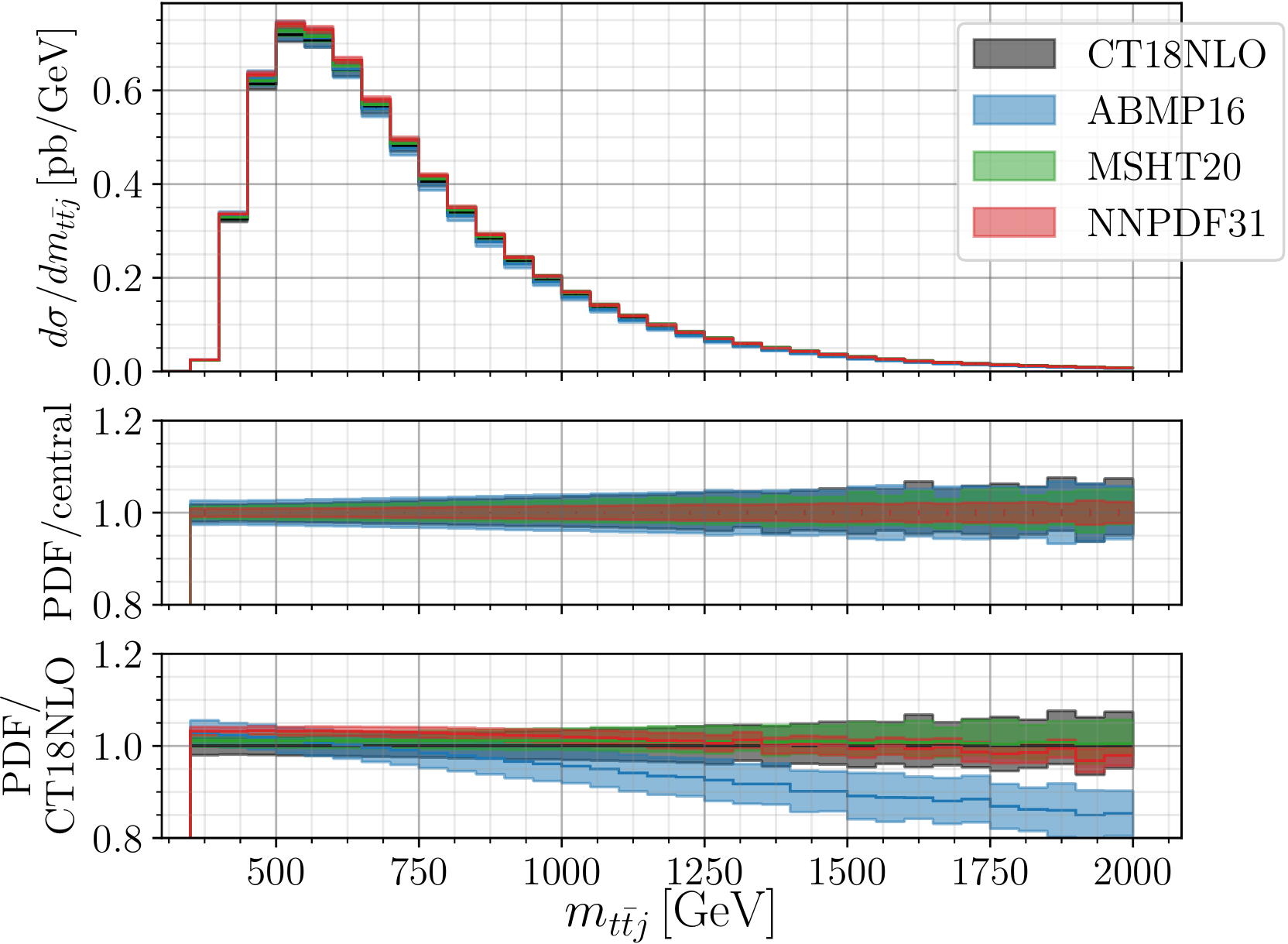}
\includegraphics[width=0.49\textwidth]{\main/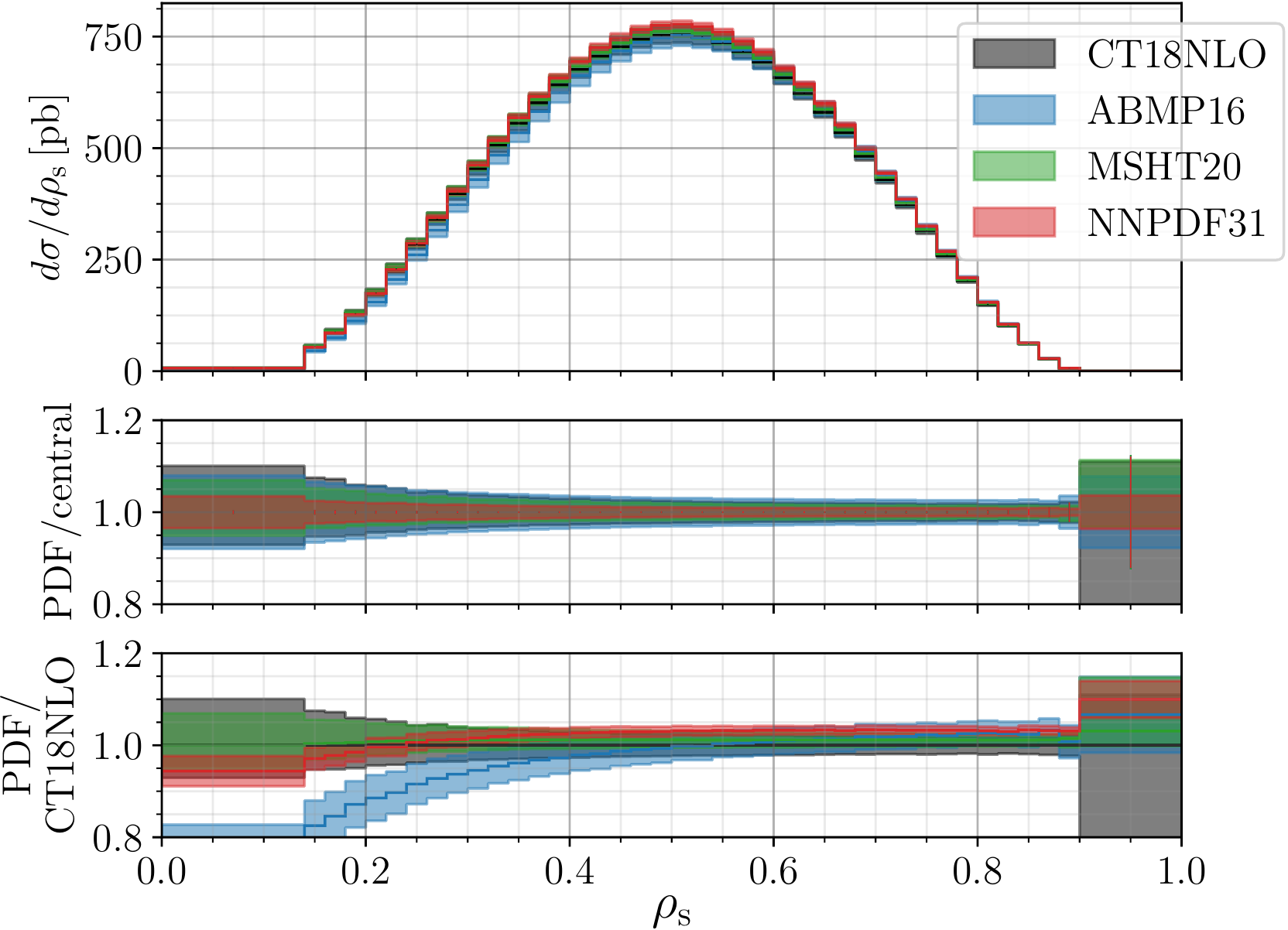}
\caption{Predictions for the $m_{t\bar{t}j}$ (left) and $\rho_s$ (right)
  distributions in a LO computation with scale $\mu_0 = H_T^B/4$, including
  PDF uncertainties of the ABMP16, CT18, MSHT20 and NNPDF3.1 NLO PDF sets. PDF
  uncertainty bands have been computed according to the prescriptions of the
  various PDF collaborations, which differ among each other. 
 The CT18 NLO uncertainty band is rescaled to 68\% C.L. 
 The $\alpha_s(M_Z)$ value is fixed according to the prescription associated
 to each PDF set and two-loop $\alpha_s$ evolution is considered. For  $\rho_s
 < 0.85$, corresponding to  $m_{t\bar{t}j} > 400\,$GeV, the percentage size of
 the PDF uncertainty band in a full NLO computation would be approximately the
 same as the one shown in these plots (see also \Fig{fig:PDF_NLOLO_rho}). 
\label{fig:PDF_NLOpdfsets_LOME}
}
\end{figure}

\begin{figure}
\includegraphics[width=\textwidth]{\main/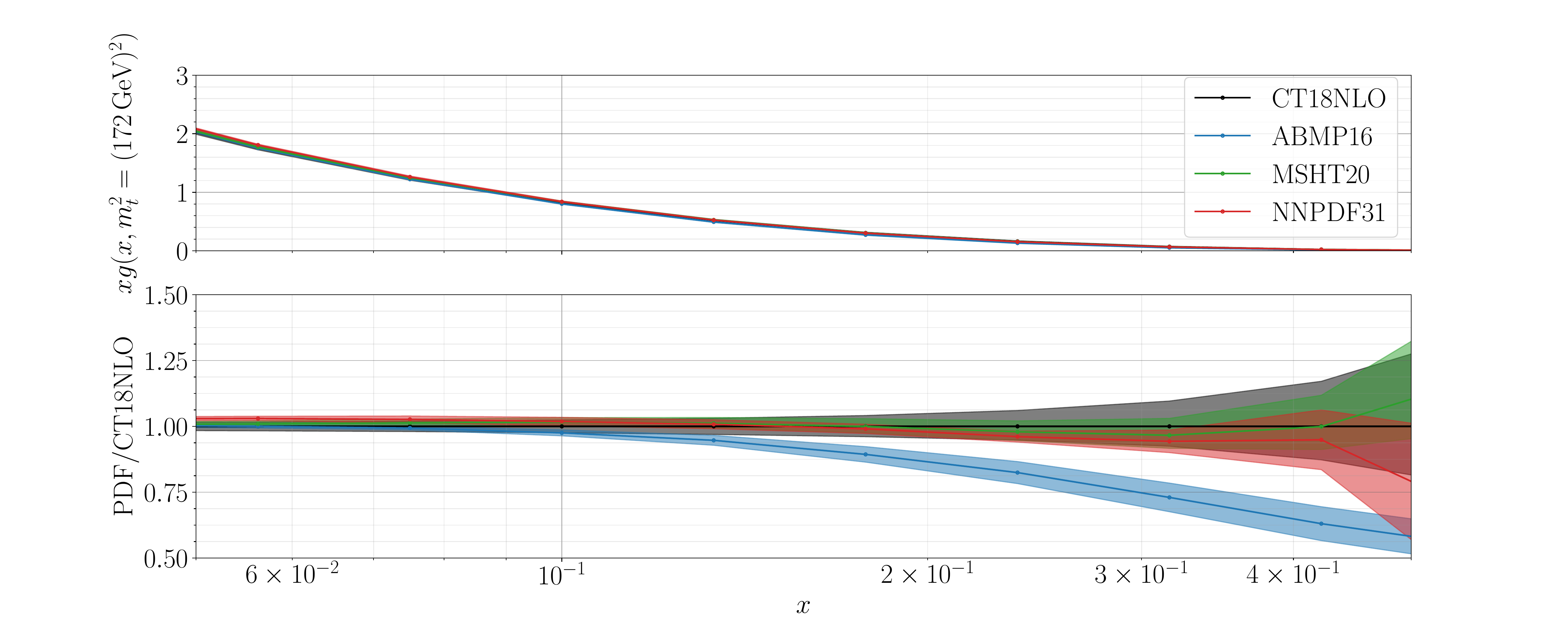}
\caption{Gluon PDFs obtained with the NLO PDF sets ABMP16, CT18NLO, MSHT20 and NNPDF3.1 as a function of the
momentum fraction $x$ for $Q^2 = m_t^2 = (172$\,{GeV})$^2$.
\label{fig:PDF_gluonpdfs}
}
\end{figure}

\begin{figure}
\includegraphics[width=0.49\textwidth]{\main/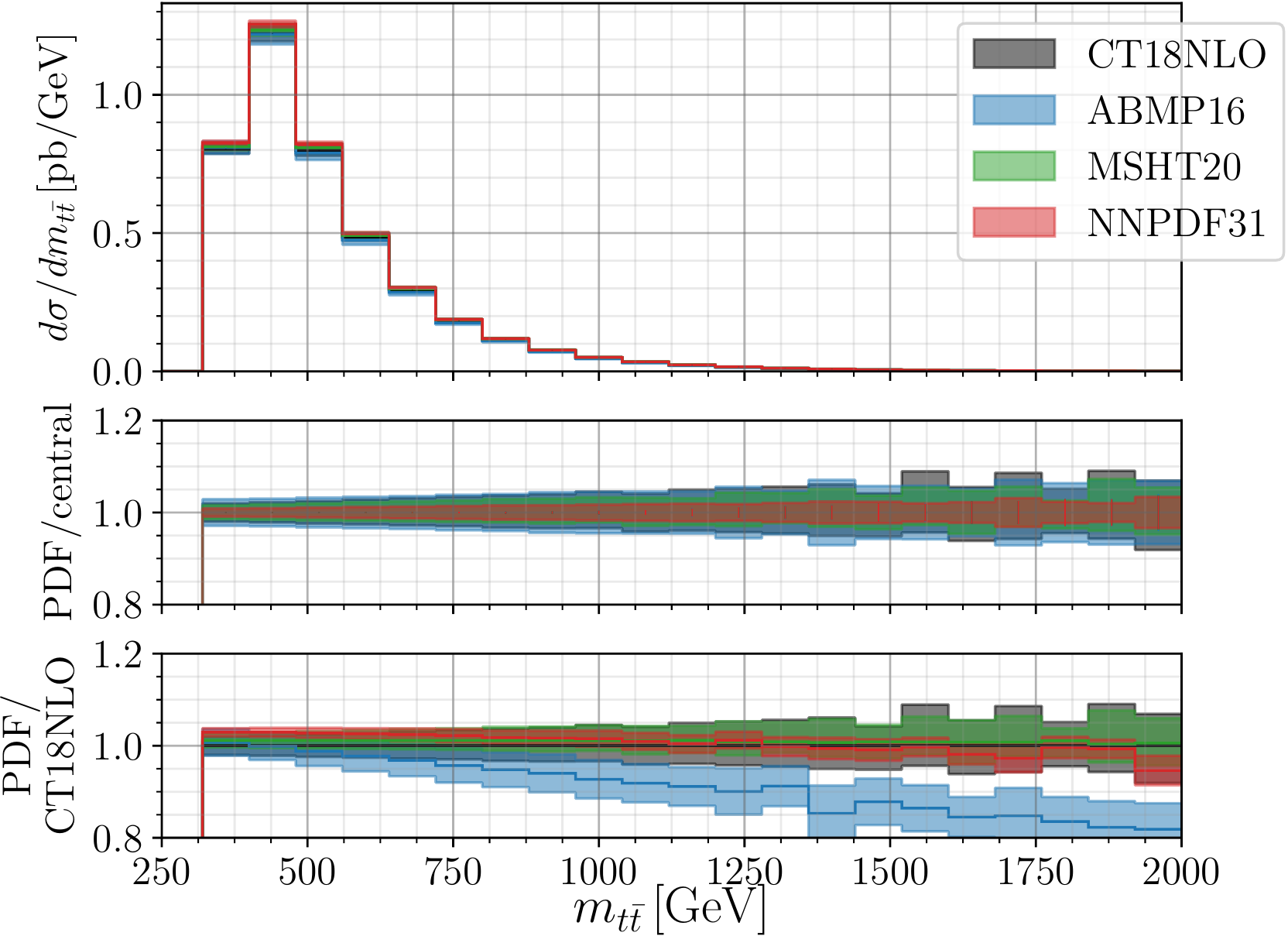}
\includegraphics[width=0.49\textwidth]{\main/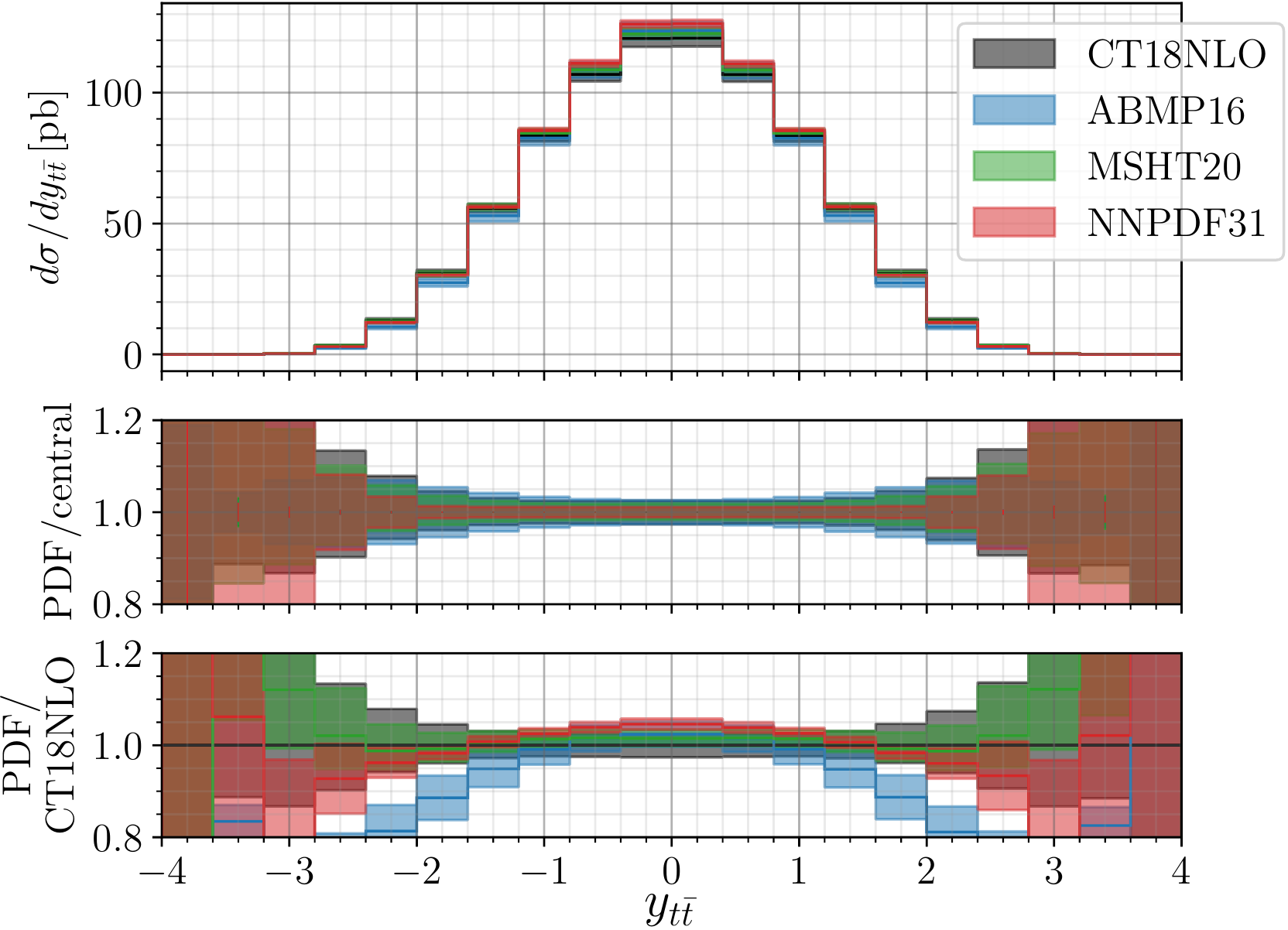}
\caption{Same as \Fig{fig:PDF_NLOpdfsets_LOME}, but for the $m_{t\bar{t}}$ (left) and the $y_{t\bar{t}}$ (right) distributions.
\label{fig:PDF_NLOpdfsets_ttbar_LOME}
}
\end{figure}

\begin{figure}
\includegraphics[width=0.49\textwidth]{\main/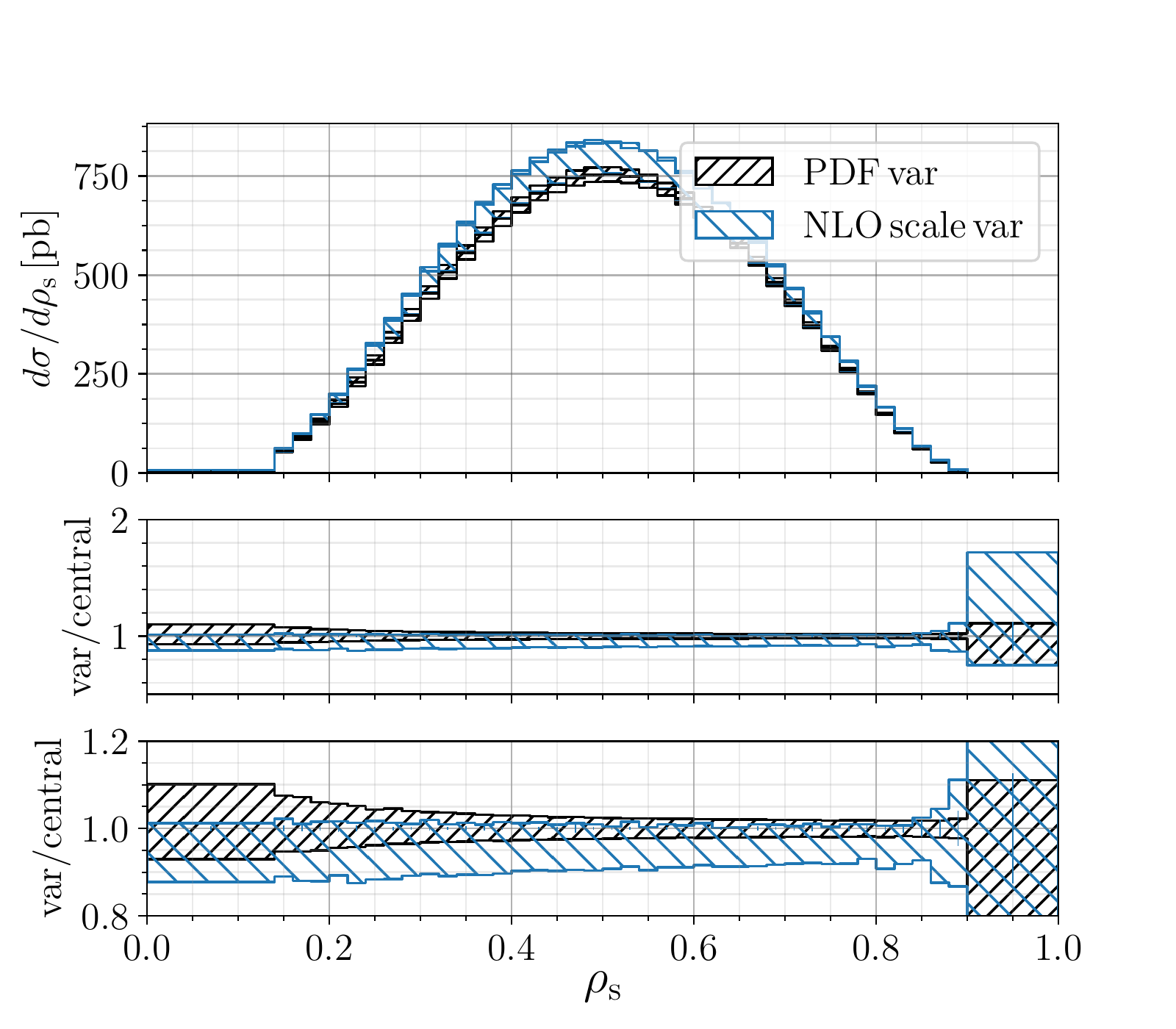}
\includegraphics[width=0.49\textwidth]{\main/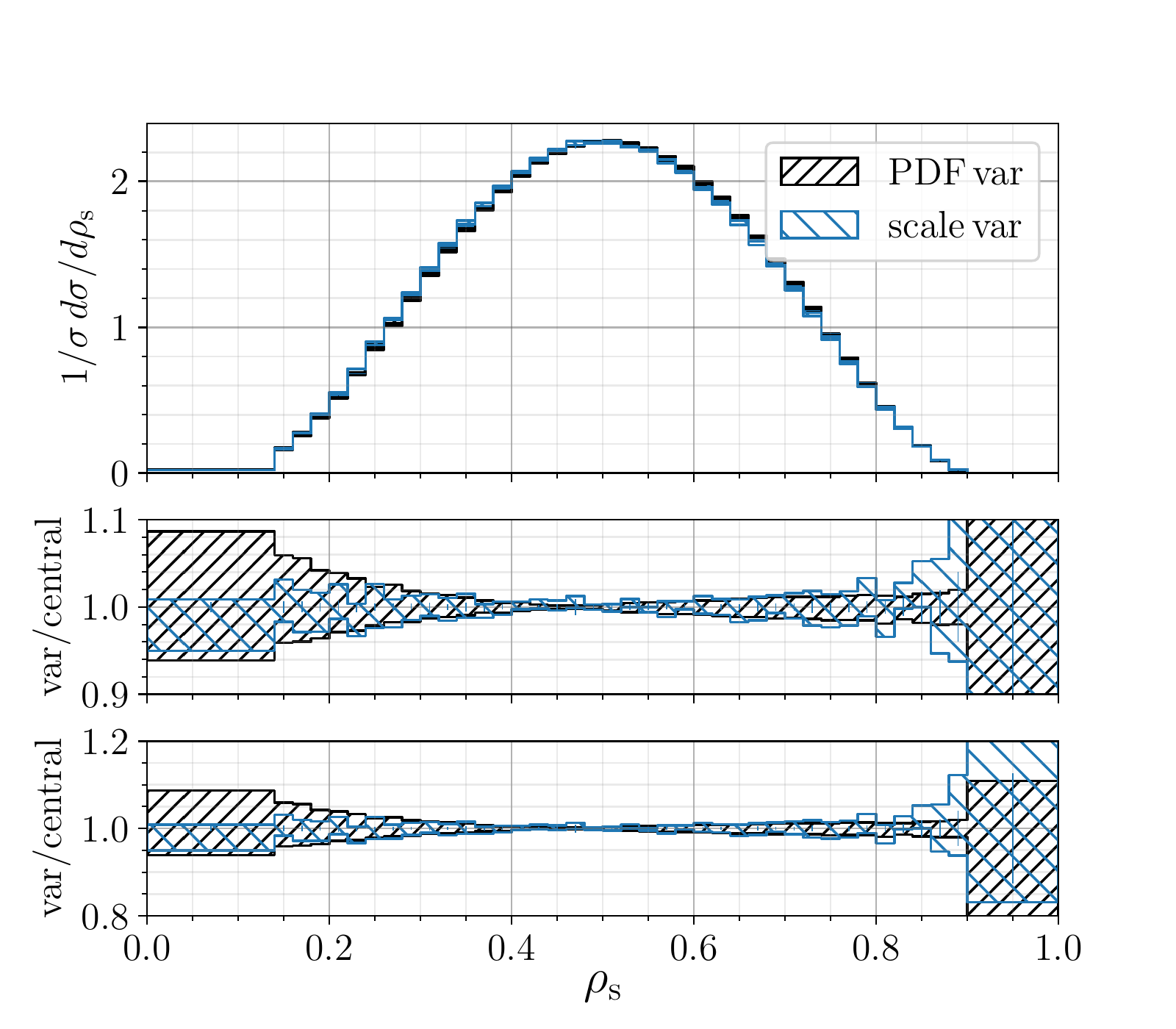}
\caption{Predictions for the 
absolute (left) and normalized (right) $\rho_s$ distribution in a LO 
computation with scale $\mu_0 = H_T^B/4$, including PDF uncertainties computed
with the CT18 NLO PDF sets and seven-point scale variation uncertainties. The
CT18 NLO uncertainty band is rescaled to 68\% C.L. The $\alpha_s(M_Z)$ value is
fixed according to the prescription associated to each PDF set and two-loop
$\alpha_s$ evolution is considered. The computation of the PDF uncertainty was
done by using LO partonic cross sections, while the scale uncertainty was
computed with the NLO partonic cross sections.
\label{fig:PDF_SCALE_HT4}
}
\end{figure}

\begin{figure}
\includegraphics[width=0.49\textwidth]{\main/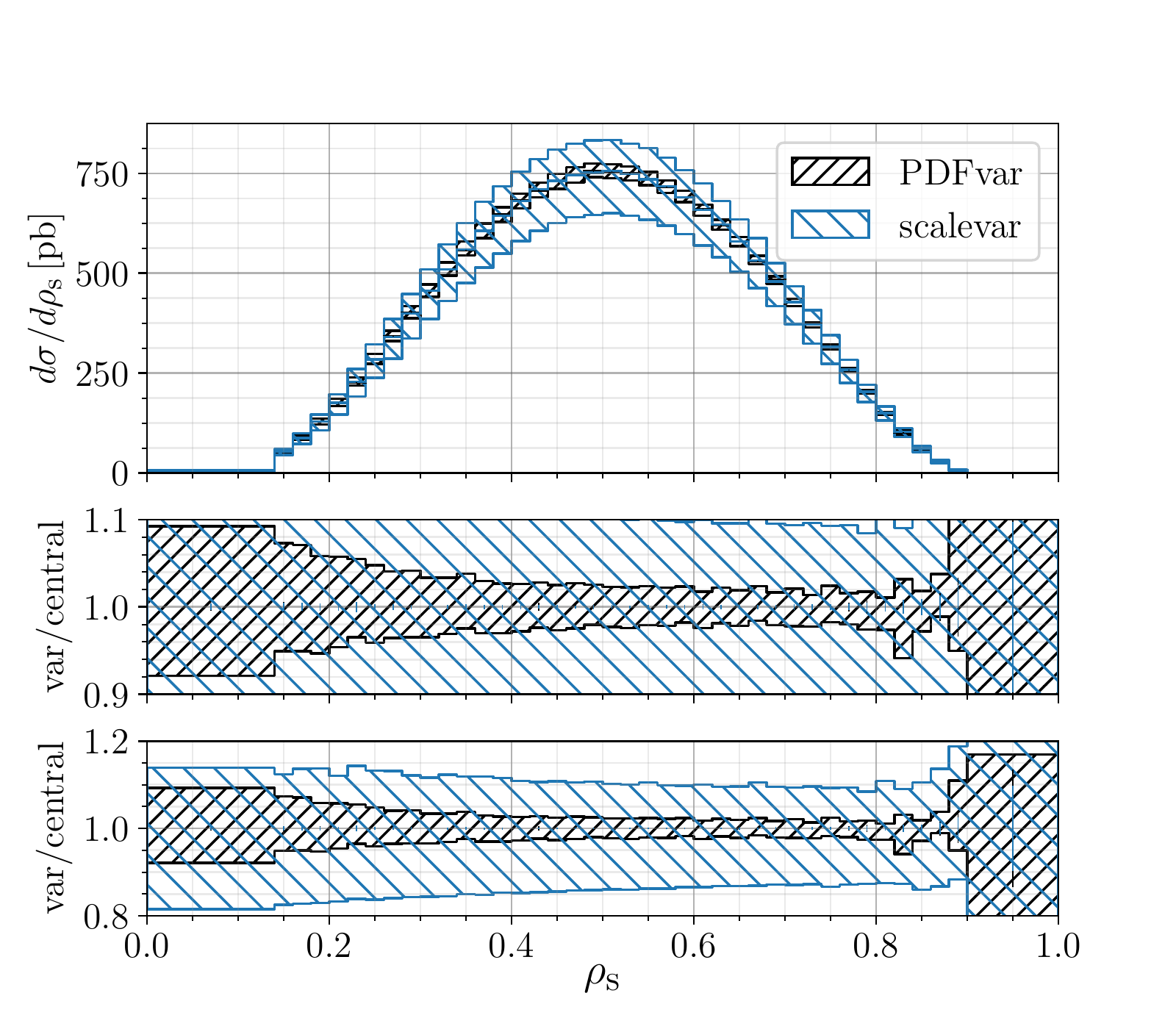}
\includegraphics[width=0.49\textwidth]{\main/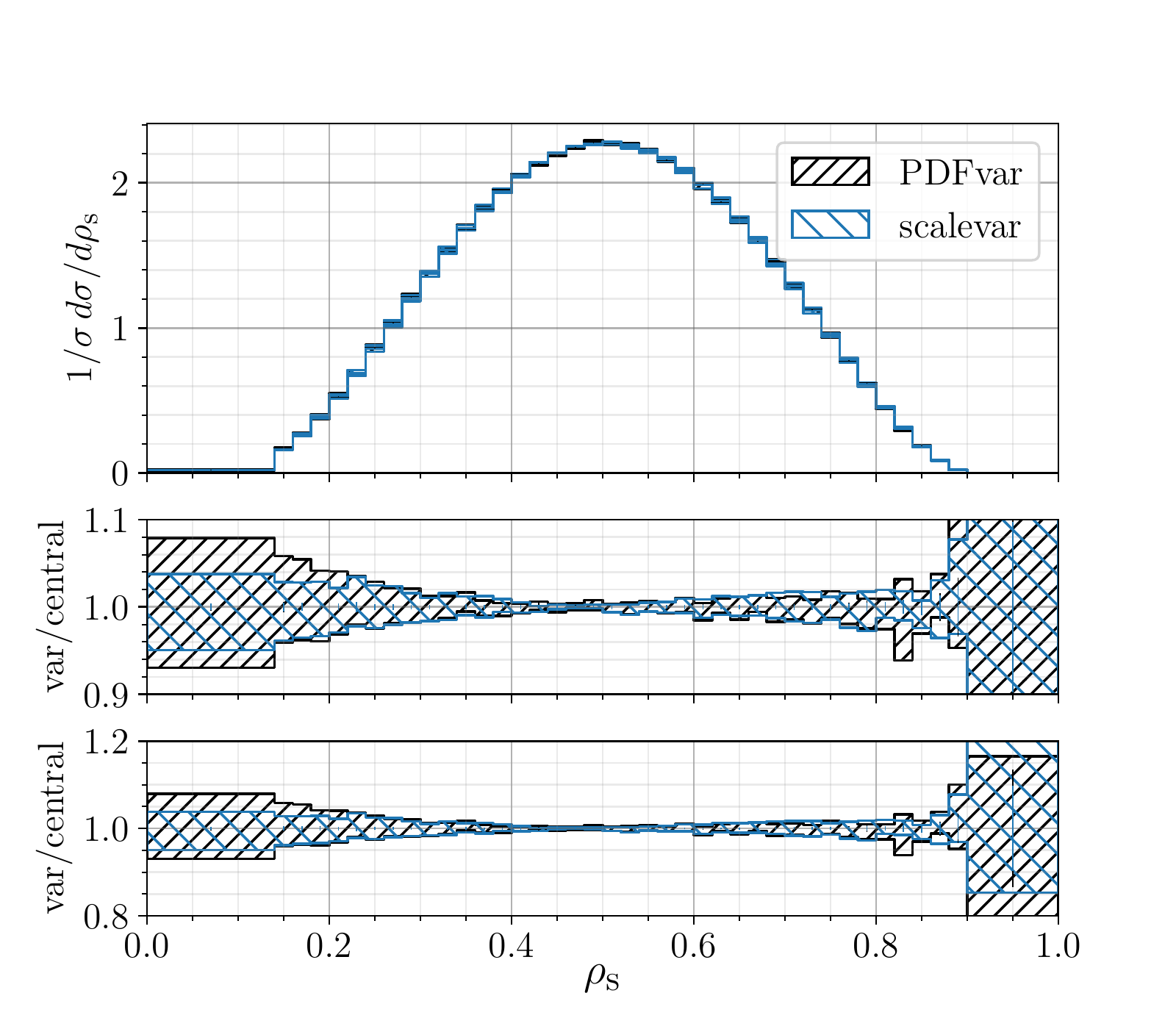}
\caption{Same as \Fig{fig:PDF_SCALE_HT4}, but for predictions obtained
  with scale $\mu_0 = H_T^B/2$. In this case both the PDF uncertainty and the
  scale variation uncertainty were obtained using the NLO partonic cross sections.
\label{fig:PDF_SCALE_HT2}
}
\end{figure}


In this Subsection we discuss the effect of PDF + $\alpha_s(M_Z)$ variations. 
We take as default the CT18 NLO PDF set, with their associated $\alpha_s(M_Z)$ value, 
the latter being subject to two-loop evolution in QCD as provided by the
\textsc{Lhapdf} interface.
We compute central predictions using the set 0 (best fit) and we determine the PDF
uncertainty band from the 58 available eigenvectors, according to the
prescription provided by the CT18 collaboration~\footnote{
    The PDF uncertainties quoted by CT18 denote the 90\% confidence level
    (C.L.) and have to be reduced by a factor of 1.645 for comparison 
    with the 68\% C.L. uncertainties quoted by other groups.}. 
First of all we observe that the relative  size of the PDF uncertainty band for many distributions remains
approximately the same when making a LO computation with NLO PDFs instead of a
full NLO computation. This is shown in \Fig{fig:PDF_NLOLO_j1}
and~\ref{fig:PDF_NLOLO_ttbar}  in case of various representative
distributions, i.e. $p_T^{j_1}$, $y_{j_1}$, $m_{t\bar{t}}$ and $y_{t\bar{t}}$,
by using the central scale $\mu_0 = H_T^B/2$ for all predictions. This is also
true for the $m_{t\bar{t}j}$ and $\rho_s$ distributions for $\rho_s$ values
not too large (corresponding to large enough $m_{t\bar{t}j}$ values), as shown
in \Fig{fig:PDF_NLOLO_rho}.  
On the other hand, for very large $\rho_s$ values ($\rho_s > 0.82$),
corresponding to small enough $m_{t\bar{t}j}$ 
(approximately, $m_{t\bar{t}j} < 400$ GeV), 
the PDF uncertainty band obtained in the LO computation is smaller
than the one from the full NLO computation. We observe, however, that the
description of this  region of phase-space is statistically limited and
differences in the PDF uncertainty bands near $\rho_s \simeq 1$, 
being subject to statistical fluctuations, 
cannot be reliably predicted with the limited statistics of our computations.
The experimental analyses aiming at the extraction of $m_t$ 
from the $\rho_s$ distribution focus on $\rho_s$ values $\sim$ 0.6 - 0.9, 
avoiding larger values of $\rho_s$ close to threshold, 
for which the statistical uncertainties of the experimental data become large, 
and improved theory descriptions beyond fixed order might be required. 
Thus, we conclude that, for the range of $\rho_s$ values corresponding to the
bulk of the available expe\-ri\-mental data, 
it is possible to determine with good accuracy the relative  size of the PDF
uncertainty bands by performing LO computations with NLO PDFs. 

This is, by far, less CPU intensive than running full NLO
computations. Therefore, in the following we use this approximation, applying
it in the computation of PDF uncertainties from a range of modern PDF sets,
including the ABMP16, CT18, MMHT20 and NNPDF3.1 NLO PDF sets using $\mu_0=H_T^B/4$.  
We also observe that, for the range of $\rho_s$ va\-lues
the experimentalists
are interested in, PDF uncertainties do not vary in a considerable way when
modifying the
central scale from $\mu_0 = H_T^B/2$ to $\mu_0 = H_T^B/4$. This is
shown for the LO computation with NLO PDFs for the $m_{t\bar{t}j}$ and
$\rho_s$ distributions in \Fig{fig:PDF_TWOSCALES_rho}, using as input the
CT18 NLO PDFs. Similar considerations are expected to be applicable also for
other PDF sets. 

With the procedure outlined above we investigated
the effect of different PDF sets on the $\rho_s$ distribution using as input the scale $\mu_0 = H_T^B/4$. 
As visible in
\Fig{fig:PDF_NLOpdfsets_LOME}, in the bulk of the $\rho_s$ distribution,
the predictions obtained with the aforementioned different PDF +
$\alpha_s(M_Z)$  sets are in good agreement among each other. On the other
hand, in the large $\rho_s$ tails, differences are observed, which are not
covered by the uncertainties of the involved PDF sets. In order to investigate
this aspect further, we show in \Fig{fig:PDF_gluonpdfs} the gluon
distribution from the considered PDF sets as a function of the longitudinal
momentum fraction $x$ retained by the partons from the incoming protons
involved in the hard scattering, for a $Q^2$ value fixed to $Q^2= m_t^2 =
(172$~{GeV})$^2$. Looking at the $x$ distributions of the incoming partons for
events leading to $\rho_s$ values within the smallest bin ($\rho_s \in [0, 0.14]$), 
we found that the 
$x_{\text{min}} = {\text{min}}(x_1, x_2)$ 
distribution is peaked around $x_{\text{min}} = 0.15$, whereas the 
$x_{\text{max}} = {\text{max}}(x_1,x_2)$ 
one is peaked around $x_{\text{max}}=0.25$. This is a region where the
gluon distributions from different PDF fits differ noticeably among each
other. On the other hand, looking at $\rho_s$ values in the bulk of the
distribution ($\rho_s \in [0.14, 0.65]$), we found that the $x_{\text{min}} $
and $x_{\text{max}}$ distributions are peaked at $x_{\text{min}} = 0.02$ and
$x_{\text{max}} = 0.07$, respectively. In this range of $x$ values, the gluon
distributions from different PDF sets are in much better agreement among each
other (see again \Fig{fig:PDF_gluonpdfs}) than in the case of small
$\rho_s$.  
We conclude that, at small $\rho_s$, the sensitivity to large $x$ PDFs, which
are not yet very well constrained by the available experimental data, would
hamper the possibility of using the $\rho_s$ distribution to extract precise
values of $m_t$. 
On the other hand, for the typical $\rho_s$ values considered by the
experimentalists to extract $m_t$, PDF uncertainties are smaller and 
we expect that the choice of a particular PDF set has a lesser impact 
on the $m_t$ value extracted from the ana\-ly\-sis of the $\rho_s$ distribution. 
In order to use the full $\rho_s$ distribution, it is very
important that uncertainties on the gluon PDF in the large $x$ tail get reduced. 
Including already available LHC differential data on
bottom hadroproduction in PDF fits could also help constraining large $x$
gluons, as pointed out in Ref.~\cite{PROSA:2015yid}, whereas current 
top-quark hadroproduction data are known to put constraints especially in the
region $x \lesssim 5 \cdot 10^{-2}$, see, e.g., ~\Ref{Alekhin:2013nda}.

For illustrative purpose, we also show PDF uncertainties on different
distributions in \Fig{fig:PDF_NLOpdfsets_ttbar_LOME}. From the left
panel, it is evident that considerations
similar to those that we discussed in the previous paragraphs
 for the $m_{t\bar{t}j}$ distribution, also apply to the
$m_{t\bar{t}}$ distribution, i.e. at large $m_{t\bar{t}}$ values this
distribution shows a large sensitivity to large-$x$ gluons. 
On the other hand, from the $y_{t\bar{t}}$ distribution in the right panel, 
it is clearly visible how PDF uncertainties blow up for large absolute values
of rapidities, corresponding to parton kinematics with more extreme $x_1$ and $x_2$ values. 
Thus, experimental data on these two distributions can be used to further
constrain PDF fits, following, for example, the analysis of the CMS
collaboration in \Ref{CMS:2019esx} based on data for the $t\bar t$ 
normalized multi-differential cross sections, which has also 
kept the full correlations between the PDFs, $\alpha_S(M_Z)$ and the top-quark mass. 

Finally, one can observe that PDF uncertainties are smaller than scale
uncertainties when looking at the absolute $\rho_s$ distribution, whereas they
are similar in size (or even larger at smaller $\rho_s$) 
when looking at the normalized $\rho_s$ distribution, 
as shown in the left and right panels of \Fig{fig:PDF_SCALE_HT4}, respectively. 
This conclusion remains valid even when changing the central scale from
$\mu_0=H_T^B/4$ to $\mu_0=H_T^B/2$, as follows from comparing
\Figs{fig:PDF_SCALE_HT4} and~\ref{fig:PDF_SCALE_HT2}.  
While in the computation using $\mu_0=H_T^B/4$ only the scale variation
uncertainty was computed using the NLO partonic cross sections and the NLO PDF
uncertainty approximated, as justified above, with the LO partonic cross
sections, in the case of the computation using  $\mu_0=H_T^B/2$ both
uncertainties were obtained with the NLO partonic cross sections. 


\begin{figure}
  \begin{center}
\includegraphics[width=0.49\textwidth]{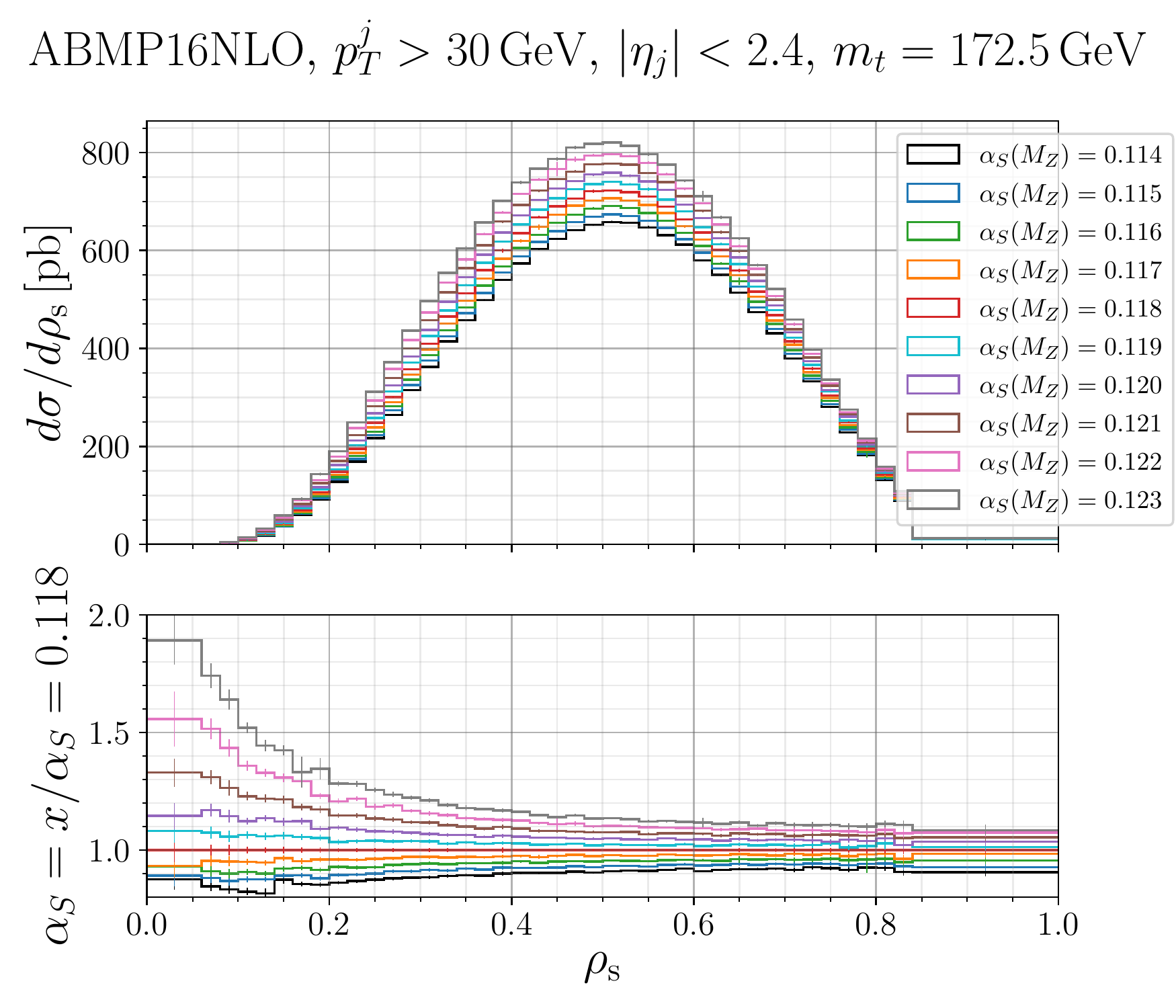}
\includegraphics[width=0.49\textwidth]{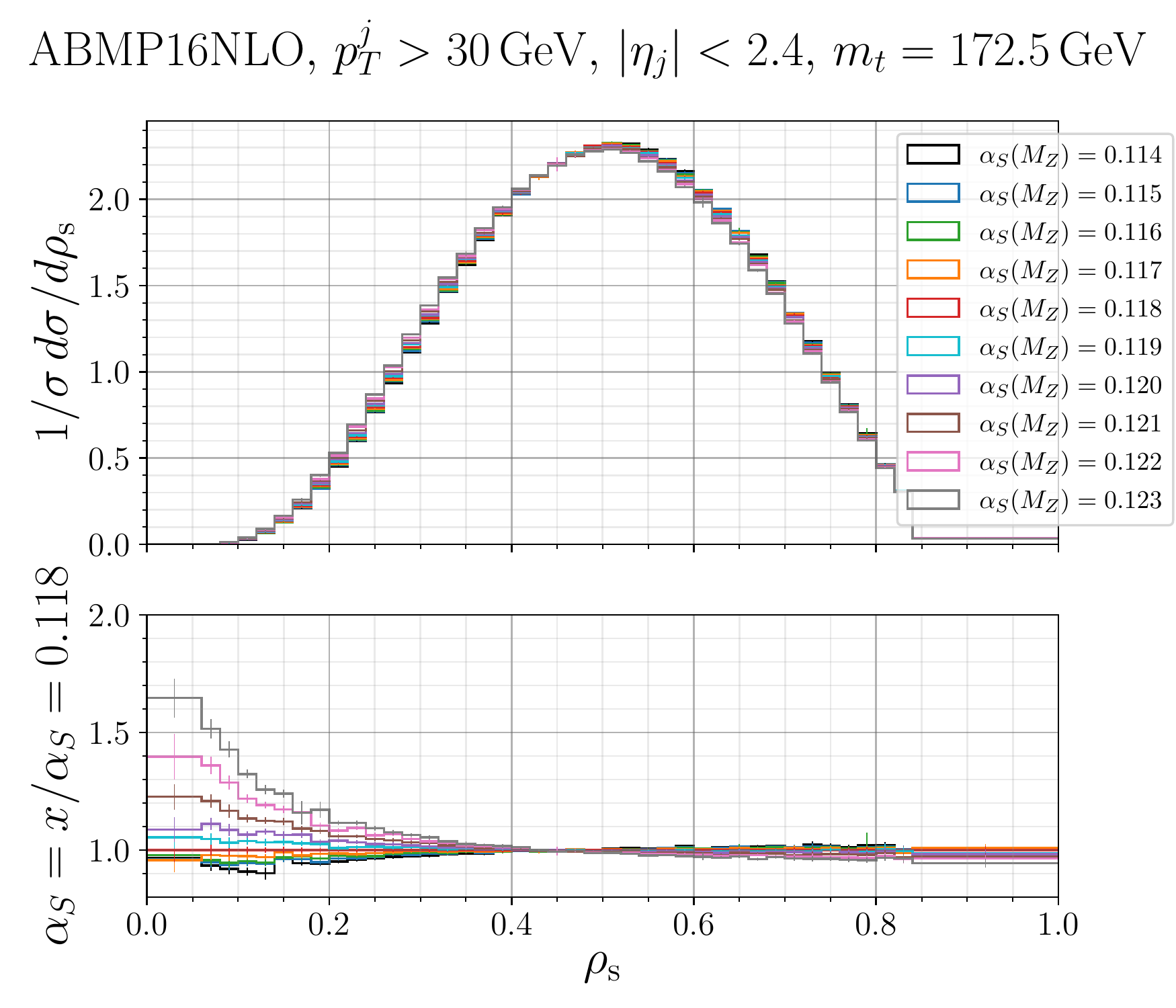}
\caption{Predictions for the $\rho_s$ (left)
  and ${\cal R}(m_t, \rho_s)$ (right)
  distributions in a
  full NLO computation with scale $\mu_0 = H_T^B/2$, using different $\alpha_S(M_Z)$ + PDF values, as extracted in the NLO ABMP16
  $\alpha_S(M_Z)$ + PDF fits. Two-loop $\alpha_s$ evolution is considered.
  The top-quark mass value $m_t$ is fixed to 172.5 GeV for all curves.
  In the lower inset, the ratio of predictions using different $\alpha_s(M_Z)$
  + PDF values with respect to the case where $\alpha_s(M_Z)=0.118$ is
  presented.  
  \label{fig:alphas}
}
\end{center}
\end{figure}


In the following of this Section, we present some further considerations on
the joined effect of $\alpha_s (M_Z)$~+~PDF variation, applied to the $\rho_s$ distribution. 
A more extensive study, that considers the prescriptions and results for $\alpha_s$ variation 
accompanying the different available PDF fits on a broad number of differential
di\-stri\-butions, is beyond the scope of this work. 
To get a first idea on the main effects of $\alpha_s(M_Z)$~+~PDF variation, and on the size of the
differences generated in the $\rho_s$ distribution, 
we use the different NLO PDF~+~$\alpha_s(M_z)$ combinations provided by
the ABMP16 collaboration. 
In particular, the ABMP16 set accounts for the correlations between PDFs and $\alpha_s$ (and heavy-quark masses). 
This implies that each fit characterized by a different $\alpha_s(M_Z)$
value, with the latter ranging in the interval from 0.114 to 0.123, is also
characterized by different PDF values. 
As shown in the left panel of \Fig{fig:alphas}, the effect appears to be particularly
sizable  at low $\rho_s$, i.e. for $\rho_s < $  0.2, 
where the most extreme $\alpha_s(M_Z)$ + PDF combinations can produce
predictions for the $\rho_s$ distribution which may differ by even a factor of two with respect to the
default choice $\alpha_s(M_z) = 0.118$. 
Differences are instead much smaller in the $\rho_s$ region currently explored
in $t\bar{t}j$ experimental analyses aiming to $m_t$ extraction, where they amount
to approximately a maximum of (-10, +20)\%. One should also observe that
predictions with different $\alpha_s(M_Z)$~+~PDFs do not cross among each
other, meaning that the effect of this variation and the corresponding
uncertainty on ${\cal R}(m_t, \rho_s)$, the variable used in the experimental
analyses, is much smaller, as shown in the right panel of \Fig{fig:alphas}. The effect is limited
to very few percent in the region of interest for the experimental analyses and becomes
increasingly sizable only for $\rho_s < 0.4$, as $\rho_s \rightarrow 0$.


\subsection{Effects of variation of the $R$ parameter used in jet reconstruction}
\label{subsec:r}

\begin{figure}
\begin{center}
  \includegraphics[width=0.999\textwidth]{\main/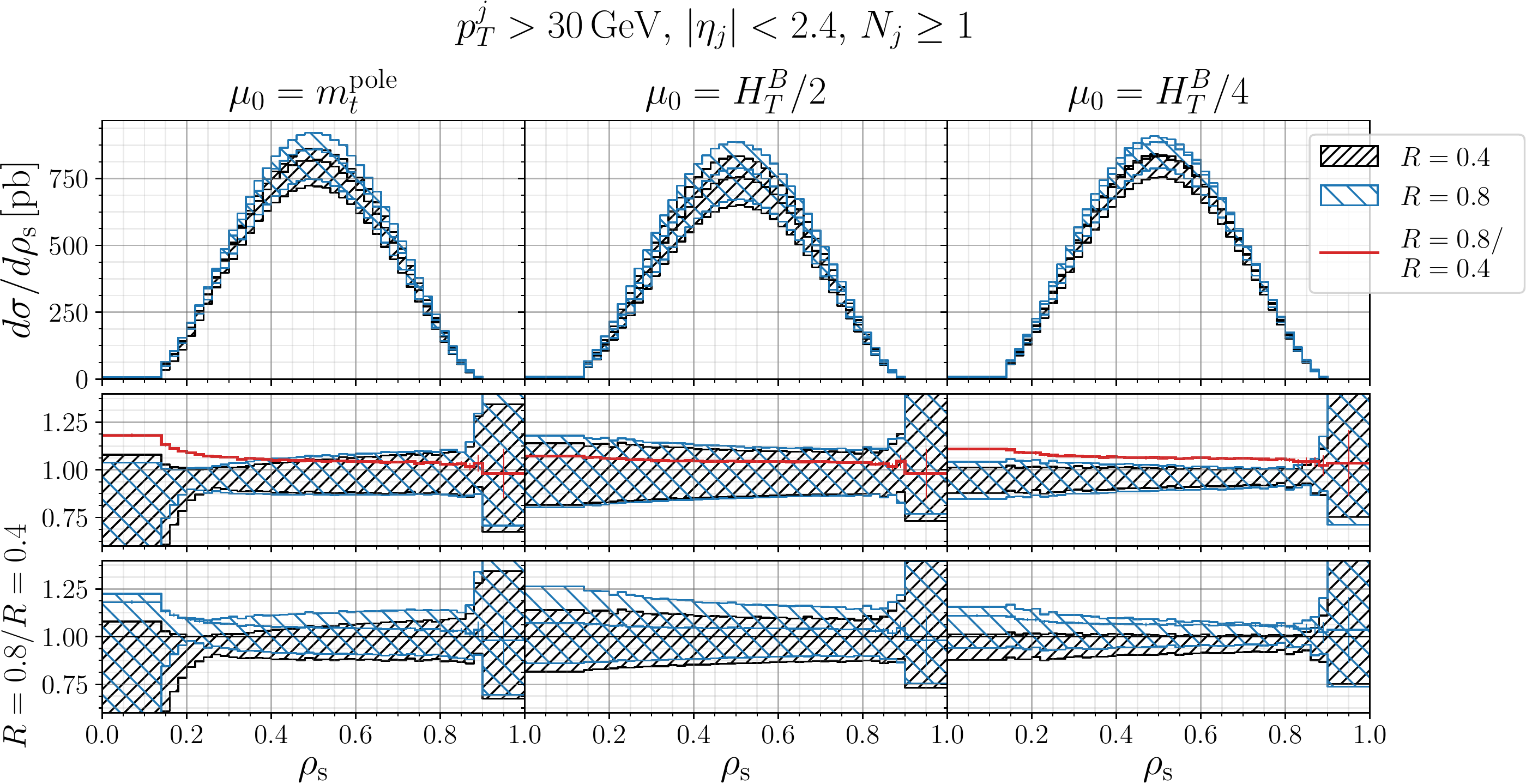}
\end{center}
  \caption{NLO differential cross section as a function of $\rho_s$ calculated
    for the $pp \rightarrow t\bar{t}j+X$ process at $\sqrt{S}=13\,$TeV using
    the static central scale $\mu_0=m_t$ (left panel) and the dynamical
    central scales $\mu_0=H_T^B/2$ (central panel) and $\mu_0 = H_T^B/4$
    (right panel), varying the $R$-value in the \mbox{anti-$k_T$} jet clustering
    algorithm from $R = 0.4$ (black) to $R= 0.8$ (blue). Thereby, besides
    central predictions, the uncertainty bands due to seven-point scale
    variation are shown. In the intermediate panel of each plot 
the relative  size of each scale uncertainty band is shown. Thereby the ratio
of the central scale prediction obtained with $R = 0.8$ and $R = 0.4$ is also
plotted (red line). In the lower panel, all ratios are computed with respect
to the central prediction with $R=0.4$.}   
  \label{fig:RHO_RVar}
\end{figure}

\begin{figure}
\begin{center}
  \includegraphics[width=0.999\textwidth]{\main/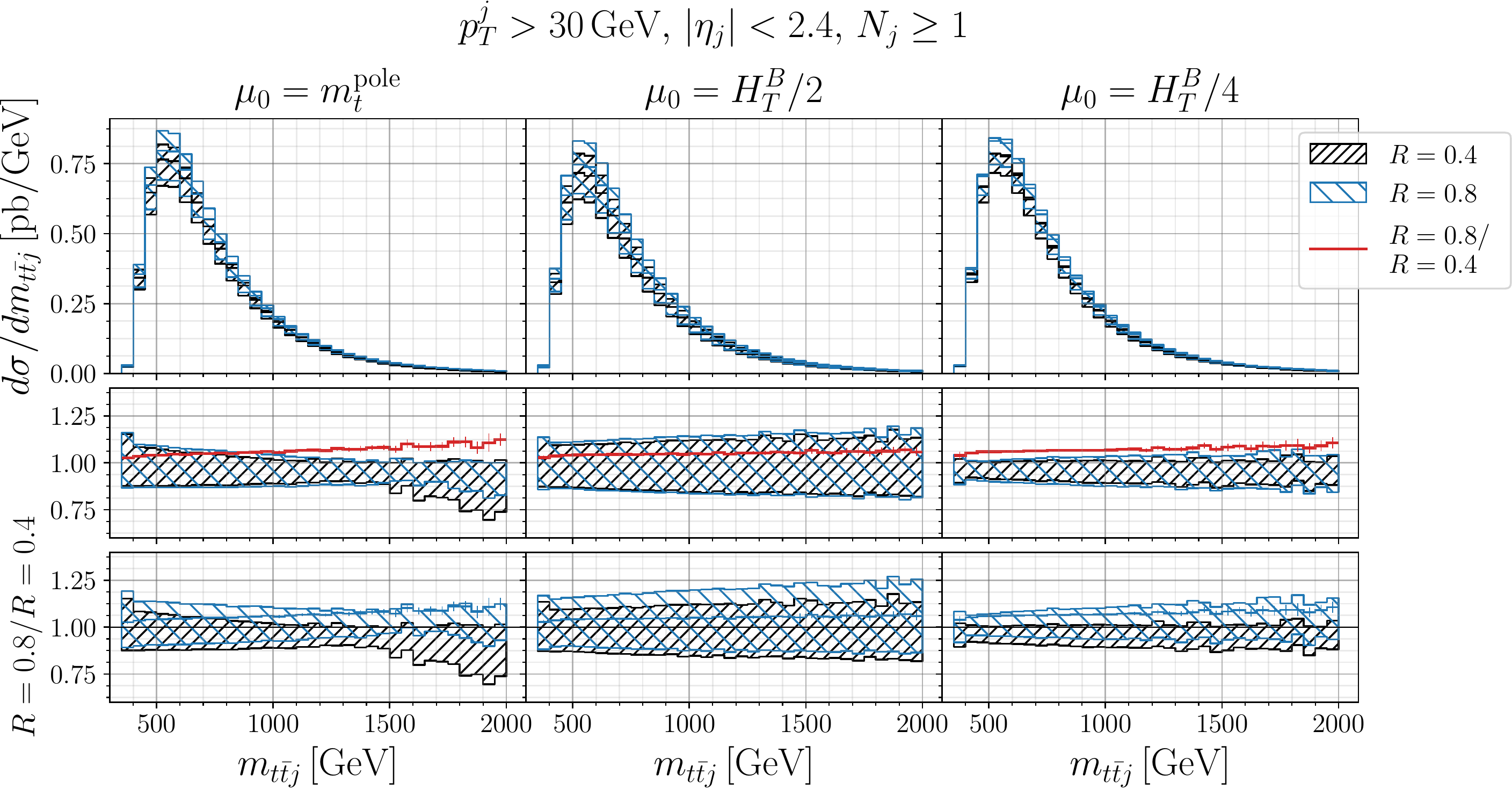}
\end{center}
\caption{
\label{fig:MTTJ_RVAR}
Same as \Fig{fig:RHO_RVar}, but for the $m_{t\bar{t}j}$ distribution.}

\end{figure}

\begin{figure}
\begin{center}
  \includegraphics[width=0.999\textwidth]{\main/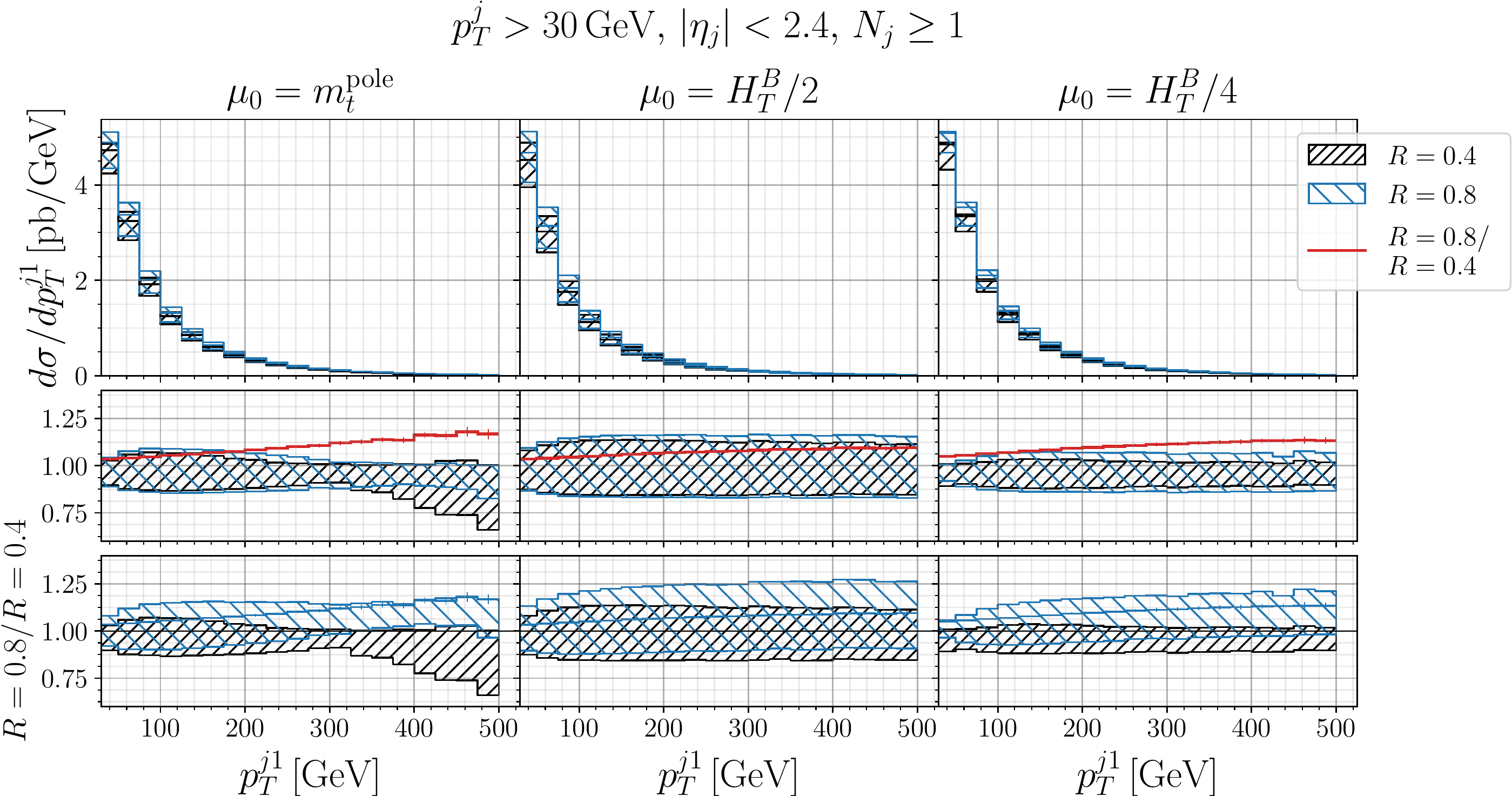}
\end{center}
\caption{
\label{fig:PTJ1_RVAR}
Same as \Fig{fig:RHO_RVar}, but for the $p_T^{j_1}$ distribution.}
\end{figure}

\begin{figure}
\begin{center}
  \includegraphics[width=0.999\textwidth]{\main/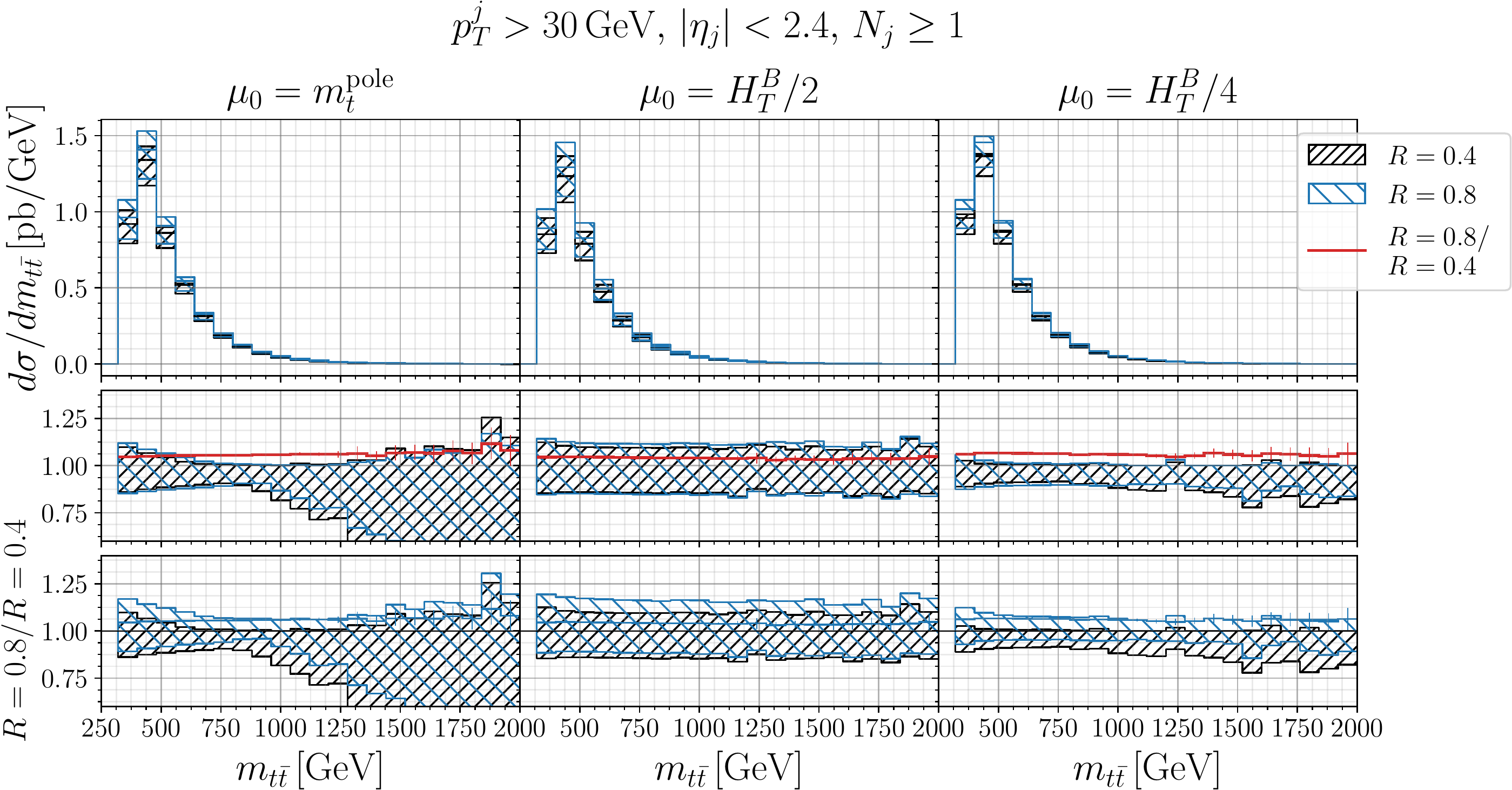}
\end{center}
\caption{
\label{fig:MTT_RVAR}
Same as \Fig{fig:RHO_RVar}, but for the $m_{t\bar{t}}$ distribution.}
\end{figure}

Additionally we have studied the influence of the variation of the $R$-parameter of
the \mbox{anti-$k_T$} jet algorithm, on the scale variation uncertainty in the case
of the static scale $\mu_0 = m_t$ and the dynamical scales $\mu_0=H_T^B/2$ and
$\mu_0 = H_T^B/4$, which have shown the most preferable behavior among the previously
discussed scale choices.
  The choice of reconstructing jets with the \mbox{anti-$k_T$} jet
algorithm using $R = 0.4$ as default in our work is motivated by the fact that
this is the present default choice of both the ATLAS and CMS collaborations. 
However, the collaborations are also exploring the possibility of larger $R$ values, 
as needed for a proper assessment of the systematic uncertainties in the experimental analyses. 
In particular, the case $R = 0.8$ is motivated by the fact that the CMS collaboration includes this 
value for $R$ in their official jet reconstruction procedure. Therefore, all the
calibration and correction factors needed to assess the systematic uncertainties associated
to the jet reconstruction are now validated and available for the experimental analysis.

The comparison of the results adopting as input the value $R = 0.4$
(black), already used to obtain the distributions presented in the previous
Subsections, and the value $R = 0.8$ (blue), is shown in
\Fig{fig:RHO_RVar} for the $\rho_s$ distribution. Thereby the fixed
central scale $\mu_0=m_t$ was used to calculate the distributions shown in the
left panel, while the central 
scale definition $\mu_0=H_T^B/2$ and $\mu_0 = H_T^B/4$  were applied to obtain those in the central and right panels, respectively. 
For all central scale definitions the differential cross section is larger for
larger values of the $R$-parameter, as expected on the basis of phase-space
considerations. The shape of central predictions is also affected, with the
$R=0.8$ choice populating especially the low $\rho_s$ (high $m_{t\bar{t}j}$) tails.   
For the considered scale choices, the scale variation uncertainty bands
applying either $R$-parameter value are very similar among each other over the
whole range of the distribution, with a slightly reduced size when using $R =
0.4$ with respect to $R = 0.8$.  Analogous considerations can be drawn when
looking at \Figs{fig:MTTJ_RVAR} and~\ref{fig:PTJ1_RVAR}, showing the
$m_{t\bar{t}j}$ and the $p_T^{j_1}$ distributions.  
Our conclusion is that the size of the scale uncertainty bands is almost
insensitive to the $R$ variation in the range [0.4 - 0.8].
On the other hand, the effect of a change in the $R$ value on observables
which only depend on the top quarks just amounts to a rescaling of the
cross~section, with no impact on the shape of the corresponding distributions,
as shown e.g. in \Fig{fig:MTT_RVAR} for the case of the $m_{t\bar{t}}$ distribution.  

Reassured by the fact that no significant shape distortion caused by the different value of the jet radius $R$ 
appears in the NLO distributions we are interested in, we want however to remind the reader that  some caution is required
when extracting
from fixed-order calculations
the final physical dependence of the cross section and of inclusive distributions on  $R$.
This is due to the fact that the $R$-dependence of the cross section is not influenced by virtual corrections, but only by the additional real
radiation. We thus expect it to display, at the NLO level, an unphysical behavior, in the form of a logarithmic divergence at small $R$. 
Only after adding shower and hadronization effects one obtains a more physical description, usually resulting in
an increase of the slope of the 
$R$ dependence of the $d\sigma/dR$ cross~section, compared to fixed-order calculations. This is a known effect, since hadron formation  further randomizes the particles’ momenta, driving
even more energy out of the cone. It is also known that, on the other hand, 
the underlying event counteracts the effect of hadronization, since it generates soft hadrons that bring more energy into the jet cone, with a probability proportional to its area.

%
\section{Theoretical predictions for top-quark mass measurements using
the ${\cal R}$ distribution}\label{sec:mass}

Some results for the ${\cal R}$~distribution have already been presented
in Section~\ref{sec:pheno}. However, for systematic extractions
of the top-quark mass using as a basis the $t\bar{t}j~+~X$ samples of events
collected at the LHC at $\sqrt{S}=$ 13 TeV, or future samples at higher
center-of-mass energies like e.g. $\sqrt{S}=$~14 TeV, many more predictions are required, using different PDF sets and covering a large top-quark mass range in small mass steps. The purpose of this Section is to provide extensive results which can be used by the  ATLAS and CMS experimental collaborations to infer
the top-quark mass values in different mass renormalization schemes and allow also to study
various syste\-matic uncertainties. In the experimental analyses the binning of the
${\cal R}$
di\-stri\-bution has been optimized to minimize systematic and
statistical uncertainties. For the exemplary results shown in the following,
we use the current binning and analysis setups as provided by ATLAS and CMS. As the analysis setup differs slightly from the one adopted in Section~\ref{sec:pheno}, we
collect the details in the next Subsection.

\subsection{Setup of the calculation}
The results presented in
Section~\ref{n3-results} have NLO QCD accuracy, refer to stable top quarks and
were obtained by properly extending the framework of
\Refs{Dittmaier:2007wz,Dittmaier:2008uj}, after having ve\-ri\-fied that,
for the case of top-quark mass renormalized in the on-shell scheme, it gives
predictions compatible with the \textsc{Powheg-Box} implementation used in
Section~\ref{sec:pheno}, as already discussed there. 
As in Section~\ref{sec:pheno}, the \mbox{anti-$k_T$} jet algorithm
\cite{Cacciari:2008gp}, as implemented in \textsc{FastJet} \cite{Cacciari:2011ma}, is used with 
$R$ set to 0.4 and employing the $E$-scheme for parton recombination into jets. 
Additional cuts are applied to the jets as returned by
\textsc{FastJet}. First of all, a mi\-ni\-mal transverse momentum $p_T^{\mbox{\scriptsize min}}$
is required for jet detection. 
We have produced results for different choices of 
$p_T^{\mbox{\scriptsize min}}$: $ p_T^{\mbox{\scriptsize min}} \in$ \{30 GeV, 50 GeV, 75 GeV, 100 GeV\}. 
This set of $p_T^{\mbox{\scriptsize min}}$ values extends the one considered in Section~\ref{sec:pheno}. 
In addition, in order to be consistent with the most recent experimental setups, we have computed predictions for two different choices of the maximal value of the pseudorapidity $\eta_j$: results using $|\eta_j|< 2.5$ as well as results using $|\eta_j|< 2.4$ have been produced (in Section~\ref{sec:pheno} only  $|\eta_j|< 2.4$ was used). 
If more than one jet satisfies these cuts (the top quarks are assumed to be always detected and not passed to the jet-algorithm), the highest energetic
jet is used to calculate $\rho_s$.

For the binning in $\rho_s$ we have investigated different choices, as provided by the ex\-pe\-ri\-ments. In the
following, predictions are presented using
\begin{displaymath}
  \{0., 0.18, 0.22, 0.27, 0.32, 0.38,
				   0.45, 0.53, 0.62, 0.71, 1.\}
\end{displaymath}
for the bin boundaries.

For the hadronic center of mass energies we used both $\sqrt{S} = 13$~TeV and
$\sqrt{S} = 14$~TeV, in view of Run 3 and the High Luminosity phase at the LHC.
Further predictions for other center-of-mass energies can be provided if required by the experiments.

Despite the slightly enhanced mass sensitivity of the ${\cal R}$ distribution compared to the
inclusive cross section, the mass effects are
still rather small and high precision in the numerical Monte Carlo
integration is required to make them visible. 
For the virtual corrections to the partonic process $gg\to t\bar{t} g$ 
additional runs were required to achieve a precision at the
sub-percent level in the combined result. To allow interpolations in the mass, results for
different top-quark mass values were produced using a step size of $1$~GeV.

For the numerical evaluation of the derivative of the leading
order cross section required in the conversion to the ${\overline{\mbox{MS}}}$ or MSR heavy-quark mass renormalization scheme
(see Section \ref{MassConversion}) a step size of $0.5$~GeV is
used. Employing different methods to calculate the derivative, we have 
checked that the step size of $0.5$~GeV leads to negligible uncertainties
for the derivative. As the pole
mass $m_t$ and the ${\overline{\mbox{MS}}}$ mass $m(m)$
differ by about $10$~GeV (doing the conversion at four loops),
we produced results 
for mass values between 158 and 178~GeV, allowing to cover not only the
relevant range in the pole mass but also in the ${\overline{\mbox{MS}}}$ and MSR mass.
The MSR mass depends on the choice of the $R$ scale. 
For typical $R$ values between $\mathcal{O}$(GeV) and the ${\overline{\mbox{MS}}}$ mass $m(m)$, the value of the MSR mass lies
between the ${\overline{\mbox{MS}}}$ mass $m(m)$ ($R=m(m)$) and the pole mass ($R\sim 1$~GeV).   

We used the (in alphabetic order)
%
ABMP16\_5\_nlo
\cite{Alekhin:2018pai},
CT18NLO
\cite{Hou:2019qau}, 
MMHT2014-nlo68cl
\cite{Harland-Lang:2014zoa}, as well as
MSHT20nlo\_as118
\cite{Bailey:2020ooq},
and
NNPDF31\_nlo\_as\_0118
\cite{NNPDF:2017mvq}
PDF sets.
The latter two fix $\alpha_s(M_Z)$ to 0.118 and 
in the evaluation of the NLO predictions we have always used 
the $\alpha_s(M_Z)$ value as provided by the corresponding PDF set. 
In addition, we also produced results always using a same fixed $\alpha_s (M_Z)$ value, following the prescription applied in \Refs{Dittmaier:2007wz,Dittmaier:2008uj}.  

\subsection{Conversion to the ${\overline{\mbox{MS}}}$
  and MSR mass scheme}
\label{MassConversion}
As mentioned before, the ${\cal R}$ distribution can also be expressed using
renormalization schemes different from the pole mass
scheme usually applied. In the following, we briefly describe how the predictions obtained
in the pole mass scheme can be translated to the ${\overline{\mbox{MS}}}$ or to the
MSR mass scheme. We follow the method outlined in \Ref{Langenfeld:2009wd} and
present only the main steps as the details can be found in
\Ref{Langenfeld:2009wd}. We start with the discussion of the ${\overline{\mbox{MS}}}$
mass. In perturbation theory the pole mass and ${\overline{\mbox{MS}}}$ mass
  are related
through a finite renormalization and the mass given in one scheme can be
expressed in terms of the mass given in a second scheme. For the
concrete case the relation between the top-quark pole mass $m_t$ and 
${\overline{\mbox{MS}}}$ mass $m(\mu)$ reads:
\begin{equation}
  m_t = m(\mu)\bigg( 1 - c_1(L_M) a^{({n_f})}(\mu) + \left(c_1(L_M)^2
    - c_2(L_M)\right)
  {a^{({n_f})}}^2(\mu)+\ldots \bigg).
\end{equation}
with
\begin{equation}
  a^{({n_f})}(\mu) = {\alpha_s^{({n_f})}(\mu)\over \pi}
  \quad \mbox{and}\quad L_M= \ln\left({\mu^2\over m_t^2}\right).
\end{equation}
The coefficients $c_i$ are known up to four-loop order
\cite{Marquard:2015qpa} and can be found conveniently collected for example in
\Ref{Chetyrkin:2000yt,Herren:2017osy}. At NLO accuracy only $c_1$ is
required which reads:
\begin{equation}
  c_1(L) =-{4\over 3} - L.
\end{equation}
At the order we are working, $L_M$ can be replaced with
$L_M= \ln\left({\mu^2\over m(\mu)^2}\right)$. The difference between 
using $m(\mu)$ or $m_t$ is of higher order in the coupling constant and does not contribute at the order we are working here.
 Using in addition $m(m)$
instead of $m(\mu)$ leads to $L_M=0$.  In
\Refs{Dittmaier:2007wz,Dittmaier:2008uj} the top-quark loops in the
gluon self energy are subtracted at finite momentum. At NLO accuracy
the strong coupling defined in this way is equivalent to
$\alpha_s^{(n_f-1)}(\mu) $, i.e. $\alpha_s$ renormalized in the
${\overline{\mbox{MS}}}$
scheme with $n_f-1$ active flavors. Note that all quarks except the top one
are treated as massless.  As the difference
between $\alpha_s^{(n_f)}(\mu) $ and $\alpha_s^{(n_f-1)}(\mu) $ is of
order $({\alpha_s^{(n_f-1)}(\mu)})^2$,  one finally obtains
\begin{equation}
  m_t = m(m)\bigg( 1 - c_1(0) a^{({n_f-1})}(m(m))+\ldots \bigg)
  = m(m)\bigg( 1 - c_1(0) a^{({n_f-1})}(\mu)+\ldots \bigg).
\end{equation}
Starting with the perturbative expansion of the cross section using
the pole mass
\begin{equation}
  \sigma = (a^{(5)}(\mu))^3 \sigma^{(0)}(m_t) + (a^{(5)}(\mu))^4
  \sigma^{(1)}(m_t)
  + \ldots,
\end{equation}
where we have set $n_f=6$, and using the above relation between $m_t$ and
$m(m)$, one obtains  after expanding in $\alpha_s$
\begin{equation}
  \label{Master1}
  \sigma = (a^{(5)}(\mu))^3 \sigma^{(0)}(m(m))
  + (a^{(5)}(\mu))^4 \left(\sigma^{(1)}(m(m))
     - c_1(0) \left.{d\sigma^{(0)}(m_t)\over dm_t}\right|_{m_t=m(m)} m(m)
  \right)
  + \ldots\, .
\end{equation}
We note that this formula can be applied to the inclusive cross
section but also to individual bins of differential distributions. 

The definition of the MSR mass \cite{Hoang:2008yj,Hoang:2017suc}
employs the fact that the ${\overline{\mbox{MS}}}$
mass is free of renormalon
ambiguities and that the renormalon ambiguity does not depend on the
mass of the heavy quark. 
Accordingly the MSR mass is defined through
\begin{equation}
  {m_{\mbox{\scriptsize MSR}}} = m_t - R\bigg( d_1(0) a^{(n_f)}(R) + d_2(0)
  (a^{(n_f)}(R))^2 +\ldots\bigg),
\end{equation}
with the expansion coefficients $d_i$ implicitly given by 
\begin{displaymath}
  m_t = m(m)\bigg( 1 + d_1(0) a^{(n_f)}(m)
  + d_2(0) (a^{(n_f)}(m))^2 +\ldots\bigg).
\end{displaymath}
Note that the coefficients $d_i$ can be
expressed in terms of the $c_i$ introduced above, e.g. $d_1(0) = -c_1(0)$.
Using again $a^{(n_f-1)}(R)$ instead of
$a^{(n_f)}(R)$ and expressing
the cross section in terms of the MSR mass leads to
\begin{equation}
  \label{Master2}
   \sigma = (a^{(5)}(\mu))^3 \sigma^{(0)}({m_{\mbox{\scriptsize MSRp}}})
  + (a^{(5)}(\mu))^4 \left(\sigma^{(1)}({m_{\mbox{\scriptsize MSRp}}})
     - c_1(0) \left.{d\sigma^{(0)}(m_t)\over dm_t}\right|_{m_t={m_{\mbox{\scriptsize MSRp}}}} R
  \right)
  + \ldots .
\end{equation}
We note that there are two variants of the
MSR scheme, leading respectively to the
\textit{natural} MSR mass and the \textit{practical} MSR mass, differing according to the way the transition from $n_f$ active flavors to $n_f-1$ active flavors is done. Expressing $\alpha_s^{(n_f)}(\mu) $ in terms
of $\alpha_s^{(n_f-1)}(\mu) $ corresponds to the practical MSR mass
scheme. This is indicated by the additional ``p'' in the subscript of the
MSR mass ${m_{\mbox{\scriptsize MSRp}}}$. For the present application, the difference between
the natural and the practical scheme is negligible, as has been shown in
\Refs{Hoang:2017suc,Garzelli:2020fmd,Hoang:2021fhn}.

Using \Eqs{Master1} and (\ref{Master2}) it is straightforward to convert
the results using the pole mass scheme into the
${\overline{\mbox{MS}}}$ or MSR scheme.

\subsection{Exemplary results for the ${\cal R}$
  distribution using different
  parton distribution function sets and different renormalization schemes}
\label{n3-results}
\begin{figure}[htbp]
  \begin{center} 
  \includegraphics[width=0.98\textwidth]{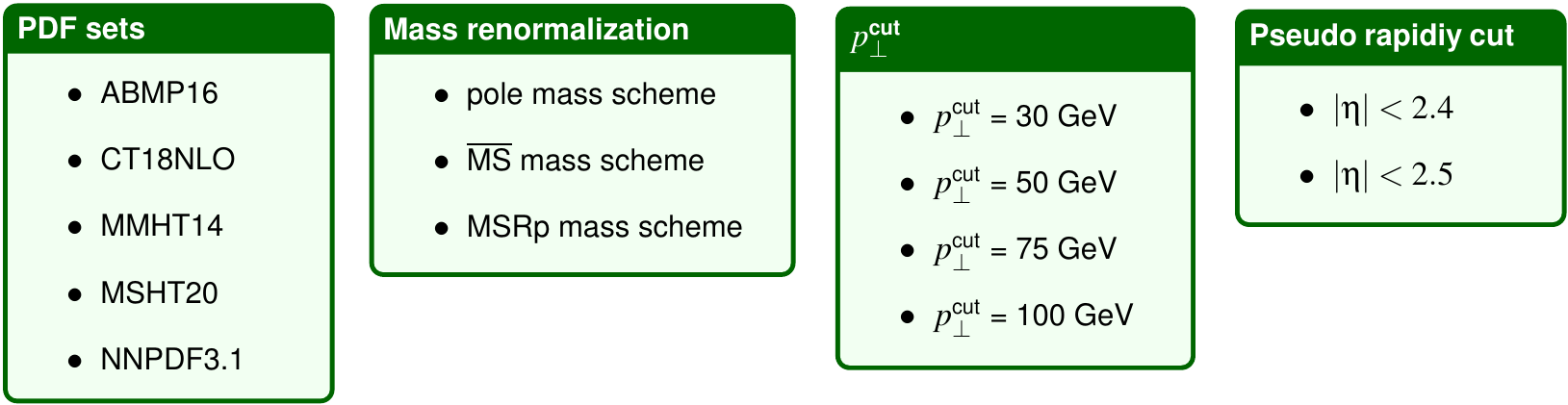}
  \end{center}
  \caption{Overview of the different settings used as input in the calculation.}
  \label{AvailableSettings}
\end{figure}

In this Section we provide 
theoretical predictions using the setup described above. The results 
allow for top-quark mass determinations from the measured ${\cal R}$
distribution. \Fig{AvailableSettings} gives an overview of the different
settings used in the calculation. 
In addition, we have considered different bin choices.
Here we only show selected results. Some reference cross sections are
collected in the Appendix~\ref{sec:appA}
and the complete set of predictions is a\-vai\-la\-ble online at \cite{onlinerepository}.

\begin{figure}[htbp]
  \begin{center}
    \includegraphics[scale=0.35]{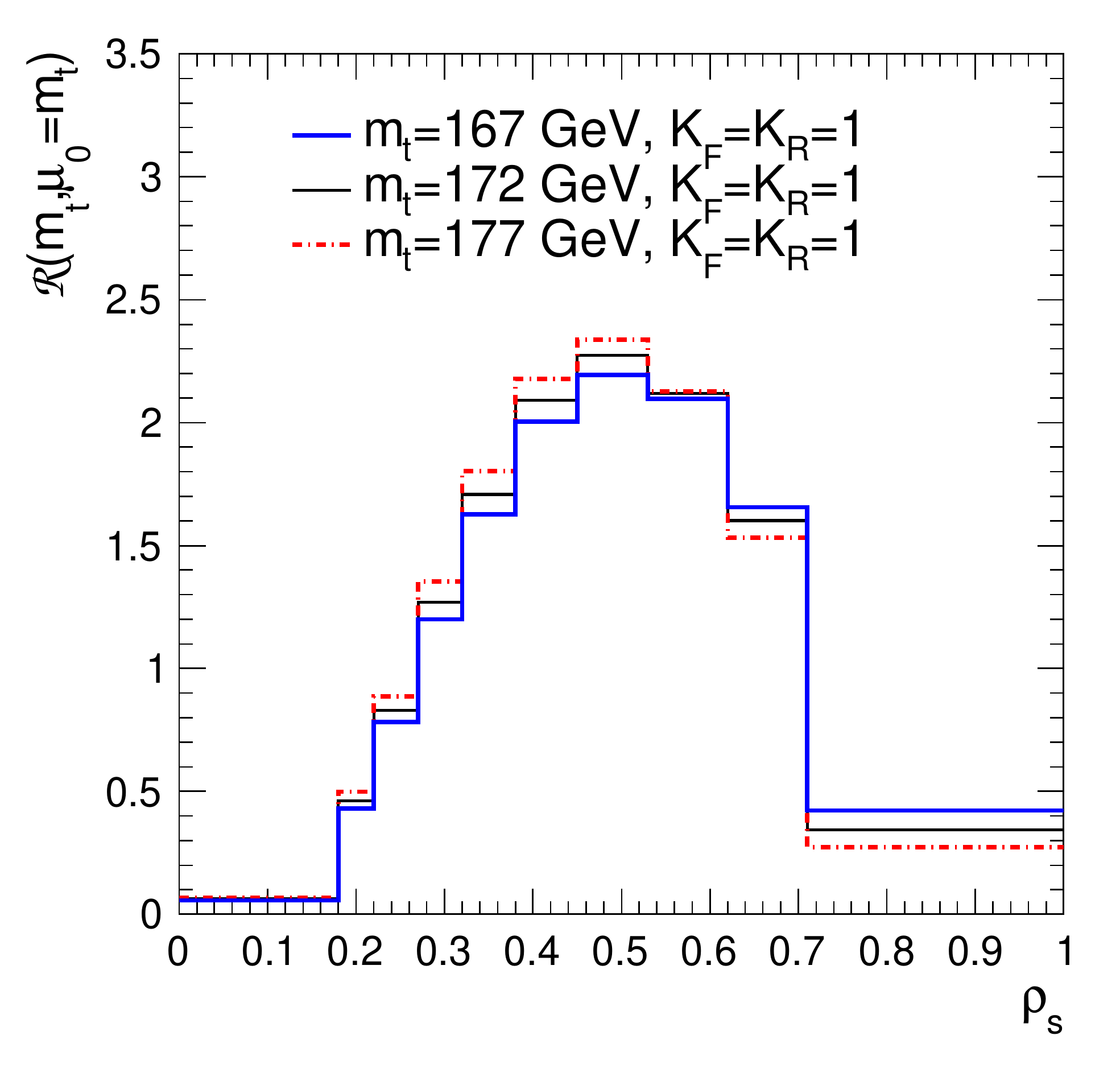}
    \includegraphics[scale=0.35]{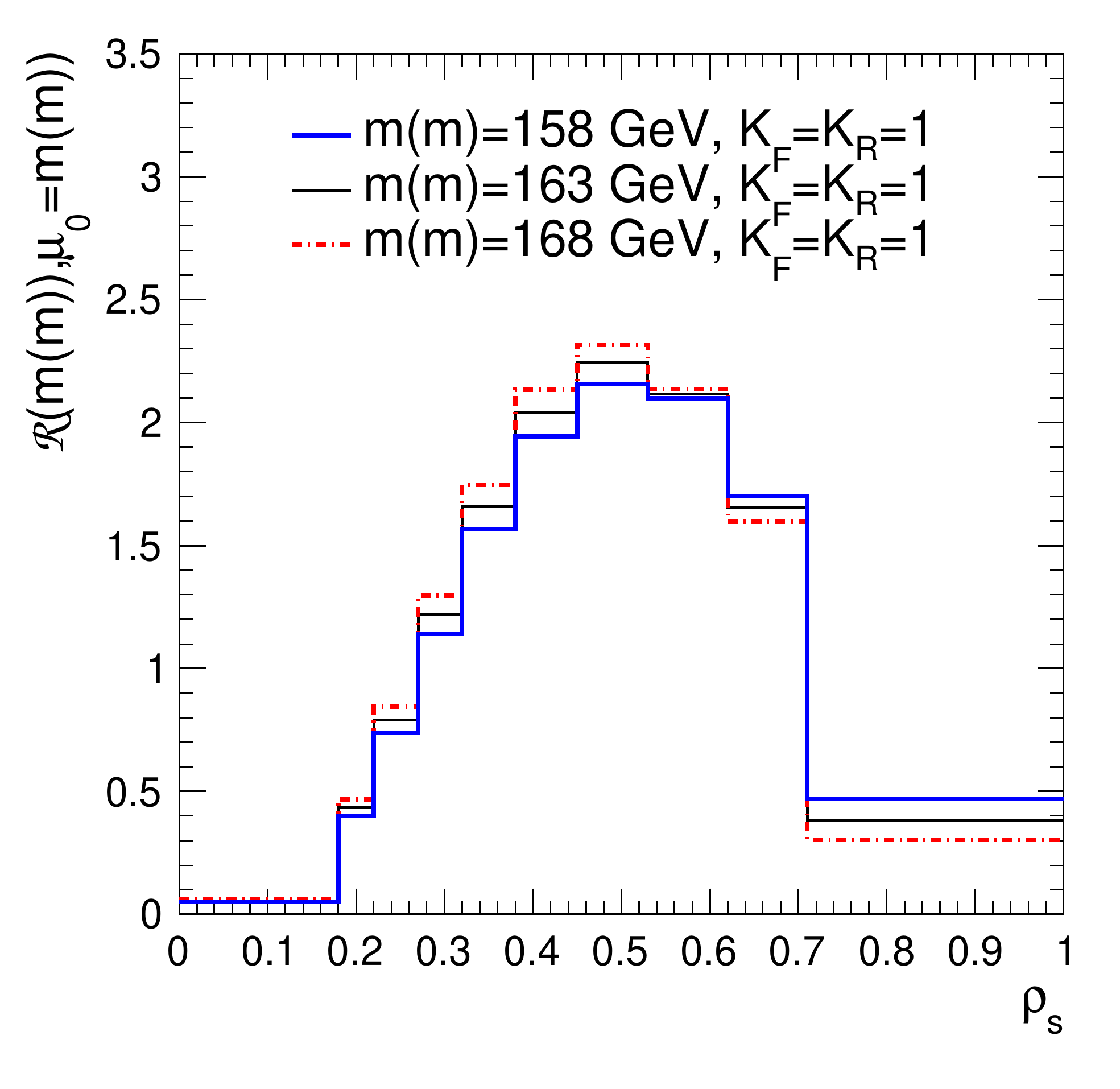}
    \includegraphics[scale=0.35]{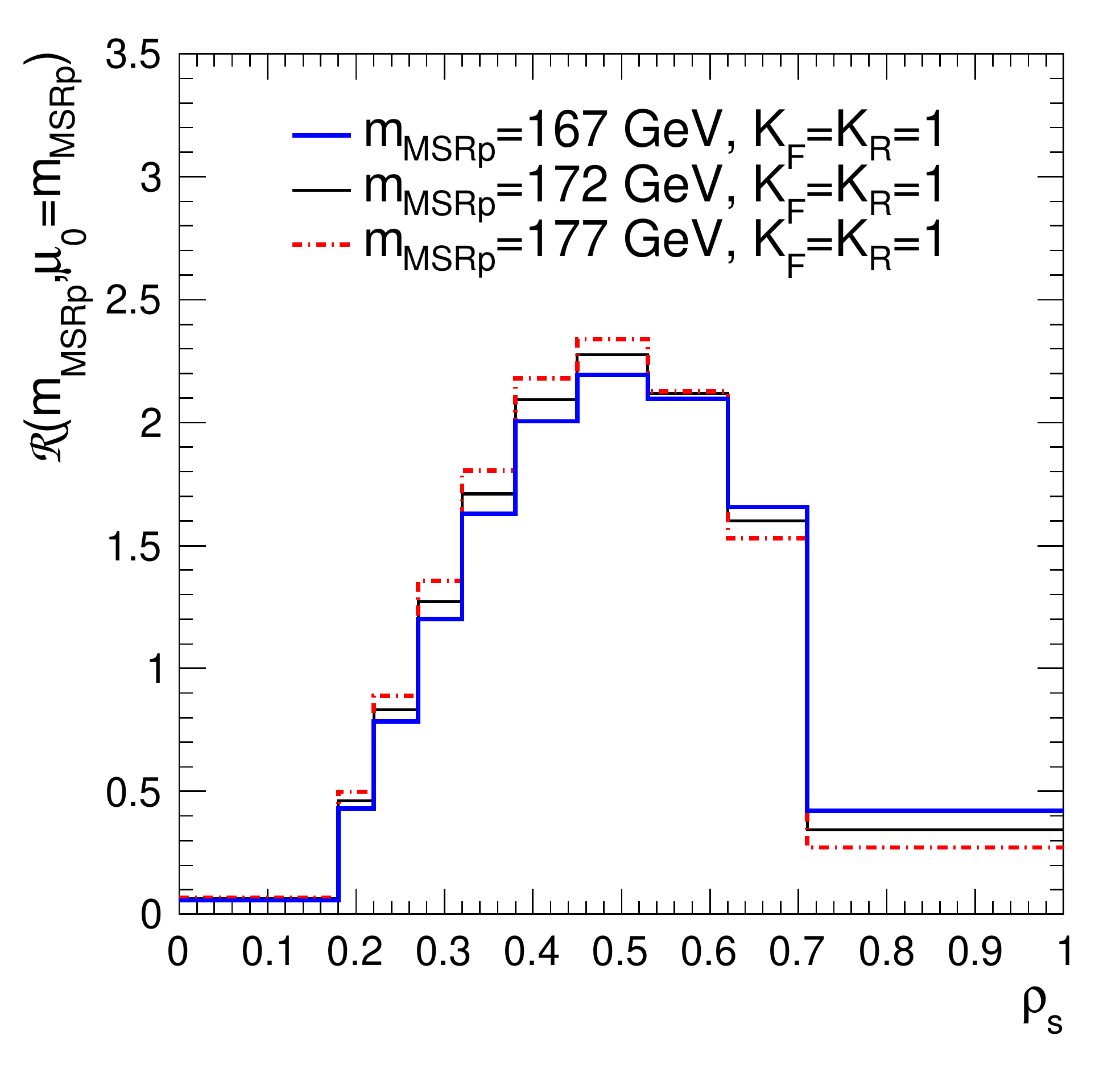}\includegraphics[scale=0.35]{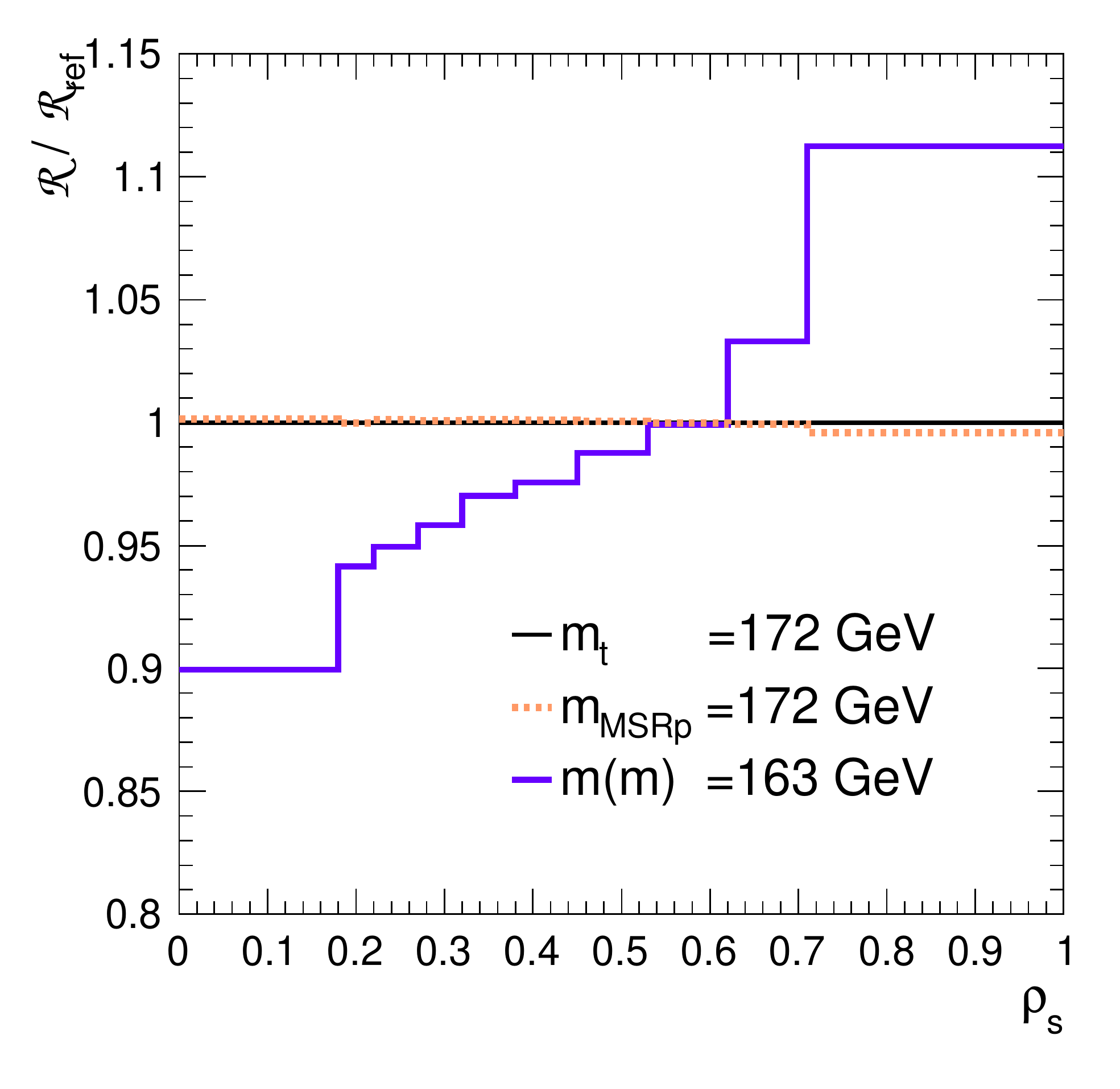}
  \end{center}
  \caption{${\cal R}$ distribution using different re\-nor\-ma\-lization schemes for the
    top-quark mass, i.e. the pole mass $m_t$, the
    ${\overline{\mbox{MS}}}$ mass $m(m)$ and the MSR
    mass ${m_{\mbox{\scriptsize MSRp}}}$. 
    Different lines correspond to different input masses, 
    implying different choices for the renormalization and the factorization
    scales (assumed to be equal).
    In the last panel, ratios of predictions with the
    $m(m)$ and ${m_{\mbox{\scriptsize MSRp}}}$ masses with respect to predictions with the on-shell mass are
    shown. 
    The central CT18NLO PDF set is used as input in all panels.} 
  \label{n3-mass-dependence}
\end{figure}

\begin{figure}[htbp]
  \begin{center}
  \includegraphics[scale=0.35]{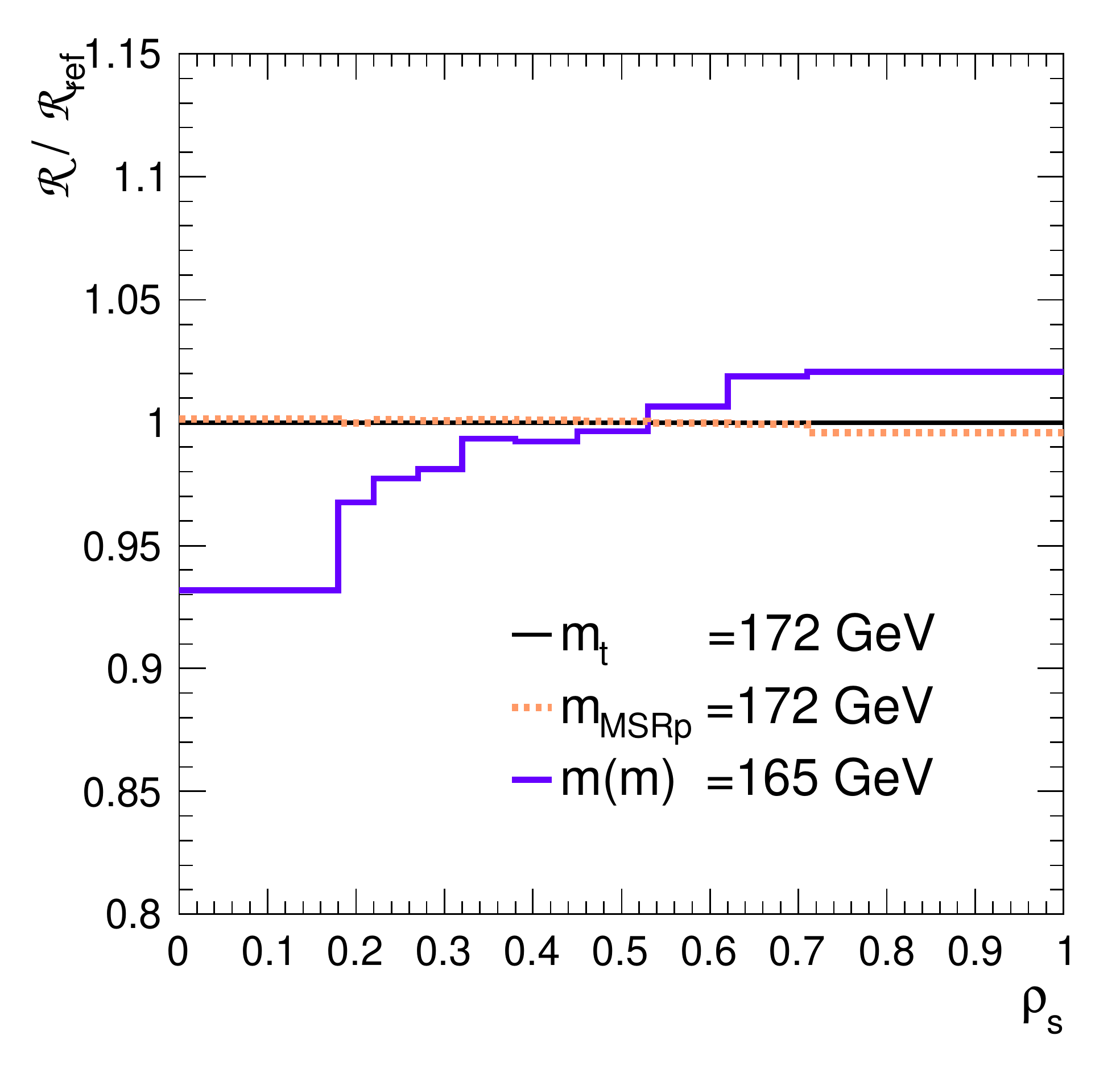}
  \end{center}
  \caption{Ratio of $\mathcal{R}$ in the ${\overline{\mbox{MS}}}$ and MSR scheme to $\mathcal{R}_{\text{\,\, ref}}$ in the pole scheme, similar to the fourth plot in
    \Fig{n3-mass-dependence}, however using $m(m)=165$~GeV instead of 163~GeV.} 
\label{Ratio2}
\end{figure}

\begin{figure}[htbp]
  \begin{center}
    \includegraphics[scale=0.35]{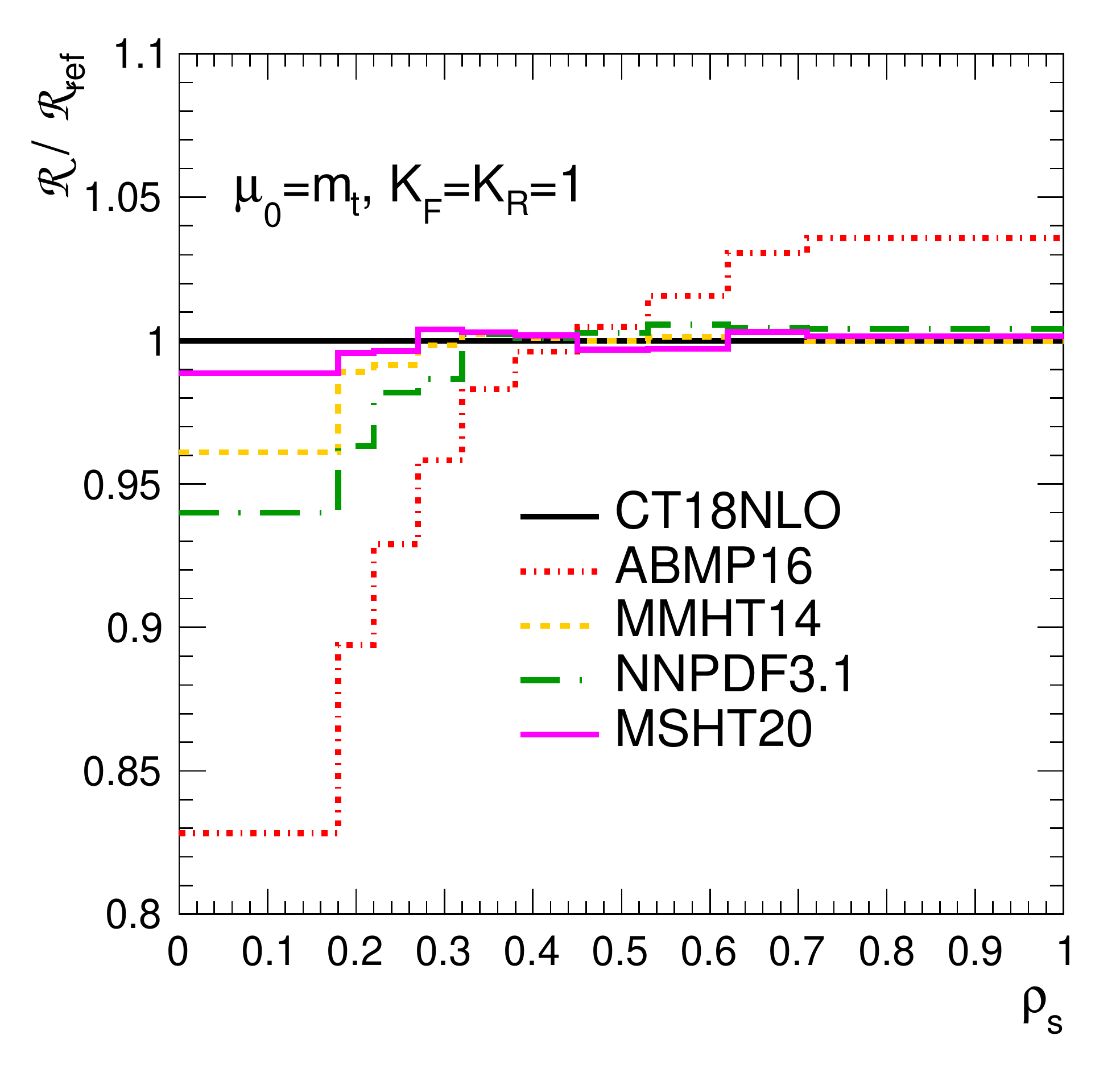}
    \includegraphics[scale=0.35]{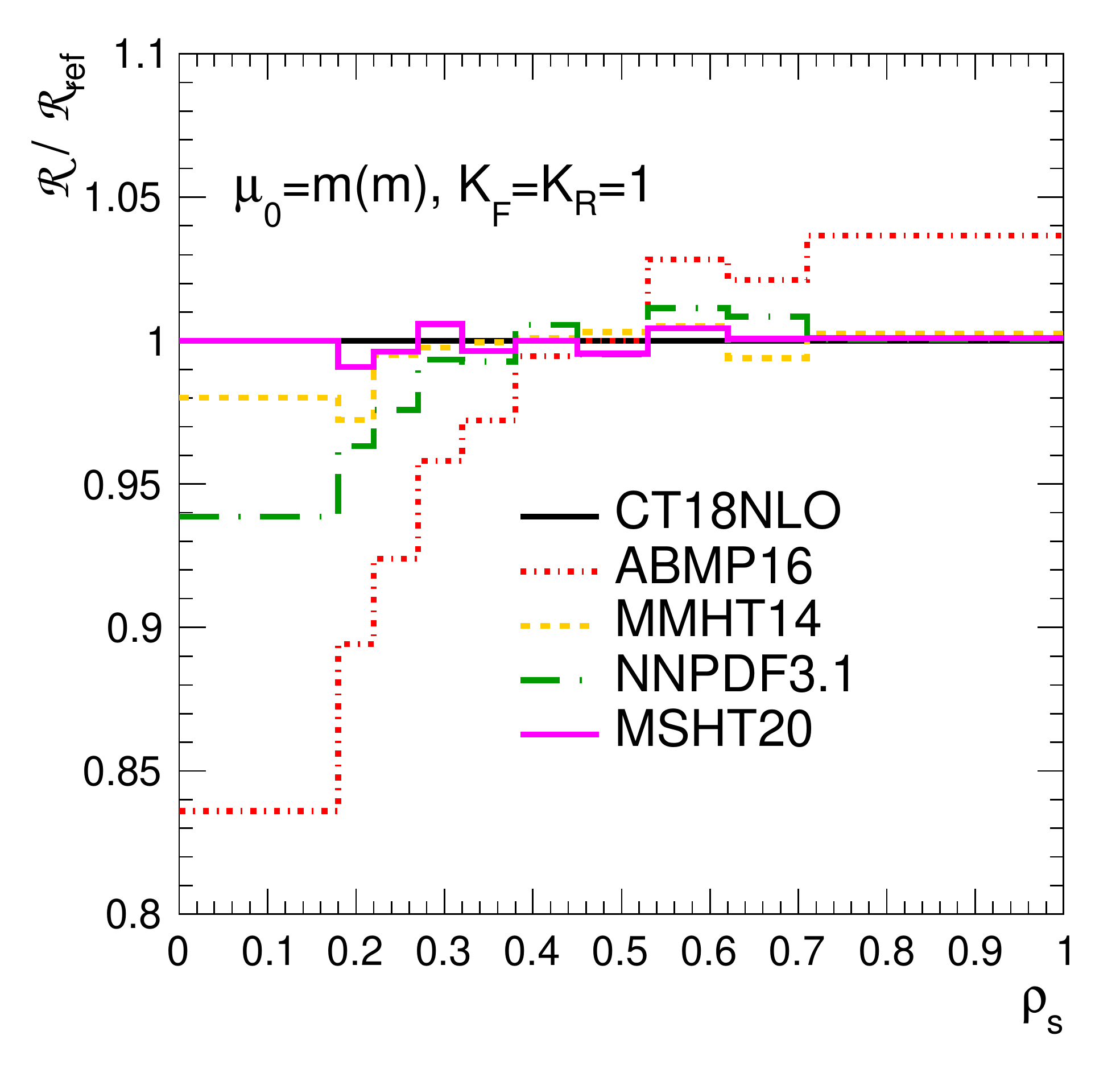}
    \includegraphics[scale=0.35]{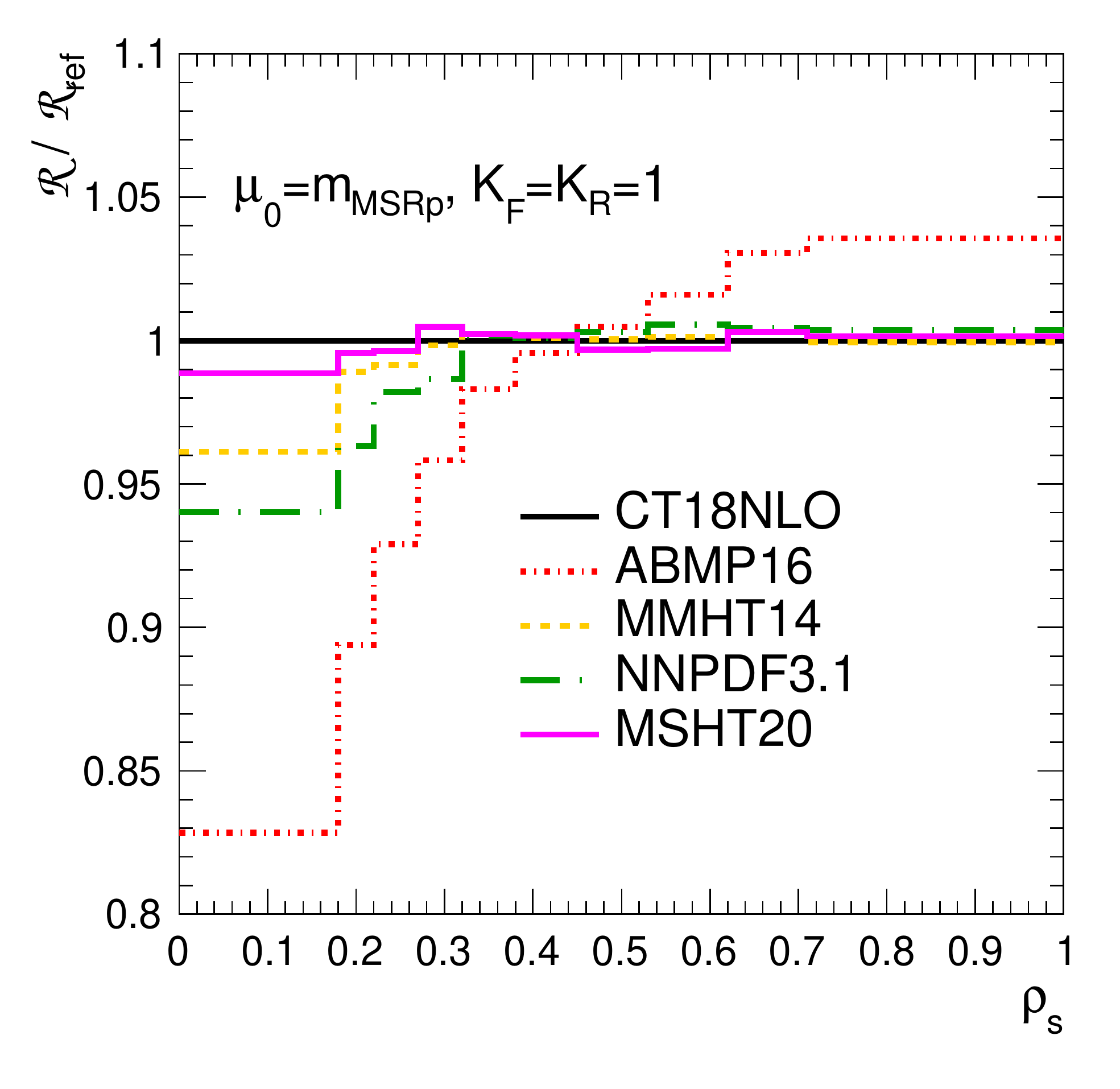}
  \end{center}
  \caption{${\cal R}$ distribution using different PDF sets. We chose again
    $\mu_F=\mu_R=\mu_0$ where $\mu_0$ is set equal to the respective mass
    value.} 
\label{R-PDF-dep}
\end{figure}

The first three panels of \Fig{n3-mass-dependence} show the ${\cal R}$
distribution using different schemes for renormalizing the top-quark mass: the on-shell scheme, the
${\overline{\mbox{MS}}}$ scheme, as well as the MSR scheme.
In each panel, the ${\cal R}$ distribution is depicted for a central
mass value together with the variation by $\pm$~5~GeV. We fix the
renormalization and factorization scale to
$\mu_F = \mu_R = m$, where $m$ is the top-quark
mass in the considered scheme.  The central
mass values in the three different schemes are chosen such
that they roughly correspond to each other after doing the
conversion at 4-loop accuracy. As we used a step size of $1$~GeV for the masses in the
calculation, this correspondence is obviously not exact.  Although
small, the effect of mass variation on the ${\cal R}$ distribution
is clearly visible within each of the three considered schemes.
In the fourth plot in \Fig{n3-mass-dependence}, we show
the predictions using the ${\overline{\mbox{MS}}}$ and MSR
masses divided by the predictions obtained in the pole mass scheme. The MSR masses were computed at a scale $R=3\,$GeV.
Note that because of
the aforementioned step size of $1$~GeV we do not expect perfect agreement. The
predictions using 
either the pole mass or the MSR mass agree with each other. However, 
a sizeable difference is observed using the ${\overline{\mbox{MS}}}$ mass.
This is a consequence of the large impact of the NNLO and N$^3$LO corrections
in the conversion formulas. Using an ${\overline{\mbox{MS}}}$ mass value of 165~GeV, which is close to
the value obtained using the one-loop conversion formula,
the agreement is much better, as expected. 
The corresponding ratio plot is shown in \Fig{Ratio2}. 
In \Fig{R-PDF-dep} we show the ${\cal R}$ distribution for different PDF choices  divided by the results for the CT18NLO set. Apart from the
first bin, where the differences are larger, the CT18NLO, MMHT14-nlo68cl, MSHT20nlo\_as118 and NNPDF3.1\_nlo\_as\_0118 central PDF sets give similar results at the level of a few percent. We note that in all results shown in \Fig{R-PDF-dep} the $\alpha_s(M_Z)$ value and evolution is used as provided by the PDF set. 
The differences in the first bin which
corresponds to high energetic events and is thus sensitive to the large
$x$ values of the PDFs is irrelevant for practical applications due to
the small number of high energetic events. The {ABMP16\_5\_nlo} PDF set shows slightly larger
deviations reaching up to 5\% if one excludes the three highest
energetic bins ($\rho_s<0.3$).
\begin{figure}[htbp]
  \begin{center}
    \includegraphics[scale=0.35]{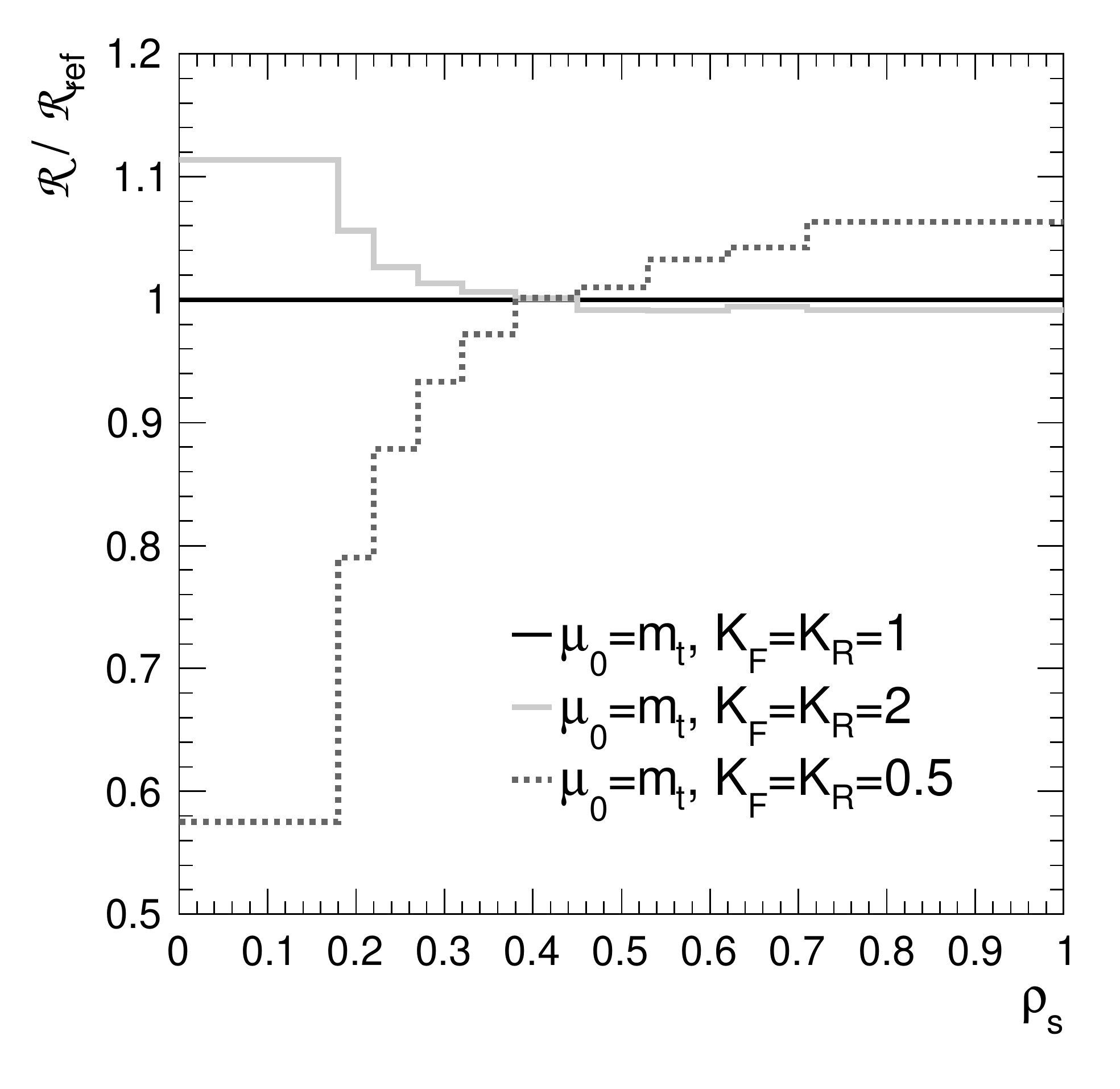}
    \includegraphics[scale=0.35]{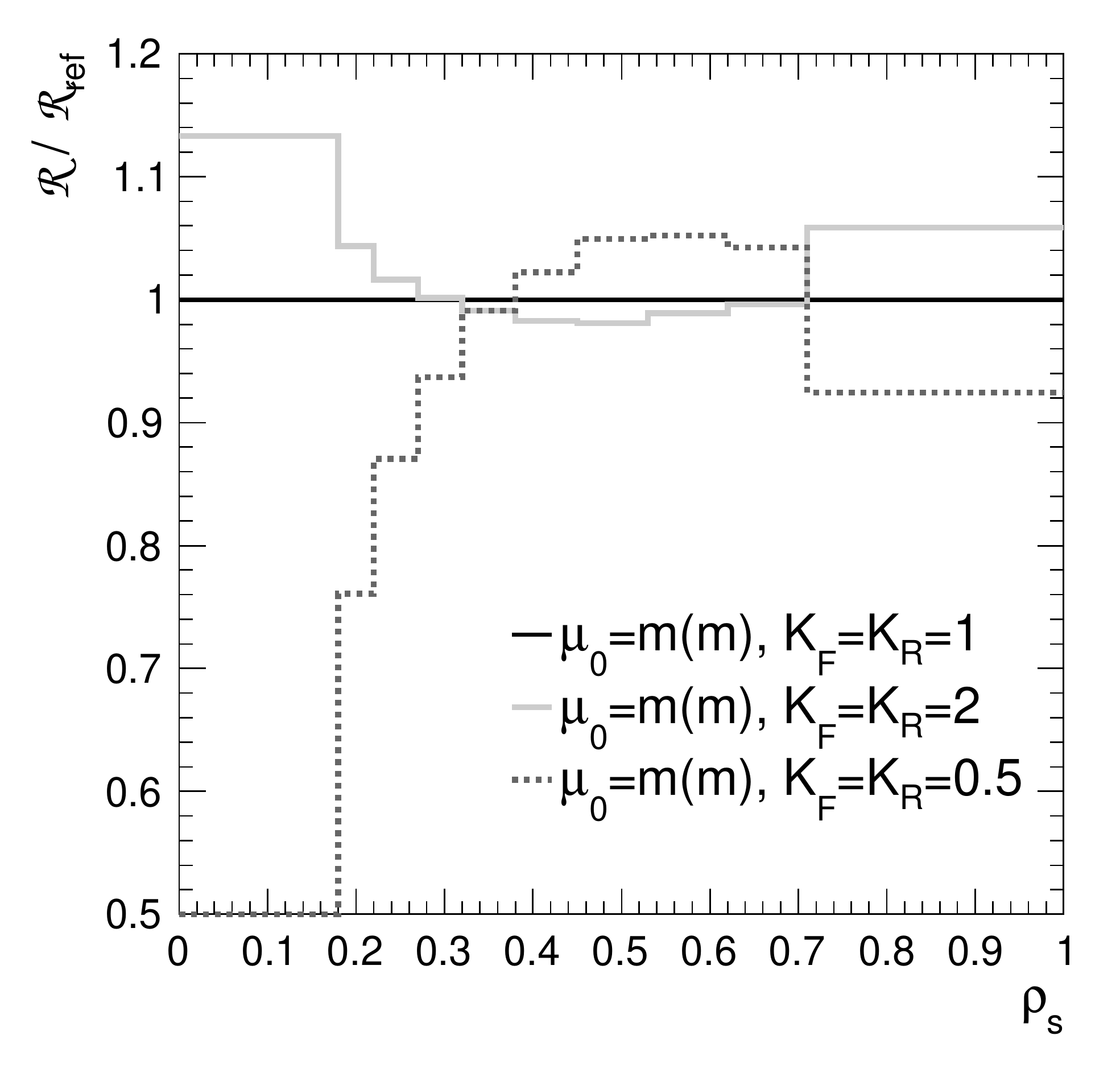}
    \includegraphics[scale=0.35]{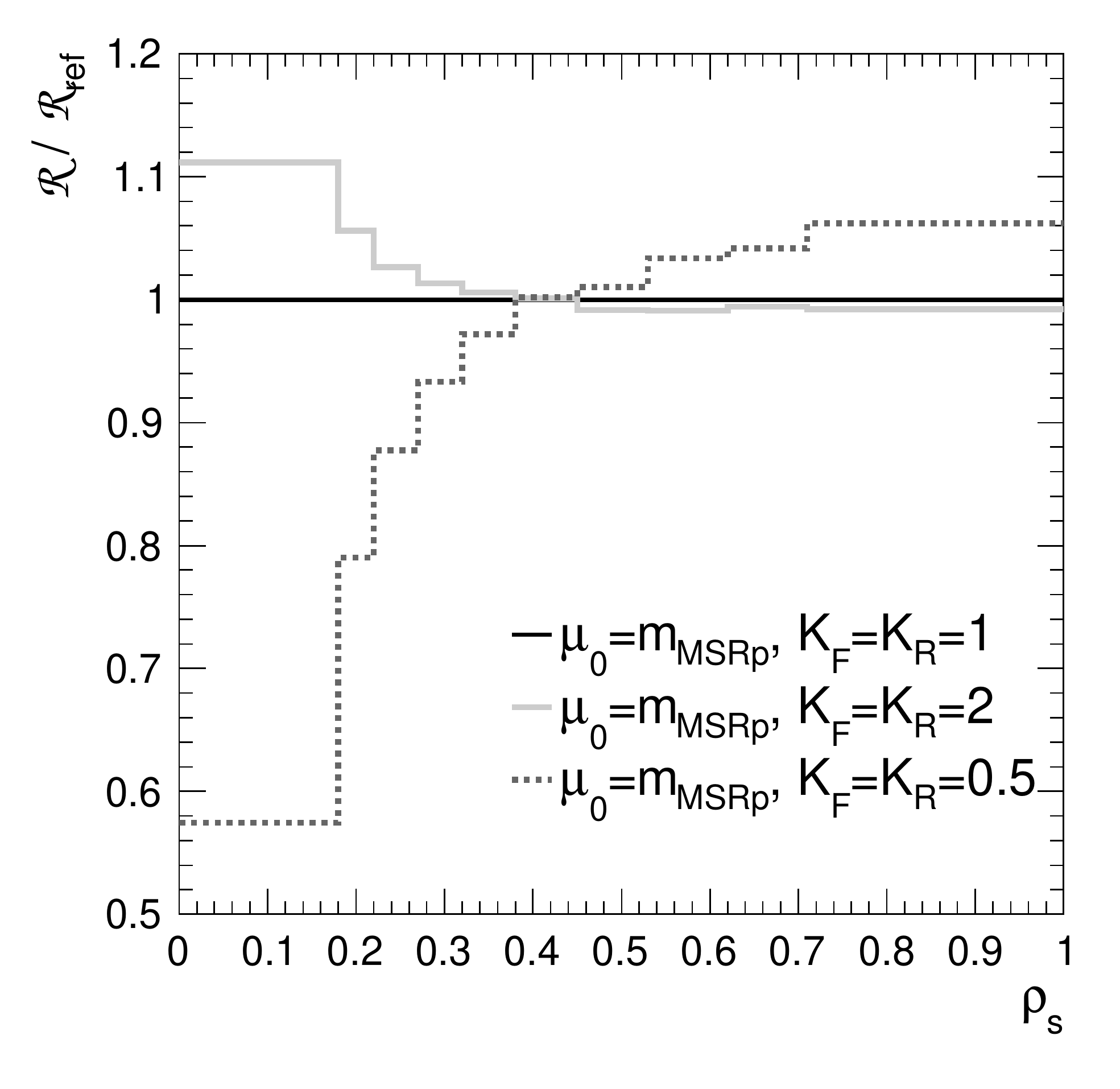}
  \end{center}
  \caption{Ratios of ${\cal R}$ distributions using different scale choices,
    i.e. $\mu_R = K_R\,\mu_0$,  $\mu_F = K_F\,\mu_0$ with 
    $\mu_0 = m$ and $K_F = K_R = 0.5$, 1, 2. Different panels refer to
    different renormalization schemes for the top-quark mass, i.e. the
    pole scheme (upper left), the ${\overline{\mbox{MS}}}$ scheme (upper right) and the MSR one (lower). }
\label{R-scale-dep}
\end{figure}
In \Fig{R-scale-dep} we show the scale dependence of the
${\cal R}$ distribution using different renorma\-li\-zation schemes for the top-quark mass. We used $\mu_F=\mu_R=\mu_0$ together with $\mu_0 = m$, $m/2$, $2m$,
where $m$ denotes the respective mass. For the pole mass and the MSR
mass the behavior is very similar. A major difference is only
observed in the highest energetic bin, where the MSR mass leads to a
larger scale dependence of the ${\cal R}$ distribution.
Again this is most likely of no practical relevance.
In case of the MSR mass one could use a larger value for the $R$-scale
bringing the MSR mass closer to the ${\overline{\mbox{MS}}}$  mass. However, the
corresponding plot for the ${\overline{\mbox{MS}}}$ mass suggests that in this case the
scale variation would become even larger.
In the range $0.2 < \rho_s < 0.7$ the scale variation band using the ${\overline{\mbox{MS}}}$  mass is similar to the one obtained by working in the pole-
and the MSR-scheme.
However, in the bin close to the threshold one obtains
very large scale variation effects using the ${\overline{\mbox{MS}}}$ mass. The origin of this effect is the strong mass dependence in the threshold region which leads to large contribution when using \Eq{Master1} to convert from the pole mass to the ${\overline{\mbox{MS}}}$ mass. 
\begin{figure}[htbp]
  \begin{center}
    \includegraphics[scale=0.35]{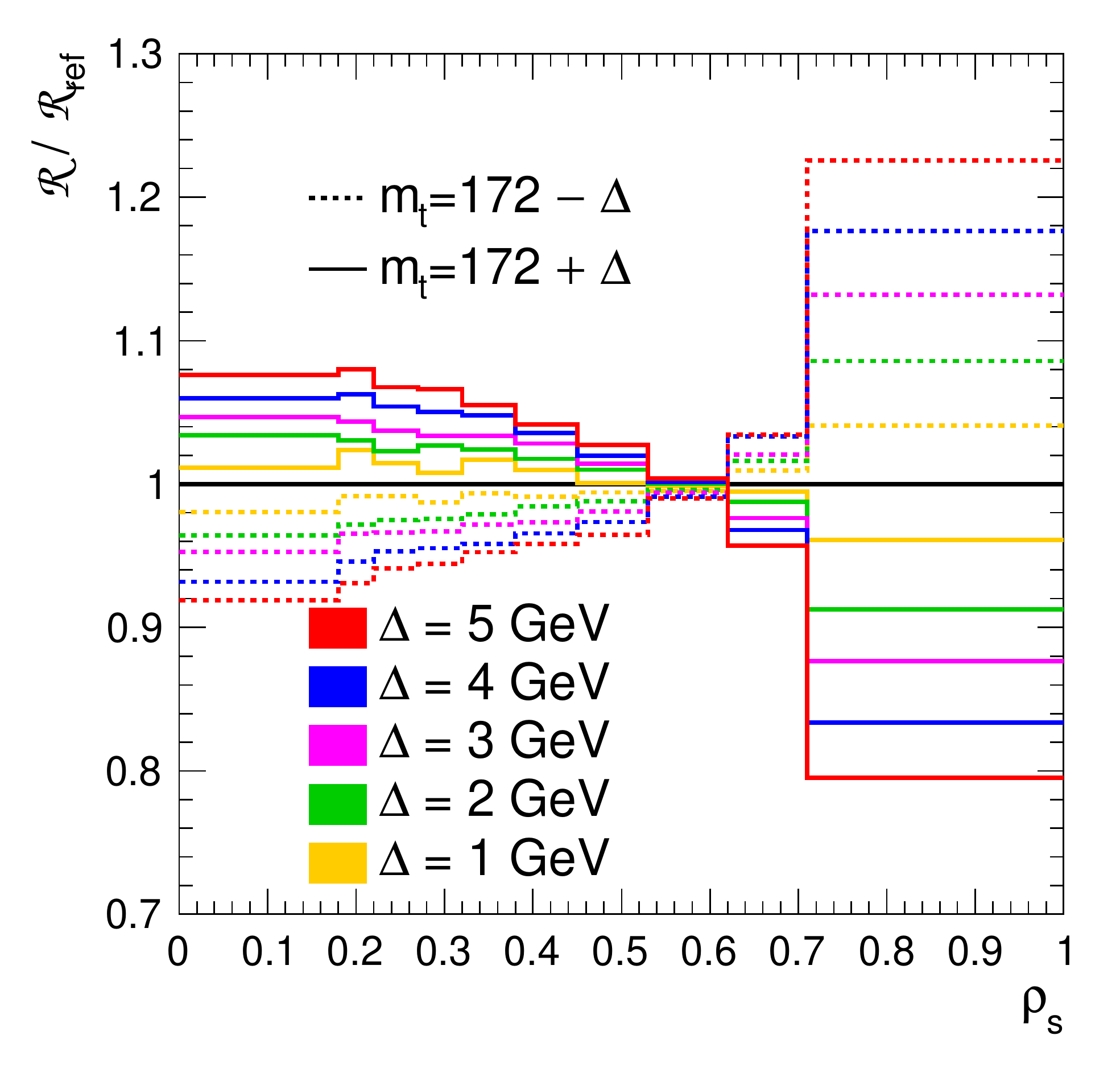}
    \includegraphics[scale=0.35]{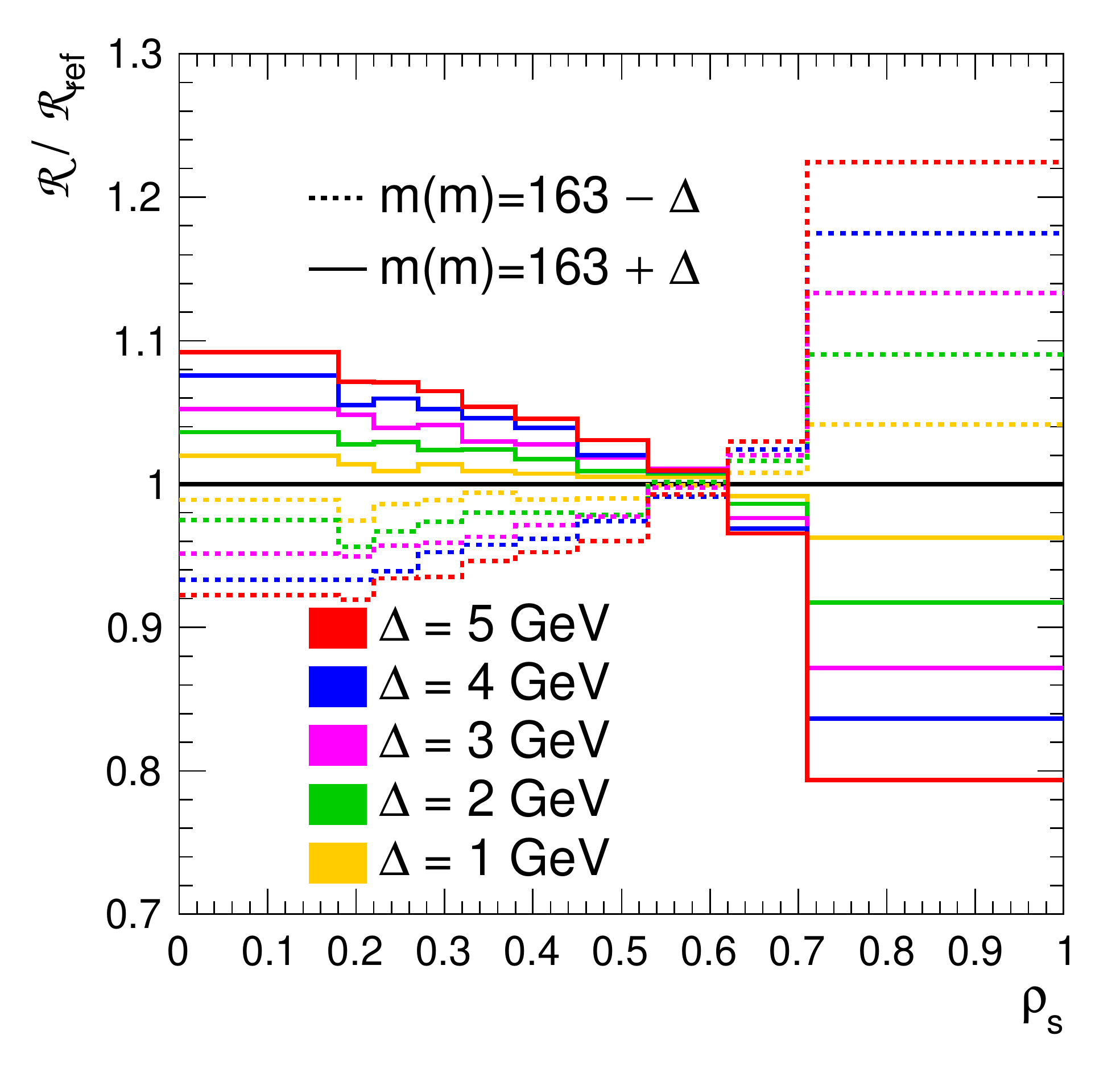}
    \includegraphics[scale=0.35]{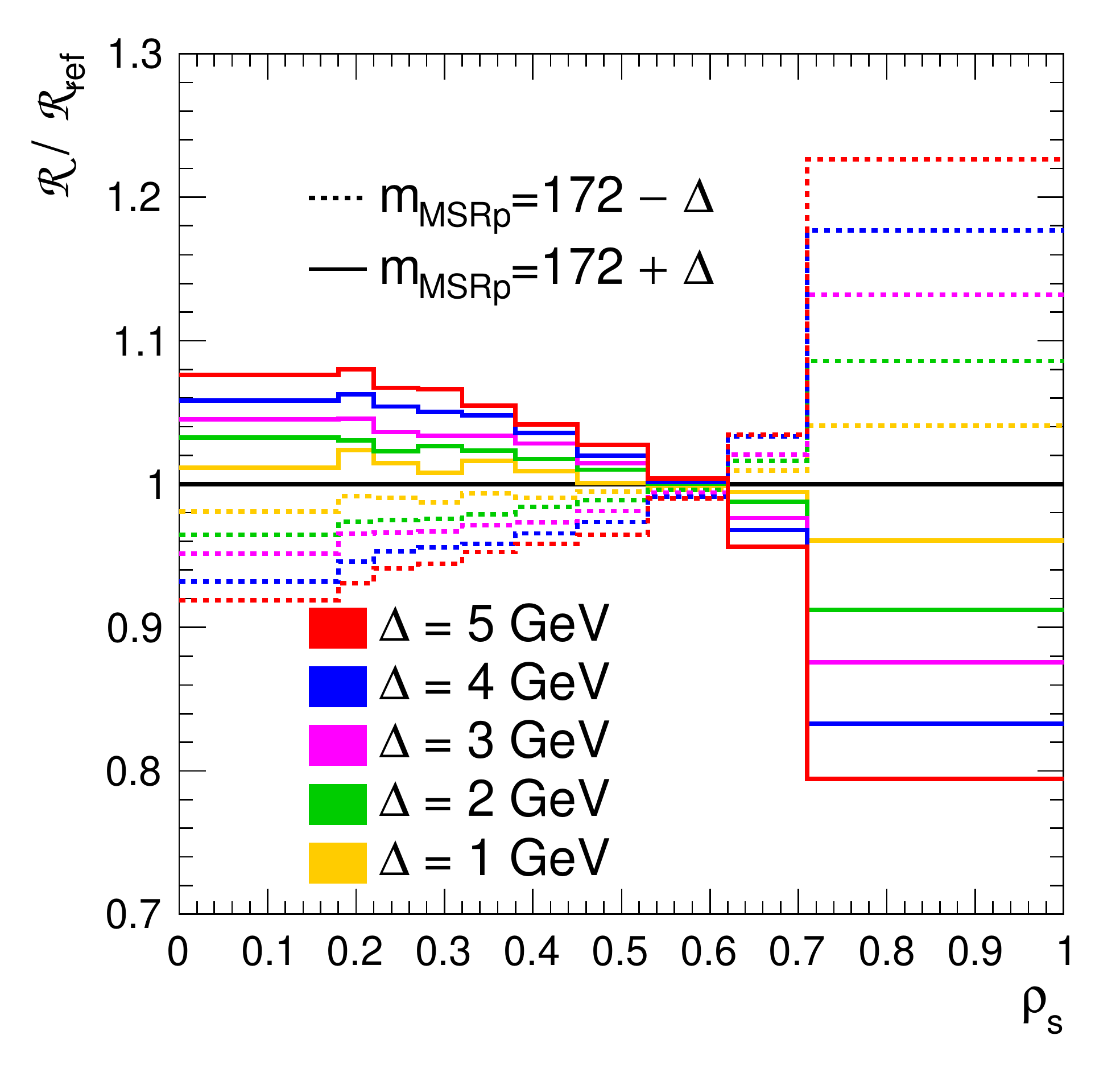}
  \end{center}
  \caption{Ratios of ${\cal R}$~distributions using as input different 
  top-quark mass values and the ${\cal R}$~distribution obtained with a fixed central top-quark mass value (assumed to be equal to 173~GeV in
all renormalization schemes). Different panels refer to different renormalization
schemes for the top-quark mass.}
\label{LinearDep}
\end{figure}
\Fig{LinearDep} is similar to \Fig{n3-mass-dependence} using however a step size of $1$~GeV instead of $5$~GeV. 
From the approximately constant step size one can see that in all bins the mass dependence
is to good approximation linear, allowing a linear interpolation when performing a $\chi^2$ fit to determine the top-quark mass from the measured distributions.  


\subsection{Off-shell effects and non-resonant/non-factorizable contributions}
As mentioned already, the results presented in previous Sections 
can only be compared with data which have been unfolded to
the \textit{parton level}. The unfolding procedure accounts for
effects due to the top-quark decay, additional gluon radiation from
top-quark decay products as well as hadronization and experimental
detector effects. Unfolding is a well established method and has been
applied in the past to a variety of different observables. 
Obviously the unfolding procedure introduces additional uncertainties which need to be taken into account.
In \Ref{Bevilacqua:2017ipv} an alternative approach has been
investigated. Instead of unfolding the measured data to the level of
the intermediate top quarks, the observable is computed using the
momenta of the final state particles in
$pp \to e^+\nu_e \, \mu^- \bar\nu_\mu \, b\bar b j+X$ events, which is a typical
final state for ${\ensuremath{t\bar t j + X}\xspace}$ assuming that both the top and antitop quarks decay
leptonically.  In principle, this allows to incorporate some of the
aforementioned effects like the top-quark decay or radiation from
top-quark decay products into the theoretical prediction instead of
modeling them with Monte Carlo (MC) programs. In particular,
spin effects like for example spin correlations are automatically taken into account and passed on to the top-quark decay products.
To compare with the results of \Ref{Bevilacqua:2017ipv}
we consider \textit{parton level with decay} of the top quarks instead of using the intermediate top quarks to define the final state.

An expected advantage of using the results of \Ref{Bevilacqua:2017ipv} is that unfolding uncertainties should be reduced as less modeling is introduced in the partonic description.
The calculation in
\Ref{Bevilacqua:2017ipv} was performed at NLO accuracy in QCD keeping
not only finite width effects for the top quark but also the
corresponding effects for the intermediate $W$-bosons. Furthermore, non-resonant/non-factorizable
contributions were also included in the prediction. The theoretical
description thus includes many physical effects. 
In the following we will refer to it as \textit{full calculation} (\textit{Full NLO}).
However, this calculation has not yet been matched to a parton shower, therefore shower effects are not 	presently accounted for, and hadronization and further soft physics effects are also absent. 
In addition, two
further approximations to model off-shell effects are studied in
\Ref{Bevilacqua:2017ipv}. First,
the approximation in which the
two decay chains
\begin{displaymath}
  pp\to{\ensuremath{t\bar t j + X}\xspace}\to e^+\nu_e \mu^- \bar\nu_\mu b\bar b j+X,
  \quad pp\to t\bar{t} \to e^+\nu_e \mu^- \bar\nu_\mu b\bar b j+X
\end{displaymath}
are calculated in the narrow width approximation,
called \textit{NWA} in \Ref{Bevilacqua:2017ipv}.
The production cross sections as well as the decay is calculated at NLO accuracy. This approach thus includes spin effects while 
non-resonant/non-factorizable contributions are neglected.
Please note that one should not naively think that these chains are calculated separately and added together, as this would lead to a double counting. Ref.~\cite{Melnikov:2011qx}, on which the discussion in Ref.~\cite{Bevilacqua:2017ipv} relies, first classifies the corrections depending on whether the additional jet is radiated in the top-quark pair production phase or from the top-quark decay products. Then it incorporates radiative corrections in a consistent way. While some contributions may be attributed to either one or the other aforementioned production chain, considered as standalone, Ref.~\cite{Melnikov:2011qx} also includes mixed contributions leading to a much more complex picture than adding the chains would suggest, avoiding double counting without the need of a merging prescription.
In a
second approximation which is called $\mbox{\textit{NWA}}_{Prod}$ in
\Ref{Bevilacqua:2017ipv}, the process $pp\to{\ensuremath{t\bar t j + X}\xspace}$ at NLO QCD accuracy is combined with the leading-order decay of the top quarks
following the approach presented in \Ref{Melnikov:2010iu}.
It should be stressed that neither the \textit{full} calculation nor the two
approximations include parton-shower effects.  In
\Ref{Bevilacqua:2017ipv} it has been argued, that employing the
aforementioned different theoretical prescriptions can lead to
significant differences in the extracted mass values. However, the interpretation of these
results need further clarifications and the significance of the observed mass shift
remains unclear as no uncertainty was
calculated for it. Furthermore, in the analysis different scale settings were
applied. While for the full calculation a dynamical scale ($H_T$) was used, a static scale with 
	$\mu_R=\mu_F=m_t$ was used for the $NWA$ and $NWA_{Prod}$ approximations. It is well known
\cite{Alioli:2013mxa,Denner:2012yc,ATLAS:2015pfy} that scale
variations can lead to sizable effects on the extracted top-quark
mass. In addition, the findings of \Ref{Bevilacqua:2017ipv}
are not directly applicable to the experimental analyses 
as the experimental approach taken by
ATLAS and CMS is rather different from the procedure applied in
\Ref{Bevilacqua:2017ipv}. In the experimental analysis the un\-fol\-ding is done to the level of stable tops. Gluon emission from top-quark decay products is thus accounted for through the unfolding. In this work we argue that in fact the bulk of
the effects leading to the aforementioned shift is,  
within the uncertainties associated to the modeling of the partonic sub-processes and the hadronization,
well described by MC tools, and is thus taken into account doing a proper
unfolding as it is done in the analyses performed by ATLAS and CMS. To show
this explicitly, the results of \Ref{Bevilacqua:2017ipv} are compared
with a Monte Carlo simulation including NLO QCD corrections in the
production. More precisely, events for the process
$pp\rightarrow t\bar{t}j \rightarrow W^+W^-b\bar{b}j$ are generated through
a MC simulation, using the parameters shown in \Tab{table:MCsettings}
matching those used in \Ref{Bevilacqua:2017ipv}.
\begin{table}[ht]\centering
\begin{tabular}{| c | c |}
\hline
 Parameter & Value \\
\hline
$\alpha_s(M_Z)$ & 0.118 \\
$G_\mu$ &  $1.6637 \times 10^{-5}~{\mbox{GeV}}^{-2}$ \\
$M_Z$ & 91.1876~{\mbox{GeV}} \\
$M_W$ & 80.3990~{\mbox{GeV}} \\
$\Gamma_Z$ & 2.50848~{\mbox{GeV}} \\
$\Gamma_W$ & 2.09875~{\mbox{GeV}} \\
 \hline
\end{tabular}
\caption{Parameters used in the Monte Carlo generation of
  $pp\rightarrow t\bar t j\rightarrow W^+ W^- b\bar{b} j$.  The
  electromagnetic coupling  $\alpha$ and the weak mixing angle
  $\theta_W$ are computed in the $G_\mu$ scheme using
  $\sin^2\theta_W = 1 - \frac{M_W^2}{M_Z^2}$ and
  $\alpha=\frac{\sqrt{2}}{\pi} G_\mu \frac{M_W^2}{M_Z^2} \left(M_Z^2 -
    M_W^2\right)$, where $M_Z$ and $M_W$ are the gauge boson masses
  and $G_\mu$ denotes the Fermi constant.
  The decay widths of the massive
  gauge bosons are defined as $\Gamma_W$ and $\Gamma_Z$.  }
\label{table:MCsettings}
\end{table}
Top quarks are produced on-shell in association with an extra jet at
NLO in QCD, using the $ttj$ package in the ${\textsc{Powheg-Box}\xspace}$ \cite{Alioli:2010xd,
  Frixione:2007vw, Nason:2004rx, Alioli:2011as}, with the CT14nlo
PDFs \cite{Dulat:2015mca} and the
$\mu_F$ and $\mu_R$ scales fixed to
the value of the top-quark mass $m_t$, renormalized in the on-shell scheme. 
The ${\textsc{Powheg-Box}\xspace}$ code is interfaced to
the {\textsc{Pythia}\xspace}8 \cite{Sjostrand:2006za} code which simulates the parton showering beyond first radiation emission and the top-quark decay. Spin correlations could be included via the ${\textsc{Powheg-Box}\xspace}$. However, in this case also the decay of the top quarks would be handled by the ${\textsc{Powheg-Box}\xspace}$. 
As the distribution we are interested in is not particularly sensitive to the decay, we simplify the 
approach by allowing {\textsc{Pythia}\xspace}8 to do the top-quark decay, without any specific optimization or mo\-di\-fi\-cation of its decay routines.
The top quark is decayed into an on-shell $W$-boson and a $b$-quark,
with decay width $\Gamma_t^{\text{on-shell}}=1.37~{\mbox{GeV}}$ for a top-quark pole mass value amounting to $m_t=173.2~{\mbox{GeV}}$ \cite{Jezabek:1988iv, Denner:2012yc}.  The lighter quarks are considered massless.  Only leptonic decays of the $W$-bosons are included in the simulation. No special settings have been used in the simulation with {\textsc{Pythia}\xspace}. The leptons are further restricted to electrons and muons, i.e. $W$-bosons decaying into $\tau$ leptons are
excluded. The MC
simulation is stopped before hadronization occurs and the
objects entering the analysis as well as the
fiducial volume are defined as in
Ref.~\cite{Bevilacqua:2017ipv}, in order to make the comparison
between the two approaches as close as possible.

\begin{figure}
\centering
\includegraphics[width=0.49\textwidth]{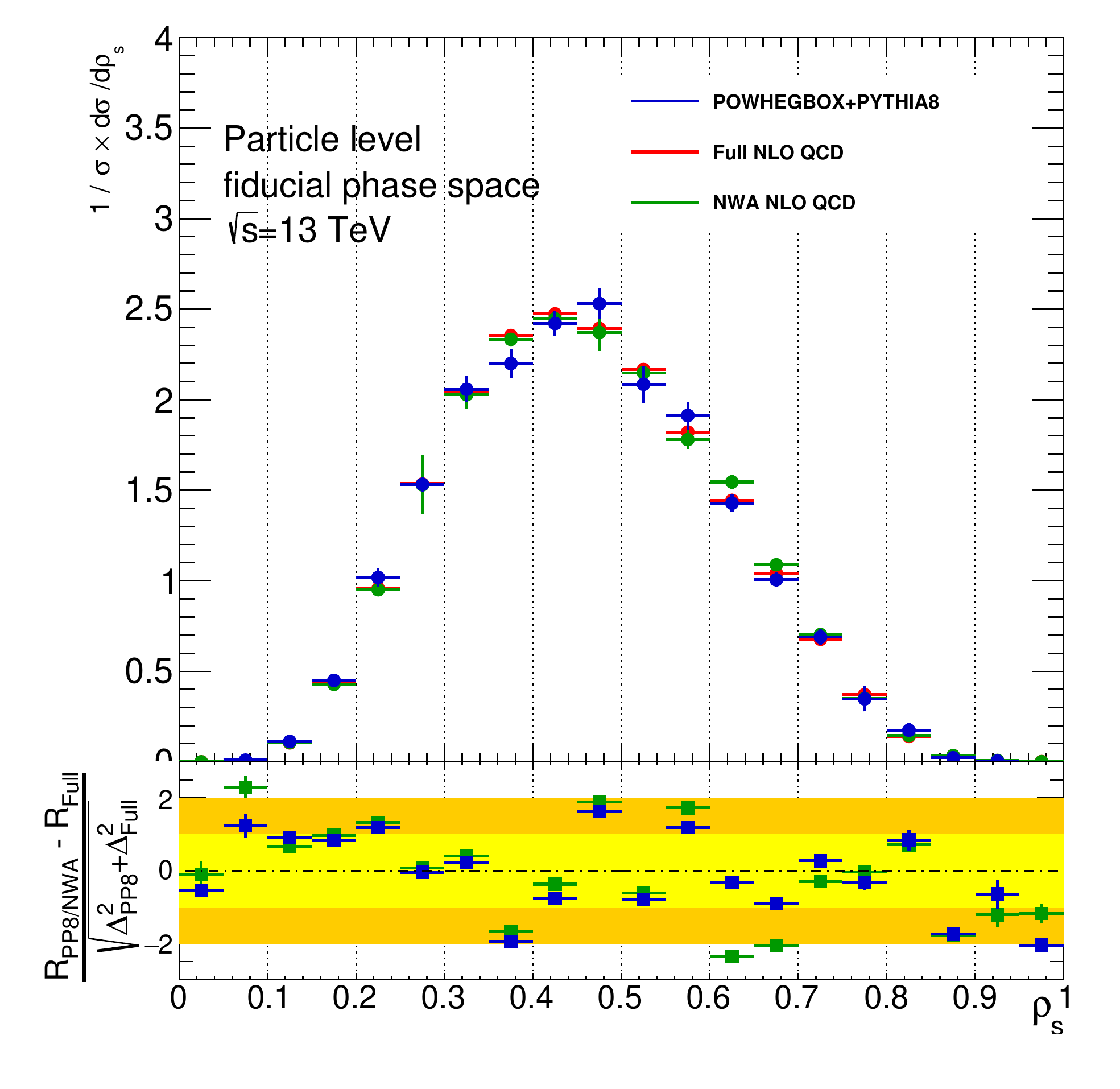}
\includegraphics[width=0.49\textwidth]{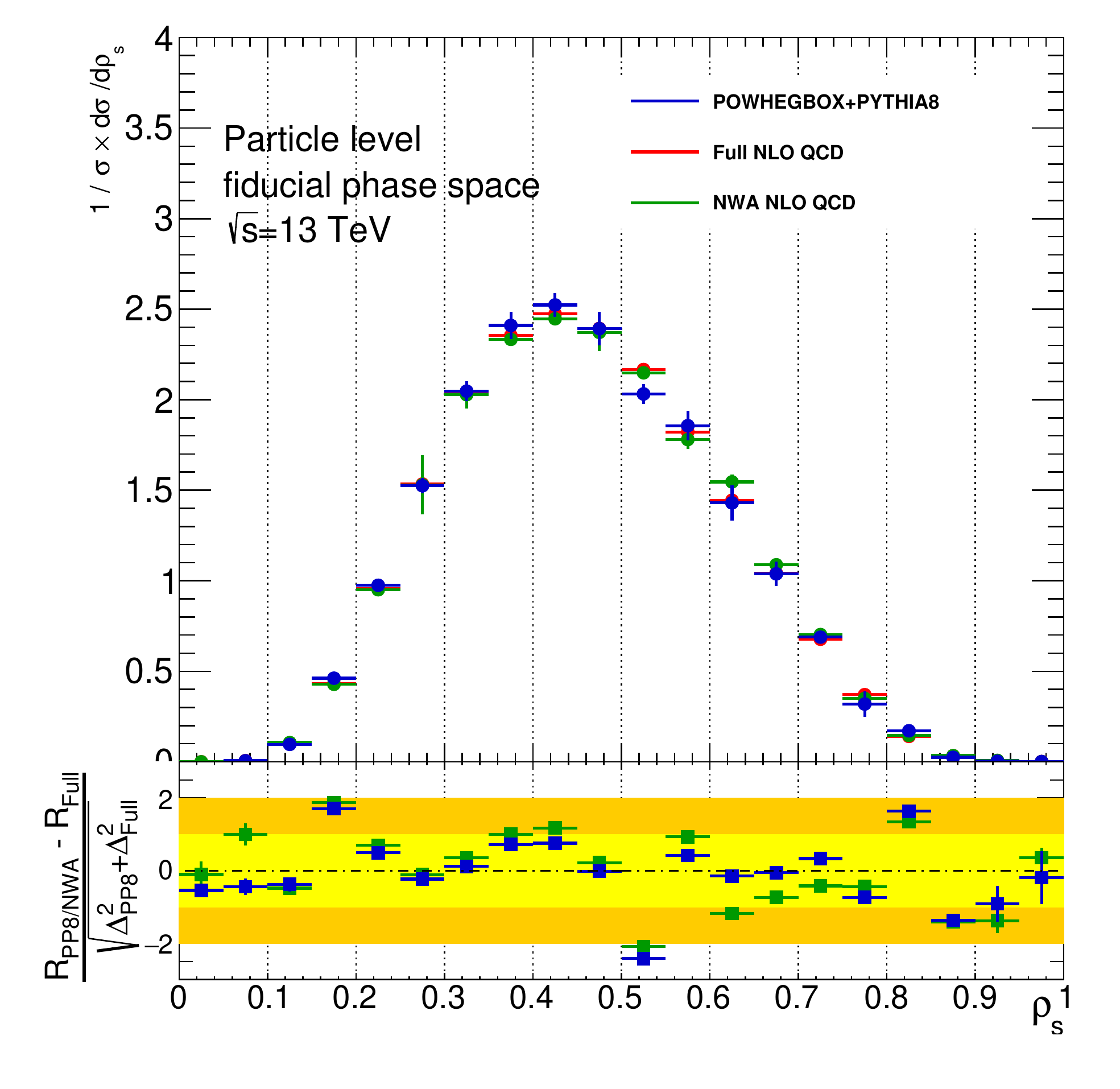}
\caption{Comparison between the ${\cal R}$ distributions computed using
  ${\textsc{Powheg-Box}\xspace}$~{+}~{\textsc{Pythia}\xspace}8 (blue), the calculation including off-shell
  and non-resonant contributions as well as finite width
  effects (\textit{Full NLO}) \cite{Bevilacqua:2017ipv} (red) and the calculation in the NWA \cite{Bevilacqua:2017ipv} (green).
  Error bars refer to  statistical uncertainties of the MC simulation.
  The lower plots show the differences between predictions of the
   ${\textsc{Powheg-Box}\xspace}$~{+}~{\textsc{Pythia}\xspace}8 or NWA
   and those of the full
  calculation, 
  normalized to the statistical uncertainty. The yellow
  (orange) area represents the 1$\sigma$ (2$\sigma$) band, where the standard deviation is calculated as the quadratic sum of the uncertainties of the two calculations, the statistical uncertainty  of ${\textsc{Powheg-Box}\xspace}$~{+}~{\textsc{Pythia}\xspace}8 prediction due to a finite number of simulated events, and the uncertainty from the numerical MC integration of the \textit{Full NLO} calculation. The uncertainties of the   {\textsc{Powheg-Box}\xspace}~{+}~{\textsc{Pythia}\xspace}8 computation dominates the total uncertainty.
    The  distribution on the left (right) is computed setting the parameter
  \texttt{hdamp} of {\textsc{Powheg-Box}\xspace} to $m_t$ (infinity).  }
\label{fig:Robs_compare}
\end{figure}

All the final-state partons with pseudorapidity $|\eta|<5$ are recombined
into jets using the \mbox{anti-$k_T$} algorithm \cite{Cacciari:2008gp} with
radius parameter $R=0.5$.  Jets are required to have a transverse
momentum $p_{T}^j$ larger than 40~{\mbox{GeV}}, absolute pseudorapidity $|\eta_j|$ smaller
than 2.5, and an angular separation satisfying
$\Delta R_{j_1j_2} = \sqrt{ (\phi_{j_1}-\phi_{j_2})^2 +
  (\eta_{j_1}-\eta_{j_2})^2} >0.5$.  A jet is called $b$-jet if it
originated from a $b$-quark.  Charged leptons are defined by
recombining electrons or muons with photons via the
\mbox{anti-$k_T$} algorithm
with $R=0.1$. Only leptons with $p_T^\ell>30~{\mbox{GeV}}$ and $|\eta| <2.5$
are considered.  Charged leptons are also required to be well
separated from other leptons ($\Delta R_{\ell\ell}>0.4$) and from jets
($\Delta R_{\ell j}>0.4$). The four momenta of the neutrinos are summed
together into a missing momentum vector $p^{\text{miss}}$.  Only
events with two $b$-jets, at least one light jet~\footnote{A light jet
  is defined as a jet originated from a $u$-, $d$-,
  $c$-, $s$-quark or a gluon.}, 
two oppositely charged leptons $\ell^+$ and $\ell^-$ and $p_T^{\text{miss}}>40~{\mbox{GeV}}$ are accepted. 
These cuts match precisely the ones used in \Ref{Bevilacqua:2017ipv}.
Events passing these cuts are used to calculate the
${\cal R}$ distribution. The
quantity $s_{t\bar tj}$ is calculated using
\begin{displaymath}
  s_{t\bar tj}=\left(p_{j_{b1}}+p_{j_{b2}}+p_{\ell^-}+p_{\ell^+}+p_{j}
    +p^{\text{miss}}\right)^2.
\end{displaymath}
In \Fig{fig:Robs_compare} the ${\cal R}$ distribution obtained from the MC
simulation (${\cal R}^{{\textsc{Powheg-Box}\xspace}{+}{\textsc{Pythia}\xspace}8}$) is presented and compared to
the result ${\cal R}^{\text{Full}}$ obtained in \Ref{Bevilacqua:2017ipv}.
For these computations {\textsc{Pythia}\xspace}8 (version 8.1.5.0) has been used. Furthermore, the top-quark mass used in {\textsc{Pythia}\xspace}8 is set equal to the mass 
used in ${\textsc{Powheg-Box}\xspace}$.
Both distributions have been generated using the same top-quark
mass $m_t=173.2~{\mbox{GeV}}$. In the
{\textsc{Powheg-Box}\xspace}~{+}~{\textsc{Pythia}\xspace}8
computation we set the factorization and renormalization
scale to $\mu_R=\mu_F=m_t$ . From the differences in terms of the total
statistical uncertainty (lower plots in \Fig{fig:Robs_compare}) one
concludes that the two calculations give consistent results within
the uncertainty. Already at this point one would expect similar
top-quark mass values treating the simulated events as pseudo-data and
using either \Ref{Bevilacqua:2017ipv} or {\textsc{Powheg-Box}\xspace}~{+}~{\textsc{Pythia}\xspace}8 
to extract the corresponding top-quark mass.
To turn this qualitative statement into a quantitative one, we have set up a $\chi^2$ minimization test to infer the $m_t$ values and quantify the difference using either \Ref{Bevilacqua:2017ipv} or {\textsc{Powheg-Box}\xspace}~{+}~{\textsc{Pythia}\xspace}8 as pseudo-data.
To do so we generated MC event samples using top-quark masses ranging from $165~{\mbox{GeV}}$ to $180~{\mbox{GeV}}$ using again the input parameters quoted in
\Tab{table:MCsettings}. For each mass value the corresponding event
sample is used to determine the ${\cal R}$ distribution. Using a second order
polynomial interpolation in the top-quark mass a continuous
parameterization for each bin is derived. The  binning choice 
follows \Refs{ATLAS:2015pfy} and~\cite{ATLAS:2019guf}. In case of the choices with 5 or 6 bins, the setting is motivated by a detailed experimental analysis done by ATLAS and CMS to optimize the bins with respect to statistical and systematic uncertainties. 
In case of the choices with 10 or 20 bins, a uniform binning is used.
 Comparing the
results obtained using different binnings, serves as check  that the
bin choice does not introduce a bias in the extracted mass value.  
The
top-quark mass is extracted minimizing the quantity:
\begin{equation}
\chi^2 \,=\, 
\sum_{i \in \text{bins}} \frac{\left[{\cal R}_{\,\,\,i}^O -  {\cal R}_{\,\,\,i}(m_t) \right]^2}{\sigma^2[{\cal R}_{\,\,\,i}^O]+\sigma^2[{\cal R}_{\,\,\,i}(m_t)] }
\label{eq:chi2}
\end{equation}
where ${\cal R}_{\,\,\,i}^O$ is one of the approximations calculated in
\Ref{Bevilacqua:2017ipv}. The corresponding  uncertainty is given by $\sigma^2[{\cal R}_{\,\,\,i}^O]$.  
We include an additional uncertainty $\sigma^2[{\cal R}_{\,\,\,i}(m_t)]$ 
for the continuous parametrization. In all quantities the index $i$ labels the respective bin.
\begin{figure}
\centering
\includegraphics[width=0.49\textwidth]{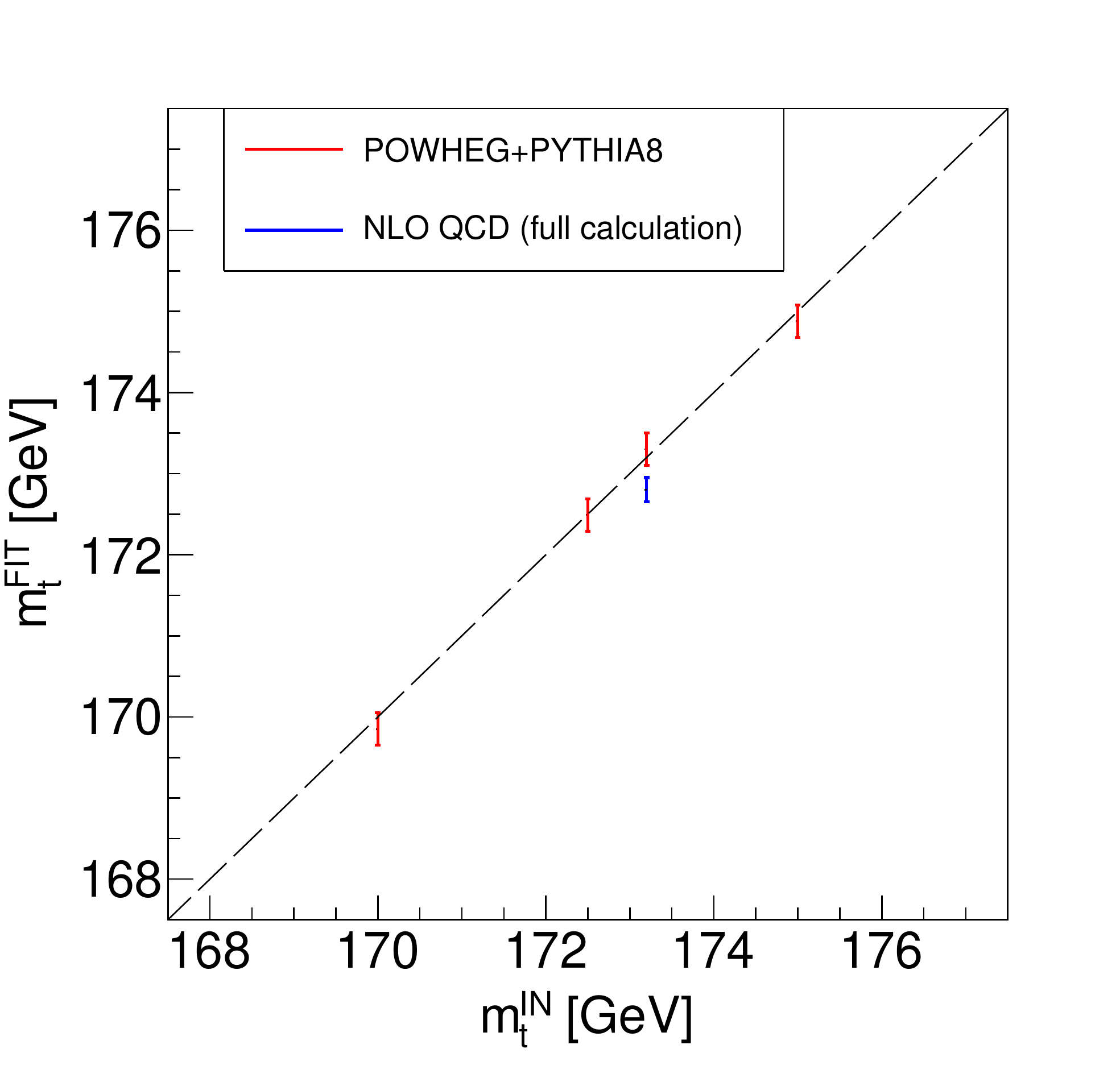}
\includegraphics[width=0.49\textwidth]{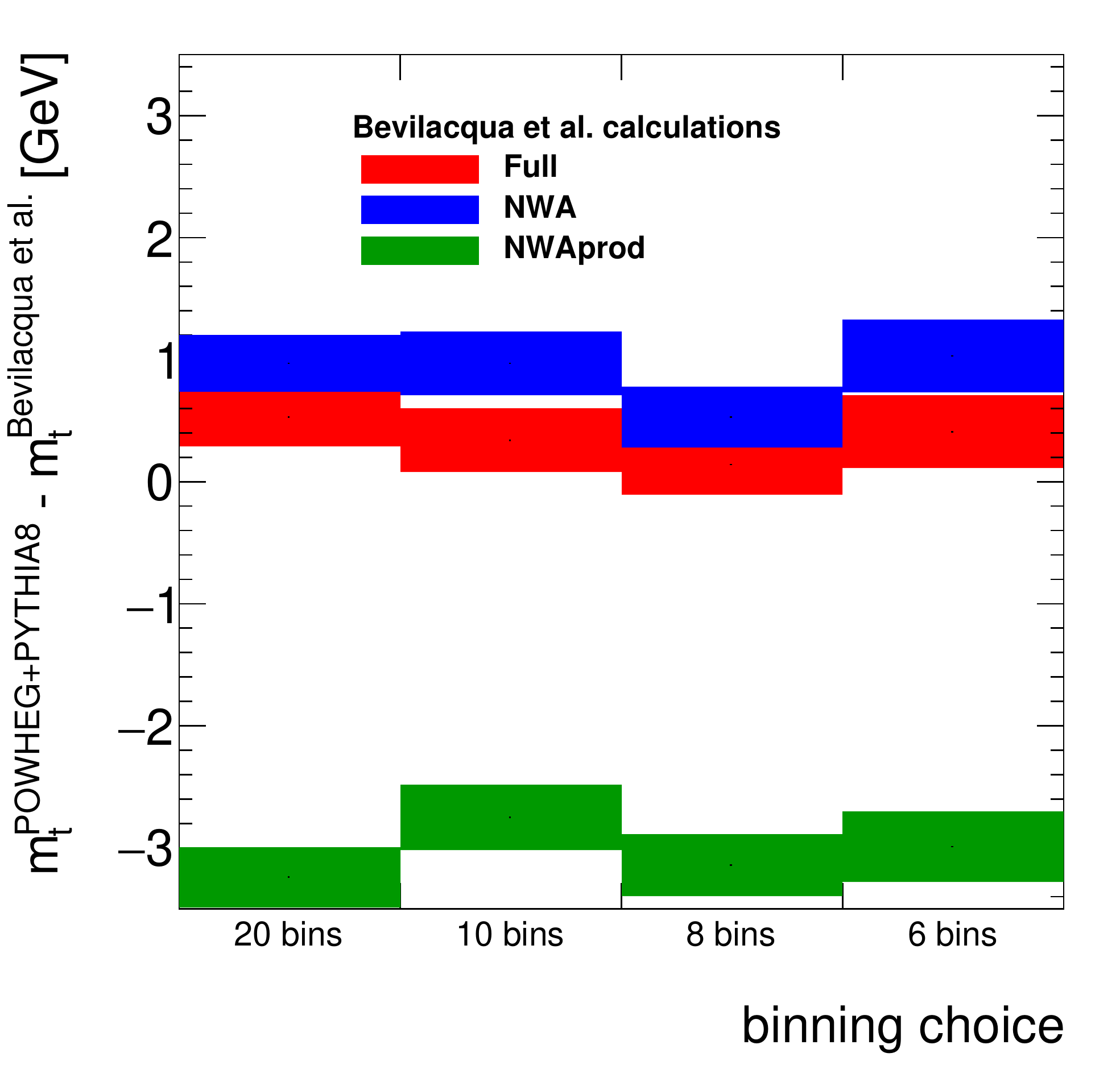}
\caption{Linearity of the fit (left) and differences between top-quark
  mass values extracted from ${\cal R}$ distributions calculated with different
  approaches (right).  Different binning choices have been tested: 20
  and 10 bins of equal size (respectively with widths 0.05 and 0.1),
  the 6 bins choice of Ref.~\cite{Fuster:2017rev, ATLAS:2015pfy} and the 8
  bins choice of Ref.~\cite{ATLAS:2019guf}.  Results are found to be
  consistent across the binning choices.}
\label{fig:MassDiff}
\end{figure}
The results are shown in \Fig{fig:MassDiff}. As a consistency check
we applied the fit procedure also to the distributions generated with {\textsc{Powheg-Box}\xspace}~{+}~{\textsc{Pythia}\xspace}8.
The left-hand plot shows the reconstructed top-quark mass 
as a function of the true top-quark mass used in generating the event
sample. The
fit reproduces the true value without an additional offset. In the
right hand plot the same procedure is applied treating the different
approximations from \Ref{Bevilacqua:2017ipv} as pseudo-data. Different
binnings using 6, 8, 10 or 20 bins have been used. For each bin
choice the result of the fit including its uncertainty is shown as a
band. As one can see using ${\cal R}^{\text{Full}}$ as pseudo-data, the
true mass value is recovered within an uncertainty of about 500
${\mbox{MeV}}$, which is below the sum of modeling uncertainties on $m_t$ reported
in \Refs{Bevilacqua:2017ipv,ATLAS:2015pfy,ATLAS:2019guf}.
Furthermore, this finding is, within the uncertainties, independent on the binning of ${\cal R}$. One can
thus conclude that {\textsc{Powheg-Box}\xspace}~{+}~{\textsc{Pythia}\xspace}8 gives a reliable description of the ${\cal R}$~distribution.
As a consequence
{\textsc{Powheg-Box}\xspace}~{+}~{\textsc{Pythia}\xspace}8 can be used to unfold the data to the parton-level.
Using instead the $\mbox{\textit{NWA}}_{Prod}$  approximation from 
\Ref{Bevilacqua:2017ipv} we reproduce the sizable shift of about 3~{\mbox{GeV}} as observed already in \Ref{Bevilacqua:2017ipv}. This is not
surprising as the  $\mbox{\textit{NWA}}_{Prod}$  approximation
neglects NLO corrections in the decay stage. The bulk of the effect
has thus nothing to do with off-shell effects but is a
consequence of missing higher-order corrections, also considering that  
for the $\mbox{\textit{NWA}}$  approximation we do not observe the
sizable shift reported in \Ref{Bevilacqua:2017ipv}. In fact inspecting
Fig.~3 of \Ref{Bevilacqua:2017ipv} the size of the shift observed in
\Ref{Bevilacqua:2017ipv} is surprising: in the range $0.3<\rho_s<0.6$
the two approximations \textit{Full NLO} and \textit{NWA} give consistent
results. For larger $\rho_s$ one can observe differences. However, 
the results fluctuate indicating that statistical uncertainties are
important. For lower $\rho_s$ there is a systematic difference. However,
this corresponds to the high-energy region which should have only a marginal
effect on the mass extraction. In fact in this region the different
scale setting procedures applied in  \textit{Full NLO} and \textit{NWA}
lead to rather different values for the renormalization scale which could at least partially explain
the differences in that region. As the \textit{NWA} approximation as
quoted in \Ref{Bevilacqua:2017ipv} is not used in the aforementioned
experimental analysis we have not tried to pin down the shift observed
in \Fig{fig:MassDiff}. 

\section{Conclusions}
\label{sec:conclu}

The main aim of this work is investigating the effects of different inputs on
theoretical predictions for $t\bar{t}j$ hadroproduction at the LHC and
providing reference cross sections with NLO QCD accuracy to the experimental collaborations,
with particular emphasis on the distributions used for top-quark mass extractions on the basis of experimental data concerning this process.

Among the different renormalization and factorization scale choices that we inve\-sti\-gated, including both the static scale
$m_t$ and dynamical scales based on the $m_{t\bar{t}j}$ and $H_T$ quantities, 
we find that the $\mu_R = \mu_F = H_T/4$ dynamical scale provides the smallest NLO$\,$/$\,$LO
$\mathcal{K}$-factor and minimizes shape variation effects when comparing predictions at
these two orders for a wide selection of differential distributions, under the typical,
quite inclusive, sets of analysis cuts currently adopted by the experimental collaborations. 
Therefore, we recommend the use of this scale, especially in case of absolute differential fiducial cross sections. On the other hand, in case of total fiducial cross sections, we find that even
the static scale still represents a competitive option.
This applies also to the
normalized distribution ${\cal R}(m_t,\rho_s)$ used in present
ATLAS and CMS analyses aiming at top-quark mass extraction, in the region $\rho_s \gtrsim 0.3$. 
If future analyses will start to make use of the shape of the ${\cal R}(m_t,\rho_s)$ distribution
for smaller values of $\rho_s$,
it will become mandatory to use dynamical scales for a more reliable theoretical description.
It will also be very important to use PDF fits with decreased uncertainties at large $x$,
considering that, for both the $\rho_s$ and the
${\cal R}(m_t,\rho_s)$
distributions, we have found that PDF uncertainties
are particularly large in the small $\rho_s$ region, that is mostly populated by hard-scattering inte\-rac\-tions involving
at least one large-$x$ parton. 
While present experimental analyses do not particularly focus on the small $\rho_s$ region, 
the sensitivity may increase with the increasing statistics of the data
collected, as expected during the high luminosity (HL) phase of the LHC.
Additionally, in the tails of the $\rho_s$ distribution, 
fixed-order NLO calculations may not provide an accurate enough description. 
Further higher-order perturbative corrections beyond NLO for the hard-scattering process can be crucial to reduce uncertainties there.

Furthermore, we have investigated the dependence of QCD uncertainties on the jet radius para\-me\-ter $R$ used
to construct jets according to the \mbox{anti-$k_T$} jet clustering algorithm.
Comparing predictions for $R=0.4$ and $R=0.8$, we found that the size of the scale uncertainty bands is almost independent of the choice of the $R$ value.
A slight advantage of using $R=0.8$ instead of the currently adopted $R=0.4$ value would be offered by the larger absolute cross sections, allowing to better exploit low-statistics kinematical regions. It will be interesting to repeat such an investigation in presence of full shower Monte Carlo effects, which affect the jet formation process.  

For the time being, we list some of our reference cross sections at $\sqrt{S} = 13$ and 14~TeV
in Appendix~\ref{sec:appA} 
and make available an exhaustive set of predictions in the online repository~\cite{onlinerepository},
confident that they might be very useful for top-quark mass extractions on the
basis of the data already collected at the LHC during Run 2 and the new data expected from the forthcoming Run~3.

These predictions, beside referring to different modern PDF sets, different scales and different analysis cuts on the jets,
cover a wide range of top-quark mass values, re\-nor\-ma\-li\-zed in different schemes.
In particular, in this work we show for the first time predictions for the $t\bar{t}j$ hadroproduction
process using the MSR top-quark mass renormalization scheme, providing an
alternative to the on-shell and the $\overline{\mathrm{MS}}$ schemes already
used for top-quark mass extraction in previous $t\bar{t}j$ phenomenological studies and experimental analyses. 
We found that in the $0.2 < \rho_s < 0.7$ region the size of scale uncertainty bands using these three different
top-quark mass renormalization schemes is approximately the same. On the other hand, close to threshold,
the $\overline{\mathrm{MS}}$ mass does not provide a good description, as
expected on the basis of theory considerations and as proven in practice by
the fact that scale uncertainties become very large. The mass dependence of
the cross~sections turns out to be linear, which allows to good approximation a linear interpolation between theory predictions for different $m_t$
values, when building the $\chi^2$ for performing the top-quark mass fits. 

We finally observe that the \mbox{${\cal R}$~distribution} in the $\rho_s$ region of largest experimental interest can be decently modeled by using simulations with \textsc{Powheg-box}~+~\textsc{Pythia8}, making use of the POWHEG NLO matching algorithm and of \textsc{Pythia8} to decay and shower
$t\bar{t}j$ Les-Houches events with stable top-quarks and at most one extra parton. As far as the \mbox{${\cal R}$~distribution} is concerned, the predictions from this setup turn out to agree within the uncertainties with the calculation of \Ref{Bevilacqua:2016jfk}, including full off-shellness and spin correlation effects, but missing parton shower and soft physics effects. 
Of course, using the full calculation can be very relevant for an accurate description of other observables, like e.g. angular
correlation observables or $t$ and $W$ invariant mass distributions, which
are, however, outside the present scope of $t\bar{t}j$ experimental
analyses aiming at top-quark mass extraction. Once a calculation like the most sophisticated  one in \Ref{Bevilacqua:2016jfk} will be combined with Shower Monte Carlo codes, it may also allow to further reduce modeling uncertainties in the top-quark mass measurements.

In conclusion, considering, on the one hand, the advanced techniques applied in the experimental analysis
to reconstruct top quarks from their decay
products and, on the other hand, the present level of experimental
uncertainties, we are confident that the predictions presented in this work
and in the associated repository are robust enough for reliable extractions of
the top-quark mass parameter using LHC Run 2 and future Run 3 data, while
hoping for future theoretical and experimental developments which might
further increase the accuracy of these extractions and decrease the
associated systematic uncertainties, du\-ring the HL-LHC phase.

\acknowledgments
We would like to thank Malgorzata Worek for providing us with numerical input for the study of off-shell effects. We are grateful to Giuseppe Bevilacqua for further clarifications and to Andr\'e Hoang, Katerina Lipka, Marcel Vos, and Sebastian Wuchterl for useful discussions.

The work of S.A. and A.G. is supported by the ERC Starting Grant REINVENT-714788. S.A. also acknowledges funding from Fondazione Cariplo and Regione Lombardia, grant 2017-2070 and from MIUR through the FARE grant R18ZRBEAFC. 
J.F. and A.I. acknowledge support from projects PGC2018-094856-B-100 (MICINN/FEDER), PROMETEO-2018/060 and CIDEGENT/2020/21 (Generalitat Valenciana), and the iLINK grant LINKB20065 (CSIC).
The work of M.V.G., S.M. and P.U. was supported in part by the
Bundesministerium f\"ur Bildung and Forschung under contracts 05H21GUCCA and 05H18KHCA1.

\appendix
\section{Numerical results for cross sections for different scales, top-quark masses and analysis cuts}
\label{sec:appA}
In the following we present predictions for cross~sections 
for $t\bar{t}j$ production in $pp$ collisions at
$\sqrt{S}$~=~13~TeV at the LHC for future reference.
In Tables~\ref{tab:RHORES_differential-mu0}, 
\ref{tab:RHORES_differential-HTB2}, 
\ref{tab:RHORES_differential-HTB4} and
\ref{tab:RHORES_differential-mttbj}, using as input $m_t^{\rm{pole}} = 172$~GeV and the CT18 NLO PDF central set, 
we provide predictions for the $\rho_s$ distribution under the default system of cuts described at the beginning of
Section~\ref{sec:pheno}, that is  
$N_j\geq 1$, $p_T^j>30$~GeV, $|\eta_j|<2.4$, 
for the different central scale choices: 
$\mu_R = \mu_F = \mu_0 = m_t$, $H_T^B/2$, $H_T^B/4$ and $m_{t\bar{t}j}^B$, as defined
in Section~\ref{sec:compu}. Scale variation effects, computed by varying
simultaneously $\mu_R$ and $\mu_F$ by a factor of two around the central value
are also shown. 
\begin{table}[h!]
  \begin{center}
\begin{tabular}{c|c|c|c|c}
\hline
$x_l$ & $x_r$ & $K_R=K_F=0.5$ & $K_R=K_F=1$ & $K_R=K_F=2$ \\
\hline
0.0	& 0.18 & 12.6$\pm$0.2 & 22.0$\pm$0.1 & 21.70$\pm$0.07 \\
0.18	 & 0.22	& 136$\pm$1 & 166.7$\pm$0.6 & 154.7$\pm$0.4 \\
0.22	 & 0.27	& 269$\pm$1 & 300.1$\pm$0.9 & 271.5$\pm$0.9 \\
0.27	 & 0.32 & 437$\pm$1 & 457.3$\pm$0.9 & 409.8$\pm$0.6 \\
0.32	 & 0.38	& 609$\pm$2 & 616$\pm$1 & 546.6$\pm$0.8 \\
0.38 & 0.45	& 765$\pm$2 & 752$\pm$1 & 663.6$\pm$0.6 \\
0.45 & 0.53	& 845$\pm$1 & 814$\pm$1 & 717$\pm$2 \\
0.53	 & 0.62 & 800$\pm$1 & 759.8$\pm$0.9 & 667.8$\pm$0.6 \\
0.62 & 0.71 & 615$\pm$1 & 576.2$\pm$0.9 & 505.6$\pm$0.6 \\
0.71 & 1.0	& 133.1$\pm$0.3 & 123.8$\pm$0.2 & 108.5$\pm$0.1
      \\
      \hline
\end{tabular}
\caption{
  \label{tab:RHORES_differential-mu0}
  NLO predictions for the $\rho_s$ distribution [pb] for $t\bar{t}j + X$ hadroproduction at $\sqrt{S} =$~13~TeV
  using as input the CT18 NLO central PDF set, $m_t = 172$ GeV and $\mu_0=m_t$,
  for $N_j\geq 1$, $p_T^j>30$~GeV, $|\eta_j|<2.4$. See Section~\ref{sec:pheno} for more detail.}
\end{center}
\end{table}

\begin{table}[h!]
  \begin{center}
\begin{tabular}{c|c|c|c|c}
\hline
$x_l$ & $x_r$ & $K_R=K_F=0.5$ & $K_R=K_F=1$ & $K_R=K_F=2$ \\
\hline
0.0	& 0.18 & 23.91$\pm$0.09 & 21.10$\pm$0.06 & 17.38$\pm$0.05\\
0.18	 & 0.22	& 171.9$\pm$0.8 & 152.4$\pm$0.5 & 126.7$\pm$0.3\\
0.22	 & 0.27	& 306$\pm$1 & 269.9$\pm$0.8 & 226.6$\pm$0.4 \\
0.27	 & 0.32 &  464$\pm$1 & 413.9$\pm$0.7 & 348.6$\pm$0.5 \\
0.32	 & 0.38	&  626$\pm$2 & 558.7$\pm$0.9 & 473.5$\pm$0.6 \\
0.38 & 0.45	&  760$\pm$1 & 684.9$\pm$0.8 & 584.5$\pm$0.6 \\
0.45 & 0.53	&  828$\pm$2 & 749.5$\pm$0.8 &  643.4$\pm$0.9\\
0.53	 & 0.62 &  7745$\pm$1 & 706.1$\pm$0.7 & 609.6$\pm$0.6 \\
0.62 & 0.71 &  592$\pm$1 & 541.1$\pm$0.6 &  470.1$\pm$0.5 \\
0.71 & 1.0	&  127.9$\pm$0.3 & 117.6$\pm$0.2 & 102.5$\pm$0.2
      \\
      \hline
\end{tabular}
\caption{
  \label{tab:RHORES_differential-HTB2}
  Same as Table~\ref{tab:RHORES_differential-mu0}, but for $\mu_0=H_T^B/2$.}
\end{center}
\end{table}

\begin{table}[h!]
  \begin{center}
\begin{tabular}{c|c|c|c|c}
\hline
$x_l$ & $x_r$ & $K_R=K_F=0.5$ & $K_R=K_F=1$ & $K_R=K_F=2$ \\
\hline
0.0	& 0.18 & 22.23$\pm$0.1 &  23.9$\pm$0.1 & 21.1$\pm$0.1\\
0.18	 & 0.22	& 164$\pm$1 & 171.9$\pm$0.8 & 152.4$\pm$0.5\\
0.22	 & 0.27	& 294$\pm$1 & 306$\pm$1 &  269.9$\pm$0.8\\
0.27	 & 0.32 & 446$\pm$2 & 464$\pm$1 & 413.9$\pm$0.7 \\
0.32	 & 0.38	& 601$\pm$2 &  626$\pm$2  & 558.7$\pm$0.9\\
0.38 & 0.45	& 732$\pm$2 & 760$\pm$1 & 684.9$\pm$0.8\\
0.45 & 0.53	& 797$\pm$2 &  828$\pm$2 & 749.5$\pm$0.8 \\
0.53	 & 0.62 &  743$\pm$2 & 775$\pm$1 & 706.1$\pm$0.7 \\
0.62 & 0.71 &  567$\pm$2 &  592$\pm$1 &  541.1$\pm$0.6\\
0.71 & 1.0	& 123.4$\pm$0.8 & 127.9$\pm$0.3 & 117.6$\pm$0.2
      \\
      \hline
\end{tabular}
\caption{
  \label{tab:RHORES_differential-HTB4}
  Same as Table~\ref{tab:RHORES_differential-mu0}, but for $\mu_0=H_T^B/4$.}
\end{center}
\end{table}

\begin{table}
  \begin{center}
\begin{tabular}{c|c|c|c|c}
\hline
$x_l$ & $x_r$ & $K_R=K_F=0.5$ & $K_R=K_F=1$ & $K_R=K_F=2$ \\
\hline
0.0	& 0.18 & 19.59$\pm$0.06 &  16.26$\pm$0.04 & 13.17$\pm$0.03 \\
0.18	 & 0.22	&  146.2$\pm$0.4 & 123.1$\pm$0.3 & 100.8$\pm$0.3 \\
0.22	 & 0.27	& 265.2$\pm$0.6 &  226.0$\pm$0.5 &  186.5$\pm$0.4\\
0.27	 & 0.32 &  412.5$\pm$0.7 & 355.0$\pm$0.5 & 296.2$\pm$0.5\\
0.32	 & 0.38	& 565.3$\pm$0.8 & 492.5$\pm$0.6 & 413.9$\pm$0.5\\
0.38 & 0.45	&  698$\pm$1 &  618.1$\pm$0.6 & 523.4$\pm$0.5\\
0.45 & 0.53	&  768$\pm$1 & 689.2$\pm$0.6 & 590.5$\pm$0.7\\
0.53	 & 0.62 &  725.9$\pm$0.9 & 659.1$\pm$0.6 & 570.4$\pm$0.5 \\
0.62 & 0.71 &  554.0$\pm$0.9 & 510.3$\pm$0.8 & 446.1$\pm$0.4\\
0.71 & 1.0	& 119.9$\pm$0.4 & 111.9$\pm$0.2 & 98.6$\pm$0.1
      \\
      \hline
\end{tabular}
\caption{
  \label{tab:RHORES_differential-mttbj}
  Same as Table~\ref{tab:RHORES_differential-mu0}, but for $\mu_0=m_{t\bar{t}j}^B/2$.}
\end{center}
\end{table}

Covering a range of top-quark mass values, 
a series of slightly different cuts $p_T^j~>~30,\,$ $50, \, 75 \,$ and $100$ GeV 
along with $|\eta_j|<2.4$, and using as input the \texttt{ABMP16\_5\_nlo} PDF set, Tables~\ref{tab:LHC13_mpole_ABMP16_n3_CMS13TeV}--\ref{tab:LHC14_mpole_ABMP16_n3_CMS13TeV_pt100}
provide further examples for reference cross sections, the complete set of which
is collected in the online repository~\cite{onlinerepository}.

\begin{landscape}
\begin{table}[!ht]
\centering
\begin{adjustbox}{width=1.4\textwidth}
\small
\begin{tabular}{c|c|ccccccccccccc}
  \hline
&  & \multicolumn{13}{|c}{ $\frac{d\sigma_{t\bar{t}j + X}}{d\rho_s}$ distribution with $N_{j}\geq1,\,p_{T,j}>30$ GeV, $|\eta_{j}|<2.4$, $K_{R}=K_{F}=1$}  \\ 
 \hline 
 $x_{l}$  & $x_{r}$  & 165 GeV & 166 GeV & 167 GeV & 168 GeV & 169 GeV & 170 GeV & 171 GeV & 172 GeV & 173 GeV & 174 GeV & 175 GeV & 176 GeV & 177 GeV \\ 
 \hline 
0.00 & 0.18 & 19.49(6) & 19.32(6) & 19.23(6) & 18.91(6) & 18.79(6) & 18.70(6) & 18.48(6) & 18.16(5) & 18.13(5) & 17.97(5) & 17.72(5) & 17.64(5) & 17.49(5) \\ 
0.18 & 0.22 & 159.1(6) & 158.6(5) & 156.4(5) & 153.7(5) & 152.7(5) & 151.2(5) & 149.6(5) & 146.6(5) & 146.0(5) & 143.9(5) & 142.1(4) & 142.3(4) & 139.6(4) \\ 
0.22 & 0.27 & 299.0(8) & 295.4(7) & 291.3(7) & 287.4(7) & 285.6(7) & 281.2(7) & 278.4(7) & 274.0(7) & 271.8(6) & 269.0(6) & 264.4(6) & 262.4(6) & 259.4(6) \\ 
0.27 & 0.32 & 475.0(1) & 469.0(1) & 461.0(1) & 457.0(1) & 452.0(1) & 443.2(1) & 439.6(1) & 432.2(9) & 426.1(9) & 421.2(9) & 414.9(9) & 409.4(9) & 401.9(9) \\ 
0.32 & 0.38 & 658.0(1) & 650.0(1) & 639.0(1) & 632.0(1) & 619.0(1) & 612.0(1) & 604.0(1) & 596.0(1) & 585.0(1) & 576.0(1) & 568.8(1) & 559.7(1) & 553.0(1) \\ 
0.38 & 0.45 & 830.0(1) & 813.0(1) & 803.0(1) & 788.0(1) & 776.0(1) & 762.0(1) & 749.0(1) & 739.0(1) & 728.0(1) & 716.0(1) & 703.0(1) & 692.0(1) & 679.0(1) \\ 
0.45 & 0.53 & 931.0(1) & 914.0(1) & 894.0(1) & 878.0(1) & 861.0(1) & 843.0(1) & 829.0(1) & 811.0(1) & 796.0(1) & 781.0(1) & 768.0(1) & 750.0(1) & 736.0(1) \\ 
0.53 & 0.62 & 899.0(1) & 883.0(1) & 858.0(1) & 840.0(1) & 823.0(1) & 806.0(1) & 783.0(1) & 763.0(1) & 746.0(1) & 731.0(1) & 711.0(1) & 695.0(1) & 678.0(1) \\ 
0.62 & 0.71 & 732.0(1) & 709.0(1) & 690.0(1) & 667.0(1) & 644.0(1) & 625.0(1) & 605.0(1) & 586.0(1) & 566.1(1) & 546.6(1) & 529.1(9) & 511.9(9) & 493.7(9) \\ 
0.71 & 1.00 & 200.0(3) & 188.2(3) & 176.5(3) & 165.3(3) & 155.2(3) & 144.9(3) & 135.4(3) & 126.6(2) & 118.3(2) & 109.8(2) & 103.0(2) & 95.4(2) & 88.8(2) \\ 
 \hline
&  & \multicolumn{13}{|c}{ $\frac{d\sigma_{t\bar{t}j + X}}{d\rho_s}$ distribution with $N_{j}\geq1,\,p_{T,j}>30$ GeV, $|\eta_{j}|<2.4$, $K_{R}=K_{F}=2$}  \\ 
 \hline 
 $x_{l}$  & $x_{r}$  & 165 GeV & 166 GeV & 167 GeV & 168 GeV & 169 GeV & 170 GeV & 171 GeV & 172 GeV & 173 GeV & 174 GeV & 175 GeV & 176 GeV & 177 GeV \\ 
 \hline 
0.00 & 0.18 & 19.30(4) & 19.24(4) & 18.88(4) & 18.67(4) & 18.52(4) & 18.36(4) & 18.19(3) & 18.01(3) & 17.80(3) & 17.58(3) & 17.41(3) & 17.32(3) & 17.13(3) \\ 
0.18 & 0.22 & 149.4(3) & 146.4(3) & 145.4(3) & 143.8(3) & 142.0(3) & 140.0(3) & 138.3(3) & 136.6(3) & 135.2(3) & 133.7(3) & 131.6(3) & 130.5(3) & 128.6(3) \\ 
0.22 & 0.27 & 271.3(5) & 267.4(5) & 264.4(5) & 260.4(5) & 258.0(4) & 254.3(4) & 251.1(4) & 248.5(4) & 245.8(4) & 242.4(4) & 240.0(4) & 235.7(4) & 232.8(4) \\ 
0.27 & 0.32 & 424.4(7) & 418.4(7) & 411.7(7) & 407.7(7) & 401.0(6) & 395.1(7) & 389.8(6) & 383.6(6) & 380.1(6) & 374.7(6) & 369.3(6) & 364.2(6) & 359.7(6) \\ 
0.32 & 0.38 & 584.2(8) & 575.0(8) & 564.6(8) & 556.4(8) & 548.0(7) & 540.0(7) & 532.2(7) & 524.3(7) & 517.6(7) & 510.0(7) & 501.3(7) & 494.4(6) & 487.6(6) \\ 
0.38 & 0.45 & 729.0(8) & 717.1(8) & 703.8(8) & 693.8(8) & 680.9(8) & 669.6(8) & 659.3(8) & 649.2(8) & 637.6(7) & 627.1(7) & 615.7(7) & 606.6(7) & 596.3(7) \\ 
0.45 & 0.53 & 812.8(9) & 796.3(9) & 782.0(9) & 767.1(9) & 752.1(8) & 740.2(8) & 723.9(8) & 710.6(8) & 696.5(8) & 683.7(8) & 670.1(7) & 657.6(7) & 645.1(7) \\ 
0.53 & 0.62 & 789.9(9) & 770.3(9) & 752.3(9) & 735.4(9) & 717.9(8) & 700.5(8) & 683.0(8) & 669.9(8) & 652.9(7) & 637.1(7) & 622.8(7) & 606.5(7) & 592.1(7) \\ 
0.62 & 0.71 & 636.7(8) & 618.8(8) & 598.9(8) & 581.7(8) & 561.3(7) & 546.1(7) & 527.9(7) & 508.9(7) & 492.6(7) & 477.1(6) & 460.4(6) & 445.8(6) & 430.7(6) \\ 
0.71 & 1.00 & 174.0(2) & 163.0(2) & 153.6(2) & 143.8(2) & 134.8(2) & 125.7(2) & 117.8(2) & 110.2(2) & 103.1(2) & 95.9(2) & 89.8(1) & 83.0(1) & 77.1(1) \\ 
 \hline
&  & \multicolumn{13}{|c}{ $\frac{d\sigma_{t\bar{t}j + X}}{d\rho_s}$ distribution with $N_{j}\geq1,\,p_{T,j}>30$ GeV, $|\eta_{j}|<2.4$, $K_{R}=K_{F}=0.5$}  \\ 
 \hline 
 $x_{l}$  & $x_{r}$  & 165 GeV & 166 GeV & 167 GeV & 168 GeV & 169 GeV & 170 GeV & 171 GeV & 172 GeV & 173 GeV & 174 GeV & 175 GeV & 176 GeV & 177 GeV \\ 
 \hline 
0.00 & 0.18 & 10.5(1) & 10.9(1) & 10.7(1) & 10.5(1) & 10.4(1) & 10.7(1) & 10.7(1) & 10.6(1) & 10.58(9) & 10.59(9) & 10.59(9) & 10.53(9) & 10.38(9) \\ 
0.18 & 0.22 & 125.9(9) & 124.6(9) & 124.1(9) & 122.9(8) & 123.3(9) & 120.7(9) & 121.3(8) & 117.1(9) & 118.1(8) & 115.9(8) & 116.0(7) & 115.3(8) & 114.8(8) \\ 
0.22 & 0.27 & 265.0(1) & 260.0(1) & 259.0(1) & 258.0(1) & 255.0(1) & 252.0(1) & 246.0(1) & 246.0(1) & 242.0(1) & 240.0(1) & 237.0(1) & 236.0(1) & 232.6(1) \\ 
0.27 & 0.32 & 446.0(2) & 443.0(2) & 439.0(2) & 434.0(2) & 423.0(2) & 422.0(2) & 418.0(2) & 410.0(2) & 405.0(2) & 402.0(2) & 397.0(2) & 388.0(1) & 387.0(1) \\ 
0.32 & 0.38 & 646.0(2) & 643.0(2) & 636.0(2) & 623.0(2) & 615.0(2) & 610.0(2) & 596.0(2) & 588.0(2) & 584.0(2) & 569.0(2) & 565.0(2) & 554.0(2) & 548.0(2) \\ 
0.38 & 0.45 & 846.0(2) & 829.0(2) & 815.0(2) & 805.0(2) & 789.0(2) & 779.0(2) & 764.0(2) & 755.0(2) & 738.0(2) & 727.0(2) & 715.0(2) & 707.0(2) & 691.0(2) \\ 
0.45 & 0.53 & 966.0(2) & 946.0(2) & 933.0(2) & 913.0(2) & 896.0(2) & 875.0(2) & 862.0(2) & 843.0(2) & 828.0(2) & 811.0(2) & 799.0(2) & 782.0(2) & 766.0(2) \\ 
0.53 & 0.62 & 955.0(2) & 933.0(2) & 910.0(2) & 889.0(2) & 866.0(2) & 844.0(2) & 826.0(2) & 809.0(2) & 789.0(2) & 770.0(2) & 751.0(2) & 735.0(2) & 719.0(2) \\ 
0.62 & 0.71 & 780.0(2) & 757.0(2) & 733.0(2) & 710.0(2) & 688.0(2) & 664.0(2) & 643.0(2) & 626.0(2) & 604.0(2) & 586.0(2) & 564.0(2) & 544.0(1) & 528.0(1) \\ 
0.71 & 1.00 & 215.2(6) & 203.0(5) & 192.2(5) & 178.3(5) & 167.9(5) & 157.8(5) & 146.3(4) & 137.0(4) & 128.3(4) & 119.2(4) & 111.4(4) & 103.0(3) & 96.3(3) \\ 
 \hline
 \hline
 \end{tabular}
 \end{adjustbox}
\caption{NLO predictions for the $\rho_s$ distribution in the pole mass scheme for $\sqrt{S}=13$~TeV using as input the static scale $\mu_0 = m_t$ and the \text{ABMP16} NLO PDF set. 
At least one jet with  $|\eta_j| < 2.4$ and $p_T^j > 30$ GeV is required. \label{tab:LHC13_mpole_ABMP16_n3_CMS13TeV}}
 \end{table}
\end{landscape}
\begin{landscape}
\begin{table}[!ht]
\centering
\begin{adjustbox}{width=1.4\textwidth}
\small
\begin{tabular}{c|c|ccccccccccccc}
  \hline
&  & \multicolumn{13}{|c}{ $\frac{d\sigma_{t\bar{t}j + X}}{d\rho_s}$ distribution with $N_{j}\geq1,\,p_{T}^j>50$ GeV, $|\eta_{j}|<2.4$, $K_{R}=K_{F}=1$}  \\ 
 \hline 
 $x_{l}$  & $x_{r}$  & 165 GeV & 166 GeV & 167 GeV & 168 GeV & 169 GeV & 170 GeV & 171 GeV & 172 GeV & 173 GeV & 174 GeV & 175 GeV & 176 GeV & 177 GeV \\ 
 \hline 
0.00 & 0.18 & 19.23(6) & 19.09(6) & 18.94(5) & 18.63(6) & 18.52(6) & 18.42(5) & 18.19(5) & 17.89(5) & 17.86(5) & 17.67(5) & 17.46(5) & 17.38(5) & 17.22(5) \\ 
0.18 & 0.22 & 152.3(5) & 152.1(5) & 149.7(4) & 147.3(5) & 146.3(5) & 145.0(4) & 143.4(4) & 139.9(4) & 139.6(4) & 137.5(4) & 136.7(4) & 135.9(4) & 133.6(4) \\ 
0.22 & 0.27 & 276.7(7) & 273.0(6) & 268.6(6) & 265.6(6) & 263.1(6) & 259.1(6) & 255.7(6) & 253.3(6) & 249.9(5) & 247.4(5) & 244.0(5) & 240.7(5) & 237.7(5) \\ 
0.27 & 0.32 & 420.1(9) & 412.0(9) & 405.6(8) & 401.0(9) & 396.9(8) & 388.2(8) & 386.0(8) & 378.3(8) & 373.8(7) & 368.8(7) & 362.2(7) & 357.7(7) & 350.9(7) \\ 
0.32 & 0.38 & 547.6(1) & 541.6(1) & 532.0(9) & 525.8(9) & 514.7(9) & 508.5(9) & 499.1(8) & 494.1(8) & 483.8(8) & 474.8(8) & 469.0(8) & 462.2(8) & 456.1(7) \\ 
0.38 & 0.45 & 647.5(1) & 634.7(1) & 623.2(9) & 612.8(9) & 601.0(9) & 591.3(9) & 579.8(8) & 570.5(8) & 560.7(8) & 551.9(8) & 541.3(7) & 531.9(8) & 522.6(8) \\ 
0.45 & 0.53 & 666.3(9) & 653.1(9) & 638.4(9) & 624.0(9) & 612.5(9) & 597.9(8) & 586.8(8) & 571.3(8) & 561.9(8) & 550.7(8) & 538.9(7) & 527.1(7) & 516.6(7) \\ 
0.53 & 0.62 & 573.9(9) & 561.5(9) & 546.1(8) & 531.4(8) & 519.5(7) & 504.3(8) & 491.6(7) & 477.7(7) & 465.2(7) & 454.9(7) & 440.1(6) & 428.3(7) & 415.8(6) \\ 
0.62 & 0.71 & 393.1(7) & 379.4(7) & 366.0(7) & 353.0(7) & 337.4(6) & 325.8(6) & 312.2(6) & 301.3(6) & 290.5(6) & 276.9(6) & 265.5(5) & 255.4(5) & 244.4(5) \\ 
0.71 & 1.00 & 66.2(2) & 61.5(2) & 56.8(1) & 52.4(1) & 48.7(1) & 44.7(1) & 41.3(1) & 37.7(1) & 34.7(1) & 31.7(1) & 29.28(9) & 26.71(9) & 24.27(8) \\ 
 \hline
&  & \multicolumn{13}{|c}{ $\frac{d\sigma_{t\bar{t}j + X}}{d\rho_s}$ distribution with $N_{j}\geq1,\,p_{T}^j>50$ GeV, $|\eta_{j}|<2.4$, $K_{R}=K_{F}=2$}  \\ 
 \hline 
 $x_{l}$  & $x_{r}$  & 165 GeV & 166 GeV & 167 GeV & 168 GeV & 169 GeV & 170 GeV & 171 GeV & 172 GeV & 173 GeV & 174 GeV & 175 GeV & 176 GeV & 177 GeV \\ 
 \hline 
0.00 & 0.18 & 18.83(4) & 18.77(4) & 18.41(4) & 18.22(4) & 18.09(3) & 17.93(3) & 17.74(3) & 17.53(3) & 17.36(3) & 17.16(3) & 16.96(3) & 16.89(3) & 16.66(3) \\ 
0.18 & 0.22 & 142.0(3) & 139.0(3) & 138.0(3) & 136.7(3) & 134.6(3) & 133.1(3) & 131.0(3) & 129.3(3) & 128.2(3) & 126.8(3) & 124.7(3) & 123.6(3) & 122.0(3) \\ 
0.22 & 0.27 & 248.4(4) & 245.0(4) & 242.1(4) & 238.3(4) & 235.9(4) & 233.1(4) & 229.9(4) & 227.0(4) & 223.9(3) & 221.0(4) & 218.0(4) & 215.1(4) & 212.2(3) \\ 
0.27 & 0.32 & 370.6(6) & 365.5(6) & 359.0(6) & 355.4(6) & 348.9(5) & 343.8(5) & 339.0(5) & 333.9(5) & 329.8(5) & 325.0(5) & 320.4(5) & 316.2(5) & 312.0(4) \\ 
0.32 & 0.38 & 482.6(6) & 473.6(6) & 466.6(6) & 459.3(6) & 451.7(5) & 444.4(5) & 438.5(5) & 431.2(5) & 424.8(5) & 418.3(5) & 411.8(5) & 405.3(5) & 397.9(5) \\ 
0.38 & 0.45 & 562.1(6) & 553.4(6) & 544.0(6) & 534.0(6) & 523.5(6) & 514.5(5) & 505.7(6) & 497.0(5) & 487.3(5) & 479.2(5) & 469.5(5) & 461.7(5) & 453.5(5) \\ 
0.45 & 0.53 & 577.4(6) & 565.8(6) & 552.3(6) & 541.8(6) & 530.8(5) & 519.7(5) & 508.1(5) & 497.9(5) & 487.6(5) & 476.7(5) & 467.5(5) & 457.0(5) & 447.2(5) \\ 
0.53 & 0.62 & 499.3(6) & 485.6(5) & 473.4(5) & 461.5(5) & 449.0(5) & 436.3(5) & 423.8(5) & 414.4(5) & 401.7(5) & 391.3(4) & 380.0(4) & 370.7(4) & 360.3(4) \\ 
0.62 & 0.71 & 340.3(5) & 327.7(5) & 316.0(5) & 303.2(4) & 291.5(4) & 281.0(4) & 269.9(4) & 259.1(4) & 249.1(4) & 239.5(4) & 229.2(3) & 219.8(3) & 211.8(3) \\ 
0.71 & 1.00 & 57.0(1) & 52.7(1) & 48.8(1) & 45.12(9) & 41.78(8) & 38.57(8) & 35.36(8) & 32.54(7) & 29.92(7) & 27.41(7) & 25.07(6) & 22.92(6) & 20.79(6) \\ 
 \hline
&  & \multicolumn{13}{|c}{ $\frac{d\sigma_{t\bar{t}j + X}}{d\rho_s}$ distribution with $N_{j}\geq1,\,p_{T}^j>50$ GeV, $|\eta_{j}|<2.4$, $K_{R}=K_{F}=0.5$}  \\ 
 \hline 
 $x_{l}$  & $x_{r}$  & 165 GeV & 166 GeV & 167 GeV & 168 GeV & 169 GeV & 170 GeV & 171 GeV & 172 GeV & 173 GeV & 174 GeV & 175 GeV & 176 GeV & 177 GeV \\ 
 \hline 
0.00 & 0.18 & 11.1(1) & 11.3(1) & 11.2(1) & 10.99(9) & 10.98(9) & 11.24(9) & 11.03(9) & 10.96(9) & 11.02(9) & 10.99(8) & 11.04(8) & 10.89(8) & 10.85(8) \\ 
0.18 & 0.22 & 124.1(9) & 123.6(8) & 122.9(8) & 121.5(8) & 121.1(8) & 119.9(8) & 119.7(7) & 115.6(8) & 116.9(7) & 114.7(7) & 114.4(7) & 113.8(7) & 113.2(7) \\ 
0.22 & 0.27 & 252.0(1) & 247.0(1) & 244.0(1) & 244.3(1) & 240.0(1) & 238.0(1) & 232.7(1) & 233.0(1) & 229.1(1) & 227.0(9) & 223.6(9) & 222.8(9) & 220.3(8) \\ 
0.27 & 0.32 & 403.0(1) & 396.0(2) & 395.0(1) & 390.0(1) & 381.0(1) & 377.0(1) & 376.0(1) & 367.0(1) & 363.0(1) & 360.0(1) & 355.0(1) & 347.0(1) & 347.0(1) \\ 
0.32 & 0.38 & 553.0(2) & 546.0(2) & 538.0(2) & 526.0(1) & 522.0(2) & 515.0(1) & 501.0(1) & 499.0(1) & 493.0(1) & 481.0(1) & 477.0(1) & 470.0(1) & 460.0(1) \\ 
0.38 & 0.45 & 670.0(2) & 656.0(2) & 644.0(2) & 636.0(1) & 623.0(2) & 616.0(1) & 604.0(1) & 593.0(1) & 581.0(1) & 575.0(1) & 561.0(1) & 554.0(1) & 542.0(1) \\ 
0.45 & 0.53 & 705.0(2) & 691.0(2) & 677.0(2) & 662.0(1) & 645.0(1) & 634.0(1) & 621.0(1) & 607.0(1) & 596.0(1) & 584.0(1) & 574.0(1) & 560.0(1) & 549.0(1) \\ 
0.53 & 0.62 & 623.0(1) & 607.0(1) & 588.0(1) & 572.0(1) & 560.0(1) & 542.0(1) & 528.0(1) & 515.0(1) & 498.0(1) & 489.0(1) & 476.0(1) & 463.0(1) & 452.0(1) \\ 
0.62 & 0.71 & 430.0(1) & 413.0(1) & 400.0(1) & 385.0(1) & 367.0(1) & 356.0(1) & 341.9(1) & 329.3(1) & 315.1(1) & 305.9(9) & 289.2(9) & 279.8(8) & 268.2(8) \\ 
0.71 & 1.00 & 73.1(3) & 67.7(2) & 63.0(2) & 58.1(2) & 54.0(2) & 49.7(2) & 45.8(2) & 42.2(2) & 38.9(2) & 35.5(2) & 32.7(1) & 29.7(1) & 27.1(1) \\ 
 \hline
 \hline
 \end{tabular}
 \end{adjustbox}
\caption{
Same as~\ref{tab:LHC13_mpole_ABMP16_n3_CMS13TeV} 
but for $p_T^j > 50$ GeV.
\label{tab:LHC13_mpole_ABMP16_n3_CMS13TeV_pt50}}
 \end{table}
\end{landscape}

\begin{landscape}
\begin{table}[!ht]
\centering
\begin{adjustbox}{width=1.4\textwidth}
\small
\begin{tabular}{c|c|ccccccccccccc}
  \hline
&  & \multicolumn{13}{|c}{ $\frac{d\sigma_{t\bar{t}j + X}}{d\rho_s}$ distribution with $N_{j}\geq1,\,p_{T}^j>75$ GeV, $|\eta_{j}|<2.4$, $K_{R}=K_{F}=1$}  \\ 
 \hline 
 $x_{l}$  & $x_{r}$  & 165 GeV & 166 GeV & 167 GeV & 168 GeV & 169 GeV & 170 GeV & 171 GeV & 172 GeV & 173 GeV & 174 GeV & 175 GeV & 176 GeV & 177 GeV \\ 
 \hline 
0.00 & 0.18 & 18.69(6) & 18.48(6) & 18.39(5) & 18.13(5) & 17.98(5) & 17.87(5) & 17.65(5) & 17.38(4) & 17.37(5) & 17.16(5) & 16.98(4) & 16.87(4) & 16.74(5) \\ 
0.18 & 0.22 & 142.0(5) & 141.3(4) & 139.4(4) & 137.4(4) & 135.5(4) & 134.7(4) & 133.1(4) & 130.4(4) & 129.7(4) & 127.5(4) & 126.8(4) & 125.7(4) & 123.7(4) \\ 
0.22 & 0.27 & 245.7(5) & 242.3(5) & 239.0(5) & 234.9(5) & 232.5(5) & 229.4(5) & 226.3(5) & 223.9(5) & 220.7(5) & 218.4(4) & 215.0(4) & 212.5(4) & 209.7(5) \\ 
0.27 & 0.32 & 351.3(7) & 345.6(7) & 341.1(6) & 336.5(7) & 332.9(7) & 325.9(6) & 323.1(6) & 317.5(6) & 312.9(6) & 308.2(6) & 304.3(6) & 299.2(6) & 294.2(6) \\ 
0.32 & 0.38 & 435.7(8) & 429.4(7) & 422.4(7) & 417.0(7) & 407.6(7) & 402.3(6) & 396.1(6) & 390.1(6) & 382.2(6) & 376.9(6) & 369.6(6) & 364.4(6) & 359.4(6) \\ 
0.38 & 0.45 & 478.6(7) & 469.1(7) & 459.0(7) & 452.0(7) & 443.3(7) & 434.6(6) & 427.2(6) & 419.6(6) & 411.7(6) & 403.4(6) & 396.2(6) & 388.7(6) & 382.2(6) \\ 
0.45 & 0.53 & 449.1(7) & 438.3(7) & 428.6(6) & 419.4(7) & 410.1(6) & 399.4(6) & 392.0(6) & 381.6(5) & 373.3(5) & 364.6(5) & 356.7(5) & 347.7(5) & 339.7(5) \\ 
0.53 & 0.62 & 337.1(6) & 328.1(6) & 316.9(5) & 307.5(5) & 299.4(5) & 289.6(5) & 280.9(5) & 271.6(4) & 262.7(4) & 255.1(4) & 246.8(4) & 239.6(4) & 231.1(4) \\ 
0.62 & 0.71 & 179.7(4) & 171.5(4) & 163.0(4) & 155.5(4) & 147.8(4) & 140.6(3) & 133.5(3) & 127.0(3) & 121.0(3) & 114.1(3) & 107.6(3) & 102.3(3) & 96.6(3) \\ 
0.71 & 1.00 & 14.44(6) & 12.98(6) & 11.64(5) & 10.37(5) & 9.28(5) & 8.29(4) & 7.39(4) & 6.58(4) & 5.83(4) & 5.10(3) & 4.46(3) & 3.87(3) & 3.34(3) \\ 
 \hline
&  & \multicolumn{13}{|c}{ $\frac{d\sigma_{t\bar{t}j + X}}{d\rho_s}$ distribution with $N_{j}\geq1,\,p_{T}^j>75$ GeV, $|\eta_{j}|<2.4$, $K_{R}=K_{F}=2$}  \\ 
 \hline 
 $x_{l}$  & $x_{r}$  & 165 GeV & 166 GeV & 167 GeV & 168 GeV & 169 GeV & 170 GeV & 171 GeV & 172 GeV & 173 GeV & 174 GeV & 175 GeV & 176 GeV & 177 GeV \\ 
 \hline 
0.00 & 0.18 & 18.22(4) & 18.14(3) & 17.79(3) & 17.64(3) & 17.49(3) & 17.27(3) & 17.12(3) & 16.94(3) & 16.75(3) & 16.58(3) & 16.39(3) & 16.29(3) & 16.09(3) \\ 
0.18 & 0.22 & 131.7(3) & 128.8(3) & 128.0(3) & 126.8(3) & 125.0(3) & 123.1(2) & 121.3(2) & 119.5(2) & 118.7(2) & 117.7(2) & 115.2(2) & 114.2(2) & 113.1(2) \\ 
0.22 & 0.27 & 219.7(3) & 217.0(3) & 213.9(3) & 210.7(3) & 208.2(3) & 205.6(3) & 203.0(3) & 200.1(3) & 197.1(3) & 195.2(3) & 192.5(3) & 189.6(3) & 187.3(3) \\ 
0.27 & 0.32 & 311.7(5) & 306.2(5) & 301.8(4) & 298.3(5) & 293.0(4) & 288.7(4) & 284.2(4) & 279.6(4) & 275.9(4) & 271.6(4) & 267.4(4) & 264.0(4) & 260.4(4) \\ 
0.32 & 0.38 & 382.7(5) & 376.5(5) & 370.1(5) & 363.6(5) & 357.6(4) & 351.7(4) & 346.0(4) & 340.4(4) & 335.3(4) & 329.6(4) & 324.3(4) & 318.6(4) & 312.9(4) \\ 
0.38 & 0.45 & 415.6(5) & 409.0(4) & 400.0(4) & 393.3(4) & 385.2(4) & 378.0(4) & 371.9(4) & 364.9(4) & 356.7(4) & 350.8(4) & 343.6(4) & 337.4(4) & 331.4(4) \\ 
0.45 & 0.53 & 388.5(4) & 379.9(4) & 370.2(4) & 362.0(4) & 354.2(4) & 346.3(4) & 338.1(4) & 330.1(4) & 323.1(4) & 315.1(4) & 308.0(3) & 300.5(3) & 294.3(3) \\ 
0.53 & 0.62 & 291.5(4) & 282.7(4) & 274.1(4) & 265.9(4) & 257.9(3) & 250.3(3) & 241.4(3) & 234.4(3) & 226.7(3) & 220.1(3) & 212.4(3) & 206.3(3) & 199.2(3) \\ 
0.62 & 0.71 & 154.4(3) & 147.2(3) & 139.9(3) & 133.6(3) & 126.7(2) & 120.8(2) & 114.8(2) & 109.0(2) & 103.2(2) & 97.9(2) & 92.8(2) & 87.7(2) & 82.8(2) \\ 
0.71 & 1.00 & 12.27(4) & 11.01(4) & 9.89(4) & 8.88(4) & 7.94(3) & 6.97(3) & 6.25(3) & 5.51(3) & 4.90(2) & 4.32(2) & 3.77(2) & 3.28(2) & 2.78(2) \\ 
 \hline
&  & \multicolumn{13}{|c}{ $\frac{d\sigma_{t\bar{t}j + X}}{d\rho_s}$ distribution with $N_{j}\geq1,\,p_{T}^j>75$ GeV, $|\eta_{j}|<2.4$, $K_{R}=K_{F}=0.5$}  \\ 
 \hline 
 $x_{l}$  & $x_{r}$  & 165 GeV & 166 GeV & 167 GeV & 168 GeV & 169 GeV & 170 GeV & 171 GeV & 172 GeV & 173 GeV & 174 GeV & 175 GeV & 176 GeV & 177 GeV \\ 
 \hline 
0.00 & 0.18 & 11.1(1) & 11.31(9) & 11.12(9) & 10.94(8) & 10.97(9) & 11.13(9) & 10.99(8) & 10.91(9) & 10.93(8) & 10.94(8) & 10.98(7) & 10.77(8) & 10.75(8) \\ 
0.18 & 0.22 & 116.7(8) & 115.9(7) & 115.3(7) & 113.7(7) & 114.8(7) & 111.8(7) & 112.5(7) & 109.1(7) & 109.3(6) & 108.1(6) & 107.1(6) & 106.3(6) & 106.1(6) \\ 
0.22 & 0.27 & 222.4(1) & 220.2(9) & 217.2(9) & 216.7(9) & 215.3(9) & 211.1(9) & 208.0(8) & 206.6(8) & 204.6(8) & 203.0(8) & 199.0(8) & 198.1(7) & 195.2(7) \\ 
0.27 & 0.32 & 341.0(1) & 334.0(1) & 332.0(1) & 327.0(1) & 321.0(1) & 317.0(1) & 315.0(1) & 310.0(1) & 306.0(1) & 304.4(1) & 299.7(1) & 292.1(9) & 291.3(9) \\ 
0.32 & 0.38 & 440.0(1) & 434.0(1) & 428.0(1) & 419.0(1) & 413.0(1) & 409.0(1) & 400.0(1) & 394.0(1) & 392.0(1) & 381.0(1) & 378.0(1) & 371.3(9) & 365.2(9) \\ 
0.38 & 0.45 & 495.0(1) & 487.0(1) & 479.0(1) & 471.0(1) & 463.0(1) & 456.0(1) & 446.0(1) & 438.0(1) & 428.0(1) & 423.1(1) & 413.5(1) & 407.1(9) & 399.1(9) \\ 
0.45 & 0.53 & 478.0(1) & 469.0(1) & 455.0(1) & 449.0(1) & 435.0(1) & 424.0(1) & 417.0(1) & 405.8(1) & 397.6(9) & 389.4(9) & 383.5(9) & 371.9(8) & 363.9(8) \\ 
0.53 & 0.62 & 366.6(1) & 356.7(9) & 344.0(9) & 332.4(8) & 325.1(9) & 314.0(8) & 304.3(8) & 296.0(8) & 287.1(8) & 278.6(7) & 269.7(7) & 261.3(7) & 252.5(7) \\ 
0.62 & 0.71 & 197.2(7) & 187.9(7) & 181.0(7) & 172.5(6) & 164.2(6) & 156.2(6) & 147.5(6) & 140.0(5) & 132.1(5) & 126.1(5) & 119.6(5) & 112.6(4) & 107.2(4) \\ 
0.71 & 1.00 & 16.2(1) & 14.7(1) & 12.91(9) & 11.87(8) & 10.67(8) & 9.47(7) & 8.59(7) & 7.60(6) & 6.56(6) & 5.76(5) & 5.13(5) & 4.45(4) & 3.81(4) \\ 
 \hline
 \hline
 \end{tabular}
 \end{adjustbox}
\caption{
Same as~\ref{tab:LHC13_mpole_ABMP16_n3_CMS13TeV}  
but for $p_T^j > 75$ GeV.
\label{tab:LHC13_mpole_ABMP16_n3_CMS13TeV_pt75}}
 \end{table}
\end{landscape}
\begin{landscape}
\begin{table}[!ht]
\centering
\begin{adjustbox}{width=1.4\textwidth}
\small
\begin{tabular}{c|c|ccccccccccccc}
  \hline
&  & \multicolumn{13}{|c}{ $\frac{d\sigma_{t\bar{t}j + X}}{d\rho_s}$ distribution with $N_{j}\geq1,\,p_{T}^j>100$ GeV, $|\eta_{j}|<2.4$, $K_{R}=K_{F}=1$}  \\ 
 \hline 
 $x_{l}$  & $x_{r}$  & 165 GeV & 166 GeV & 167 GeV & 168 GeV & 169 GeV & 170 GeV & 171 GeV & 172 GeV & 173 GeV & 174 GeV & 175 GeV & 176 GeV & 177 GeV \\ 
 \hline 
0.00 & 0.18 & 17.87(5) & 17.62(5) & 17.61(5) & 17.33(5) & 17.13(5) & 17.14(5) & 16.87(5) & 16.62(4) & 16.57(4) & 16.43(4) & 16.27(4) & 16.13(4) & 16.03(4) \\ 
0.18 & 0.22 & 128.7(4) & 128.7(4) & 126.6(4) & 124.6(4) & 123.8(4) & 122.8(4) & 121.2(4) & 118.9(3) & 117.8(3) & 116.0(3) & 115.5(3) & 114.4(3) & 112.5(3) \\ 
0.22 & 0.27 & 214.3(5) & 211.6(5) & 208.2(4) & 205.2(5) & 202.5(4) & 199.7(4) & 196.8(4) & 194.6(4) & 192.2(4) & 189.7(4) & 187.5(4) & 184.7(4) & 182.3(4) \\ 
0.27 & 0.32 & 293.0(6) & 287.3(6) & 284.0(5) & 280.5(6) & 276.6(5) & 272.1(5) & 268.5(5) & 264.2(5) & 259.8(5) & 255.7(5) & 252.5(5) & 248.7(5) & 244.0(5) \\ 
0.32 & 0.38 & 346.4(6) & 340.2(6) & 333.7(6) & 329.8(6) & 323.0(5) & 317.9(5) & 312.8(5) & 308.8(5) & 302.4(5) & 297.7(5) & 291.6(5) & 287.7(5) & 283.2(5) \\ 
0.38 & 0.45 & 356.1(6) & 348.3(6) & 342.1(5) & 335.2(5) & 328.7(5) & 322.7(5) & 315.8(5) & 308.7(5) & 303.2(5) & 298.3(4) & 292.0(4) & 285.9(5) & 281.1(4) \\ 
0.45 & 0.53 & 303.7(5) & 296.3(5) & 290.2(5) & 281.8(5) & 275.4(5) & 267.7(4) & 261.7(4) & 255.4(4) & 247.6(4) & 241.9(4) & 235.9(4) & 229.9(4) & 224.9(4) \\ 
0.53 & 0.62 & 194.3(4) & 187.5(4) & 180.2(4) & 174.5(4) & 168.0(3) & 162.4(3) & 156.0(3) & 151.2(3) & 144.9(3) & 140.0(3) & 134.4(3) & 129.5(3) & 124.5(3) \\ 
0.62 & 0.71 & 71.1(2) & 66.8(2) & 62.4(2) & 58.2(2) & 54.6(2) & 50.7(2) & 47.7(2) & 44.4(2) & 41.3(2) & 38.1(2) & 35.3(1) & 32.8(1) & 29.9(1) \\ 
0.71 & 1.00 & 1.73(2) & 1.51(2) & 1.27(1) & 1.03(1) & 0.87(1) & 0.69(1) & 0.560(8) & 0.458(7) & 0.356(6) & 0.281(5) & 0.215(4) & 0.161(3) & 0.116(2) \\ 
 \hline
&  & \multicolumn{13}{|c}{ $\frac{d\sigma_{t\bar{t}j + X}}{d\rho_s}$ distribution with $N_{j}\geq1,\,p_{T}^j>100$ GeV, $|\eta_{j}|<2.4$, $K_{R}=K_{F}=2$}  \\ 
 \hline 
 $x_{l}$  & $x_{r}$  & 165 GeV & 166 GeV & 167 GeV & 168 GeV & 169 GeV & 170 GeV & 171 GeV & 172 GeV & 173 GeV & 174 GeV & 175 GeV & 176 GeV & 177 GeV \\ 
 \hline 
0.00 & 0.18 & 17.44(3) & 17.38(3) & 17.03(3) & 16.90(3) & 16.72(3) & 16.53(3) & 16.41(3) & 16.23(3) & 16.05(3) & 15.87(3) & 15.70(3) & 15.56(3) & 15.37(3) \\ 
0.18 & 0.22 & 120.0(2) & 117.8(2) & 116.6(3) & 115.7(2) & 114.2(2) & 112.3(2) & 110.6(2) & 109.2(2) & 108.6(2) & 107.3(2) & 105.4(2) & 104.3(2) & 103.0(2) \\ 
0.22 & 0.27 & 192.4(3) & 189.6(3) & 187.1(3) & 184.5(3) & 181.5(3) & 179.8(3) & 177.5(3) & 174.7(3) & 172.2(3) & 170.2(3) & 168.0(3) & 165.4(2) & 163.4(2) \\ 
0.27 & 0.32 & 260.5(4) & 255.9(4) & 252.3(4) & 249.5(4) & 244.4(3) & 240.9(3) & 237.4(3) & 233.5(3) & 229.9(3) & 226.7(3) & 222.7(3) & 220.1(3) & 216.9(3) \\ 
0.32 & 0.38 & 305.0(4) & 299.5(4) & 294.1(4) & 288.5(4) & 284.1(3) & 278.7(3) & 274.8(3) & 270.0(3) & 265.8(3) & 260.7(3) & 257.0(3) & 252.0(3) & 246.8(3) \\ 
0.38 & 0.45 & 310.2(4) & 304.5(3) & 298.0(3) & 292.2(3) & 286.3(3) & 280.6(3) & 275.3(3) & 270.0(3) & 263.8(3) & 259.2(3) & 253.3(3) & 248.8(3) & 244.2(3) \\ 
0.45 & 0.53 & 263.5(3) & 256.9(3) & 249.7(3) & 244.6(3) & 238.1(3) & 232.4(3) & 226.4(3) & 220.5(3) & 215.2(3) & 209.4(3) & 204.5(3) & 199.2(2) & 194.0(2) \\ 
0.53 & 0.62 & 167.2(3) & 161.3(2) & 155.9(2) & 150.4(2) & 145.0(2) & 139.9(2) & 134.3(2) & 130.0(2) & 124.7(2) & 120.5(2) & 115.8(2) & 111.6(2) & 107.3(2) \\ 
0.62 & 0.71 & 60.7(2) & 57.2(2) & 53.7(1) & 50.3(1) & 46.5(1) & 43.6(1) & 40.8(1) & 37.9(1) & 35.2(1) & 32.5(1) & 30.1(1) & 27.75(9) & 25.73(9) \\ 
0.71 & 1.00 & 1.42(1) & 1.22(1) & 0.99(1) & 0.836(9) & 0.689(7) & 0.554(7) & 0.443(6) & 0.355(5) & 0.277(4) & 0.215(3) & 0.153(3) & 0.116(2) & 0.083(2) \\ 
 \hline
&  & \multicolumn{13}{|c}{ $\frac{d\sigma_{t\bar{t}j + X}}{d\rho_s}$ distribution with $N_{j}\geq1,\,p_{T}^j>100$ GeV, $|\eta_{j}|<2.4$, $K_{R}=K_{F}=0.5$}  \\ 
 \hline 
 $x_{l}$  & $x_{r}$  & 165 GeV & 166 GeV & 167 GeV & 168 GeV & 169 GeV & 170 GeV & 171 GeV & 172 GeV & 173 GeV & 174 GeV & 175 GeV & 176 GeV & 177 GeV \\ 
 \hline 
0.00 & 0.18 & 10.67(9) & 10.79(9) & 10.69(9) & 10.36(8) & 10.45(8) & 10.65(8) & 10.60(8) & 10.43(8) & 10.48(8) & 10.48(7) & 10.50(7) & 10.37(7) & 10.35(7) \\ 
0.18 & 0.22 & 105.5(7) & 104.8(6) & 104.3(6) & 103.7(6) & 103.7(6) & 101.3(6) & 101.8(6) & 98.7(6) & 98.9(6) & 97.9(6) & 97.7(5) & 96.8(5) & 96.2(5) \\ 
0.22 & 0.27 & 192.6(8) & 190.7(8) & 187.9(8) & 188.3(7) & 185.9(8) & 182.4(7) & 179.7(7) & 180.1(7) & 177.2(7) & 176.2(7) & 172.2(7) & 171.6(6) & 168.8(6) \\ 
0.27 & 0.32 & 283.0(1) & 277.7(1) & 273.6(1) & 270.6(9) & 264.8(9) & 262.6(9) & 259.6(9) & 256.5(9) & 253.5(9) & 249.9(8) & 248.7(8) & 241.3(8) & 241.1(8) \\ 
0.32 & 0.38 & 347.6(1) & 341.9(1) & 336.7(1) & 329.0(9) & 325.2(1) & 322.2(9) & 315.2(8) & 310.1(8) & 307.1(9) & 299.6(9) & 296.1(8) & 292.4(7) & 286.4(8) \\ 
0.38 & 0.45 & 368.3(1) & 360.3(9) & 356.8(9) & 348.0(9) & 342.8(9) & 336.3(8) & 329.1(8) & 322.8(8) & 316.7(8) & 310.1(8) & 303.8(8) & 300.3(7) & 294.0(7) \\ 
0.45 & 0.53 & 323.3(8) & 315.7(8) & 308.4(8) & 301.9(8) & 291.5(8) & 286.0(8) & 279.8(7) & 271.4(7) & 264.3(7) & 259.0(7) & 254.0(7) & 245.9(6) & 239.5(6) \\ 
0.53 & 0.62 & 210.4(7) & 203.4(6) & 196.5(6) & 189.0(6) & 183.1(6) & 176.3(6) & 170.6(5) & 164.6(6) & 158.2(5) & 152.0(5) & 147.6(5) & 142.0(5) & 136.1(4) \\ 
0.62 & 0.71 & 78.3(4) & 73.9(4) & 69.4(3) & 65.6(3) & 61.0(3) & 56.8(3) & 53.0(3) & 48.4(3) & 46.2(3) & 42.7(3) & 39.2(2) & 36.3(2) & 34.0(2) \\ 
0.71 & 1.00 & 2.17(3) & 1.83(3) & 1.49(2) & 1.31(2) & 1.09(2) & 0.89(2) & 0.75(1) & 0.58(1) & 0.484(9) & 0.366(8) & 0.293(6) & 0.233(5) & 0.182(4) \\ 
 \hline
 \hline
 \end{tabular}
 \end{adjustbox}
\caption{
Same as~\ref{tab:LHC13_mpole_ABMP16_n3_CMS13TeV} 
but for $p_T^j > 100$ GeV.
\label{tab:LHC13_mpole_ABMP16_n3_CMS13TeV_pt100}}
 \end{table}
\end{landscape}

\begin{landscape}
\begin{table}[!ht]
\centering
\begin{adjustbox}{width=1.4\textwidth}
\small
\begin{tabular}{c|c|ccccccccccccc}
  \hline
&  & \multicolumn{13}{|c}{ $\frac{d\sigma_{t\bar{t}j + X}}{d\rho_s}$ distribution with $N_{j}\geq1,\,p_{T}^j>30$ GeV, $|\eta_{j}|<2.4$, $K_{R}=K_{F}=1$}  \\ 
 \hline 
 $x_{l}$  & $x_{r}$  & 165 GeV & 166 GeV & 167 GeV & 168 GeV & 169 GeV & 170 GeV & 171 GeV & 172 GeV & 173 GeV & 174 GeV & 175 GeV & 176 GeV & 177 GeV \\ 
 \hline 
0.00 & 0.18 & 26.70(8) & 26.61(8) & 26.24(7) & 26.18(8) & 25.77(8) & 25.57(8) & 25.25(7) & 25.09(7) & 24.94(7) & 24.68(7) & 24.50(7) & 24.39(6) & 24.03(6) \\ 
0.18 & 0.22 & 208.5(7) & 206.6(7) & 203.9(6) & 201.7(7) & 199.9(6) & 197.8(7) & 195.0(6) & 192.9(6) & 190.5(6) & 189.5(6) & 187.3(6) & 185.5(6) & 182.5(5) \\ 
0.22 & 0.27 & 381.3(9) & 376.4(9) & 371.6(9) & 366.9(1) & 362.8(9) & 358.3(9) & 353.8(9) & 349.6(8) & 344.2(8) & 340.5(8) & 335.0(8) & 332.5(8) & 327.8(8) \\ 
0.27 & 0.32 & 593.0(1) & 582.0(1) & 575.0(1) & 569.0(1) & 560.0(1) & 551.0(1) & 544.0(1) & 539.0(1) & 530.0(1) & 522.0(1) & 516.0(1) & 512.0(1) & 505.0(1) \\ 
0.32 & 0.38 & 807.0(1) & 799.0(1) & 786.0(1) & 775.0(2) & 762.0(1) & 750.0(1) & 740.0(1) & 727.0(1) & 716.0(1) & 709.0(1) & 698.0(1) & 687.0(1) & 676.0(1) \\ 
0.38 & 0.45 & 1002.0(2) & 984.0(2) & 968.0(1) & 951.0(2) & 936.0(2) & 922.0(1) & 907.0(1) & 890.0(1) & 876.0(1) & 860.0(1) & 849.0(1) & 832.0(1) & 820.0(1) \\ 
0.45 & 0.53 & 1110.0(2) & 1083.0(2) & 1064.0(2) & 1044.0(2) & 1025.0(2) & 1003.0(2) & 983.0(2) & 969.0(1) & 948.0(1) & 929.0(1) & 913.0(1) & 896.0(1) & 880.0(1) \\ 
0.53 & 0.62 & 1063.0(2) & 1040.0(2) & 1017.0(1) & 993.0(2) & 968.0(1) & 947.0(1) & 925.0(1) & 901.0(1) & 881.0(1) & 862.0(1) & 839.0(1) & 819.0(1) & 801.0(1) \\ 
0.62 & 0.71 & 853.0(1) & 831.0(1) & 804.0(1) & 780.0(1) & 754.0(1) & 731.0(1) & 707.0(1) & 685.0(1) & 664.0(1) & 640.0(1) & 617.0(1) & 598.0(1) & 577.1(1) \\ 
0.71 & 1.00 & 232.0(4) & 218.9(4) & 204.1(3) & 192.0(4) & 179.0(3) & 168.4(3) & 156.8(3) & 146.7(3) & 137.1(3) & 127.8(3) & 119.2(2) & 111.3(2) & 103.0(2) \\ 
 \hline
&  & \multicolumn{13}{|c}{ $\frac{d\sigma_{t\bar{t}j + X}}{d\rho_s}$ distribution with $N_{j}\geq1,\,p_{T}^j>30$ GeV, $|\eta_{j}|<2.4$, $K_{R}=K_{F}=2$}  \\ 
 \hline 
 $x_{l}$  & $x_{r}$  & 165 GeV & 166 GeV & 167 GeV & 168 GeV & 169 GeV & 170 GeV & 171 GeV & 172 GeV & 173 GeV & 174 GeV & 175 GeV & 176 GeV & 177 GeV \\ 
 \hline 
0.00 & 0.18 & 26.63(5) & 26.29(5) & 25.96(5) & 25.77(5) & 25.63(5) & 25.23(5) & 25.08(5) & 24.79(4) & 24.35(5) & 24.28(4) & 24.06(4) & 23.78(4) & 23.53(4) \\ 
0.18 & 0.22 & 194.1(4) & 191.4(4) & 188.9(4) & 187.6(4) & 184.7(4) & 182.6(4) & 181.3(4) & 177.8(4) & 176.9(4) & 174.0(4) & 171.6(4) & 169.6(4) & 168.2(4) \\ 
0.22 & 0.27 & 345.6(6) & 341.2(6) & 336.0(6) & 333.1(6) & 329.4(6) & 324.0(6) & 319.3(6) & 316.5(5) & 311.9(5) & 306.5(5) & 304.0(5) & 299.9(5) & 296.6(5) \\ 
0.27 & 0.32 & 530.5(9) & 521.4(9) & 514.7(8) & 507.1(9) & 498.5(9) & 493.2(8) & 487.4(8) & 479.2(8) & 475.3(8) & 467.5(7) & 460.5(7) & 456.3(7) & 447.3(7) \\ 
0.32 & 0.38 & 715.1(1) & 705.9(1) & 695.4(1) & 683.9(1) & 676.7(9) & 664.2(9) & 651.3(9) & 644.0(9) & 633.1(9) & 626.4(8) & 615.9(8) & 607.5(8) & 599.0(8) \\ 
0.38 & 0.45 & 882.0(1) & 867.0(1) & 852.6(1) & 837.0(1) & 823.8(9) & 813.3(9) & 799.5(9) & 783.5(9) & 773.0(9) & 758.0(8) & 746.0(9) & 734.0(8) & 724.5(8) \\ 
0.45 & 0.53 & 973.0(1) & 953.0(1) & 937.0(1) & 919.0(1) & 903.0(1) & 882.8(9) & 866.8(9) & 849.4(9) & 831.9(9) & 816.8(8) & 802.1(9) & 785.6(9) & 769.5(8) \\ 
0.53 & 0.62 & 932.0(1) & 912.0(1) & 891.4(1) & 869.1(1) & 849.1(9) & 830.8(9) & 809.8(9) & 790.0(9) & 771.2(9) & 754.5(8) & 735.2(8) & 718.8(8) & 701.4(7) \\ 
0.62 & 0.71 & 750.6(1) & 727.0(9) & 704.0(9) & 682.6(9) & 658.4(9) & 639.4(9) & 619.2(8) & 598.5(8) & 579.0(8) & 560.4(7) & 541.8(7) & 524.8(7) & 506.4(7) \\ 
0.71 & 1.00 & 202.6(3) & 190.5(2) & 178.3(2) & 167.4(2) & 156.8(2) & 147.1(2) & 137.4(2) & 128.5(2) & 120.1(2) & 111.9(2) & 104.0(2) & 96.9(2) & 90.1(1) \\ 
 \hline
&  & \multicolumn{13}{|c}{ $\frac{d\sigma_{t\bar{t}j + X}}{d\rho_s}$ distribution with $N_{j}\geq1,\,p_{T}^j>30$ GeV, $|\eta_{j}|<2.4$, $K_{R}=K_{F}=0.5$}  \\ 
 \hline 
 $x_{l}$  & $x_{r}$  & 165 GeV & 166 GeV & 167 GeV & 168 GeV & 169 GeV & 170 GeV & 171 GeV & 172 GeV & 173 GeV & 174 GeV & 175 GeV & 176 GeV & 177 GeV \\ 
 \hline 
0.00 & 0.18 & 15.0(1) & 14.8(1) & 14.8(1) & 15.0(1) & 15.0(1) & 14.7(1) & 14.7(1) & 14.8(1) & 14.8(1) & 14.5(1) & 14.6(1) & 14.7(1) & 14.5(1) \\ 
0.18 & 0.22 & 165.0(1) & 163.0(1) & 163.0(1) & 164.0(1) & 159.0(1) & 159.0(1) & 159.0(1) & 155.0(1) & 155.0(1) & 153.9(9) & 153.3(1) & 151.4(9) & 149.5(9) \\ 
0.22 & 0.27 & 338.0(2) & 332.0(2) & 331.0(2) & 328.0(2) & 325.0(2) & 322.0(1) & 318.0(1) & 312.0(1) & 310.0(1) & 306.0(1) & 305.0(1) & 300.0(1) & 298.0(1) \\ 
0.27 & 0.32 & 560.0(2) & 555.0(2) & 548.0(2) & 535.0(2) & 530.0(2) & 522.0(2) & 518.0(2) & 514.0(2) & 506.0(2) & 503.0(2) & 495.0(2) & 486.0(2) & 485.0(2) \\ 
0.32 & 0.38 & 798.0(3) & 787.0(3) & 775.0(3) & 764.0(3) & 754.0(2) & 743.0(2) & 734.0(2) & 721.0(2) & 712.0(2) & 696.0(2) & 695.0(2) & 684.0(2) & 672.0(2) \\ 
0.38 & 0.45 & 1017.0(3) & 998.0(3) & 981.0(3) & 968.0(3) & 952.0(3) & 938.0(2) & 922.0(2) & 912.0(2) & 893.0(2) & 879.0(2) & 860.0(2) & 853.0(2) & 836.0(2) \\ 
0.45 & 0.53 & 1151.0(3) & 1129.0(3) & 1105.0(3) & 1086.0(3) & 1061.0(3) & 1046.0(2) & 1025.0(3) & 1005.0(2) & 984.0(2) & 972.0(2) & 949.0(2) & 935.0(2) & 910.0(2) \\ 
0.53 & 0.62 & 1119.0(3) & 1095.0(3) & 1073.0(3) & 1045.0(2) & 1019.0(2) & 996.0(2) & 972.0(2) & 952.0(2) & 928.0(2) & 907.0(2) & 883.0(2) & 859.0(2) & 845.0(2) \\ 
0.62 & 0.71 & 912.0(2) & 886.0(2) & 858.0(2) & 827.0(2) & 803.0(2) & 776.0(2) & 751.0(2) & 728.0(2) & 704.0(2) & 680.0(2) & 660.0(2) & 638.0(2) & 616.0(2) \\ 
0.71 & 1.00 & 250.0(6) & 234.8(6) & 219.3(6) & 205.9(6) & 192.9(6) & 180.6(5) & 169.5(5) & 157.8(4) & 147.6(4) & 138.2(4) & 127.9(4) & 118.8(4) & 111.4(4) \\ 
 \hline
 \hline
 \end{tabular}
 \end{adjustbox}
\caption{
NLO predictions for the $\rho_s$ distribution in the pole mass scheme for $\sqrt{S}=14$~TeV using as input the static scale $\mu_0 = m_t$  and the \text{ABMP16} NLO PDF set. 
At least one jet with $|\eta_j| < 2.4$ and $p_T^j > 30$~GeV is required.
 \label{tab:LHC14_mpole_ABMP16_n3_CMS13TeV}}
 \end{table}
\end{landscape}
\begin{landscape}
\begin{table}[!ht]
\centering
\begin{adjustbox}{width=1.4\textwidth}
\small
\begin{tabular}{c|c|ccccccccccccc}
  \hline
&  & \multicolumn{13}{|c}{ $\frac{d\sigma_{t\bar{t}j + X}}{d\rho_s}$ distribution with $N_{j}\geq1,\,p_{T}^j>50$ GeV, $|\eta_{j}|<2.4$, $K_{R}=K_{F}=1$}  \\ 
 \hline 
 $x_{l}$  & $x_{r}$  & 165 GeV & 166 GeV & 167 GeV & 168 GeV & 169 GeV & 170 GeV & 171 GeV & 172 GeV & 173 GeV & 174 GeV & 175 GeV & 176 GeV & 177 GeV \\ 
 \hline 
0.00 & 0.18 & 26.39(7) & 26.18(7) & 25.91(7) & 25.82(7) & 25.41(7) & 25.23(7) & 24.91(7) & 24.72(7) & 24.56(6) & 24.31(7) & 24.13(6) & 23.99(6) & 23.60(6) \\ 
0.18 & 0.22 & 199.9(6) & 197.9(6) & 194.2(6) & 193.4(6) & 191.3(6) & 189.0(6) & 186.9(5) & 184.1(5) & 181.7(5) & 180.2(5) & 178.8(5) & 176.4(5) & 174.4(5) \\ 
0.22 & 0.27 & 351.2(8) & 346.0(8) & 341.7(8) & 336.8(8) & 332.9(8) & 330.0(7) & 324.7(7) & 320.1(7) & 317.0(7) & 312.8(7) & 307.3(7) & 306.1(7) & 300.9(7) \\ 
0.27 & 0.32 & 521.0(1) & 510.0(1) & 505.0(1) & 499.0(1) & 490.0(1) & 482.8(1) & 477.0(1) & 470.5(9) & 463.8(9) & 457.2(1) & 450.9(9) & 445.5(9) & 440.8(9) \\ 
0.32 & 0.38 & 672.0(1) & 663.0(1) & 652.0(1) & 643.0(1) & 632.0(1) & 622.0(1) & 612.0(1) & 602.0(1) & 592.0(1) & 584.9(1) & 575.7(1) & 566.7(9) & 556.9(9) \\ 
0.38 & 0.45 & 778.0(1) & 767.0(1) & 751.0(1) & 739.0(1) & 723.0(1) & 713.0(1) & 701.0(1) & 687.7(1) & 675.9(1) & 663.9(1) & 651.6(9) & 641.1(9) & 630.1(9) \\ 
0.45 & 0.53 & 794.0(1) & 776.0(1) & 759.0(1) & 744.0(1) & 731.0(1) & 713.0(1) & 699.0(1) & 685.8(9) & 669.3(9) & 655.8(9) & 642.4(9) & 629.5(9) & 615.8(8) \\ 
0.53 & 0.62 & 681.8(1) & 665.5(1) & 647.7(9) & 628.0(1) & 612.1(1) & 595.8(9) & 580.5(9) & 564.9(8) & 549.5(8) & 535.1(8) & 519.4(8) & 504.0(8) & 491.2(7) \\ 
0.62 & 0.71 & 462.6(8) & 445.6(8) & 427.9(8) & 411.9(8) & 397.0(8) & 382.3(8) & 365.9(7) & 353.3(7) & 339.5(6) & 325.5(6) & 312.3(6) & 300.6(6) & 286.7(6) \\ 
0.71 & 1.00 & 77.2(2) & 71.4(2) & 66.5(2) & 61.4(2) & 56.5(2) & 52.0(1) & 48.1(1) & 44.1(1) & 40.4(1) & 37.1(1) & 34.1(1) & 30.9(1) & 28.35(9) \\ 
 \hline
&  & \multicolumn{13}{|c}{ $\frac{d\sigma_{t\bar{t}j + X}}{d\rho_s}$ distribution with $N_{j}\geq1,\,p_{T}^j>50$ GeV, $|\eta_{j}|<2.4$, $K_{R}=K_{F}=2$}  \\ 
 \hline 
 $x_{l}$  & $x_{r}$  & 165 GeV & 166 GeV & 167 GeV & 168 GeV & 169 GeV & 170 GeV & 171 GeV & 172 GeV & 173 GeV & 174 GeV & 175 GeV & 176 GeV & 177 GeV \\ 
 \hline 
0.00 & 0.18 & 25.97(5) & 25.68(5) & 25.33(5) & 25.13(5) & 24.99(4) & 24.60(4) & 24.45(4) & 24.15(4) & 23.73(4) & 23.65(4) & 23.42(4) & 23.16(4) & 22.90(4) \\ 
0.18 & 0.22 & 184.1(4) & 181.5(4) & 179.2(4) & 177.8(4) & 175.2(4) & 173.0(4) & 171.7(4) & 168.7(3) & 167.3(4) & 164.7(3) & 162.4(3) & 160.3(3) & 158.8(3) \\ 
0.22 & 0.27 & 315.6(5) & 311.7(5) & 306.9(5) & 303.8(5) & 300.0(5) & 295.5(5) & 291.1(5) & 288.0(5) & 284.5(5) & 279.2(4) & 277.2(4) & 273.0(4) & 269.3(4) \\ 
0.27 & 0.32 & 462.4(7) & 454.5(7) & 448.8(7) & 440.6(7) & 434.2(7) & 429.0(7) & 423.3(6) & 416.1(6) & 412.1(6) & 405.6(6) & 398.6(6) & 394.7(6) & 387.9(6) \\ 
0.32 & 0.38 & 590.9(8) & 582.0(7) & 573.8(7) & 564.1(7) & 555.1(7) & 545.1(7) & 535.5(7) & 529.1(7) & 519.8(6) & 513.3(6) & 504.4(6) & 496.3(6) & 488.8(6) \\ 
0.38 & 0.45 & 681.6(8) & 669.5(8) & 656.2(7) & 643.1(8) & 631.8(7) & 623.6(7) & 611.8(7) & 599.6(7) & 590.8(6) & 578.9(6) & 570.1(6) & 558.5(6) & 549.8(6) \\ 
0.45 & 0.53 & 691.6(7) & 675.0(7) & 661.9(7) & 649.2(7) & 634.5(7) & 620.4(7) & 607.5(6) & 596.5(6) & 583.2(6) & 570.2(6) & 558.0(6) & 546.4(6) & 534.2(5) \\ 
0.53 & 0.62 & 589.0(7) & 575.5(6) & 560.5(6) & 546.1(6) & 531.1(6) & 517.7(6) & 504.1(6) & 489.4(5) & 475.2(6) & 464.1(5) & 449.9(5) & 438.4(5) & 426.5(5) \\ 
0.62 & 0.71 & 400.6(6) & 385.2(6) & 370.6(5) & 357.4(5) & 342.7(5) & 331.1(5) & 318.0(5) & 305.7(4) & 293.5(4) & 282.2(4) & 270.4(4) & 259.7(4) & 248.2(4) \\ 
0.71 & 1.00 & 66.5(1) & 61.6(1) & 57.0(1) & 52.6(1) & 48.7(1) & 44.76(9) & 41.32(9) & 38.08(9) & 34.83(8) & 31.95(8) & 29.28(7) & 26.73(7) & 24.33(7) \\ 
 \hline
&  & \multicolumn{13}{|c}{ $\frac{d\sigma_{t\bar{t}j + X}}{d\rho_s}$ distribution with $N_{j}\geq1,\,p_{T}^j>50$ GeV, $|\eta_{j}|<2.4$, $K_{R}=K_{F}=0.5$}  \\ 
 \hline 
 $x_{l}$  & $x_{r}$  & 165 GeV & 166 GeV & 167 GeV & 168 GeV & 169 GeV & 170 GeV & 171 GeV & 172 GeV & 173 GeV & 174 GeV & 175 GeV & 176 GeV & 177 GeV \\ 
 \hline 
0.00 & 0.18 & 15.6(1) & 15.5(1) & 15.4(1) & 15.6(1) & 15.6(1) & 15.3(1) & 15.4(1) & 15.3(1) & 15.3(1) & 15.1(1) & 15.1(1) & 15.3(1) & 15.2(1) \\ 
0.18 & 0.22 & 163.0(1) & 161.0(1) & 162.0(1) & 162.0(1) & 157.0(1) & 156.0(1) & 156.6(1) & 153.6(1) & 153.0(1) & 152.0(8) & 151.5(8) & 150.5(8) & 147.3(8) \\ 
0.22 & 0.27 & 319.0(1) & 314.0(1) & 314.0(1) & 309.0(1) & 306.0(1) & 303.0(1) & 300.0(1) & 294.0(1) & 293.0(1) & 290.0(1) & 286.0(1) & 283.0(1) & 282.0(1) \\ 
0.27 & 0.32 & 503.0(2) & 500.0(2) & 492.0(2) & 480.0(2) & 478.0(2) & 470.0(2) & 464.0(2) & 459.0(2) & 450.0(2) & 449.0(1) & 441.0(1) & 434.0(1) & 432.0(1) \\ 
0.32 & 0.38 & 680.0(2) & 666.0(2) & 657.0(2) & 646.0(2) & 637.0(2) & 627.0(2) & 620.0(2) & 609.0(2) & 601.0(2) & 591.0(2) & 584.0(2) & 574.0(2) & 566.0(1) \\ 
0.38 & 0.45 & 808.0(2) & 791.0(2) & 777.0(2) & 766.0(2) & 751.0(2) & 739.0(2) & 728.0(2) & 714.0(2) & 703.0(2) & 690.0(2) & 678.0(2) & 668.0(1) & 657.0(1) \\ 
0.45 & 0.53 & 840.0(2) & 823.0(2) & 805.0(2) & 792.0(2) & 771.0(2) & 757.0(2) & 741.0(2) & 725.0(2) & 711.0(2) & 699.0(2) & 683.0(1) & 670.0(1) & 653.0(1) \\ 
0.53 & 0.62 & 731.0(2) & 714.0(2) & 696.0(2) & 676.0(2) & 662.0(2) & 641.0(1) & 627.0(1) & 608.0(1) & 592.0(1) & 576.0(1) & 560.0(1) & 542.0(1) & 531.0(1) \\ 
0.62 & 0.71 & 504.0(1) & 487.0(1) & 467.0(1) & 448.0(1) & 433.0(1) & 416.0(1) & 399.0(1) & 384.0(1) & 370.0(1) & 353.0(1) & 340.7(1) & 326.9(1) & 315.6(9) \\ 
0.71 & 1.00 & 85.6(3) & 78.7(3) & 73.2(3) & 68.2(3) & 62.5(3) & 57.8(2) & 53.3(2) & 48.8(2) & 45.2(2) & 41.1(2) & 37.6(2) & 34.4(2) & 31.3(1) \\ 
 \hline
 \hline
 \end{tabular}
 \end{adjustbox}
\caption{
Same as Fig.~\ref{tab:LHC14_mpole_ABMP16_n3_CMS13TeV} but for $p_T^j > 50$~GeV.
 \label{tab:LHC14_mpole_ABMP16_n3_CMS13TeV_pt50}}
 \end{table}
\end{landscape}
\begin{landscape}
\begin{table}[!ht]
\centering
\begin{adjustbox}{width=1.4\textwidth}
\small
\begin{tabular}{c|c|ccccccccccccc}
  \hline
&  & \multicolumn{13}{|c}{ $\frac{d\sigma_{t\bar{t}j + X}}{d\rho_s}$ distribution with $N_{j}\geq1,\,p_{T}^j>75$ GeV, $|\eta_{j}|<2.4$, $K_{R}=K_{F}=1$}  \\ 
 \hline 
 $x_{l}$  & $x_{r}$  & 165 GeV & 166 GeV & 167 GeV & 168 GeV & 169 GeV & 170 GeV & 171 GeV & 172 GeV & 173 GeV & 174 GeV & 175 GeV & 176 GeV & 177 GeV \\ 
 \hline 
0.00 & 0.18 & 25.61(7) & 25.42(7) & 25.11(6) & 25.06(7) & 24.63(7) & 24.45(7) & 24.13(6) & 24.00(6) & 23.79(6) & 23.61(6) & 23.32(6) & 23.22(6) & 22.88(5) \\ 
0.18 & 0.22 & 185.5(6) & 183.8(5) & 180.1(5) & 178.8(6) & 176.8(5) & 175.1(5) & 173.5(5) & 170.0(5) & 168.5(5) & 166.9(5) & 165.8(5) & 163.5(5) & 161.3(4) \\ 
0.22 & 0.27 & 309.4(7) & 305.6(7) & 302.5(6) & 297.7(7) & 295.0(7) & 290.6(6) & 286.5(6) & 282.8(6) & 279.4(6) & 275.1(6) & 271.3(6) & 268.2(6) & 264.5(5) \\ 
0.27 & 0.32 & 437.3(9) & 429.0(8) & 423.8(8) & 418.7(9) & 410.7(8) & 403.9(8) & 397.5(8) & 393.9(7) & 387.6(7) & 382.7(8) & 377.5(7) & 372.1(7) & 367.3(7) \\ 
0.32 & 0.38 & 534.6(9) & 525.0(9) & 516.9(8) & 509.5(9) & 500.3(8) & 492.8(8) & 483.5(8) & 474.9(7) & 467.1(8) & 461.6(8) & 454.0(7) & 446.0(7) & 437.8(7) \\ 
0.38 & 0.45 & 574.9(9) & 566.9(9) & 555.5(8) & 545.6(9) & 534.3(8) & 525.0(8) & 515.2(8) & 506.5(7) & 495.8(7) & 486.0(7) & 476.5(7) & 469.2(7) & 460.2(6) \\ 
0.45 & 0.53 & 535.5(8) & 523.1(8) & 512.3(7) & 499.8(8) & 489.5(7) & 476.7(7) & 465.4(7) & 455.3(7) & 445.8(7) & 435.1(6) & 424.7(6) & 415.7(6) & 406.0(6) \\ 
0.53 & 0.62 & 400.0(7) & 387.7(7) & 375.4(6) & 364.2(7) & 352.8(6) & 341.8(6) & 332.5(6) & 320.8(5) & 310.7(5) & 300.9(5) & 292.3(5) & 281.7(5) & 273.8(5) \\ 
0.62 & 0.71 & 210.4(5) & 201.3(5) & 191.3(4) & 181.7(5) & 173.8(4) & 165.4(4) & 157.1(4) & 149.2(4) & 141.4(4) & 133.7(4) & 127.2(3) & 120.5(3) & 112.9(3) \\ 
0.71 & 1.00 & 16.76(7) & 15.24(7) & 13.69(6) & 12.33(7) & 10.93(6) & 9.68(5) & 8.67(5) & 7.70(5) & 6.78(4) & 5.87(4) & 5.16(4) & 4.50(3) & 3.85(3) \\ 
 \hline
&  & \multicolumn{13}{|c}{ $\frac{d\sigma_{t\bar{t}j + X}}{d\rho_s}$ distribution with $N_{j}\geq1,\,p_{T}^j>75$ GeV, $|\eta_{j}|<2.4$, $K_{R}=K_{F}=2$}  \\ 
 \hline 
 $x_{l}$  & $x_{r}$  & 165 GeV & 166 GeV & 167 GeV & 168 GeV & 169 GeV & 170 GeV & 171 GeV & 172 GeV & 173 GeV & 174 GeV & 175 GeV & 176 GeV & 177 GeV \\ 
 \hline 
0.00 & 0.18 & 25.04(5) & 24.76(5) & 24.43(4) & 24.25(4) & 24.10(4) & 23.74(4) & 23.59(4) & 23.31(4) & 22.88(4) & 22.81(4) & 22.62(4) & 22.35(4) & 22.05(4) \\ 
0.18 & 0.22 & 170.5(4) & 168.4(4) & 166.0(4) & 164.0(4) & 161.6(3) & 159.7(3) & 158.3(3) & 155.6(3) & 154.3(3) & 152.1(3) & 150.0(3) & 147.9(3) & 146.5(3) \\ 
0.22 & 0.27 & 278.6(5) & 275.2(5) & 270.9(4) & 267.7(4) & 264.1(4) & 260.4(4) & 256.6(4) & 253.5(4) & 250.2(4) & 245.6(4) & 243.6(4) & 240.0(4) & 236.4(4) \\ 
0.27 & 0.32 & 387.5(6) & 380.5(6) & 375.3(5) & 369.0(6) & 363.3(5) & 358.7(5) & 353.5(5) & 348.1(5) & 343.6(5) & 338.4(5) & 333.1(5) & 328.8(5) & 323.2(5) \\ 
0.32 & 0.38 & 468.0(6) & 461.3(6) & 453.1(5) & 445.7(6) & 439.4(5) & 430.4(5) & 423.1(5) & 417.2(5) & 409.1(5) & 403.5(5) & 397.3(5) & 389.9(5) & 384.4(4) \\ 
0.38 & 0.45 & 503.3(6) & 494.4(6) & 484.1(5) & 473.4(6) & 465.7(5) & 457.6(5) & 449.1(5) & 440.0(5) & 431.7(5) & 423.8(5) & 415.8(5) & 407.8(4) & 399.8(4) \\ 
0.45 & 0.53 & 465.3(5) & 453.9(5) & 444.1(5) & 434.2(5) & 425.6(5) & 413.4(5) & 404.2(4) & 395.4(4) & 386.5(4) & 376.4(4) & 368.3(4) & 359.7(4) & 350.4(4) \\ 
0.53 & 0.62 & 344.1(4) & 333.7(4) & 323.8(4) & 314.4(4) & 304.5(4) & 295.8(4) & 286.9(4) & 277.4(4) & 269.0(3) & 260.7(3) & 252.2(3) & 244.7(3) & 236.6(3) \\ 
0.62 & 0.71 & 181.4(3) & 173.5(3) & 164.8(3) & 156.9(3) & 149.5(3) & 142.4(3) & 134.9(3) & 128.5(2) & 121.5(2) & 115.0(2) & 108.8(2) & 103.5(2) & 97.1(2) \\ 
0.71 & 1.00 & 14.35(5) & 12.90(5) & 11.59(4) & 10.39(4) & 9.21(4) & 8.27(4) & 7.34(3) & 6.52(3) & 5.71(3) & 4.99(3) & 4.39(2) & 3.82(2) & 3.30(2) \\ 
 \hline
&  & \multicolumn{13}{|c}{ $\frac{d\sigma_{t\bar{t}j + X}}{d\rho_s}$ distribution with $N_{j}\geq1,\,p_{T}^j>75$ GeV, $|\eta_{j}|<2.4$, $K_{R}=K_{F}=0.5$}  \\ 
 \hline 
 $x_{l}$  & $x_{r}$  & 165 GeV & 166 GeV & 167 GeV & 168 GeV & 169 GeV & 170 GeV & 171 GeV & 172 GeV & 173 GeV & 174 GeV & 175 GeV & 176 GeV & 177 GeV \\ 
 \hline 
0.00 & 0.18 & 15.6(1) & 15.4(1) & 15.3(1) & 15.5(1) & 15.5(1) & 15.3(1) & 15.3(1) & 15.3(1) & 15.3(1) & 15.0(1) & 15.0(1) & 15.2(1) & 15.0(1) \\ 
0.18 & 0.22 & 152.5(1) & 150.8(1) & 151.7(1) & 150.7(9) & 147.4(9) & 146.5(9) & 146.6(9) & 144.4(9) & 143.1(9) & 142.3(8) & 141.0(8) & 140.3(8) & 138.3(8) \\ 
0.22 & 0.27 & 284.0(1) & 279.0(1) & 278.0(1) & 274.0(1) & 271.0(1) & 269.0(1) & 266.0(1) & 261.0(1) & 260.0(1) & 256.2(9) & 254.4(1) & 250.0(1) & 250.0(1) \\ 
0.27 & 0.32 & 421.0(2) & 421.0(2) & 414.0(1) & 405.0(1) & 401.0(1) & 395.0(1) & 389.0(1) & 385.0(1) & 379.0(1) & 377.0(1) & 371.0(1) & 365.0(1) & 362.0(1) \\ 
0.32 & 0.38 & 541.0(2) & 530.0(2) & 521.0(2) & 513.0(2) & 508.0(1) & 497.0(1) & 492.0(1) & 484.0(1) & 476.0(1) & 467.0(1) & 462.0(1) & 455.0(1) & 450.0(1) \\ 
0.38 & 0.45 & 600.0(1) & 588.0(1) & 580.0(1) & 568.0(1) & 558.0(1) & 548.0(1) & 541.0(1) & 529.0(1) & 518.0(1) & 510.0(1) & 498.0(1) & 490.0(1) & 483.0(1) \\ 
0.45 & 0.53 & 571.0(1) & 559.0(1) & 545.0(1) & 532.0(1) & 520.0(1) & 509.0(1) & 497.0(1) & 486.0(1) & 475.0(1) & 467.0(1) & 454.5(1) & 445.4(1) & 434.7(1) \\ 
0.53 & 0.62 & 431.0(1) & 420.0(1) & 408.0(1) & 396.0(1) & 385.0(1) & 371.7(1) & 361.9(9) & 349.9(9) & 339.4(9) & 328.1(8) & 318.8(8) & 306.4(8) & 297.1(8) \\ 
0.62 & 0.71 & 232.2(8) & 222.1(8) & 209.8(8) & 199.9(7) & 190.7(7) & 181.7(7) & 172.2(6) & 164.2(6) & 157.1(6) & 148.2(6) & 140.2(5) & 132.2(5) & 125.5(5) \\ 
0.71 & 1.00 & 19.0(1) & 17.0(1) & 15.4(1) & 13.9(1) & 12.46(9) & 11.11(9) & 9.86(8) & 8.65(7) & 7.68(7) & 6.71(6) & 5.95(6) & 5.21(5) & 4.43(5) \\ 
 \hline
 \hline
 \end{tabular}
 \end{adjustbox}
\caption{
Same as Fig.~\ref{tab:LHC14_mpole_ABMP16_n3_CMS13TeV} but for $p_T^j > 75$~GeV.
\label{tab:LHC14_mpole_ABMP16_n3_CMS13TeV_pt75}}
 \end{table}
\end{landscape}
\begin{landscape}
\begin{table}[!ht]
\centering
\begin{adjustbox}{width=1.4\textwidth}
\small
\begin{tabular}{c|c|ccccccccccccc}
  \hline
&  & \multicolumn{13}{|c}{ $\frac{d\sigma_{t\bar{t}j + X}}{d\rho_s}$ distribution with $N_{j}\geq1,\,p_{T}^j>100$ GeV, $|\eta_{j}|<2.4$, $K_{R}=K_{F}=1$}  \\ 
 \hline 
 $x_{l}$  & $x_{r}$  & 165 GeV & 166 GeV & 167 GeV & 168 GeV & 169 GeV & 170 GeV & 171 GeV & 172 GeV & 173 GeV & 174 GeV & 175 GeV & 176 GeV & 177 GeV \\ 
 \hline 
0.00 & 0.18 & 24.43(7) & 24.23(6) & 24.04(6) & 23.89(6) & 23.57(6) & 23.43(6) & 23.02(6) & 22.90(6) & 22.71(6) & 22.60(6) & 22.32(6) & 22.14(5) & 21.91(5) \\ 
0.18 & 0.22 & 168.6(5) & 167.4(5) & 164.2(4) & 162.4(5) & 161.0(5) & 159.0(5) & 157.4(4) & 154.7(4) & 153.3(4) & 151.6(4) & 149.9(4) & 147.8(4) & 146.5(4) \\ 
0.22 & 0.27 & 268.8(6) & 266.0(6) & 262.6(5) & 258.4(6) & 256.4(6) & 253.0(6) & 248.9(5) & 246.2(5) & 242.7(5) & 239.8(5) & 235.6(5) & 233.1(5) & 230.2(5) \\ 
0.27 & 0.32 & 364.3(7) & 356.5(7) & 352.9(7) & 348.3(8) & 342.7(7) & 335.9(7) & 330.6(7) & 326.6(6) & 322.8(6) & 317.6(6) & 312.6(6) & 309.8(6) & 303.9(6) \\ 
0.32 & 0.38 & 423.4(7) & 415.4(7) & 408.6(7) & 402.9(7) & 395.9(7) & 388.9(6) & 382.1(7) & 375.6(6) & 368.8(6) & 363.9(6) & 357.6(6) & 352.6(6) & 345.8(6) \\ 
0.38 & 0.45 & 428.3(7) & 421.7(7) & 411.8(6) & 404.0(7) & 395.3(6) & 389.1(6) & 381.3(6) & 374.0(6) & 366.1(6) & 358.2(5) & 351.4(5) & 344.0(5) & 337.8(5) \\ 
0.45 & 0.53 & 363.2(6) & 353.4(6) & 345.9(6) & 335.6(6) & 328.9(6) & 320.0(5) & 311.5(5) & 303.8(5) & 296.8(5) & 289.5(5) & 282.2(5) & 274.4(4) & 268.0(4) \\ 
0.53 & 0.62 & 228.8(5) & 221.5(5) & 213.4(4) & 206.2(5) & 198.7(4) & 191.5(4) & 185.2(4) & 177.9(4) & 171.8(3) & 165.1(4) & 159.9(3) & 152.1(3) & 147.7(3) \\ 
0.62 & 0.71 & 83.6(3) & 78.2(3) & 73.9(2) & 68.9(3) & 64.3(2) & 59.7(2) & 55.8(2) & 52.1(2) & 48.4(2) & 44.8(2) & 41.3(2) & 38.6(2) & 35.3(2) \\ 
0.71 & 1.00 & 2.08(2) & 1.76(2) & 1.45(2) & 1.21(2) & 1.01(1) & 0.82(1) & 0.68(1) & 0.524(8) & 0.419(7) & 0.332(6) & 0.245(5) & 0.181(4) & 0.143(3) \\ 
 \hline
&  & \multicolumn{13}{|c}{ $\frac{d\sigma_{t\bar{t}j + X}}{d\rho_s}$ distribution with $N_{j}\geq1,\,p_{T}^j>100$ GeV, $|\eta_{j}|<2.4$, $K_{R}=K_{F}=2$}  \\ 
 \hline 
 $x_{l}$  & $x_{r}$  & 165 GeV & 166 GeV & 167 GeV & 168 GeV & 169 GeV & 170 GeV & 171 GeV & 172 GeV & 173 GeV & 174 GeV & 175 GeV & 176 GeV & 177 GeV \\ 
 \hline 
0.00 & 0.18 & 23.99(4) & 23.67(4) & 23.40(4) & 23.17(4) & 23.02(4) & 22.69(4) & 22.54(4) & 22.28(4) & 21.88(4) & 21.79(4) & 21.60(3) & 21.36(3) & 21.08(4) \\ 
0.18 & 0.22 & 155.6(3) & 153.6(3) & 151.6(3) & 149.6(3) & 147.3(3) & 145.6(3) & 144.0(3) & 141.8(3) & 140.2(3) & 138.7(3) & 136.7(3) & 134.9(3) & 133.0(3) \\ 
0.22 & 0.27 & 243.5(4) & 240.4(4) & 236.7(4) & 233.8(4) & 230.4(4) & 227.2(3) & 223.7(3) & 220.7(3) & 218.2(3) & 214.3(3) & 211.9(3) & 209.2(3) & 206.0(3) \\ 
0.27 & 0.32 & 322.7(5) & 317.7(5) & 313.1(5) & 308.2(5) & 303.7(4) & 298.7(4) & 294.3(4) & 289.9(4) & 286.2(4) & 282.2(4) & 276.8(4) & 273.5(4) & 268.9(4) \\ 
0.32 & 0.38 & 371.9(5) & 366.3(5) & 359.6(4) & 354.4(5) & 347.8(4) & 341.8(4) & 335.8(4) & 329.9(4) & 323.7(4) & 319.3(4) & 313.8(4) & 308.7(4) & 303.2(4) \\ 
0.38 & 0.45 & 375.3(4) & 367.7(4) & 360.4(4) & 352.8(4) & 345.1(4) & 339.6(4) & 332.7(4) & 325.4(4) & 319.3(4) & 313.1(3) & 305.8(4) & 300.7(3) & 294.7(3) \\ 
0.45 & 0.53 & 315.5(4) & 306.7(4) & 300.0(4) & 292.5(4) & 285.7(3) & 277.7(3) & 270.6(3) & 264.3(3) & 257.4(3) & 250.7(3) & 244.6(3) & 238.0(3) & 231.4(3) \\ 
0.53 & 0.62 & 197.4(3) & 191.1(3) & 184.2(3) & 178.3(3) & 172.1(3) & 165.2(3) & 160.0(3) & 153.5(2) & 148.0(2) & 142.4(2) & 137.3(2) & 132.1(2) & 127.4(2) \\ 
0.62 & 0.71 & 72.0(2) & 67.5(2) & 62.9(2) & 58.9(2) & 55.1(2) & 51.6(2) & 47.6(1) & 44.6(1) & 41.4(1) & 38.2(1) & 35.4(1) & 32.8(1) & 30.1(1) \\ 
0.71 & 1.00 & 1.68(1) & 1.43(1) & 1.19(1) & 0.97(1) & 0.808(9) & 0.655(8) & 0.514(7) & 0.413(6) & 0.320(5) & 0.242(4) & 0.183(3) & 0.141(3) & 0.094(2) \\ 
 \hline
&  & \multicolumn{13}{|c}{ $\frac{d\sigma_{t\bar{t}j + X}}{d\rho_s}$ distribution with $N_{j}\geq1,\,p_{T}^j>100$ GeV, $|\eta_{j}|<2.4$, $K_{R}=K_{F}=0.5$}  \\ 
 \hline 
 $x_{l}$  & $x_{r}$  & 165 GeV & 166 GeV & 167 GeV & 168 GeV & 169 GeV & 170 GeV & 171 GeV & 172 GeV & 173 GeV & 174 GeV & 175 GeV & 176 GeV & 177 GeV \\ 
 \hline 
0.00 & 0.18 & 14.9(1) & 14.8(1) & 14.7(1) & 14.9(1) & 14.8(1) & 14.6(1) & 14.6(1) & 14.7(1) & 14.5(1) & 14.50(9) & 14.36(9) & 14.49(9) & 14.39(9) \\ 
0.18 & 0.22 & 137.6(9) & 136.3(9) & 137.3(9) & 136.8(8) & 133.5(8) & 132.7(8) & 131.9(8) & 130.7(8) & 130.0(8) & 128.4(7) & 127.8(7) & 127.1(7) & 125.8(7) \\ 
0.22 & 0.27 & 247.0(1) & 240.0(1) & 240.0(1) & 237.5(1) & 234.3(1) & 232.7(9) & 229.2(9) & 226.5(9) & 225.4(9) & 221.2(8) & 219.5(8) & 216.6(8) & 216.5(8) \\ 
0.27 & 0.32 & 350.0(1) & 348.0(1) & 341.0(1) & 336.0(1) & 333.0(1) & 328.0(1) & 323.0(1) & 318.0(1) & 313.0(1) & 312.0(1) & 305.9(1) & 302.5(1) & 299.3(1) \\ 
0.32 & 0.38 & 426.0(1) & 418.0(1) & 412.0(1) & 403.0(1) & 400.0(1) & 391.0(1) & 387.0(1) & 379.0(1) & 377.0(1) & 368.8(1) & 364.7(1) & 358.5(9) & 353.8(9) \\ 
0.38 & 0.45 & 444.0(1) & 436.0(1) & 428.0(1) & 422.0(1) & 415.0(1) & 405.0(1) & 398.0(1) & 389.8(9) & 381.8(9) & 374.2(9) & 365.6(9) & 360.7(9) & 355.4(8) \\ 
0.45 & 0.53 & 388.0(1) & 376.8(1) & 366.9(1) & 358.8(1) & 349.6(9) & 341.4(9) & 332.5(9) & 325.8(8) & 316.6(8) & 311.2(8) & 301.9(8) & 294.9(7) & 289.0(7) \\ 
0.53 & 0.62 & 248.0(8) & 240.8(8) & 232.5(8) & 224.2(7) & 217.0(7) & 208.7(7) & 202.8(6) & 194.0(6) & 188.6(6) & 180.6(6) & 174.5(5) & 167.7(5) & 161.5(5) \\ 
0.62 & 0.71 & 93.1(5) & 87.2(4) & 82.0(4) & 76.4(4) & 71.2(4) & 66.4(4) & 61.8(3) & 57.5(3) & 53.6(3) & 49.7(3) & 46.4(3) & 42.6(3) & 39.5(2) \\ 
0.71 & 1.00 & 2.45(3) & 2.12(3) & 1.82(3) & 1.52(3) & 1.29(2) & 1.08(2) & 0.83(2) & 0.70(1) & 0.56(1) & 0.443(9) & 0.346(7) & 0.291(6) & 0.201(4) \\ 
 \hline
 \hline
 \end{tabular}
 \end{adjustbox}
\caption{ 
Same as Fig.~\ref{tab:LHC14_mpole_ABMP16_n3_CMS13TeV} but for $p_T^j > 100$~GeV.
\label{tab:LHC14_mpole_ABMP16_n3_CMS13TeV_pt100}}
 \end{table}
\end{landscape}

{\small 
\providecommand{\href}[2]{#2}\begingroup\raggedright\endgroup

}

\end{document}